%% file: Planck_Constraints_on_primordial_non-Gaussianity.tex
\documentclass[a4paper, traditabstract,longauth]{aa} 
\usepackage{graphicx,amsmath}
\usepackage{epsf}
\usepackage{ifthen}
\usepackage[table,usenames,dvipsnames]{xcolor}
\usepackage{fixltx2e}

\usepackage{color}
\usepackage{epsfig}
\usepackage[breaklinks, colorlinks, citecolor=blue]{hyperref} 
\usepackage[nonamebreak]{natbib}

\usepackage[varg]{txfonts}

\usepackage{etex}
\reserveinserts{28}

\pdfoutput=1
\pdfmapfile{+txfonts.map}

\bibpunct{(}{)}{;}{a}{}{,} 

\input Planck 
\input macros

\begin{document}
\title{ \textit{Planck} 2015 results. XVII. Constraints on primordial non-Gaussianity}

\input{A19_Primordial_non_gaussianity_authors_and_institutes.tex}
\date{Received xxxx, Accepted xxxxx}


\input{A19_abstract}


\authorrunning{Planck Collaboration}
\titlerunning{ \textit{Planck} 2015 Results. Constraints on primordial NG}
\maketitle

\section {Introduction}
\label{sec:intro}

\input{A19_Section1}

\section {Models}
\label{sec:models}

\input{A19_Section2}

\section{Statistical estimation of the CMB bispectrum for polarized maps}
\label{sec:SEP}
\input{A19_Section3}

\section{Non-primordial contributions to the CMB bispectrum}
\label{sec:npNG}

\input{A19_Section4}

\section{Validation tests}
\label{sec:Validation}

\input{A19_Section5}

\section{Results}
\label{sec:Results}

\input{A19_Section6}

\section{Validation of \textit{Planck} results}
\label{sec:Sec_valid_data}

\input{A19_Section7}

\section{Other non-Gaussianity shapes for $\fnl$}
\label{sec:Other_shapes}

\input{A19_Section8}

\section{Limits on the primordial trispectrum}
\label{sec:tau_gnl}

\input{A19_Section9}

\section{Minkowski functionals results} 
\label{sec:Minkowski}

\input{A19_Section10}

\section{Implications for early Universe physics}
\label{sec:Implications}

\input{A19_Section11}

\section{Conclusions}
\label{sec:Conc}

\input{A19_Section12}


\input{A19_acknowledgements.tex}

\bibliographystyle{aat}
\bibliography{Planck_bib,A19_bib}

\appendix

\input{A19_AppendixA}

\label{App:A}

\input{A19_AppendixB}
\label{App:B}

\raggedright
\end{document}

%% file: Planck.tex
\def\setsymbol#1#2{\expandafter\def\csname #1\endcsname{#2}}
\def\getsymbol#1{\csname #1\endcsname}

\def\Planck{\textit{Planck}}

\def\alltwentythirteenresultspapers{\nocite{planck2013-p01, planck2013-p02, planck2013-p02a, planck2013-p02d, planck2013-p02b, planck2013-p03, planck2013-p03c, planck2013-p03f, planck2013-p03d, planck2013-p03e, planck2013-p01a, planck2013-p06, planck2013-p03a, planck2013-pip88, planck2013-p08, planck2013-p11, planck2013-p12, planck2013-p13, planck2013-p14, planck2013-p15, planck2013-p05b, planck2013-p17, planck2013-p09, planck2013-p09a, planck2013-p20, planck2013-p19, planck2013-pipaberration, planck2013-p05, planck2013-p05a, planck2013-pip56, planck2013-p06b, planck2013-p01a}}

\newbox\tablebox    \newdimen\tablewidth
\def\leaderfil{\leaders\hbox to 5pt{\hss.\hss}\hfil}
\def\endPlancktable{\tablewidth=\columnwidth 
    $$\hss\copy\tablebox\hss$$
    \vskip-\lastskip\vskip -2pt}
\def\endPlancktablewide{\tablewidth=\textwidth 
    $$\hss\copy\tablebox\hss$$
    \vskip-\lastskip\vskip -2pt}
\def\tablenote#1 #2\par{\begingroup \parindent=0.8em
    \abovedisplayshortskip=0pt\belowdisplayshortskip=0pt
    \noindent
    $$\hss\vbox{\hsize\tablewidth \hangindent=\parindent \hangafter=1 \noindent
    \hbox to \parindent{$^#1$\hss}\strut#2\strut\par}\hss$$
    \endgroup}
\def\doubleline{\vskip 3pt\hrule \vskip 1.5pt \hrule \vskip 5pt}

\def\L2{\ifmmode L_2\else $L_2$\fi}

\def\DeltaT{\ifmmode \Delta T\else $\Delta T$\fi}
\def\deltat{\ifmmode \Delta t\else $\Delta t$\fi}
\def\fknee{\ifmmode f_{\rm knee}\else $f_{\rm knee}$\fi}
\def\Fmax{\ifmmode F_{\rm max}\else $F_{\rm max}$\fi}
\def\solar{\ifmmode{\rm M}_{\mathord\odot}\else${\rm M}_{\mathord\odot}$\fi}
\def\Msolar{\ifmmode{\rm M}_{\mathord\odot}\else${\rm M}_{\mathord\odot}$\fi}
\def\Lsolar{\ifmmode{\rm L}_{\mathord\odot}\else${\rm L}_{\mathord\odot}$\fi}
\def\inv{\ifmmode^{-1}\else$^{-1}$\fi}
\def\mo{\ifmmode^{-1}\else$^{-1}$\fi}
\def\sup#1{\ifmmode ^{\rm #1}\else $^{\rm #1}$\fi}
\def\expo#1{\ifmmode \times 10^{#1}\else $\times 10^{#1}$\fi}
\def\,{\thinspace}
\def\lsim{\mathrel{\raise .4ex\hbox{\rlap{$<$}\lower 1.2ex\hbox{$\sim$}}}}
\def\gsim{\mathrel{\raise .4ex\hbox{\rlap{$>$}\lower 1.2ex\hbox{$\sim$}}}}

\def\simprop{\mathrel{\raise .4ex\hbox{\rlap{$\propto$}\lower 1.2ex\hbox{$\sim$}}}}
\def\deg{\ifmmode^\circ\else$^\circ$\fi}
\def\pdeg{\ifmmode $\setbox0=\hbox{$^{\circ}$}\rlap{\hskip.11\wd0 .}$^{\circ}
          \else \setbox0=\hbox{$^{\circ}$}\rlap{\hskip.11\wd0 .}$^{\circ}$\fi}
\def\arcs{\ifmmode {^{\scriptstyle\prime\prime}}
          \else $^{\scriptstyle\prime\prime}$\fi}
\def\arcm{\ifmmode {^{\scriptstyle\prime}}
          \else $^{\scriptstyle\prime}$\fi}
\newdimen\sa  \newdimen\sb
\def\parcs{\sa=.07em \sb=.03em
     \ifmmode \hbox{\rlap{.}}^{\scriptstyle\prime\kern -\sb\prime}\hbox{\kern -\sa}
     \else \rlap{.}$^{\scriptstyle\prime\kern -\sb\prime}$\kern -\sa\fi}
\def\parcm{\sa=.08em \sb=.03em
     \ifmmode \hbox{\rlap{.}\kern\sa}^{\scriptstyle\prime}\hbox{\kern-\sb}
     \else \rlap{.}\kern\sa$^{\scriptstyle\prime}$\kern-\sb\fi}
\def\ra[#1 #2 #3.#4]{#1\sup{h}#2\sup{m}#3\sup{s}\llap.#4}
\def\dec[#1 #2 #3.#4]{#1\deg#2\arcm#3\arcs\llap.#4}
\def\deco[#1 #2 #3]{#1\deg#2\arcm#3\arcs}
\def\rra[#1 #2]{#1\sup{h}#2\sup{m}}

\def\dots{\relax\ifmmode \ldots\else $\ldots$\fi}
\def\WHzsr{\ifmmode $W\,Hz\mo\,sr\mo$\else W\,Hz\mo\,sr\mo\fi}
\def\mHz{\ifmmode $\,mHz$\else \,mHz\fi}
\def\GHz{\ifmmode $\,GHz$\else \,GHz\fi}
\def\mKs{\ifmmode $\,mK\,s$^{1/2}\else \,mK\,s$^{1/2}$\fi}
\def\muKs{\ifmmode \,\mu$K\,s$^{1/2}\else \,$\mu$K\,s$^{1/2}$\fi}
\def\muKRJs{\ifmmode \,\mu$K$_{\rm RJ}$\,s$^{1/2}\else \,$\mu$K$_{\rm RJ}$\,s$^{1/2}$\fi}
\def\muKHz{\ifmmode \,\mu$K\,Hz$^{-1/2}\else \,$\mu$K\,Hz$^{-1/2}$\fi}
\def\MJysr{\ifmmode \,$MJy\,sr\mo$\else \,MJy\,sr\mo\fi}
\def\MJysrmK{\ifmmode \,$MJy\,sr\mo$\,mK$_{\rm CMB}\mo\else \,MJy\,sr\mo\,mK$_{\rm CMB}\mo$\fi}
\def\microns{\ifmmode \,\mu$m$\else \,$\mu$m\fi}

\def\muK{\ifmmode \,\mu$K$\else \,$\mu$\hbox{K}\fi}
\def\microK{\ifmmode \,\mu$K$\else \,$\mu$\hbox{K}\fi}
\def\muW{\ifmmode \,\mu$W$\else \,$\mu$\hbox{W}\fi}
\def\kms{\ifmmode $\,km\,s$^{-1}\else \,km\,s$^{-1}$\fi}
\def\kmsMpc{\ifmmode $\,\kms\,Mpc\mo$\else \,\kms\,Mpc\mo\fi}
\providecommand{\sorthelp}[1]{}

%% file: macros.tex
\newcommand{\fnl}{f_{\rm{NL}}}
\newcommand{\fnlloc}{f^{\rm local}_{\rm NL}}

\newcommand{\gnl}{g_{\rm{NL}}}

\def\fnllocal{f_\textrm {NL}^\textrm{local}}
\def\fnlortho{f_\textrm {NL}^\textrm{ortho}}
\def\fnlequi{f_\textrm {NL}^\textrm{equi}}

\providecommand{\planck}{\textit{Planck}}

\def\WMAP{{WMAP}}

\newcommand\ba{\begin{eqnarray}}
\newcommand\ea{\end{eqnarray}}
\newcommand\bea{\begin{eqnarray}}
\newcommand\eea{\end{eqnarray}}

\newcommand\be{\begin{equation}}
\newcommand\ee{\end{equation}}

\newcommand{\boldvec}[1]{{{\vec{#1}}}}

\newcommand{\vk}{\boldvec{k}}

\newcommand{\vn}{\boldvec{n}}

\newcommand{\vx}{\boldvec{x}}

\newcommand{\vnhat}{\hat{\vn}}

\newcommand{\vkhat}{\hat{\vk}}

\newcommand{\itT}{\textit{T}}
\newcommand{\itQ}{\textit{Q}}
\newcommand{\itU}{\textit{U}}
\newcommand{\itE}{\textit{E}}

\newcommand{\itTpE}{\textit{T+E}}

\def\Commander{\texttt{Commander}}

\def\NILC{\texttt{NILC}}
\def\SMICA{\texttt{SMICA}}
\def\SEVEM{\texttt{SEVEM}}

\def\={\nonumber &=}
\def\nn{\nonumber}

\def\({\left(}
\def\){\right)}
\def\[{\left[}
\def\]{\right]}
\def\<{\left\langle}
\def\>{\right\rangle}

\def\curl{\mathcal}

\def\eq{\begin{eqnarray}}
\def\qe{\end{eqnarray}}

\def\and{\quad \mbox{and} \quad}

\def\fnl{f_\textrm{NL}}

\def\fnllocal{f_\textrm {NL}^\textrm{local}}

\def\bfnl{\kern2pt\overline{\kern-2ptf}_\textrm{NL}}

\def\lmax{\ell_\textrm{max}}

\def\exp{\textrm{exp}}

\def\kall{k_1,k_2,k_3}

\def\alm{a_{\ell m}}

\def\Ylm{Y_{\ell m}}

\def\barQ{\kern2pt\overline{\kern-2pt\curl{Q}}}

\def\bargamma{\kern2pt\overline{\kern-2pt\gamma}}
\def\barzeta{\kern2pt\overline{\kern-2pt\zeta}}

\def\barR{\kern2pt\overline{\kern-2pt\curl{R}}}

\def\nmax{n_\textrm{max}}

\def\klist{k_1,k_2,k_3}
\def\eqref#1{(\ref{#1})}

\def\leaderfi1{\leaders\hbox to 5pt{\hss.\hss}\hfil}

\def\setsize{\csname @setfontsize\endcsname \setsize}

\newcommand{\C}{C}

\def\gnlloc{g_{\rm NL}^{\rm local}}
\def\gnldotpi4{g_{\rm NL}^{\dot\sigma^4}}
\def\gnldpi4{g_{\rm NL}^{(\partial\sigma)^4}}
\def\gnlB{g_{\rm NL}^{\dot\sigma^2 (\partial\sigma)^2}}

\def\Nin{N_{\rm in}}
\def\Nout{N_{\rm out}}

%% file: A19_Primordial_non_gaussianity_authors_and_institutes.tex
\author{\small
Planck Collaboration: P.~A.~R.~Ade\inst{97}
\and
N.~Aghanim\inst{63}
\and
M.~Arnaud\inst{79}
\and
F.~Arroja\inst{71, 85}
\and
M.~Ashdown\inst{75, 6}
\and
J.~Aumont\inst{63}
\and
C.~Baccigalupi\inst{95}
\and
M.~Ballardini\inst{51, 53, 34}
\and
A.~J.~Banday\inst{109, 10}
\and
R.~B.~Barreiro\inst{70}
% ...
}
\institute{\small
APC, AstroParticule et Cosmologie, Universit\'{e} Paris Diderot, CNRS/IN2P3, CEA/lrfu, Observatoire de Paris, Sorbonne Paris Cit\'{e}, 10, rue Alice Domon et L\'{e}onie Duquet, 75205 Paris Cedex 13, France\goodbreak
\and
Aalto University Mets\"{a}hovi Radio Observatory and Dept of Radio Science and Engineering, P.O. Box 13000, FI-00076 AALTO, Finland\goodbreak
\and
African Institute for Mathematical Sciences, 6-8 Melrose Road, Muizenberg, Cape Town, South Africa\goodbreak
\and
Agenzia Spaziale Italiana Science Data Center, Via del Politecnico snc, 00133, Roma, Italy\goodbreak
\and
Aix Marseille Universit\'{e}, CNRS, LAM (Laboratoire d'Astrophysique de Marseille) UMR 7326, 13388, Marseille, France\goodbreak
\and
Astrophysics Group, Cavendish Laboratory, University of Cambridge, J J Thomson Avenue, Cambridge CB3 0HE, U.K.\goodbreak
\and
Astrophysics \& Cosmology Research Unit, School of Mathematics, Statistics \& Computer Science, University of KwaZulu-Natal, Westville Campus, Private Bag X54001, Durban 4000, South Africa\goodbreak
\and
CGEE, SCS Qd 9, Lote C, Torre C, 4$^{\circ}$ andar, Ed. Parque Cidade Corporate, CEP 70308-200, Bras\'{i}lia, DF, Brazil\goodbreak
\and
CITA, University of Toronto, 60 St. George St., Toronto, ON M5S 3H8, Canada\goodbreak
\and
CNRS, IRAP, 9 Av. colonel Roche, BP 44346, F-31028 Toulouse cedex 4, France\goodbreak
\and
CRANN, Trinity College, Dublin, Ireland\goodbreak
\and
California Institute of Technology, Pasadena, California, U.S.A.\goodbreak
\and
Centre for Theoretical Cosmology, DAMTP, University of Cambridge, Wilberforce Road, Cambridge CB3 0WA, U.K.\goodbreak
\and
Centro de Estudios de F\'{i}sica del Cosmos de Arag\'{o}n (CEFCA), Plaza San Juan, 1, planta 2, E-44001, Teruel, Spain\goodbreak
\and
Computational Cosmology Center, Lawrence Berkeley National Laboratory, Berkeley, California, U.S.A.\goodbreak
\and
Consejo Superior de Investigaciones Cient\'{\i}ficas (CSIC), Madrid, Spain\goodbreak
\and
DSM/Irfu/SPP, CEA-Saclay, F-91191 Gif-sur-Yvette Cedex, France\goodbreak
\and
DTU Space, National Space Institute, Technical University of Denmark, Elektrovej 327, DK-2800 Kgs. Lyngby, Denmark\goodbreak
\and
D\'{e}partement de Physique Th\'{e}orique, Universit\'{e} de Gen\`{e}ve, 24, Quai E. Ansermet,1211 Gen\`{e}ve 4, Switzerland\goodbreak
\and
Departamento de Astrof\'{i}sica, Universidad de La Laguna (ULL), E-38206 La Laguna, Tenerife, Spain\goodbreak
\and
Departamento de F\'{\i}sica, Universidad de Oviedo, Avda. Calvo Sotelo s/n, Oviedo, Spain\goodbreak
\and
Department of Astronomy and Astrophysics, University of Toronto, 50 Saint George Street, Toronto, Ontario, Canada\goodbreak
\and
Department of Astrophysics/IMAPP, Radboud University Nijmegen, P.O. Box 9010, 6500 GL Nijmegen, The Netherlands\goodbreak
\and
Department of Physics \& Astronomy, University of British Columbia, 6224 Agricultural Road, Vancouver, British Columbia, Canada\goodbreak
\and
Department of Physics and Astronomy, Dana and David Dornsife College of Letter, Arts and Sciences, University of Southern California, Los Angeles, CA 90089, U.S.A.\goodbreak
\and
Department of Physics and Astronomy, University College London, London WC1E 6BT, U.K.\goodbreak
\and
Department of Physics and Astronomy, University of Sussex, Brighton BN1 9QH, U.K.\goodbreak
\and
Department of Physics, Florida State University, Keen Physics Building, 77 Chieftan Way, Tallahassee, Florida, U.S.A.\goodbreak
\and
Department of Physics, Gustaf H\"{a}llstr\"{o}min katu 2a, University of Helsinki, Helsinki, Finland\goodbreak
\and
Department of Physics, Princeton University, Princeton, New Jersey, U.S.A.\goodbreak
\and
Department of Physics, University of California, Santa Barbara, California, U.S.A.\goodbreak
\and
Department of Physics, University of Illinois at Urbana-Champaign, 1110 West Green Street, Urbana, Illinois, U.S.A.\goodbreak
\and
Dipartimento di Fisica e Astronomia G. Galilei, Universit\`{a} degli Studi di Padova, via Marzolo 8, 35131 Padova, Italy\goodbreak
\and
Dipartimento di Fisica e Astronomia, ALMA MATER STUDIORUM, Universit\`{a} degli Studi di Bologna, Viale Berti Pichat 6/2, I-40127, Bologna, Italy\goodbreak
\and
Dipartimento di Fisica e Scienze della Terra, Universit\`{a} di Ferrara, Via Saragat 1, 44122 Ferrara, Italy\goodbreak
\and
Dipartimento di Fisica, Universit\`{a} La Sapienza, P. le A. Moro 2, Roma, Italy\goodbreak
\and
Dipartimento di Fisica, Universit\`{a} degli Studi di Milano, Via Celoria, 16, Milano, Italy\goodbreak
\and
Dipartimento di Fisica, Universit\`{a} degli Studi di Trieste, via A. Valerio 2, Trieste, Italy\goodbreak
\and
Dipartimento di Matematica, Universit\`{a} di Roma Tor Vergata, Via della Ricerca Scientifica, 1, Roma, Italy\goodbreak
\and
Discovery Center, Niels Bohr Institute, Blegdamsvej 17, Copenhagen, Denmark\goodbreak
\and
Discovery Center, Niels Bohr Institute, Copenhagen University, Blegdamsvej 17, Copenhagen, Denmark\goodbreak
\and
European Space Agency, ESAC, Planck Science Office, Camino bajo del Castillo, s/n, Urbanizaci\'{o}n Villafranca del Castillo, Villanueva de la Ca\~{n}ada, Madrid, Spain\goodbreak
\and
European Space Agency, ESTEC, Keplerlaan 1, 2201 AZ Noordwijk, The Netherlands\goodbreak
\and
Gran Sasso Science Institute, INFN, viale F. Crispi 7, 67100 L'Aquila, Italy\goodbreak
\and
HGSFP and University of Heidelberg, Theoretical Physics Department, Philosophenweg 16, 69120, Heidelberg, Germany\goodbreak
\and
Helsinki Institute of Physics, Gustaf H\"{a}llstr\"{o}min katu 2, University of Helsinki, Helsinki, Finland\goodbreak
\and
ICTP South American Institute for Fundamental Research, Instituto de F\'{\i}sica Te\'{o}rica, Universidade Estadual Paulista, S\~{a}o Paulo, Brazil\goodbreak
\and
INAF - Osservatorio Astronomico di Padova, Vicolo dell'Osservatorio 5, Padova, Italy\goodbreak
\and
INAF - Osservatorio Astronomico di Roma, via di Frascati 33, Monte Porzio Catone, Italy\goodbreak
\and
INAF - Osservatorio Astronomico di Trieste, Via G.B. Tiepolo 11, Trieste, Italy\goodbreak
\and
INAF/IASF Bologna, Via Gobetti 101, Bologna, Italy\goodbreak
\and
INAF/IASF Milano, Via E. Bassini 15, Milano, Italy\goodbreak
\and
INFN, Sezione di Bologna, Via Irnerio 46, I-40126, Bologna, Italy\goodbreak
\and
INFN, Sezione di Roma 1, Universit\`{a} di Roma Sapienza, Piazzale Aldo Moro 2, 00185, Roma, Italy\goodbreak
\and
INFN, Sezione di Roma 2, Universit\`{a} di Roma Tor Vergata, Via della Ricerca Scientifica, 1, Roma, Italy\goodbreak
\and
INFN/National Institute for Nuclear Physics, Via Valerio 2, I-34127 Trieste, Italy\goodbreak
\and
IPAG: Institut de Plan\'{e}tologie et d'Astrophysique de Grenoble, Universit\'{e} Grenoble Alpes, IPAG, F-38000 Grenoble, France, CNRS, IPAG, F-38000 Grenoble, France\goodbreak
\and
IUCAA, Post Bag 4, Ganeshkhind, Pune University Campus, Pune 411 007, India\goodbreak
\and
Imperial College London, Astrophysics group, Blackett Laboratory, Prince Consort Road, London, SW7 2AZ, U.K.\goodbreak
\and
Infrared Processing and Analysis Center, California Institute of Technology, Pasadena, CA 91125, U.S.A.\goodbreak
\and
Institut N\'{e}el, CNRS, Universit\'{e} Joseph Fourier Grenoble I, 25 rue des Martyrs, Grenoble, France\goodbreak
\and
Institut Universitaire de France, 103, bd Saint-Michel, 75005, Paris, France\goodbreak
\and
Institut d'Astrophysique Spatiale, CNRS, Univ. Paris-Sud, Universit\'{e} Paris-Saclay, B\^{a}t. 121, 91405 Orsay cedex, France\goodbreak
\and
Institut d'Astrophysique de Paris, CNRS (UMR7095), 98 bis Boulevard Arago, F-75014, Paris, France\goodbreak
\and
Institut f\"ur Theoretische Teilchenphysik und Kosmologie, RWTH Aachen University, D-52056 Aachen, Germany\goodbreak
\and
Institute for Space Sciences, Bucharest-Magurale, Romania\goodbreak
\and
Institute of Astronomy, University of Cambridge, Madingley Road, Cambridge CB3 0HA, U.K.\goodbreak
\and
Institute of Theoretical Astrophysics, University of Oslo, Blindern, Oslo, Norway\goodbreak
\and
Instituto de Astrof\'{\i}sica de Canarias, C/V\'{\i}a L\'{a}ctea s/n, La Laguna, Tenerife, Spain\goodbreak
\and
Instituto de F\'{\i}sica de Cantabria (CSIC-Universidad de Cantabria), Avda. de los Castros s/n, Santander, Spain\goodbreak
\and
Istituto Nazionale di Fisica Nucleare, Sezione di Padova, via Marzolo 8, I-35131 Padova, Italy\goodbreak
\and
Jet Propulsion Laboratory, California Institute of Technology, 4800 Oak Grove Drive, Pasadena, California, U.S.A.\goodbreak
\and
Jodrell Bank Centre for Astrophysics, Alan Turing Building, School of Physics and Astronomy, The University of Manchester, Oxford Road, Manchester, M13 9PL, U.K.\goodbreak
\and
Kavli Institute for Cosmological Physics, University of Chicago, Chicago, IL 60637, USA\goodbreak
\and
Kavli Institute for Cosmology Cambridge, Madingley Road, Cambridge, CB3 0HA, U.K.\goodbreak
\and
Kazan Federal University, 18 Kremlyovskaya St., Kazan, 420008, Russia\goodbreak
\and
LAL, Universit\'{e} Paris-Sud, CNRS/IN2P3, Orsay, France\goodbreak
\and
LERMA, CNRS, Observatoire de Paris, 61 Avenue de l'Observatoire, Paris, France\goodbreak
\and
Laboratoire AIM, IRFU/Service d'Astrophysique - CEA/DSM - CNRS - Universit\'{e} Paris Diderot, B\^{a}t. 709, CEA-Saclay, F-91191 Gif-sur-Yvette Cedex, France\goodbreak
\and
Laboratoire Traitement et Communication de l'Information, CNRS (UMR 5141) and T\'{e}l\'{e}com ParisTech, 46 rue Barrault F-75634 Paris Cedex 13, France\goodbreak
\and
Laboratoire de Physique Subatomique et Cosmologie, Universit\'{e} Grenoble-Alpes, CNRS/IN2P3, 53, rue des Martyrs, 38026 Grenoble Cedex, France\goodbreak
\and
Laboratoire de Physique Th\'{e}orique, Universit\'{e} Paris-Sud 11 \& CNRS, B\^{a}timent 210, 91405 Orsay, France\goodbreak
\and
Lawrence Berkeley National Laboratory, Berkeley, California, U.S.A.\goodbreak
\and
Lebedev Physical Institute of the Russian Academy of Sciences, Astro Space Centre, 84/32 Profsoyuznaya st., Moscow, GSP-7, 117997, Russia\goodbreak
\and
Leung Center for Cosmology and Particle Astrophysics, National Taiwan University, Taipei 10617, Taiwan\goodbreak
\and
Max-Planck-Institut f\"{u}r Astrophysik, Karl-Schwarzschild-Str. 1, 85741 Garching, Germany\goodbreak
\and
McGill Physics, Ernest Rutherford Physics Building, McGill University, 3600 rue University, Montr\'{e}al, QC, H3A 2T8, Canada\goodbreak
\and
National University of Ireland, Department of Experimental Physics, Maynooth, Co. Kildare, Ireland\goodbreak
\and
Nicolaus Copernicus Astronomical Center, Bartycka 18, 00-716 Warsaw, Poland\goodbreak
\and
Niels Bohr Institute, Blegdamsvej 17, Copenhagen, Denmark\goodbreak
\and
Niels Bohr Institute, Copenhagen University, Blegdamsvej 17, Copenhagen, Denmark\goodbreak
\and
Nordita (Nordic Institute for Theoretical Physics), Roslagstullsbacken 23, SE-106 91 Stockholm, Sweden\goodbreak
\and
Optical Science Laboratory, University College London, Gower Street, London, U.K.\goodbreak
\and
Perimeter Institute for Theoretical Physics, Waterloo ON N2L 2Y5, Canada\goodbreak
\and
SISSA, Astrophysics Sector, via Bonomea 265, 34136, Trieste, Italy\goodbreak
\and
SMARTEST Research Centre, Universit\`{a} degli Studi e-Campus, Via Isimbardi 10, Novedrate (CO), 22060, Italy\goodbreak
\and
School of Physics and Astronomy, Cardiff University, Queens Buildings, The Parade, Cardiff, CF24 3AA, U.K.\goodbreak
\and
School of Physics and Astronomy, University of Nottingham, Nottingham NG7 2RD, U.K.\goodbreak
\and
Sorbonne Universit\'{e}-UPMC, UMR7095, Institut d'Astrophysique de Paris, 98 bis Boulevard Arago, F-75014, Paris, France\goodbreak
\and
Space Research Institute (IKI), Russian Academy of Sciences, Profsoyuznaya Str, 84/32, Moscow, 117997, Russia\goodbreak
\and
Space Sciences Laboratory, University of California, Berkeley, California, U.S.A.\goodbreak
\and
Special Astrophysical Observatory, Russian Academy of Sciences, Nizhnij Arkhyz, Zelenchukskiy region, Karachai-Cherkessian Republic, 369167, Russia\goodbreak
\and
Stanford University, Dept of Physics, Varian Physics Bldg, 382 Via Pueblo Mall, Stanford, California, U.S.A.\goodbreak
\and
Sub-Department of Astrophysics, University of Oxford, Keble Road, Oxford OX1 3RH, U.K.\goodbreak
\and
Sydney Institute for Astronomy, School of Physics A28, University of Sydney, NSW 2006, Australia\goodbreak
\and
The Oskar Klein Centre for Cosmoparticle Physics, Department of Physics,Stockholm University, AlbaNova, SE-106 91 Stockholm, Sweden\goodbreak
\and
Theory Division, PH-TH, CERN, CH-1211, Geneva 23, Switzerland\goodbreak
\and
UPMC Univ Paris 06, UMR7095, 98 bis Boulevard Arago, F-75014, Paris, France\goodbreak
\and
Universit\'{e} de Toulouse, UPS-OMP, IRAP, F-31028 Toulouse cedex 4, France\goodbreak
% ...
}

%% file: A19_abstract.tex
\abstract{
The \Planck\ full mission cosmic microwave background (CMB)
temperature and $E$-mode polarization maps are analysed to
obtain constraints on primordial non-Gaussianity (NG).  Using three
classes of optimal bispectrum estimators --- separable template-fitting (KSW),
binned, and modal --- we obtain consistent values for the primordial
local, equilateral, and orthogonal bispectrum amplitudes, quoting as
our final result from temperature alone 
$f_{\rm NL}^{\rm local} = 2.5 \pm 5.7$, $f_{\rm NL}^{\rm equil} 
= - 16 \pm 70$, and $f_{\rm NL}^{\rm ortho} = - 34 \pm 33$
(68\,\%~CL, statistical).
Combining  temperature and  polarization data we obtain
$f_{\rm NL}^{\rm local} = 0.8 \pm 5.0$, $f_{\rm NL}^{\rm equil} 
= - 4 \pm 43$, and $f_{\rm NL}^{\rm ortho} = - 26 \pm 21$
(68\,\%~CL, statistical). The results are based on comprehensive
cross-validation of these estimators on Gaussian and non-Gaussian
simulations,  are stable across component separation techniques, pass
an extensive suite of tests, and are consistent with estimators based on measuring the Minkowski
functionals of the CMB. The effect of time-domain de-glitching systematics on the bispectrum is negligible. In spite of these test outcomes we conservatively label the results including polarization data as preliminary, owing to a known mismatch of the noise model in simulations and the data. Beyond estimates of individual shape
amplitudes, we present model-independent, three-dimensional
reconstructions of the \Planck\ CMB bispectrum and derive
constraints on early universe scenarios that generate primordial NG,
including general single-field models of inflation, axion inflation, initial state modifications, models producing parity-violating tensor bispectra, and directionally dependent vector models. 
We present a wide survey of
scale-dependent feature and resonance models,  
accounting for the ``look elsewhere" effect in estimating the statistical significance of features.
We also look for isocurvature NG, and find no signal, but we obtain constraints that improve
significantly with the inclusion of polarization. 
The primordial trispectrum amplitude in the local model is constrained to be
$g_{\rm NL}^{\rm local} = (-9.0 \pm 7.7) \times 10^4$
(68\,\%~CL statistical), and we perform an analysis of trispectrum
shapes beyond the local case. The global picture that emerges is one of consistency with the premises of the $\Lambda$CDM cosmology, namely that the structure we observe today was sourced by adiabatic, passive, Gaussian, and primordial seed perturbations.}

%% file: A19_Section1.tex
This paper, one of a set associated with the 2015 release of data from the 
\Planck\footnote{\Planck\ (\url{http://www.esa.int/Planck}) is a project of the European Space Agency  (ESA) with instruments provided by two scientific consortia funded by ESA member states and led by Principal Investigators from France and Italy, telescope reflectors provided through a collaboration between ESA and a scientific consortium led and funded by Denmark, and additional contributions from NASA (USA).} 
mission~\citep{planck2014-a01}, 
\alltwentythirteenresultspapers\ describes the constraints on primordial 
non-Gaussianity (NG) obtained using the cosmic microwave background (CMB) 
maps from the full \Planck\ mission, including a first analysis of some of 
the \Planck\ polarization data.

Primordial NG is one of the most powerful tests of inflation, and more generally of high-energy early Universe physics (for some reviews, see~\citealt{Bartolo:2004if},~\citealt{2010AdAst2010E..73L},~\citealt{2010AdAst2010E..72C},~\citealt{2010CQGra..27l4010K},~\citealt{2010AdAst2010E..71Y}). In fact, the simplest models of inflation (characterized by a single scalar field slowly rolling along a smooth potential) predict the generation of primordial fluctuations that are almost Gaussian distributed, with a tiny deviation from Gaussianity of the order of the slow-roll parameters~(\citealt{2003NuPhB.667..119A,2003JHEP...05..013M}). The 2013 \Planck\ results on primordial NG are consistent with such a prediction, being compatible with Gaussian primordial fluctuations: the standard scenario of single-field, slow-roll inflation has survived its most stringent test to date. For example, in~\cite{planck2013-p09a} we obtained 
$f_{\rm NL}^{\rm local} = 2.7 \pm 5.8$, $f_{\rm NL}^{\rm equil} = -42 \pm 75$, and $f_{\rm NL}^{\rm ortho} = -25 \pm 39$ for the amplitudes of three of the most well-studied shapes of primordial NG. 
On the other hand, it is well known that any deviations from the standard picture of inflation have the potential to produce distinctive NG signatures at a detectable level in the CMB anisotropies.\footnote{We refer the reader to~\cite{planck2013-p09a} and references therein for a detailed summary of the models and underlying physical mechanisms generating various types of primordial NG.} Therefore, as already shown in~\cite{planck2013-p09a} (see also~\cite{planck2013-p17}) improved NG constraints allow severe limits to be placed on various classes of inflationary models that extend the simplest paradigm, in a way that is strongly complementary to the power-spectrum constraints (i.e., scalar spectral index of curvature perturbations and tensor-to-scalar amplitude ratio). 
 
One of the main goals of this paper is to improve NG constraints using mainly the angular bispectrum of CMB anisotropies, i.e., the harmonic transform of the 3-point angular correlation function. We also investigate higher-order NG correlators like the trispectrum. 
We follow the same notation as~\cite{planck2013-p09a}. The CMB angular bispectrum is related to the primordial bispectrum
\begin{equation}
\label{bispectrumPhi}
\langle  \Phi({\vec k}_1) \Phi({\vec k}_2) \Phi({\vec k}_3) \rangle= (2 \pi)^3 \delta^{(3)}({\vec k}_1+{\vec k}_2+{\vec k}_3) B_{\rm \Phi}(k_1,k_2,k_3), 
\end{equation}
where the field $\Phi$, related to the comoving curvature perturbation $\zeta$ on super-horizon scales by $\Phi \equiv (3/5) \zeta$, is such that in the matter
era, and on super-horizon scales, it reduces to Bardeen's gauge-invariant gravitational potential (\citealt{1980PhRvD..22.1882B}).
The bispectrum  $B_{\Phi}(k_1,k_2,k_3)$ measures the correlation among three perturbation modes. If translational and rotational invariance are assumed, it depends only on the magnitude of the three wavevectors. 
In general the bispectrum can be written as 
\begin{equation}
\label{amplitudeandF}
B_{\Phi}(k_1,k_2,k_3)= f_{\rm NL} F(k_1,k_2,k_3) \, ,
\end{equation}
where we have introduced the dimensionless ``nonlinearity parameter" $f_{\rm NL}$ (\citealt{1994ApJ...430..447G,2000PhRvD..61f3504W,2001PhRvD..63f3002K,2004JCAP...08..009B}), measuring the NG amplitude. 
The bispectrum is obtained by sampling triangles in Fourier space. The dependence of the function $F(k_1,k_2,k_3)$ on the type of triangle (i.e., the configuration) formed by the three wavevectors describes the \textit{shape} (and the scale dependence) of the bispectrum~(\citealt{2004JCAP...08..009B}), which encodes much physical information. Different NG shapes are linked to distinctive physical mechanisms that can generate such NG fingerprints in the early Universe.

In this paper the limits on primordial NG are mainly improved through the use of the full mission data, as well as by exploiting the polarization information. 


\Planck\ results on primordial NG also provide a reconstruction of the full CMB bispectrum through different techniques (see Sect.~\ref{Bisp_Rec}). This complements (and adds to) the extraction of single amplitudes $f_{\rm NL}$ for specific bispectrum shapes. Such a reconstruction can point to interesting features in the bispectrum signal that go beyond the usual standard scale-invariant shapes (such as the well known ``local'' and ``equilateral'' configurations).

As we have seen, the \Planck\ 2013 NG paper \citep{planck2013-p09a} significantly improved constraints on  the standard primordial NG models with scale-invariant local, equilateral or  orthogonal shapes.  The \Planck\ NG paper also included constraints from the modal estimator on a variety of other primordial models, including DBI inflation, non-Bunch-Davies models (excited initial states), directionally-dependent vector inflation models, warm inflation, and scale-dependent feature and resonance models.   All scale-invariant bispectra 
were strongly constrained, with the possible exception of highly flattened non-Bunch-Davies models.  On the other hand, the preliminary investigation of primordial oscillatory models seemed to be more promising, in that 
two specific feature models appeared to produce fits of some significance. 
One aim of the present work is to expand the detail and scope of investigations of feature and resonant models and to examine the significance of these results with a 
more careful analysis of the ``look elsewhere" effect, through exploring multi-parameter results using large ensembles of Gaussian simulations.
Also we will thoroughly analyse or re-analyse other primordial NG signals that are theoretically well-motivated and those which have appeared in the literature since the first data release. These include primordial NG arising in the context of inflation models where vector fields play a non-negligible role or primordial NG generated in the tensor (gravitational waves) perturbations. Each of these primordial NG signals carry distinctive signatures that may have been imprinted at the inflationary epoch, thus opening up a new window into the detailed physics of inflation.  
 

The paper is organized as follows. In Sect.~\ref{sec:models} we briefly discuss the primordial NG models that we test in this paper. Section~\ref{sec:SEP} summarizes the optimal statistical estimators used to constrain the CMB bispectrum and trispectrum from \Planck\ temperature and polarization data. In Sect.~\ref{sec:npNG} we discuss the non-primordial contributions to the CMB bispectrum and trispectrum, including foreground residuals after component separation and focusing on the $f_{\rm NL}$ bias induced by the ISW-lensing bispectrum. We also analyse the impact on primordial NG estimation  from the residuals of the deglitching processing. Section~\ref{sec:Validation} describes an extensive suite of tests performed on realistic simulations to validate the different estimator pipelines, and compare their performance. Using simulations, we also quantify the impact on $f_{\rm NL}$ of using a variety of component-separation techniques. In Sect.~\ref{sec:Results} we derive constraints on $f_{\rm NL}$ for the local, equilateral, and orthogonal bispectra and present a reconstruction of the CMB bispectrum. We also present a reconstruction of the primordial curvature fluctuations. In Sect.~\ref{sec:Sec_valid_data} we validate these results by performing a series of null tests on the data to assess the robustness of our results. Section~\ref{sec:Other_shapes} investigates scale-dependent NG models and other selected bispectrum shapes. Section~\ref{sec:tau_gnl} presents the \Planck\ limits on the CMB trispectrum. In Sect.~\ref{sec:Minkowski} we provide constraints on CMB local bispectrum and trispectrum  from Minkowski functionals. In Sect.~\ref{sec:Implications} we discuss the main implications of \Planck's constraints on primordial NG for early Universe models. We conclude in Sect.~\ref{sec:Conc}. Appendix~\ref{App:A} contains  the derivation of the statistical estimator for the amplitudes characterizing a ``direction-dependent'' primordial non-Gaussianity. Appendix~\ref{App:B} contains some details about Minkowski functionals.

%% file: A19_Section2.tex
In this Section we briefly highlight the \textit{classes} of inflationary models investigated in this paper, and describe the distinctive NG they generate.  Within each class a common underlying physical process gives rise to the corresponding NG shape, illustrated by concrete realizations of inflationary models. For each class we therefore provide the explicit form of the bispectrum shapes chosen for the data analysis, emphasizing extensions with variants and distinctly new shapes beyond those already described in \cite{planck2013-p09a}.

\subsection{General single-field models of inflation}
This class of models includes inflationary models with a non-standard kinetic term (or more general higher-derivative interactions), in which the inflaton fluctuations propagate with an effective sound speed $c_{\rm s}$ which can be smaller than the speed of light. For example, models with a non-standard kinetic term are described by an inflaton Lagrangian ${\mathcal L}=P(X, \phi)$, where $X=g^{\mu \nu} \partial_\mu \phi \, \partial_\nu \phi$, with at most one derivative on $\phi$, and the sound speed is given by $c_{\rm s} ^2=(\partial P/\partial X)/(\partial P/\partial X+2 X (\partial^2 P/\partial X^2))$. 

The NG parameter space of this class of models is generically well described by two NG shapes --- ``equilateral'' and ``orthogonal''  \citep{2010JCAP...01..028S} --- since usually there are two dominant interaction terms of the inflaton field giving rise 
to the overall NG signal. One of these typically produces a bispectrum very close to the equilateral type with $f_{\rm NL} \sim c_{\rm s}^{-2}$ in the limit $c_{\rm s} \ll 1$~\citep{2007JCAP...01..002C,2010JCAP...01..028S}.

The equilateral-type NG is well approximated by the template~(\citealt{2006JCAP...05..004C})
\begin{eqnarray}
\label{equilateralBis}
\nonumber
& &	B_{\Phi}^{\rm equil}(k_1,k_2,k_3)= 6A^2 f_{\rm NL}^{\rm equil}\\
\nonumber
& \times& \left\{
-\frac1{k^{4-n_{\rm s}}_1k^{4-n_{\rm s}}_2}-\frac1{k^{4-n_{\rm s}}_2k^{4-n_{\rm s}}_3}
-\frac1{k^{4-n_{\rm s}}_3k^{4-n_{\rm s}}_1} 
-\frac2{(k_1k_2k_3)^{2(4-n_{\rm s})/3}}
\right. \\
& &
\left.+\left[\frac1{k^{(4-n_{\rm s})/3}_1k^{2(4-n_{\rm s})/3}_2k^{4-n_{\rm s}}_3}
+\mbox{(5 permutations)}\right]\right\}\, .
\end{eqnarray}
Here $P_{\Phi}(k)=A/k^{4-n_{\rm s}}$ represents  Bardeen's gravitational potential power spectrum, $A^2$ being the normalization and  $n_{\rm s}$ the scalar spectral index.
DBI inflationary models based on string theory~(\citealt{2004PhRvD..70j3505S,2004PhRvD..70l3505A}) provide physically well-motivated examples of the $P(X, \phi)$-model. They are characterized by an almost equilateral NG with $f^{\rm equil}_{\rm NL}=-(35/108)c_{\rm s}^{-2}$ for $c_{\rm s} \ll 1$, which typically is $f^{\rm equil}_{\rm NL} <-5$.

The  ``orthogonal'' shape template is~(\citealt{2010JCAP...01..028S})
\begin{eqnarray}\label{orthogonalBis}
\nonumber
& &	B^{\rm ortho}_{\Phi}(k_1,k_2,k_3)= 6A^2 f_{\rm NL}^{\rm ortho}\\
\nonumber
& \times& \left\{
-\frac3{k^{4-n_{\rm s}}_1k^{4-n_{\rm s}}_2}-\frac3{k^{4-n_{\rm s}}_2k^{4-n_{\rm s}}_3}
-\frac3{k^{4-n_{\rm s}}_3k^{4-n_{\rm s}}_1}\right.
-\frac8{(k_1k_2k_3)^{2(4-n_{\rm s})/3}}
\\
& &
\left.+\left[\frac3{k^{(4-n_{\rm s})/3}_1k^{2(4-n_{\rm s})/3}_2k^{4-n_{\rm s}}_3}
+\mbox{(5 perm.)}\right]\right\}.
\end{eqnarray}
Equilateral and orthogonal shapes emerge also in models characterized by more general higher-derivative interactions, such as ghost inflation~(\citealt{2004JCAP...04..001A}), effective field theories of inflation~(\citealt{2008JHEP...03..014C,2010JCAP...01..028S,2010JCAP...08..008B}), or the so ``Galileon-like'' models of inflation (see, e.g.,~\citealt{2011JCAP...01..014B}). The latter model is constructed starting from some specific underlying symmetry for the inflaton field, and is characterized by strongly constrained derivative interactions.  
\medskip


\subsection{Multi-field models}

This class of models is characterized by the presence of additional light scalar 
degrees of freedom besides the inflaton, whose fluctuations give rise, or 
contribute, to the final primordial curvature perturbation at the end of 
inflation. This includes the case of ``multiple-field inflation'', where 
inflation is driven by more than one scalar field, as well as scenarios in which 
additional scalar fields remain subdominant during the inflationary 
expansion.
From the point of view of primordial NG, the element in common to all these 
models is that a potentially detectable level of NG in the curvature 
perturbation is generated via a transfer of super-horizon non-Gaussian 
isocurvature perturbations in the second field (not necessarily the inflaton) to 
the adiabatic (curvature) density perturbations, accompanied by nonlinearities in the transfer mechanism. 
This process typically takes place on super-horizon scales, thus implying a 
local form of NG in real space. When going to Fourier space, this leads to a correlation between large and 
small scale modes. The bispectrum for this class of models is indeed largest on so-called ``squeezed'' triangles ($k_1 \ll k_2\simeq 
k_3$).
The local bispectrum is~(\citealt{1993ApJ...403L...1F,1994ApJ...430..447G,2000MNRAS.313..323G,2000MNRAS.313..141V,2000PhRvD..61f3504W,2001PhRvD..63f3002K})
\begin{eqnarray}
\label{localBis}
B^{\rm local}_{\Phi}(k_1,k_2,k_3)&=&2 f^{\rm local}_{\rm NL} \Big[ P_{\Phi}(k_1) P_{\Phi}(k_2)+P_{\Phi}(k_1) P_{\Phi}(k_3) \nonumber\\
&+&P_{\Phi}(k_2) P_{\Phi}(k_3) \Big] \nonumber 
\\
&=& 2 A^2  f^{\rm local}_{\rm NL} \left[  \frac{1}{k_1^{4-n_{\rm s}}k_2^{4-n_{\rm s}}} +{\rm cycl.} \right]\, .
\end{eqnarray}
There is a broad literature on examples and specific realizations of this transfer mechanism 
from isocurvature to adiabatic perturbations 
(\citealt{2002PhRvD..65j3505B,2002PhRvD..66j3506B,2006JCAP...05..019V,2006PhRvD..73h3522R,2007PhRvD..76h3512R,2005PhRvL..95l1302L,Tzavara:2010ge}; 
for a review on NG from multiple-field inflation models, see~\citealt{2010AdAst2010E..76B}). 
An alternative, important possibility is the curvaton model~(\citealt{1990PhRvD..42..313M,1997PhRvD..56..535L,2002NuPhB.626..395E,2002PhLB..524....5L,2001PhLB..522..215M}).
In this type of scenario, a second light scalar field, subdominant during inflation, decays after inflation, 
generating primordial density perturbations with a potentially high level of NG~(e.g., \citealt{2002PhLB..524....5L,2003PhRvD..67b3503L,2004PhRvD..69d3503B}).  
In the (simplest) adiabatic curvaton models, the local $\fnl$ parameter was found to be~\citep{2004PhRvD..69d3503B,2004PhRvL..93w1301B} $f_{\rm NL}^{\rm 
local}=(5/4r_\mathrm{D})-5r_\mathrm{D}/6-5/3$, when the curvaton field has a quadratic potential~(\citealt{2002PhLB..524....5L,2003PhRvD..67b3503L,2005PhRvL..95l1302L,2006JCAP...09..008M,2006PhRvD..74j3003S}). 
In the previous formula, $r_{\rm D}=[3\rho_{\rm curvaton}/(3 \rho_{\rm curvaton}+4\rho_{\rm radiation})]_{\rm D}$  is the ``curvaton decay fraction''  
evaluated at the epoch of the curvaton decay in the sudden decay approximation. It is then easy to see that, for low values of $r_{\rm D}$, a high level of NG can be 
generated.~\footnote{NG perturbations can arise also at the end of inflation, e.g., from nonlinearities during the (p)reheating phase (e.g.,~\citealt{2005PhRvL..94p1301E,2008PhRvL.100d1302C,2006PhRvD..73j6012B}; 
see also~\citealt{2009PhRvL.103g1301B}) or from fluctuations in the inflaton decay rate or interactions, as found in modulated (p)reheating and modulated hybrid inflation~(\citealt{2003astro.ph..3614K,2004PhRvD..69h3505D,2004PhRvD..69b3505D,2004PhRvD..70h3004B,Zaldarriaga:2003my,2005JCAP...11..006L,2005PhRvD..72l3516S,2006PhRvL..97l1301L,2006PhRvD..73b3522K,2012JCAP...05..039C}).}


\subsection{Isocurvature non-Gaussianity}
\label{Sec:isocurv_NG_intro}
Isocurvature NG, which was only sketched from the purely theoretical point of view in the 2013 paper, can now be analysed thanks to the polarization information.

In most of the models mentioned above, the main focus is on the level
of primordial NG in the final curvature perturbation $\zeta$. However,
in inflationary scenarios where different scalar fields play a
non-negligible role, residual isocurvature perturbation modes can remain after inflation. Isocurvature modes are usually investigated
by considering their contribution to the power spectrum. However,
if present, they would also contribute to the bispectrum,
producing in general both a pure isocurvature bispectrum and mixed
bispectra because of the cross-correlation between isocurvature and adiabatic 
perturbations~(\citealt{2002astro.ph..6039K,2002PhRvD..65j3505B,Komatsu:2003iq,
2008JCAP...11..019K,Langlois:2008vk,Kawasaki:2008pa,Hikage:2008sk,
2011JCAP...01..008L,Langlois:2011hn,2012JCAP...07..037K,Langlois:2012tm}).  
While one might expect isocurvature NG to be negligible, since both 
(linear) isocurvature modes and (adiabatic) NG appear to be very small, 
and searches for isocurvature NG using WMAP data did not lead to any 
detections~(\citealt{2013JCAP...07..007H,2013JCAP...03..020H}),
this expectation can be tested at significantly higher precision by \Planck. Moreover, there exist inflation 
models~(\citealt{2011JCAP...01..008L}) where isocurvature 
modes, while remaining a small fraction in the power spectrum, would 
dominate the bispectrum.

At the time of recombination there are in principle four possible
distinct isocurvature modes (in addition to the adiabatic mode): 
cold dark matter (CDM); baryon; neutrino density; and 
neutrino velocity isocurvature modes~(\citealt{2000PhRvD..62h3508B}).  
In this paper we will
only consider isocurvature NG of the local type and always limit
ourselves to considering the adiabatic mode together with just one
type of isocurvature mode (considering each of the four types
separately). Otherwise the number of free parameters becomes so large
that no meaningful limits can be derived. Moreover, we assume the same
spectral index for the primordial isocurvature power spectrum and the
isocurvature-adiabatic cross-power spectrum as for
the adiabatic power spectrum. Under those assumptions, as shown 
by~\cite{Langlois:2011hn}, we have in principle 
six independent $f_\mathrm{NL}$ parameters: the usual purely adiabatic one; 
a purely isocurvature one; and four correlated ones. 

The primordial shape templates are a generalization of Eq.~(\ref{localBis}), 
see~\cite{Langlois:2011hn,Langlois:2012tm}:
\begin{eqnarray}
B^{IJK}(k_1, k_2, k_3) & = &
2 f_{\rm NL}^{I, JK}  P_\Phi(k_2) P_\Phi(k_3) 
+ 2 f_{\rm NL}^{J, KI}  P_\Phi(k_1) P_\Phi(k_3) \nonumber\\
&& + 2 f_{\rm NL}^{K, IJ}  P_\Phi(k_1)P_\Phi(k_2), 
\label{localBis_isocurv}
\end{eqnarray}
where $I,J,K$ label the different modes (adiabatic and isocurvature).
The invariance under the simultaneous exchange of two of these indices
and the corresponding momenta means that $f_{\rm NL}^{I, JK} = f_{\rm NL}^{I, KJ}$,
hence reducing the number of independent parameters from eight to six, in the case
of two modes.
The different bispectra vary most importantly through the fact that different
types of radiation transfer functions $g^I_\ell(k)$ are used to project the 
primordial template onto the CMB: the reduced bispectra are of the form
\begin{equation}
\label{b_IJK_isocurv}
b_{\ell_1 \ell_2 \ell_3}^{I,JK}= 6 \int_0^\infty r^2 dr \, 
\alpha^I_{(\ell_1}(r)\beta^{J}_{\ell_2}(r)\beta^{K}_{\ell_3)}(r),
\end{equation}
with   
\begin{eqnarray}
\label{alpha_isocurv}
\alpha^I_{\ell}(r)&\equiv& \frac{2}{\pi} \int k^2 dk\,  j_\ell(kr) \, g^I_{\ell}(k),
\\
\label{beta_isocurv}
\beta^{I}_{\ell}(r)&\equiv& \frac{2}{\pi}  \int k^2 dk \,  j_\ell(kr) 
\, g^I_{\ell}(k)\,  P_\Phi(k).
\end{eqnarray}
Here $j_\ell$ is the spherical bessel function and we use the notation
$(\ell_1 \ell_2 \ell_3)\equiv [\ell_1\ell_2\ell_3+ 5\,  {\rm perm.}]/3!$.
In addition to the isocurvature index, each transfer function carries a
polarization index that we do not show here.
It is important to note that, unlike
the case of the purely adiabatic mode, the inclusion of polarization improves 
the constraints on the isocurvature NG significantly,
as predicted by~\cite{Langlois:2011hn,Langlois:2012tm}.


\subsection{Resonance and axion monodromy models}
\label{resonance}

Oscillatory models for NG are physically well-motivated.   Large-field inflation faces an inherent UV completion problem because the inflaton field is required to move over large distances in field space relative to the Planck mass $m_{\rm Pl}$.  An effective shift symmetry can enforce potential flatness and this can be naturally implemented in a string theory context with axions and a periodically modulated potential, so-called ``axion monodromy" models.  This periodicity can generate resonances in the inflationary fluctuations with logarithmically-spaced oscillations, creating imprints in the power spectrum, the bispectrum and trispectrum \citep{2008JCAP...04..010C,2010JCAP...06..009F,Hannestad:2009yx,2011JCAP...01..017F}.  On the other hand, sharp features or corners in an inflationary potential can temporarily drive the inflaton away from slow-roll; these large changes in the field and derivatives can create evenly-spaced oscillations, to be discussed in the next subsection. However, in multifield models residual oscillations after corner-turning can also lead to log-spaced oscillations, just as in the resonance models \citep{2011arXiv1104.1323C,2011JCAP...01..030A,2013JCAP...05..006B,2015JCAP...02..027C}.    A preliminary search for bispectrum resonance signals was performed in the first \Planck\ analysis  \citep{planck2013-p09a} and our purpose here is to substantially increase the frequency range and number of models investigated.   

\smallskip

\noindent \textit{Simple resonance model:}
Periodic features in the inflationary potential can induce oscillations with frequency $\omega$ that can resonate through any interactions with the inflationary fluctuations, contributing to the bispectrum.   Provided that $\omega > H$, this mode starts inside the horizon but its frequency decreases as it is stretched by inflation, until frozen when $\omega\simeq H$.   Thus periodic features introduce a driving force which can scan across a wide range of frequencies.  The simplest basic behaviour of such resonant models yields  logarithmic stretching and can be described by the non-scale-invariant shape (see \citealt{2008JCAP...04..010C,2010AdAst2010E..72C})
\eq\label{eq:resBprim}
B_\Phi^{\rm res}(\klist) = \frac{6A ^2\fnl^{\rm res} }{(k_1 k_2 k_3)^2}\sin\[C\ln({k_1+k_2+k_3}) + \phi\]\,, 
\qe
where the constant $C=1/\ln (3k_{\rm *})$, $k_*$ is a wavenumber associated with the periodicity, and $\phi$ is a phase.   These oscillations constructively and destructively interfere with the oscillations created by the CMB transfer functions, introducing additional nodal points in the CMB bispectrum.

\smallskip

\noindent \textit{Generalized resonance models:} In a more general context, it is possible to have more complicated resonant shapes and envelopes.  Resonant single-field models with varying sound speed $c_{\rm s}$ generate three  leading-order bispectrum terms \citep{Chen:2010bka}:
\eq\label{eq:resChen1Bprim}
&B^{\rm res-cs}(\kall) =\frac{6A ^2 }{(k_1 k_2 k_3)^2}\left\{\fnl^{\rm res1}   \sin\[C\ln({k_1+k_2+k_3}) + \phi\]\right.\nonumber\\
& +\,3\fnl^{\rm res2}\frac{ k_1^2+ k_2^2+ k_3^2}{(k_1+k_2+k_3)^2}\cos\[C\ln({k_1+k_2+k_3})  +\phi\] \nonumber\\
& ~~~\left.+\,27\fnl^{\rm res3}\frac{k_1k_2k_3}{(k_1+k_2+k_3)^3}  \sin\[C\ln({k_1+k_2+k_3}) + \phi\]\right\}.
\qe
The first term on the right-hand side of Eq.~\eqref{eq:resChen1Bprim} is the basic resonant shape given in Eq.~\eqref{eq:resBprim}, while the second and third terms have the same oscillatory behaviour, but modulated by a (mildly) flattened shape, and an equilateral shape respectively.   The third term is in fact the second generic shape arising in effective field theory and correlates well with the equilateral shape in Eq.~\eqref{eq:equilBprim}.   The second term  in Eq.~\eqref{eq:resChen1Bprim} weakly favours flattened triangles, but there are regimes for resonant models that can generate much stronger flat shapes.  If the resonance begins very deep inside the horizon, then the second (negative energy) mode can also make a significant contribution that is associated with enfolded or flat bispectra; this is similar to having an excited initial state or non-Bunch-Davies (NBD) vacuum.  

With these two physical motivations in mind we also investigate classes of models with resonant oscillations modulated by both the equilateral and flattened shapes, defined by
\eq\label{eq:equilBprim}
S^{\rm eq} (\klist) = \frac{\tilde k_1\tilde k_2\tilde k_3}{k_1k_2k_3}\,, ~~~~ S^{\rm flat} = 1 - S^{\rm eq}\,,
\qe
where $\tilde k_1 \equiv k_2+k_3 - k_1$ (here, for simplicity we ignore the spectral index dependence of the equilateral shape in Eq.~\ref{equilateralBis}).   The corresponding equilateral and flattened resonant bispectra ans\" atze are then
\eq\label{eq:resequilBprim}
&&B^{\rm res-eq}(\klist)\equiv S^{\rm eq} (\klist) \times B^{\rm res} (\klist)\nn\\
&&~~~~~~=~\frac{6A ^2\fnl^{\rm res-eq} }{(k_1 k_2 k_3)^2} \frac{\tilde k_1\tilde k_2\tilde k_3}{k_1k_2k_3} \sin\[C\ln({k_1+k_2+k_3}) + \phi\]\,,\\
\label{eq:resflatBprim}
&& B^{\rm res-flat}(\klist) \equiv  S^{\rm flat} (\klist) \times B^{\rm res} (\klist)\,.
\qe

We note that typically non-Bunch-Davies bispectra can be much more sharply peaked in the flattened or squeezed limits than Eq.~\eqref{eq:resflatBprim}, but our purpose here is to determine if this type of resonant model is favoured by the \Planck\ data; that is, whether Eq.~\eqref{eq:resflatBprim} warrants further investigation with other flattened profiles.

\subsection{Scale-dependent oscillatory feature models}

Temporary violations of slow-roll inflation can occur if there are sharp features in the inflationary potential \citep{2007JCAP...06..023C}, as well as changes in the sound speed $c_{\rm s}$ or sharp turns in field space in multifield inflation.   The inflaton field makes temporary departures from the attractor solution, which typically have a strong scale-dependent running modulated by a sinusoidal oscillation; there are model-dependent counterparts in the power spectrum, bispectrum, and trispectrum.   For example, sharper or narrower features induce a relatively larger signal in the bispectrum (see e.g.,~\citealt{2010AdAst2010E..72C}).  An example is the analytic envelope solutions predicted for both the power spectrum and bispectrum for the single field models with a specific inflaton feature shape \citep{Adshead:2011jq}; a search for these was presented previously in the \Planck\ Inflation paper \citep{planck2013-p17} and likewise no significant signal was found using the corresponding bispectrum envelopes at the available modal resolution \citep{planck2013-p09a}.   
In this new analysis,  we will emphasize the search for generic oscillatory behaviour in the data over a larger range in modal resolution, although we will also look for the shapes predicted for simple features in single field models. 

\smallskip

\noindent \textit{Constant feature model:} In the previous investigation of \Planck\ data using a coarse parameter grid \citep{planck2013-p09a}, we searched for the simplest ansatz for an oscillatory bispectrum signal  \citep{2007JCAP...06..023C}:
\begin{align}
\label{eq:featureBprim}
B^{\rm feat}(\kall) = \frac{6A ^2\fnl^{\rm feat} }{(k_1 k_2 k_3)^2}\sin\[{\omega(k_1+k_2+k_3)} + \phi\]\,,
\end{align}
where $\phi$ is a phase factor and $\omega$ is a frequency associated with the specific shape of the feature in the potential that disrupts the slow-roll evolution.   In the earlier analysis, we also considered a damping envelope, which slightly increased the apparent significance of the best-fit feature models, though at the cost of an additional parameter (see single-field solutions below). 

\smallskip

\noindent \textit{Generalized feature models:}   Here, we again search for oscillatory signals in a model-independent manner. We will modulate the bispectrum cross-sections with the physically motivated equilateral and flattened shapes, reflecting the physical contexts in which they could have been generated, as for the resonant models described above in Eq.~\eqref{eq:resChen1Bprim}.  
If there are potential features in a model with a varying sound speed, then we can expect there to be oscillatory contributions to the bispectrum signal with a dominant equilateral shape. Motivated by the equilateral resonance model in Eq.~\eqref{eq:resChen1Bprim}, we will search for the following {\it equilateral feature} ansatz:
\eq\label{eq:featequilBprim}
&&B^{\rm feat-eq}(\klist) \equiv  S^{\rm eq} (\klist) \times B^{\rm feat} (\klist)\\
&& ~~~~~~=~\frac{6A ^2\fnl^{\rm feat-eq} }{(k_1 k_2 k_3)^2} \frac{\tilde k_1\tilde k_2\tilde k_3}{k_1k_2k_3}  \sin\[\omega(k_1+k_2+k_3) + \phi\]\,.
\qe
For extremely sharp features, it is possible to excite the inflationary fluctuations as if there were a non-Bunch Davies vacuum:  the oscillatory signal becomes modulated with a flattened shape \citep{2007JCAP...06..023C}.  Again, motivated by the enfolded resonance model in Eq.~\eqref{eq:resflatBprim}, we take the following simple flattened ansatz:
\eq\label{eq:featflatBprim}
B^{\rm feat-flat}(\klist) &\equiv&  S^{\rm flat} (\klist) \times B^{\rm feat} (\klist)\,.
\qe
Although the exact profile of the flattened shape can be much more highly peaked on the faces in these NBD models, this ansatz should be adequate for testing whether these models are favoured.   We note that while the power spectrum is insensitive to the underlying scenario creating the features, the bispectrum shape will reveal whether features arise from varying sound speed or highly excited features in the potential. 


\smallskip
\noindent \textit{Single field feature solutions:} Here we use the full analytic bispectrum solution given by \cite{Adshead:2011jq}, but the dominant leading-order behaviour takes the form
\eq\label{eq:Adsetal1}
B^{K^2\cos}(\kall) =\frac{6A ^2 \fnl^{\rm K^2\cos} }{(k_1 k_2 k_3)^2} K^2 D(\alpha \omega K) \cos (\omega K)\,,
\qe
where $K= k_1+ k_2 +k_3$ and  $D(\alpha \omega K) =\alpha \omega/ (K \sinh (\alpha\omega K))$ 
is an envelope function, with parameter $\alpha$ setting an overall cut-off for the bispectrum at large wavenumbers or multipoles. This envelope and the overall $K^2$ scaling distinguishes this realistic case from the simple separable constant feature ansatz of Eq. (\ref{eq:featureBprim}). 
We shall allow the envelope parameter $\alpha$ to vary from $\alpha = 0$, with no envelope (the infinitely thin limit for a feature in the potential) through to large $\alpha$, with a narrow domain for the bispectrum.  
Alternative analytic solutions where the bispectrum is created by a variation in the sound speed $c_{\rm s}$ are dominated by the $K\sin (\omega K)$ term, as in
\eq\label{eq:Adsetal2}
B^{K\sin}(\kall) =\frac{6A ^2 \fnl^{\rm K\sin} }{(k_1 k_2 k_3)^2}  K \,D(\alpha \omega K) \sin (\omega K)\,.
\qe

For the simplest models there is a predicted relationship between the power spectrum and bispectrum amplitude (e.g., see also \citealt{Achucarro:2012fd} for a two-field model).   We note that typically the power spectrum has larger signal-to-noise at low frequency (i.e., below $\omega \simeq 1000$) while the bispectrum dominates at higher frequency.  


\subsection{Non-Gaussianity from excited initial states}  

It is well known that if the initial vacuum state for inflation is excited and deviates from the standard Bunch-Davies vacuum, then measurable non-Gaussianities can be produced \citep{2007JCAP...01..002C, Holman:2007na, Meerburg:2009ys, Ashoorioon:2010xg}. These models generically lead to non-Gaussianity that peaks in the flattened limit, where $k_1 + k_2 \approx k_3$, and also often has oscillatory behaviour.  Here we constrain the same selection of templates found in the 2013 \Planck\ analysis, namely the flat model in Eq.~\eqref{eq:equilBprim}, Non-Bunch-Davies (NBD) \citep{2007JCAP...01..002C}, NBD1 and NBD2 models \citep{Agullo:2011xv} (now called ``NBD1 cos" and ``NBD2 cos") and NBD3 \citep{2010AdAst2010E..72C}.  We also introduce three new templates, NBD sin which is motivated by \citep{Chen:2010bka} and takes the form
\begin{align}
&&B^{\rm NBD-sin}\left(k_1,k_2,k_3\right) = \frac{2A^2 \fnl^{\rm NBD-sin}}{\left(k_1 k_2 k_3\right)^2} \left( e^{-\omega\tilde{k}_1} + e^{-\omega\tilde{k}_2} + e^{-\omega\tilde{k}_3} \right)\cr
&&~~~~~~~\times ~ \sin \left( \omega K + \phi \right)\,,
\end{align}
where again $K = k_1+k_2+k_3$ and $\tilde{k}_i = K - 2 k_i$. The other two templates are extensions of the NBD1 cos and NBD2 cos models found in \cite{Agullo:2011xv} and take the form
\begin{align}
&&B^{\rm NBD{\it i}-sin}\left(k_1,k_2,k_3\right) = \frac{2A^2\fnl^{\rm NBD{\it i}-sin}}{(k_1 k_2 k_3)^2} \big[  f_i(k_1;k_2,k_3) \cr  
&& ~~~~~~~~~~~~~~~~\times~ {\sin(\omega \tilde{k}_1)}/{\tilde{k}_1}+ 2\,\mbox{perm.}\big]\,,
\end{align}
where $f_1(k_1;k_2,k_3) = k_1^2(k_2^2+k_3^2)/2$, which is dominated by squeezed configurations, and $f_2(k_1;k_2,k_3) = k_2^2k_3^2$, which has a flattened shape.


\subsection{Directional-dependence motivated by gauge fields}  
Some models where primordial vector fields are present during inflation predict interesting NG signatures. This is the case of a coupling of the inflaton field $\varphi$ to the kinetic term of a gauge field $A^\mu$, ${\mathcal L}$ contains $-I^2(\varphi) F^2$, where $F_{\mu\nu}=\partial_\mu A_\nu-\partial_\nu A_\mu$ and the coupling $I^2(\varphi) F^2$ is chosen so that scale invariant vector perturbations are produced on superhorizon scales~(\citealt{Barnaby:2012tk,Bartolo:2012sd}). 
The bispectrum turns out to be the sum of two contributions: one of the local shape; and another that is also enhanced in the squeezed limit ($k_1 \ll k_2\simeq k_3$), but featuring a non-trivial dependence on the angle between the small and the large wave vectors through the parameter $\mu_{12} =\hat {\vec k}_1\cdot\hat{\vec k}_2$ (where $\hat{\vec k} = {\vec k}/k$) as $\mu_{12}^2$. Also, primordial magnetic fields sourcing curvature perturbations can cause a dependence on  both  $\mu$ and $\mu^2$ \citep{Shiraishi:2012rm}. 

We can parametrize these shapes as variations on the local shape \citep{Shiraishi:2013vja}, as 
\eq\label{vectorBis}
B_\Phi(\kall) = \sum_L c_L [P_L(\mu_{12})P_{\Phi}(k_1)P_{\Phi}(k_2)+\hbox{2\,perm.}],
\qe
where $P_L(\mu)$ is the Legendre polynomial with $P_0=1, \;P_1=\mu$, and $P_2=\textstyle {\frac{1}{2} (3 \mu^2-1)}$.   For example, for $L=1$ we have the shape 
\eq\label{eq:dbiS}
 B_\Phi^{L=1}(\kall) = \frac{2A ^2\fnl^{L=1}}{(k_1 k_2 k_3)^2}\,\left[\frac{k_3^2}{k_1^2k_2^2}(k_1^2+k_2^2-k_3^2)  + \hbox{2\,perm.}\right].
 \qe
The local template corresponds to $c_i= 2 f_{\rm NL} \delta_{i0}$. Here and in the following the nonlinearity parameters $f_{\rm NL}^L$ are related to the $c_L$ coefficients by $c_0=2 f_{\rm NL}^{L=0}$, $c_1=-4 f_{\rm NL}^{L=1}$, and $c_2=-16 f_{\rm NL}^{L=2}$. The $L=1,2$ shapes exhibit sharp variations in the flattened limit, for example for $k_1+k_2\approx k_3$, while in the squeezed limit, $L=1$ is suppressed whereas $L=2$ grows like the local bispectrum shape (i.e., the $L=0$ case). The $I^2(\varphi) F^2$ models predict $c_2=c_0/2$, while primordial curvature perturbations sourced by large-scale magnetic fields generate non-vanishing $c_0$, $c_1$, and $c_2$. Quite interestingly, in the proposed ``solid inflation'' scenario (\citealt{2012arXiv1210.0569E}; see also~\citealt{2013JCAP...08..022B,2014PhRvD..90f3506E,2014PhRvD..89l3509S,2014JCAP...11..009B}) bispectra similar to Eq.~(\ref{vectorBis})  can be generated, in this case with $c_2 \gg c_0$~\citep{2012arXiv1210.0569E,2014PhRvD..90f3506E}. Therefore, measurements of the $c_i$ coefficients can be an efficient probe of some detailed aspects of the inflationary mechanism, such as the existence of primordial vector fields during inflation (or a non-trivial symmetry structure of the inflaton fields, as in solid inflation).

 \subsection{Non-Gaussianity from gauge-field production during axion inflation} \label{sec:axioninf}

The same shift symmetry that leads to axion (monodromy) models of inflation (Sect.~\ref{resonance})  naturally allows (from an effective field theory point of view) for a coupling between a pseudoscalar axion inflaton field and a gauge field of the type 
${\mathcal L} \supset\ -(\alpha/4f) \phi F^{\mu\nu}\tilde F_{\mu\nu}$, where the parameter $\alpha$ is dimensionless and $f$ is the axion decay constant ($\tilde F^{\mu\nu}=\epsilon^{\mu\nu\gamma\beta}
F_{\gamma\beta}/2$).  This scenario has a rich and interesting phenomenology both for scalar and tensor primordial fluctuations (see, e.g.,~\citealt{2011PhRvL.106r1301B,2011JCAP...06..003S,2011JCAP...04..009B,2012PhRvD..85b3525B,2013PhRvD..87j3506L,2013JCAP...02..017M,2014arXiv1409.5799F}). Gauge field quanta are produced by the background motion of the inflaton field, and these in turn source curvature perturbations through an inverse decay process of the gauge field. A bispectrum of curvature fluctuations is generated as~\citep{2011JCAP...04..009B,2013JCAP...02..017M}\footnote{For simplicity we assume a scale-invariant bispectrum.}
\begin{equation}
\label{invdec}
B^{\rm inv.dec}=6 A^2 f^{\rm inv.dec}_{\rm NL} \frac{\sum_i k_i^3}{\prod k_i^3} \frac{f_3\left(\xi_*,\frac{k_2}{k_1},\frac{k_3}{k_1}\right)}{f_3(\xi_*,1,1)}\, , 
\end{equation} 
where the exact expression for the function $f_3$ can be found in equation (3.29) of ~\cite{2011JCAP...04..009B} (see also~\citealt{2013JCAP...02..017M}). Here $\xi$ characterizes the coupling strength of the axion to the gauge field $\xi=\alpha |\dot\phi|/(2fH)$. The inverse decay bispectrum peaks for equilateral configuration, since $\delta \varphi$ is mostly sourced by the inverse decay $(\delta A +\delta A \rightarrow \delta \varphi)$, when two modes of the vector fields are of comparable magnitude (the correlation with the equilateral template is $94\,\%$  and with the orthogonal one is $4\,\%$). We do however constrain the exact shape in Eq.~(\ref{invdec}), without resorting to the equilateral template. Another interesting observational signature that can shed light on the role played by pseudo-scalars in the early Universe is provided by tensor NG, to which we turn next.  
 
\subsection{Parity-violating tensor non-Gaussianity motivated by pseudo-scalars} 

While the majority of the studies on primordial and CMB NG focus on the scalar mode, tensor-mode NG has been attracting attention as a probe of high-energy theories of gravity (e.g., \citealt{2011JHEP...09..045M, 2011JCAP...06..030M, 2011JHEP...08..067S, 2011PThPh.126..937S, 2011PhRvL.107u1301G}) or primordial magnetic fields \citep{2005PhRvD..72f3002B, Shiraishi:2012rm, 2012JCAP...06..015S}.\footnote{See \cite{planck2014-a22} for the \Planck\ constraints on magnetically-induced NG.} 

Recently, the possibility of observable tensor bispectra has been vigorously discussed in a model where the inflaton couples to a pseudoscalar field \citep{2012PhRvD..86j3508B, 2013JCAP...11..047C, 2014arXiv1409.5799F}. In this model, through the gravitational coupling to the U(1) gauge field, gravitational waves ($h_{ij} \equiv \delta g_{ij}^{TT}/a^2 = \sum_{s = \pm} h^{(s)} e_{ij}^{(s)}$) receive NG corrections, where only one of the two spin states is enhanced. The bispectrum, generally formed as 
\begin{eqnarray}
\left\langle \prod_{i=1}^3 h^{(s_i)}(\vec{k}_i) \right\rangle
&=& (2\pi)^3 \delta^{(3)}(\vec{k}_1 + \vec{k}_2 + \vec{k}_3) B_{\rm h}^{s_1 s_2 s_3}(\vec{k}_1, \vec{k}_2, \vec{k}_3) ~,
\end{eqnarray}
is accordingly polarized, with $B_{\rm h}^{+++} \gg B_{\rm h}^{++-}, B_{\rm h}^{+--}, B_{\rm h}^{---}$.  This NG enhancement is a sub-horizon effect and therefore $B_{\rm h}^{+++}$ is maximized at the equilateral limit ($k_1 \simeq k_2 \simeq k_3$) \citep{2013JCAP...11..047C}. 

A model-independent template of the equilateral-type polarized tensor bispectrum is given by \citep{2013JCAP...11..051S, 2015JCAP...01..007S} 
\begin{eqnarray}
B_{\rm h}^{+++}(\vec{k}_1, \vec{k}_2, \vec{k}_3)
&=& f_{\rm NL}^{\rm tens} F_\zeta^{\rm equil}(k_1, k_2, k_3) 
\nonumber \\
&&\times 
\frac{16\sqrt{2}}{27} e_{ij}^{(+) *}(\hat{\vec{k}}_1)
e_{jk}^{(+) *}(\hat{\vec{k}}_2)
e_{ki}^{(+) *}(\hat{\vec{k}}_3) ~, \label{eq:bis_tens_form}
\end{eqnarray}
with the polarization tensor $e_{ij}^{(s)}$ obeying $e_{ij}^{(s)}(\hat{\vec{k}}) e_{ij}^{(s')}(\hat{\vec{k}}) = 2 \delta_{s, -s'}$ and $e_{ij}^{(s) *}(\hat{\vec{k}}) = e_{ij}^{(-s)}(\hat{\vec{k}}) = e_{ij}^{(s)}(- \hat{\vec{k}})$.
We here have introduced a tensor nonlinearity parameter, by normalizing with the equilateral bispectrum template of curvature perturbations ($F_\zeta^{\rm equil} \equiv (5/3)^3 F_\Phi^{\rm equil} = (5/3)^3 B_\Phi^{\rm equil} / f_{\rm NL}^{\rm equil}$) in the equilateral limit, yielding
\begin{eqnarray}
f_{\rm NL}^{\rm tens} \equiv \lim_{k_i \to k} \frac{B_{\rm h}^{+++}(\vec{k}_1, \vec{k}_2, \vec{k}_3)}{ F_{\zeta}^{\rm equil}(k_1, k_2, k_3)} ~.
\end{eqnarray}
The template Eq.~\eqref{eq:bis_tens_form} can adequately reconstruct the tensor bispectra created in the pseudoscalar inflation 
models\footnote{The form of the tensor bispectrum is the same whether the inflaton field is identified with the pseudoscalar field or not.} \citep{2013JCAP...11..051S}, and thus the amplitude $f_{\rm NL}^{\rm tens}$ is directly connected with the model parameters, e.g., the coupling strength of the pseudoscalar field to the gauge field $\xi$ (for details see Sect.~\ref{sec:Implications}). 

The CMB temperature and $E$-mode bispectra sourced by the parity-violating tensor NG have not only the usual parity-even ($\ell_1 + \ell_2 + \ell_3 = {\rm even}$) signals but also parity-odd ($\ell_1 + \ell_2 + \ell_3 = {\rm odd}$) contributions, which cannot be sourced by known scalar bispectra \citep{2011PhRvD..83b7301K, 2011PThPh.126..937S}. Moreover, their shapes are mostly distinct from the scalar templates, due the different radiation transfer functions; hence they can be measured essentially independently of the scalar NG \citep{2013JCAP...11..051S}. The analysis of the WMAP temperature data distributed in $\ell_1 + \ell_2 + \ell_3 = {\rm odd}$ configurations leads to an observational limit $f_{\rm NL}^{\rm tens} = (0.8 \pm 1.1) \times 10^4$ \citep{2015JCAP...01..007S}. This paper updates the limit, by analysing both parity-even and parity-odd signals in the \Planck\ temperature and $E$-mode polarization data.

%% file: A19_Section3.tex
We now provide a brief overview of the main statistical techniques that we 
use to estimate the nonlinearity parameter $\fnl$ from temperature and polarization CMB data,
 followed by a description of the data set that will be used in our analysis. 

The CMB temperature and polarization fields are characterized using 
the multipoles of a spherical harmonic decomposition of the CMB maps:
\begin{align}
\frac{\Delta T}{T} (\vnhat) = & \sum_{\ell m} a_{\ell m}^T \Ylm(\vnhat) \nonumber \, ,\\
E (\vnhat) = & \sum_{\ell m} a_{\ell m}^E \Ylm(\vnhat) \ .
\end{align}
At linear order, the relation between the primordial perturbation field and the CMB multipoles is (e.g., \citealt{2001PhRvD..63f3002K})
\begin{equation}\label{eq:phi2alm}
\alm^X = 4 \pi (-i)^\ell \int \frac{d^3 k}{(2 \pi)^3} \, \Phi(\vk)  \Ylm(\vkhat) \Delta^{X}_\ell(k)  \; ,
\end{equation}
where $X=\{T,E\}$ denotes either temperature or $E$-mode polarization, $\Phi$ is the primordial 
gravitational potential, and $\Delta^{X}_\ell$ represents the linear CMB radiation transfer function. 

The CMB angular bispectrum is the three-point correlator of the $a_{\ell m}$s
\begin{equation}
B_{\ell_1 \ell_2 \ell_3}^{m_1 m_2 m_3, X_1 X_2 X_3} \equiv\langle a^{X_1}_{\ell_1 m_1}a^{X_2}_{\ell_2 m_2}a^{X_3}_{\ell_3 m_3}\rangle \, ,
\end{equation}
where $X_i=\{T,E\}$.
If the CMB sky is rotationally invariant, and the bispectra we are considering have even parity 
(which is true for combinations of $T$ and $E$),  then the angular bispectrum can be factorized as
\begin{equation}\label{eq:Bred}
\langle a^{X_1}_{\ell_1 m_1}a^{X_2}_{\ell_2 m_2}a^{X_3}_{\ell_3 m_3}\rangle = 
\curl{G}^{\ell_1 \ell_2 \ell_3}_{m_1 m_2 m_3} \,b^{X_1 X_2 X_3}_{\ell_1 \ell_2
\ell_3}\, ,
\end{equation} 
where $b^{X_1 X_2 X_3}_{\ell_1 \ell_2\ell_3}$ is the so-called {\em reduced bispectrum}, and $\curl{G}^{\ell_1 \ell_2 \ell_3}_{m_1 m_2 m_3}$ 
 is the Gaunt integral, defined as the integral over the solid angle of the product of three spherical harmonics,
\begin{equation}\label{eq:Gaunt}
\curl{G}^{\ell_1 \ell_2 \ell_3}_{m_1 m_2 m_3} \equiv\int Y_{\ell_1 m_1}(\vnhat) \, Y_{\ell_2 m_2}(\vnhat) \, Y_{\ell_3 m_3}(\vnhat) \, d^2\vnhat \; .
\end{equation}
The Gaunt integral (often written in terms of Wigner 3$j$-symbols) enforces rotational symmetry, and restricts attention to a tetrahedral domain of multipole triplets
$\{\ell_1,\ell_2,\ell_3\}$, satisfying both a triangle condition and a limit
given by some maximum resolution $\ell_\mathrm{max}$ (the latter being defined by the finite angular 
resolution of the experiment under study). 

Our goal is to extract the nonlinearity parameter $\fnl$ from the data, for different primordial shapes. To achieve this, we essentially fit a theoretical CMB bispectrum 
ansatz $b_{\ell_1 \ell_2 \ell_3}$ to the observed 3-point function. Theoretical predictions for CMB angular bispectra arising from early Universe primordial models can be obtained 
by applying Eq.~\eqref{eq:phi2alm} to the primordial bispectra of Sect.~\ref{sec:models}, (see e.g., \citealt{2001PhRvD..63f3002K}).  Optimized cubic bispectrum estimators were introduced by \cite{1998MNRAS.299..805H}, and it has been shown that for small NG the general optimal polarized $\fnl$ estimator can be written as \citep{2006JCAP...05..004C}
\begin{align}\label{eq:optimalestimator}
\hat{f}_{\textrm{NL}}  = & \frac{1}{N} \sum_{X_i, X'_i} \sum_{\ell_i,m_i} \sum_{\ell'_i, m'_i} \curl{G}^{\,\,\ell_1\; \ell_2\; \ell_3}_{m_1 m_2 m_3 } b^{X_1 X_2 X_3, \, \rm th}_{\ell_1 \ell_2 \ell_3} \nonumber \\
& \left\{ \left[ (\tens{C}^{-1}_{\ell_1 m_1, \ell_1' m_1'})^{X_1 X'_1} a^{X'_1}_{\ell_1'm_1'}\, (\tens{C}^{-1}_{\ell_2 m_2, \ell_2' m_2'} )^{X_2 X'_2} a^{X'_2}_{\ell_2'm_2'} \right. \right. \nonumber \\
& \left. (\tens{C}^{-1}_{\ell_3 m_3, \ell_3' m_3'})^{X_3 X'_3} a^{X'_3}_{\ell_3'm_3'} \right] - \nonumber \\
& \left. \left[ \,(\tens{C}^{-1}_{\ell_1 m_1, \ell_2 m_2})^{X_1 X_2} (\tens{C}^{-1}_{\ell_3 m_3,\ell_3' m_3'})^{X_3 X'_3} a^{X'_3}_{\ell_3'm_3'}  + \mathrm{cyclic} \right] \right\} \; , 
\end{align}
where $N$ is a suitable normalization chosen to produce unit response to $ b_{\ell_1 \ell_2 \ell_3}^{\rm th}$. 
Note that we are implicitly defining a suitable normalization convention so that 
$b_{\ell_1 \ell_2 \ell_3} = \fnl b_{\ell_1 \ell_2 \ell_3}^{\rm th}$, and $b_{\ell_1 \ell_2 \ell_3}^{\rm th}$ is the value of the theoretical template when $\fnl=1$.
 $\tens{C}^{-1}$ is the inverse of the block matrix:
\begin{equation}
\tens{C}=\left(
\begin{array}{c|c}
\tens{C}^{TT} & \tens{C}^{TE} \\
\hline
\tens{C}^{ET} & \tens{C}^{EE}
\end{array}\right) \, ,
\end{equation}
and the blocks represent the full \textit{TT}, \textit{TE}, and \textit{EE} covariance matrices, with $\tens{C}^{ET}$ being the transpose of $\tens{C}^{TE}$.
All quantities in the previous equation (i.e., CMB multipoles, bispectrum template and covariances matrices) are assumed to properly incorporate instrumental beam and noise.

As standard for these estimators, we note in square brackets (below) the presence of two contributions. One is cubic in the observed $\alm$s, and correlates the bispectrum of the data to the theoretical fitting template $b_{\ell_1 \ell_2 \ell_3}^{\rm th}$.
This is generally called the ``cubic term" of the estimator. The other contribution is linear in the observed $\alm$s (``linear term"). This part corrects for mean-field contributions to the error bars, introduced by rotational invariance-breaking features, such as a mask or anisotropic/correlated instrumental noise \citep{2006JCAP...05..004C,Yadav:2007ny}. 
 
The inverse covariance filtering operation implied by Eq.~\eqref{eq:optimalestimator} is a challenging numerical task, which
has been successfully performed only recently \citep{2009JCAP...09..006S,2012arXiv1211.0585E}.
This step can be avoided by working in the ``diagonal covariance approximation". In this approach, the estimator is built by neglecting off-diagonal entries of the covariance matrix in the cubic term in Eq.~\eqref{eq:optimalestimator}, 
and then finding the linear term that minimizes the variance for this specific cubic statistic. Applying such a procedure yields \citep{Yadav:2007rk}
\begin{flalign}\label{eq:diagcovestimator}
\hat{f}_{\textrm{NL}}  =  & \frac{1}{N} \sum_{X_i, X'_i} \sum_{\ell_i,m_i} \curl{G}^{\,\,\ell_1\; \ell_2\; \ell_3}_{m_1 m_2 m_3 } (\tens{C}^{-1})_{\ell_1}^{X_1 X'_1} (\tens{C}^{-1})_{\ell_2} ^{X_2 X'_2}  (\tens{C}^{-1})_{\ell_3}^{X_3 X'_3} b^{X_1 X_2 X_3, \, \rm th}_{\ell_1 \ell_2 \ell_3}
\;\;\;\; \nonumber
\\ & \left[ a^{X'_1}_{\ell_1 m_1}\, a^{X'_2}_{\ell_2 m_2}\,a^{X'_3}_{\ell_3 m_3} - \tens{C}_{\ell_1 m_1,\ell_2 m_2}^{X'_1 X'_2} a^{X'_3}_{\ell_3 m_3}  - \tens{C}_{\ell_1 m_1,\ell_3 m_3}^{X'_1 X'_3} a^{X'_2}_{\ell_2 m_2} \right. \nonumber
\\ & \left. - \tens{C}_{\ell_2 m_2,\ell_3 m_3}^{X'_2 X'_3} a^{X'_1}_{\ell_1 m_1} \right] \, ,
\end{flalign}
where $\tens{C}^{-1}_\ell$ is the inverse of the $2\times2$ matrix
\begin{equation}
\tens{C}_\ell=\left(
\begin{array}{cc}
\tens{C}^{TT}_\ell & \tens{C}_\ell^{TE} \\
\tens{C}_\ell^{ET} & \tens{C}_\ell^{EE}
\end{array}\right) \, .
\end{equation}
This expression can also be written as
\begin{equation}\label{eq:fnlestimator_alt}
\hat{f}_{\textrm{NL}}  = 
\frac{\langle b^{\rm th}, b^{\rm obs} \rangle}{\langle b^{\rm th}, b^{\rm th} \rangle},
\end{equation}
where the observed (reduced) bispectrum includes the linear correction term
and the inner product is defined as
\begin{flalign}\label{eq:bispec_innerprod}
& \langle b^A, b^B \rangle = \\
& \sum_{X_i, X'_i} \sum_{\ell_i} 
b^{X_1 X_2 X_3,\, A}_{\ell_1 \ell_2 \ell_3} h^2_{\ell_1 \ell_2 \ell_3}
(\tens{C}^{-1})_{\ell_1}^{X_1 X'_1} (\tens{C}^{-1})_{\ell_2} ^{X_2 X'_2}  (\tens{C}^{-1})_{\ell_3}^{X_3 X'_3}
b^{X'_1 X'_2 X'_3, \, B}_{\ell_1 \ell_2 \ell_3} \nonumber
\end{flalign}
with 
\begin{equation}\label{eq:numtriangles}
h_{\ell_1 \ell_2 \ell_3} = \sqrt{\frac{(2\ell_1+1)(2\ell_2+1)(2\ell_3+1)}{4\pi}}
\begin{pmatrix}
\ell _1 &\ell _2 &\ell _3\cr
0&0 &0\cr
\end{pmatrix},
\end{equation}
where the last term is the Wigner 3j symbol. The denominator in Eq. (\ref{eq:fnlestimator_alt}), $\langle b^{\rm th}, b^{\rm th} \rangle$ is the normalization
constant $N$.

The price to pay for the simplification obtained in Eq.~\eqref{eq:diagcovestimator} is, in principle, loss of optimality. However, in practice we found in our previous temperature analysis \citep{planck2013-p09a} that error bars obtained 
with this simplified procedure are very close to optimal, provided the $\alm$s are pre-filtered with a simple diffusive inpainting technique (see \citealt{planck2013-p09a} for details). We find that this still holds true when we include polarization
 and pre-inpaint the \itT,\itQ,\itU\ input maps. Given its practical advantages in terms of speed and simplicity, we adopt this method in the following analysis. 

A well-known, major issue with both Eq.~\eqref{eq:optimalestimator} and \eqref{eq:diagcovestimator} is that their direct implementation would require evaluation of all the bispectrum configurations from the data. The computational cost 
of this would scale like $\ell_{\rm max}^5$ and be totally prohibitive for  high-resolution CMB experiments like \Planck. The different bispectrum estimation techniques applied to our analysis are essentially defined by the 
approach adopted to circumvent this problem. The advantage of having multiple independent implementations of the optimal bispectrum estimator is twofold. First, by cross-validating and comparing 
 outputs of different pipelines, it strongly improves the robustness of the results. Second, different methods are complementary, in the sense that they have specific capabilities which go beyond simple $\fnl$ estimation. For example, 
the skew-$C_\ell$ method defined below facilitates the monitoring of NG foreground contamination, while the binned and modal estimators allow model-independent reconstruction of the data bispectrum, and so on. The skew-$C_\ell$ method enables the nature of any detected NG to be determined. Thus, the simultaneous 
application of all these techniques also allows us to increase the range and scope of our analysis.

In the following, we briefly outline the main features of the three optimal bispectrum estimation pipelines that are used for \Planck\ measurements of $\fnl$.  We will only provide a short summary here, focused on the extension to polarization data, referring the reader who is 
interested in more technical aspects to our previous analysis of temperature data \citep{planck2013-p09a}.

\subsection{KSW and skew-$C_\ell$ estimators} 
\label{sec:KSW_est}

KSW and skew-$C_\ell$ estimators \citep{Komatsu:2003iq,2010MNRAS.401.2406M} can be 
used for bispectrum templates that are written in {\em factorizable} 
(separable) form, i.e., as a linear combination of separate products of 
functions\footnote{We note that the local, equilateral, and orthogonal 
templates of Sect.~\ref{sec:models} are separable. In fact, while 
the theoretical local NG models are manifestly separable, the 
equilateral and orthogonal templates of Eqs.~\eqref{equilateralBis} 
and \eqref{orthogonalBis} are factorizable approximations of the 
original non-separable shapes, that were derived exactly with the 
purpose of allowing the application of  this type of estimator 
\citep{2006JCAP...05..004C,2010JCAP...01..028S}.} of $k_1$, $k_2$, and 
$k_3$. This allows reduction of the three-dimensional integration over the bispectrum configurations into a product of three separate one-dimensional sums over $\ell_1$, $\ell_2$, $\ell_3$. This leads to a massive reduction in computational time ($O(N_{\rm pix})$, where $N_{\rm pix}$ is the number of pixels in the map). The main difference between the KSW and skew-$C_\ell$ pipelines is 
that the former estimates the $\fnl$ amplitude directly, whereas the latter initially estimates the so called ``bispectrum-related power spectrum'' (in short, ``skew-$C_\ell$") function. Roughly speaking, the skew-$C_\ell$ associates, with each angular wavenumber $\ell$, the contribution to the amplitude $\fnl$ (for each given shape) extracted from all triangles with one fixed side of size $\ell$. After resumming over the contributions from each $\ell$-bin, the final point-like $\fnl$ estimate is obtained exactly as KSW. Equipping the KSW estimator with a skew-$C_\ell$ extension can be particularly useful in the presence of (expected) spurious NG contaminants in the data. The slope of the skew-$C_\ell$ statistic is in fact 
shape-dependent and can be used to separate multiple NG components in the map.

\subsection{Modal estimators}  
\label{sec:modal_est}

Modal estimators \citep{2010PhRvD..82b3502F,2012JCAP...12..032F} are based on decomposing the bispectrum 
(both from theory and from data) into a sum of uncorrelated separable templates, forming a {\em complete basis} 
in bispectrum space, and measuring the amplitude of each. The evaluation of the amplitude for each 
template can be sped up by using a KSW approach (since the templates themselves are separable by construction). 
All amplitudes form a vector, also referred to as the ``mode spectrum". 
It is then possible to measure the correlation of the observed data mode spectrum with the theoretical mode spectra 
for different primordial shapes, in order to obtain estimates of the primordial $\fnl$. Note also that 
the observed mode spectrum from data is theory-independent, and contains {\em all} the information from the data. 
Correlating the observed mode spectrum to theoretical mode vectors then allows the extraction of all the $\fnl$ amplitudes simultaneously. 
This makes modal estimators naturally suited for NG analyses, both when there are a large number of competing 
models to analyse, or when a model has free parameters through which we wish to scan 
(more than 500 shapes were analysed when applying this technique to \Planck\ data). Another advantage is that 
by expanding into separable basis templates, the modal estimator does not require separability of the 
starting theoretical shape in order to be applicable. Finally, after obtaining the data mode spectrum, 
it is possible to build a linear combination of the basis templates, using the measured amplitudes as coefficients, 
thus obtaining a model-independent full reconstruction of the bispectrum of the data. Of course the 
reconstructed bispectrum will be smoothed, as the estimator must use a finite number of basis templates. 

For this analysis, the modal method is implemented in two ways. One of them generalizes our previous temperature 
modal pipeline by expanding,  
for each shape, the corresponding \textit{TTT}, \textit{EEE}, \textit{TTE} and \textit{EET} bispectra. 
We then exploit separability to build the covariance matrix of these expanded bispectra \citep{Liguori:2014}, 
and to measure $\fnl$ efficiently using Eq.~\eqref{eq:diagcovestimator}. This modal 
pipeline will be referred to throughout the paper as the ``Modal 1'' pipeline. 

The other implementation, which we will refer to as ``Modal 2'',  utilizes a novel approach where the $\alm^T$ and $\alm^E$ are first orthogonalized to produce new uncorrelated unit variance $\hat{a}_{\ell m}$ coefficients, 
\begin{align}\label{eq:orthoalm} 
\hat{a}^T_{\ell m} &= \frac{\alm^T }{\sqrt{C^{TT}_\ell}} \\ 
\hat{a}^E_{\ell m} &= \frac{C^{TT}_\ell \alm^E - C^{TE}_\ell \alm^T}{\sqrt{C^{TT}_\ell}\sqrt{C^{TT}_\ell C^{EE}_\ell - {C^{TE}_\ell}^2}}\,. 
\end{align} 
We then decompose the new bispectra as 
\begin{align}\label{eq:orthoalm} 
\hat{b}^{X_1 X_2 X_3}_{\ell_1 \ell_2 \ell_3} = \sum_{m_i} 
\begin{pmatrix} 
\ell _1 &\ell _2 &\ell _3\cr m_1& m_2 &m_3 \cr 
\end{pmatrix}\, 
\left< \hat{a}^{X_1}_{\ell_1 m_1} \hat{a}^{X_2}_{\ell_2 m_2} \hat{a}^{X_3}_{\ell_3 m_3}\right>\,, 
\end{align} 
which can be constrained independently, since they are uncorrelated. In this case the estimator then takes on a particularly simple 
form \citep{2014arXiv1403.7949F}.  This new form is mathematically equivalent to the previous modal method, but  involves significantly 
fewer terms in the estimator. However, due to the orthogonalization procedure we cannot constrain the full \textit{EEE} bispectrum without further processing, just the additional part which is orthogonal to temperature.  For this reason, although the ``Modal 2'' \textit {T+E} results incorporate all the polarization information,  the \textit{EEE} results alone are not presented here. 

In our analysis, both 
modal techniques (together with all the other estimators described in this section) were used 
to measure $\fnl$
 for the three main shapes i.e., local, equilateral, and orthogonal. Besides this, we optimized the two pipelines 
for different purposes. The ``Modal 1'' estimator was adopted to perform a large number of robustness tests of our results, 
especially in relation to the local, equilateral and orthogonal measurements. The ``Modal 2'' pipeline 
was instead mostly used to study a large number of ``non-standard'' primordial shapes (e.g., oscillatory bispectra). 
 For this reason, each pipeline uses a different set of basis templates. The ``Modal 1'' estimator starts
from a polynomial basis with 600 modes, and includes nine more modes that are the contributions from last scattering 
to the exact radial KSW expansion of the local, equilateral and orthogonal templates. The ``Modal 2'' expansion uses a 
high-resolution basis with 2000 polynomial modes, augmented with a Sachs-Wolfe local bispectrum template, in order 
to improve convergence efficiency in the squeezed limit. In this way, the  high resolution estimator provides the ability to scan 
across a wide variety of non-separable and oscillatory shapes, while the lower resolution pipeline gives efficient convergence in the $\fnl$ measurements 
for the standard local, equilateral, and orthogonal shapes, offering rapid analysis for validation purposes.  The ``Modal 1'' pipeline can 
also be generalized for the estimation of parity-odd bispectra, which is included in our analysis of non-standard shapes.  

\subsection{Binned bispectrum estimator}
\label{sec:binned_est} 

One can also use a binned estimator \citep{Bucher:2009nm, Bucher:2015ura}, 
exploiting the fact that the theoretical
bispectra of interest are generally smooth functions in
$\ell$-space. As a result, data and templates can be binned in $\ell$
with minimal loss of information, but with large computational gains
from data compression. The data bispectrum in the binning grid is
then computed and compared to the binned primordial shapes to obtain
$\fnl$. No KSW-like approach, which requires separability and mixing
of theoretical and observational bispectra in the computation, is
required.  Instead, the binned data bispectrum and the binned
theoretical bispectrum and covariance are computed and stored
completely independently, and only combined at the very last stage in
a sum over the bins to obtain $\fnl$. This means that it is very easy
to test additional shapes or different cosmologies, and the data
bispectrum can also be studied on its own in a non-parametric
approach.  In particular the smoothed binned bispectrum approach, also
used in this paper, investigates the (smoothed) binned bispectrum of
the map divided by its expected standard deviation, to test if there is
a significant bispectral NG of any type in the
map. Another advantage of the binned bispectrum estimator is that the
dependence of $\fnl$ on $\ell$ can be investigated for free, simply by
leaving out bins from the final sum.

In more detail, the computation for the binned bispectrum estimator is
based on Eqs.~\eqref{eq:fnlestimator_alt} and \eqref{eq:bispec_innerprod}.
However, instead of using the reduced bispectrum 
$b_{\ell_1 \ell_2 \ell_3}^{X_1 X_2 X_3}$,
all expressions start from the alternative rotationally-invariant reduced
bispectrum $B_{\ell_1 \ell_2 \ell_3}^{X_1 X_2 X_3} = h^2_{\ell_1 \ell_2 \ell_3}
b_{\ell_1 \ell_2 \ell_3}^{X_1 X_2 X_3}$, where $h$ is defined in Eq.~\eqref{eq:numtriangles}.
The expression in Eq.~\eqref{eq:bispec_innerprod}
for the inner product remains the same when replacing $b$ by $B$, except
that the $h^2$ becomes $h^{-2}$. The importance of 
$B_{\ell_1 \ell_2 \ell_3}^{X_1 X_2 X_3}$ is that it can be determined directly 
from maximally-filtered maps;
\begin{equation}\label{eq:B_from_maps}
B_{\ell_1 \ell_2 \ell_3}^{X_1 X_2 X_3} = \int d^2 \vnhat \, M_{\ell_1}^{X_1}(\vnhat)
M_{\ell_2}^{X_2}(\vnhat) M_{\ell_3}^{X_3}(\vnhat),
\end{equation}
where
\begin{equation}\label{eq:filtered_map}
M_{\ell}^{X}(\vnhat) = \sum_m a_{\ell m}^X Y_{\ell m}(\vnhat).
\end{equation}
Binning is then implemented by adding a sum over all $\ell$
inside a bin to the expression for the filtered map given in Eq.~\eqref{eq:filtered_map},
thus obtaining the observed binned bispectrum of the map
$B_{i_1 i_2 i_3}^{X_1 X_2 X_3,\,\mathrm{obs}}$, with bin indices $i_1, i_2, i_3$.
The linear correction term is obtained in a similar way (and subtracted from
the cubic term), but with two of the maps in Eq.~\eqref{eq:B_from_maps} replaced 
by Gaussian simulations, and taking the average over a large number of those.

The theoretical templates are binned simply by summing the exact expression 
over the $\ell$s inside a bin,\footnote{We note that the enormous computational 
gain of the binned bispectrum estimator comes from the binned determination of 
the observed bispectrum; determining the theoretical bispectrum templates is 
fast, even when done exactly.} and the same is true for the covariance matrix 
$h^2_{\ell_1 \ell_2 \ell_3} C_{\ell_1}^{X_1 X'_1} C_{\ell_2}^{X_2 X'_2} C_{\ell_3}^{X_3 X'_3}$.
The binning is optimized in such a way as to maximize the overlap, 
defined using the inner product of Eq.~\eqref{eq:bispec_innerprod} between
the binned and the exact template for all shapes under consideration.
Finally the estimate of $f_\mathrm{NL}$ is computed using 
Eq.~\eqref{eq:fnlestimator_alt}, where the inner product now contains a sum over
bin indices $i$ instead of multipoles $\ell$, and the bispectra and 
covariance matrix are replaced by their binned versions.

\subsection{Data set and simulations}
\label{sec:dataset}

In the following, we will apply our bispectrum estimation pipelines
 both to simulations and data, and consider a large number of 
shapes, either primordial or non-primordial in origin. Simulations will 
 be used for a wide range of purposes, from comparisons of the outcomes of 
different estimators, to tests of instrumental 
systematics and foreground contamination, as well as Monte Carlo evaluation 
of error bars. For this reason, many different sets of simulated maps 
will be used, with features that will vary, depending on the specific 
application, and will be described case by case throughout the paper. 
Most of the time, however, we will use the FFP8 simulation data set 
described in \cite{planck2014-a14}, or 
mock data sets obtained by processing FFP8 maps in various ways. 
These are the most realistic 
simulations available, modelling the CMB sky and the
instrumental effects of \Planck\ to the best of our current knowledge. They have passed
through the same steps of the component separation pipelines as the real
sky map and are the same maps as used for the final validation of the 
estimators in Sect.~\ref{Sec_valid_est_3}.
 
As far as actual data are concerned~\citep{planck2014-a01,planck2014-a03,planck2014-a04,planck2014-a05,planck2014-a06,planck2014-a07,planck2014-a08,planck2014-a09} the maps analysed in this work  
are the \Planck\ 2015 sky map, both in temperature
and in \itE\ polarization, as cleaned with the four component separation methods 
\SMICA, \SEVEM, \NILC, and \Commander~\citep{planck2014-a11}. As explained in 
\cite{planck2014-a08}, the polarization map has had a high-pass filter
applied to it, since the characterization of systematics and foregrounds in
low-$\ell$ polarization is not yet satisfactory. This filter removes the 
scales below $\ell=20$ completely, and those
between $\ell=20$ and $\ell=40$ partially. In all our analyses we use
$\ell_\mathrm{min}=40$ for polarization, in order to be independent of the
details of this filter. For temperature, we use 
$\ell_\mathrm{min}=2$. All the final cleaned maps are smoothed with a 
$5\arcmin$ Gaussian beam in temperature, and a $10\arcmin$ Gaussian beam in polarization.

The maps are masked to remove the brightest parts of the Galaxy as well as 
significant point sources. The masks used are the common masks of the \Planck\ 
 2015 release in temperature and polarization, which are the union of the confidence 
masks for the four component separation methods\footnote{We note that the \Planck\ collaboration 
produced two slightly different sets of union masks (see \citealt{planck2014-a11} for details). We
choose to adopt 
the more conservative set in this paper, as we found that the agreement between different component 
separation methods significantly increases with these masks when we measure $\fnl$ of shapes 
that peak in the squeezed limit (while the differences are very small in other cases).}
 \citep{planck2014-a11}. The sky coverages are respectively $f_{\rm sky} = 0.76$ in temperature and 
$f_{\rm sky} = 0.74$ in polarization. The stability
of our results as a function of the mask is investigated in 
Sect.~\ref{sec:dep_mask}, where we show that our temperature and joint 
temperature plus polarization results do not change significantly 
 when we consider a larger sky coverage. 

In Sects. \ref{sec:dep_methods} and \ref{sec:FFP8tests} we also compare 
the performance of different component separation methods, and conclude that, 
with respect to bispectrum estimation, the most accurate results are 
obtained using the $\SMICA$ data set. As already done for the 2013 release, 
we will thus consider $\SMICA$ as our main data set, using the other methods 
for important cross-checking purposes.   

If we consider only temperature, current $\SMICA$ data become noise dominated at 
$\ell \simeq 2000$, while previous nominal mission data were noise dominated 
at $\ell \simeq 1700$. The mask used for the 2013 
release was also slightly larger than the current one ($f_{\rm sky} = 0.73$ in 
2013 vs. $f_{\rm sky} = 0.76$ in 2015). Since the $\fnl$ signal-to-noise ratio, as quantified 
by the Fisher Information Matrix, scales as (S/N) $ \propto \ell \sqrt{f_\mathrm{sky}}$ in 
the signal dominated regime and saturates in the noise-dominated regime, we expect 
an improvement in our $\fnl$ temperature constraints of about $20\, \%$ when going from 
the 2013 nominal mission release to the current results. Adding polarization  
and accounting for all possible \textit{TTE}, \textit{EET}, and \textit{EEE} bispectra produces  
further improvements. Since we are neglecting the first $40$ polarization multipoles, such improvements 
are expected to be fairly small for shapes peaking in the squeezed limit, and more 
pronounced for equilateral type bispectra. A Fisher matrix approach shows 
that error bars are expected to improve by about $10\, \%$ for the local shape and 
about $40\, \%$ for the equilateral shape. This is in good agreement with our actual measurements, 
as can be seen from the results presented in Sect.~\ref{sec:Results} onwards.

\subsection{Data analysis settings}\label{sec:settings}

Now we detail the general setup adopted for the analysis 
of \Planck\ 2015 data by the four different optimal bispectrum 
estimation pipelines, described in previous sections. 

As already explained, inpainting of the masked 
regions of the sky is a preliminary data filtering operation that all 
pipelines must perform, in order to retain optimality.  We found that the inpainting method used in 2013 \citep{planck2013-p09a} for temperature
maps still works well when polarization is included (note that it is the 
original \itT, \itQ, and \itU\ maps that should be inpainted, not the derived 
\itE\ map). We adopt a simple diffusive inpainting method. First the
masked regions of the map are filled with the average value of the rest of
the map. Then the value of each masked pixel is replaced by the average value
of its (generally eight) direct neighbour pixels. The latter step is repeated
a fixed number of times (2000).\footnote{For bispectrum purposes we found 
no difference between the results
when performing the procedure without a buffer (the so-called ``Gauss-Seidel" 
method, where amongst the neighbours will be pixels both at the current and 
at the previous iteration) and with a buffer (the so-called ``Jacobi" method, 
where all neighbour pixels will be at the previous iteration), except that
the former converges faster. We found 2000 iterations to work well in the
`Gauss-Seidel' case.} Relevant, final computations in 
map space (see e.g., Eq.~\ref{eq:B_from_maps}) are always done after remasking,
so that the inpainted regions of the map are not used directly.
The relevance of the inpainting procedure is that it reduces the effect of 
the sharp edges and the lack of large-scale power inside the mask leaking 
into the rest of the map during harmonic transforms. 
 
For the linear correction term and to determine error bars, we use
the FFP8 simulations (see \citealt{planck2014-a14}, and Sect.~\ref{sec:dataset}), 
 filtered through the different component separation pipelines, using the same weights  
as used for the actual data when co-adding frequency channels.
To compute all theoretical 
quantities (like the bispectrum templates and
the ISW-lensing bias) we use the \Planck\ 2015 best-fit cosmological 
parameters as our fiducial cosmology. However, results are quite insensitive 
to small changes in these parameters.

As pointed out in Sect.~\ref{sec:dataset}, low-$\ell$ multipoles are filtered out 
of the input polarization data set, so that all of our analyses will 
use $\ell_{\rm min}=40$ in polarization, and $\ell_{\rm min}=2$ in temperature.
The choice of $\ell_{\rm max}$ is dictated by the angular resolution of the cleaned maps, 
which is $5\arcmin$ in temperature, and $10\arcmin$ in polarization, and by the fact that the 
 temperature data become noise-dominated at $\ell \simeq 2000$, while the polarization information 
saturates around $\ell \simeq 1000$. 
The KSW and binned estimators use $\ell_\mathrm{max}=2500$ for temperature,
while the modal estimators use $\ell_\mathrm{max}=2000$. As shown 
explicitly in Sect.~\ref{sec:ldep}, results are completely stable between
$\ell=2000$ and $\ell=2500$, so that this has no impact on $\fnl$.
Similarly the binned estimator uses $\ell_\mathrm{max}=2000$ for polarization,
while the other estimators use $\ell_\mathrm{max}=1500$, but again
Sect.~\ref{sec:ldep} shows that this difference is unimportant. The estimators
also differ in the number of maps used to compute the linear correction term
and the error bars, but generally it is of the order of 200. This difference
is due to the different convergence properties of the estimators, some
converging faster than others.

The binned bispectrum estimator uses 57 bins\footnote{The boundary values 
of the bins are: 2, 4, 10, 18, 30, 40, 53, 71, 99, 126, 154, 211, 243, 281, 
309, 343, 378, 420, 445, 476, 518, 549, 591, 619, 659, 700, 742, 771, 800, 
849, 899, 931, 966, 1001, 1035, 1092, 1150, 1184, 1230, 1257, 1291, 1346, 
1400, 1460, 1501, 1520, 1540, 1575, 1610, 1665, 1725, 1795, 1846, 1897, 2001, 
2091, 2240, and 2500 (i.e., the first bin is [2, 3], the second [4, 9], etc., 
while the last one is [2240, 2500]).} for the analysis, which were 
determined by optimizing the correlation between the exact and the binned
templates for the different shapes in temperature and polarization, as well as 
the full combined case. This is equivalent to minimizing the variance
of the different $\fnl$ parameters, where we focus in particular on the 
primordial shapes. 

As previously explained, we use two different versions of the polarized modal pipelines, called 
``Modal 1'' and ``Modal 2'' in the paper. Besides technical and conceptual 
 implementation differences, the two modal estimators also use different 
 sets of basis templates. The ``Modal 1'' pipeline uses $600$ polynomial 
modes, plus 
 nine ``KSW radial modes'', computed at last scattering, while ``Modal 2'' 
has a basis formed by $2000$ polynomial modes, augmented with a Sachs-Wolfe 
local bispectrum template. 
Due to the way polarization is implemented in the ``Modal 2'' pipeline,
it cannot determine results for \itE-only.
More details and explanations of the different 
choices are provided in Sect.~\ref{sec:modal_est}.

As already stressed, the use of several independent bispectrum estimators,  
and several completely independent component separation methods
allows a remarkable level
of cross-validation of our results in order to establish their robustness.
The fact that the bispectrum estimators are statistically equivalent
and produce practically optimal results will be established in 
Sect.~\ref{sec:Validation}. The validation of the component separation methods
is described in \cite{planck2014-a11} and Sect~\ref{sec:Sec_valid_data}.

%% file: A19_Section4.tex
Here we investigate several bispectra of non-primordial origin that are expected to 
be present in the data, and quantify their impact on our $\fnl$ results. We devote 
particular attention to assessing potential biases that these NG signals might induce 
on the primordial bispectra. When forecasting such biases, we assume 
 the data analysis settings discussed in Sects.~\ref{sec:dataset} and~\ref{sec:settings}.

\subsection{Non-Gaussianity from the lensing-ISW bispectrum}\label{subsec:lensingISW} 
The correlation between the gravitational lensing of the CMB anisotropies and the integrated Sachs-Wolfe (ISW) effect gives rise to a secondary 
CMB bispectrum --- characterized by an oscillatory behaviour and peaked on squeezed configurations --- 
that is a well-known contaminant to the primordial 
NG signal \citep{Hanson:2009gu,2009PhRvD..80l3007M,2011JCAP...03..018L,2013arXiv1303.1722M}. The temperature-only 2013 \Planck\ results \citep{planck2013-p09a,planck2013-p14,planck2013-p12} showed evidence for the first time for the lensing-ISW CMB bispectrum and associated bias. Based on the same methodology used for the 2013 \Planck\ data analysis \citep{planck2013-p09a}, here we update the computation of the lensing-ISW bispectrum and its bias to include the full mission temperature and polarization data. 

As shown by 
\cite{Cooray2006}, the direct lensing-ISW correlation in \itE-polarization
due to rescattering of the temperature quadrupole generated by the
ISW effect is negligible. However, as explained in \cite{2011JCAP...03..018L},
there is an important correlation between the lensing potential and
the large-scale \itE-polarization generated by scattering at reionization.
Because the lensing potential is highly correlated with the ISW signal,
this also leads to a non-zero lensing-ISW bispectrum
in polarization.

To determine $f_\mathrm{NL}^\mathrm{LISW}$, the amplitude parameter of the 
lensing-ISW bispectrum, one simply inserts the theoretical template for this
shape into the general expression of Eq.~\eqref{eq:fnlestimator_alt}. The template
is given by \citep{2000PhRvD..62d3007H,2011JCAP...03..018L}
\begin{eqnarray}
b_{\ell_1 \ell_2 \ell_3}^{X_1 X_2 X_3,\,\mathrm{LISW}} \!\!\! & = &
C_{\ell_2}^{X_2\phi} \tilde{C}_{\ell_3}^{X_1X_3} f_{\ell_1 \ell_2 \ell_3}^{X_1}
+ C_{\ell_3}^{X_3\phi} \tilde{C}_{\ell_2}^{X_1X_2} f_{\ell_1 \ell_3 \ell_2}^{X_1} \nonumber\\
&& + C_{\ell_1}^{X_1\phi} \tilde{C}_{\ell_3}^{X_2X_3} f_{\ell_2 \ell_1 \ell_3}^{X_2}
+ C_{\ell_3}^{X_3\phi} \tilde{C}_{\ell_1}^{X_1X_2} f_{\ell_2 \ell_3 \ell_1}^{X_2} \nonumber\\
&& + C_{\ell_1}^{X_1\phi} \tilde{C}_{\ell_2}^{X_2X_3} f_{\ell_3 \ell_1 \ell_2}^{X_3}
+ C_{\ell_2}^{X_2\phi} \tilde{C}_{\ell_1}^{X_1X_3} f_{\ell_3 \ell_2 \ell_1}^{X_3}.
\label{LISW_redbisp_pol}
\end{eqnarray}
Here $C_\ell^{T\phi}$ and $C_\ell^{E\phi}$ are the temperature/polarization-lensing 
potential cross power spectra, and the tilde on $\tilde{C}_\ell^{TT}$,
$\tilde{C}_\ell^{TE}$, and $\tilde{C}_\ell^{EE}$ indicates that it is the lensed
$TT$, $TE$, or $EE$ power spectrum. The functions $f_{\ell_1 \ell_2 \ell_3}^{T,E}$ are defined
by
\begin{eqnarray}
f_{\ell_1 \ell_2 \ell_3}^T & = & 
\frac{1}{2} \left[ \ell_2 (\ell_2 + 1) + \ell_3 (\ell_3 + 1) - \ell_1 (\ell_1+1)
\right ], \nonumber\\
f_{\ell_1 \ell_2 \ell_3}^E & = & 
\frac{1}{2} \left[ \ell_2 (\ell_2 + 1) + \ell_3 (\ell_3 + 1) - \ell_1 (\ell_1+1)
\right ] \nonumber\\
&& \times 
\left(\begin{array}{ccc} \ell_1 & \ell_2 & \ell_3 \\ 2 & 0 & -2 \end{array}\right)
\left(\begin{array}{ccc} \ell_1 & \ell_2 & \ell_3 \\ 0 & 0 & 0 \end{array}\right)^{-1},
\end{eqnarray}
if $\ell_1+\ell_2+\ell_3$ is even and $\ell_1,\ell_2,\ell_3$ satisfy the
triangle inequality, and zero otherwise.

In this paper our main concern with the lensing-ISW bispectrum is not so much
to determine its amplitude (although that is also of great interest),
but to compute its influence on the primordial shapes.
The bias $\Delta f_\mathrm{NL}^{\mathrm P}$ due to the lensing-ISW bispectrum on the 
estimation of a given primordial amplitude $f^{\mathrm P}_\mathrm{NL}$ is given by
\be
\Delta f_\mathrm{NL}^{\rm P} = \frac{\langle b^\mathrm{LISW}, b^{\mathrm P} \rangle}
{\langle b^{\mathrm P}, b^{\mathrm P} \rangle},
\label{eq:biaseq}
\ee
where the inner product is defined in Eq.~\eqref{eq:bispec_innerprod}.

\begin{table}             
\begingroup
\newdimen\tblskip \tblskip=5pt
\caption{Bias in the three primordial $f_{\rm NL}$ parameters due to the
lensing-ISW signal for the four component separation methods.}
\label{tab:lisw_bias}                            
\nointerlineskip
\vskip -6mm
\footnotesize
\setbox\tablebox=\vbox{
   \newdimen\digitwidth
   \setbox0=\hbox{\rm 0}
   \digitwidth=\wd0
   \catcode`*=\active
   \def*{\kern\digitwidth}
   \newdimen\signwidth
   \setbox0=\hbox{+}
   \signwidth=\wd0
   \catcode`!=\active
   \def!{\kern\signwidth}
   \newdimen\dotwidth
   \setbox0=\hbox{.}
   \dotwidth=\wd0
   \catcode`^=\active
   \def^{\kern\dotwidth}
\halign{\hbox to 1.1in{#\leaderfil}\tabskip 1em&
\hfil#\hfil&
\hfil#\hfil&
\hfil#\hfil&
\hfil#\hfil\tabskip 0pt\cr
\noalign{\doubleline\vskip 2pt}
\omit&\multispan4\hfil lensing-ISW $f_{\rm NL}$ bias\hfil\cr
\omit&\multispan4\hrulefill\cr
Shape\hfill&\SMICA&\SEVEM&\NILC&\Commander\cr
\noalign{\vskip 4pt\hrule\vskip 6pt}
\itT\ Local & !*7.5& !*7.5& !*7.3& !*7.0 \cr
\itT\ Equilateral & !*1.1& !*1.2& !*1.3& !*1.8 \cr
\itT\ Orthogonal & $-$27^*& $-$27^*& $-$26^*& $-$26^*\cr
\noalign{\vskip 4pt\hrule\vskip 6pt}
\itE\ Local & !*1.0& !*1.1& !*1.0& !*1.1 \cr
\itE\ Equilateral & !*2.6& !*2.7& !*2.5& !*2.9 \cr
\itE\ Orthogonal & !$-$1.3& !$-$1.3& !$-$1.2& !$-$1.5 \cr
\noalign{\vskip 4pt\hrule\vskip 6pt}
\itTpE\ Local & !*5.2& !*5.5& !*5.1& !*4.9\cr
\itTpE\ Equilateral & !*3.4& !*3.4& !*3.4& !*3.6\cr
\itTpE\ Orthogonal & $-$10^*& $-$11^*& $-$10^*&$-$10^* \cr
\noalign{\vskip 3pt\hrule\vskip 4pt}}}
\endPlancktable                    

\endgroup
\end{table}                        

\begin{table}               
\begingroup
\newdimen\tblskip \tblskip=5pt
\caption{Results for the amplitude of the lensing-ISW bispectrum from the
\SMICA, \SEVEM, \NILC, and \Commander\ foreground-cleaned maps,
for different bispectrum estimators. Error bars are 68\,\%~CL; see the main
text for how they have been determined.}
\label{tab:fNL_lisw}                            
\nointerlineskip
\vskip -6mm
\footnotesize
\setbox\tablebox=\vbox{
   \newdimen\digitwidth
   \setbox0=\hbox{\rm 0}
   \digitwidth=\wd0
   \catcode`*=\active
   \def*{\kern\digitwidth}
   \newdimen\signwidth
   \setbox0=\hbox{+}
   \signwidth=\wd0
   \catcode`!=\active
   \def!{\kern\signwidth}
\halign{\hbox to 0.65in{#\leaderfil}\tabskip 1em&
\hfil#\hfil\tabskip 0.7em&
\hfil#\hfil\tabskip 0.7em&
\hfil#\hfil\tabskip 0.7em&
\hfil#\hfil\tabskip 0pt\cr
\noalign{\doubleline\vskip 2pt}
\omit&\multispan4\hfil lensing-ISW amplitude\hfil\cr
\omit&\multispan4\hrulefill\cr
Method\hfill&\SMICA&\SEVEM&\NILC&\Commander\cr
\noalign{\vskip 4pt\hrule\vskip 6pt}
\omit\hfil \itT\hfil&&\cr
KSW & 0.79 $\pm$ 0.28 & 0.78 $\pm$ 0.28 & 0.78 $\pm$ 0.28 & 0.84 $\pm$ 0.28 \cr
Binned & 0.59 $\pm$ 0.33 & 0.60 $\pm$ 0.33 & 0.68 $\pm$ 0.33 & 0.65 $\pm$ 0.36 \cr
Modal2 & 0.72 $\pm$ 0.26 & 0.73 $\pm$ 0.26 & 0.73 $\pm$ 0.26 & 0.78 $\pm$ 0.27 \cr
\noalign{\vskip 4pt\hrule\vskip 6pt}
\omit\hfil \textit{T+E}\hfil&&\cr
Binned & 0.82 $\pm$ 0.27 & 0.75 $\pm$ 0.28 & 0.85 $\pm$ 0.26 & 0.84 $\pm$ 0.27 \cr
\noalign{\vskip 3pt\hrule\vskip 4pt}}}
\endPlancktable                    
\endgroup
\end{table}                        

The values for the bias are given in Table~\ref{tab:lisw_bias}.
It should be noted that these are the results as computed exactly
with Eq.~\eqref{eq:biaseq}. They can differ slightly from the ones
used in e.g.,\ Table~\ref{tab:fNLsmicah}, where each estimator adopts values
computed using the approximations appropriate to the method. However, these differences
are completely insignificant.
As seen already in \cite{planck2013-p09a}, for \itT-only the bias is very 
significant for local
and to a lesser extent for orthogonal NG. For local NG the bias is larger than the error bars on $\fnl$. We see that for \itE-only the 
effect is non-zero but not significant.
For the full \itT+\itE\ case, the bias is smaller than for \itT-only, but large
enough that it is important to take into account.

The results for $f_\mathrm{NL}^\mathrm{LISW}$ can be found in 
Table~\ref{tab:fNL_lisw}. The polarized version of the template has only
been implemented in the binned bispectrum estimator. 
Error bars have been determined based on FFP8 
simulations as usual.\footnote{The
average value of the lensing-ISW amplitude determined from the FFP8 simulations 
is around 0.85 of the expected value. This value is very consistent across bispectrum estimators 
and component separation methods, which provides a useful consistency test in 
its own right. Except for this effect, all other tests on the temperature 
FFP8 maps show them to be very robust and
to behave as expected, for example in the determination of the lensing-ISW 
bias on the local shape. We took this effect into account by
increasing all error bars in the table by the appropriate factor 
(i.e., dividing them by $\simeq 0.85$).}
The KSW estimator implements the lensing-ISW template exactly, while the
binned and modal estimators use approximations, as explained in 
Sect.~\ref{sec:SEP}. In particular for the binned estimator the correlation 
between the binned and exact lensing-ISW template is relatively low, since
it is a difficult template to bin (unlike all the other templates considered
in this paper), which is reflected in the larger error bars. Tests performed
on FFP8, as well as other tests, demonstrate that the lower correlation does
not lead to a bias compared to the other estimators. We will 
use the KSW results to draw our conclusions.

\begin{figure}
\begin{center}
\includegraphics[width=\hsize]{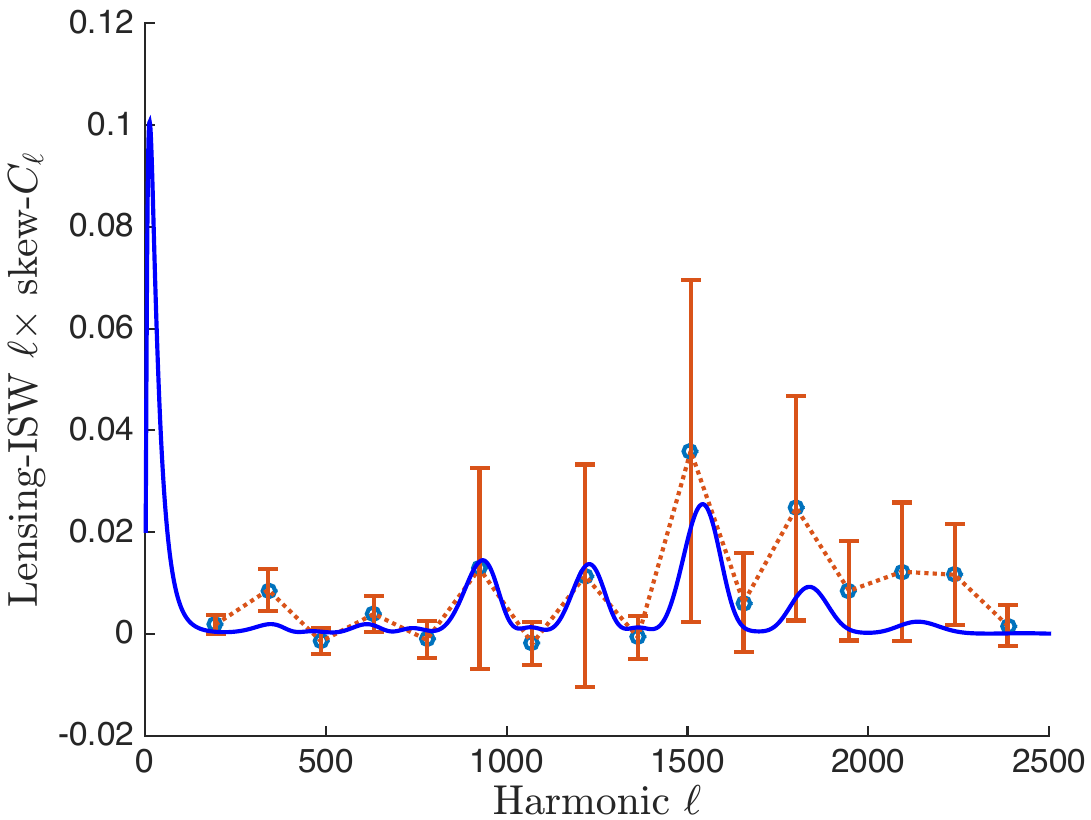}
\end{center}
\caption{The skew-$C_\ell$ spectrum for the lensing-ISW effect (red line with
data points), from the temperature map. The blue curve is the theoretically-expected spectrum.
Note that the points beyond $\ell=1500$ are significantly correlated.}
\label{fig:skewClLISW}
\end{figure}

We see that temperature results from the full mission
are consistent with the 2013 nominal mission \citep{planck2013-p09a}. 
Including polarization yields results that also appear consistent and decrease the error bars. 
However, for now the \textit{T+E} conclusions should be considered preliminary,
for the reasons related to polarization data discussed in detail in 
Sects.~\ref{sec:Results} and \ref{sec:Sec_valid_data}. The error bars will also
improve when measured with the other bispectrum estimators.
As already seen in 2013, the values for $f_\mathrm{NL}^\mathrm{LISW}$ are
slightly low compared to the expected value of 1, but not significantly so.
On the other hand, the detection of the lensing-ISW bispectrum is significant, 
even with our conservative rescaling of the error bars. The hypothesis of
having no lensing-ISW bispectrum is excluded at $2.8\,\sigma$ using temperature
alone, and improves to $3.0\,\sigma$ with the current preliminary result when
including polarization. As mentioned above, the latter result is likely to
improve with further analysis of the \Planck\ data.  
In Fig.~\ref{fig:skewClLISW} we present the results of 
the skew-$C_\ell$ analysis for lensing-ISW NG for the $T$ map, which illustrates that the instrument and data processing are not removing this expected NG signal from the data.

\subsection{Non-Gaussianity from extragalactic point sources}
\label{sec:point_sources}
The auto-bispectra of extragalactic point sources are a potential contaminant to primordial NG estimates at \Planck\ frequencies. The basic modelling and methodology of this section follows the corresponding section in \cite{planck2013-p09a}.

Extragalactic point sources are divided into populations of unclustered and clustered sources. The former are radio and late-type infrared galaxies \citep[see e.g.,][]{Toffolatti1998,Gonzalez-Nuevo2005}, while the latter are dusty star-forming galaxies constituting the cosmic infrared background \citep[CIB;][]{Lagache2005}.
The contamination due to both types of sources in NG estimators is handled via dedicated bispectrum templates which are fitted jointly with the primordial NG templates.

\begin{figure}
\begin{center}
\includegraphics[width=\hsize]{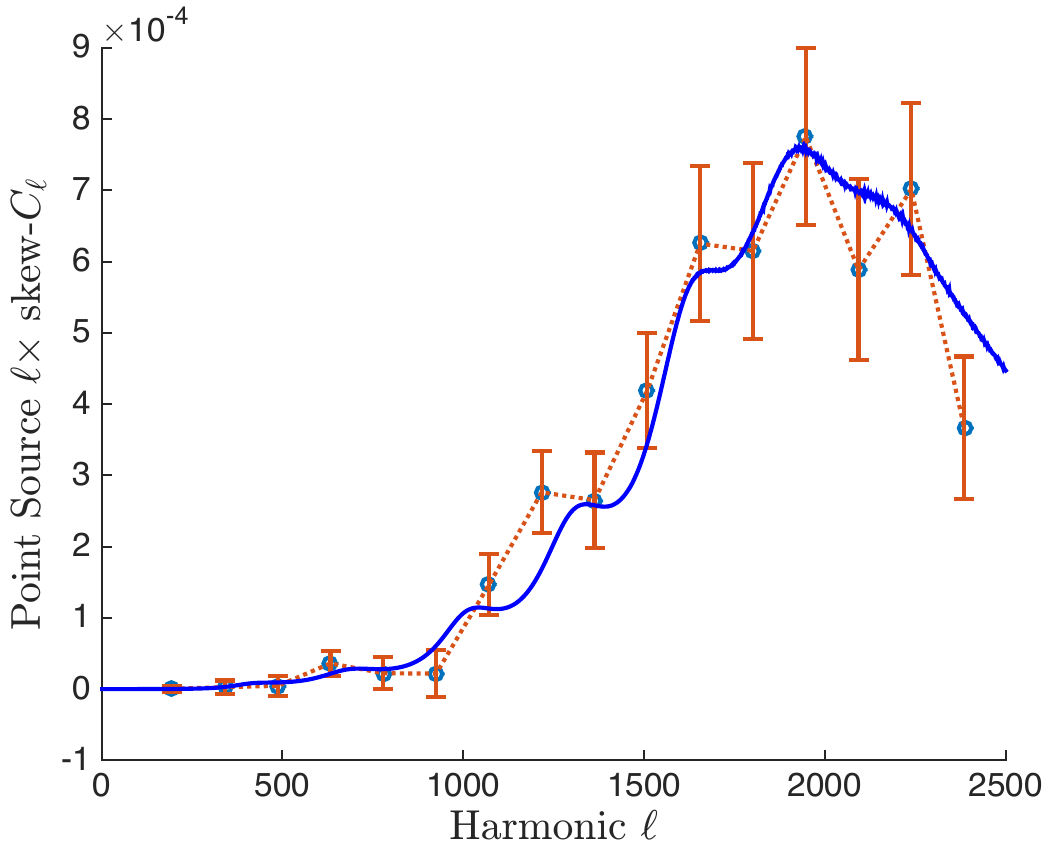}
\end{center}
\caption{The skew-$C_\ell$ spectrum for unclustered point sources (red line with data points), from the temperature map. The blue curve is the theoretical spectrum, given the amplitude determined with the KSW estimator.}
\label{fig:skewClPS}
\end{figure}

The unclustered sources have a white noise distribution, and hence
constant polyspectra. Their reduced angular bispectrum template is thus
\begin{equation}\label{Eq:unclust_template}
b_{\ell_1 \ell_2 \ell_3}^\mathrm{unclust} = \mathrm{constant.}
\end{equation}
This constant is usually noted $b_\mathrm{PS}$ or $b_\mathrm{src}$ in
the literature \citep[e.g.,][]{2001PhRvD..63f3002K}.
This constant template is valid in polarization as well as temperature, since the polarization angles of point sources are less clustered than the source density. However, since not all these point sources are polarized, we
do not measure the same sources in temperature and in polarization. In fact,
there is no detection of the bispectrum of unclustered point sources in the cleaned \Planck\ 
polarization map, unlike in the temperature map, where 
Table~\ref{Table:bps_and_ACIB} (binned bispectrum estimator) and 
Fig.~\ref{fig:skewClPS} (skew-$C_\ell$s) show a clear detection.

The clustered sources (CIB) have a more complex bispectrum in temperature, reflecting
the distribution of the large-scale structure and the clustering of galaxies in dark matter halos \citep{Argueso2003,Lacasa2012,Crawford2014}. The \Planck\ results have allowed the measurement of the CIB bispectrum at frequencies 217, 353, 545 GHz in the range $\ell\simeq 200-700$ \citep{planck2013-pip56}. In this multipole range, a power law was found to fit the measurement, with an exponent consistent between frequencies. However, at lower multipoles theoretical models for the CIB power spectrum
\citep[e.g.,][]{planck2013-pip56} and bispectrum \citep{Lacasa2014,Penin2014} predict a flattening of the CIB power. We thus take the {\it TTT} CIB bispectrum template to be a broken power law,
\begin{equation}\label{Eq:CIB_template}
b_{\ell_1 \ell_2 \ell_3}^\mathrm{CIB} \propto \left[ \frac{(1+\ell_1/\ell_\mathrm{break}) (1+\ell_2/\ell_\mathrm{break}) (1+\ell_3/\ell_\mathrm{break})}{(1+\ell_0/\ell_\mathrm{break})^3}\right]^q,
\end{equation}
where the index is $q=0.85$, the break is located at $\ell_\mathrm{break}=70$, and $\ell_0=320$  is the pivot scale for normalization.
Dusty star-forming galaxies emit with a low polarization fraction, and polarization correlates only over the smallest scales, so that the CIB is negligibly polarized. We thus take vanishing templates for its polarized bispectra
\begin{equation}
b_{\ell_1 \ell_2 \ell_3}^{\mathrm{CIB},TTE} = b_{\ell_1 \ell_2 \ell_3}^{\mathrm{CIB},TEE} = b_{\ell_1 \ell_2 \ell_3}^{\mathrm{CIB},EEE} = 0.
\end{equation}

Both point source templates, Eqs.~\eqref{Eq:unclust_template} and \eqref{Eq:CIB_template}, have been implemented in the binned bispectrum estimator described in Sect.~\ref{sec:SEP}. The results for these two templates applied to the \Planck\ temperature map, cleaned with the four component separation methods,
can be found in Table~\ref{Table:bps_and_ACIB}. Since the two templates are 
highly correlated, the results have been determined in a combined analysis. The results have also been determined jointly with the primordial local, equilateral, and orthogonal templates, and the lensing-ISW bias has been subtracted, but all of this makes a negligible difference.
Contamination from unclustered sources is detected in all component-separated maps. However, $A_\mathrm{CIB}$ is not detected.

\begin{table}               
\begingroup
\newdimen\tblskip \tblskip=5pt
\caption{Joint estimates of the bispectrum amplitudes of unclustered and 
clustered point sources in the cleaned \Planck\ temperature map, determined 
with the binned bispectrum estimator. The error bars have been determined 
using FFP8 simulations.}                          
\label{Table:bps_and_ACIB}                         
\nointerlineskip
\vskip -3mm
\footnotesize
\setbox\tablebox=\vbox{
   \newdimen\digitwidth
   \setbox0=\hbox{\rm 0}
   \digitwidth=\wd0
   \catcode`*=\active
   \def*{\kern\digitwidth}
   \newdimen\signwidth
   \setbox0=\hbox{+}
   \signwidth=\wd0
   \catcode`!=\active
   \def!{\kern\signwidth}
\halign{\hbox to 0.9in{#\leaderfil}\tabskip 2.2em&
\hfil#\hfil\tabskip 2.2em&
\hfil#\hfil\tabskip 0pt\cr
\noalign{\doubleline\vskip 2pt}
\hfill Map\hfill&$b_\mathrm{PS}/(10^{-29})$ & $A_\mathrm{CIB}/(10^{-27})$ \cr
\noalign{\vskip3pt\hrule\vskip 3pt}
\SMICA&5.6 $\pm$ 2.7&*0.4 $\pm$ 1.4\cr
\SEVEM&7.9 $\pm$ 2.8&*0.8 $\pm$ 1.4\cr
\NILC&9.3 $\pm$ 2.7&$-$0.3 $\pm$ 1.4\cr
\Commander &5.9 $\pm$ 3.2&*1.4 $\pm$ 1.6\cr
\noalign{\vskip 3pt\hrule}}}
\endPlancktable                    
\endgroup
\end{table}                        

The order of magnitude of the bispectrum amplitudes found in Table~\ref{Table:bps_and_ACIB} is consistent with expectations. Indeed, for radio sources at 217 GHz and with a flux cut based on the Planck ERCSC \citep{planck2011-1.10}, \cite{Lacasa2014b}, the forecasted $b_\mathrm{PS}$ is approximately $2 \times 10^{-28}$. For the CIB, the \Planck\ 2013 measurement \citep{planck2013-pip56} at 217~GHz gives $A_\mathrm{CIB} \simeq 6 \times 10^{-27}$ when translated into dimensionless units. The results reported in Table~\ref{Table:bps_and_ACIB} are consistent at the order-of-magnitude level with these estimates, although they are lower, because we are analysing cleaned maps.

The unclustered point source and CIB templates are highly correlated, at
93\,\%. For this reason it was not deemed a priority for the other bispectrum 
estimators to implement the CIB template as well. Moreover, both point source
templates are negligibly correlated with the primordial NG templates and
the lensing-ISW template (the maximum being the correlation between equilateral
and CIB templates at 2.7\,\%, while correlations with the unclustered point source template
are well below 1\,\%). For this reason, and despite the detection of point
sources in the cleaned maps, it makes no difference for the primordial results if
point sources are included in a joint analysis or completely neglected.

An additional contaminant to the cosmological CMB bispectrum arises
from the correlation between the gravitational lensing of the CMB
anisotropies and the CIB anisotropies.  This correlation was detected in \citet{planck2013-p13} using an optimal cross-spectrum
estimator. The CIB-lensing bispectrum might couple with any of the
primordial shapes. However, the amplitude of the CIB bispectrum is
predicted to be small in the cleaned \Planck\ maps and it has actually
not been detected (see Table~\ref{Table:bps_and_ACIB}). The
CIB-lensing bispectrum signal is frequency dependent, and it is mostly
dominant in the very high \Planck\ frequencies (see
e.g., \citealt{curto14}).

\subsection{Non-Gaussianity from residuals of the deglitching processing}

Cosmic rays interacting with the cryogenic detectors induce spikes in timelines. These high-amplitude, fast-rising signals are followed by a decay tail. We observe three families of glitches, characterized by their temporal shape. The amplitude and time constants of the decays depend on which part of the satellite is hit \citep{2014arXiv1403.6592C, planck2013-p03e}. 
These random events are Poisson-distributed in time and produce highly non-Gaussian systematics. 

A method has been developed to remove them directly at the time-ordered information (TOI) level. This process is performed iteratively, and is described in detail in \citet{planck2013-p03e}.  The short glitches are just flagged in the data, whereas for the long ones only the fast part is flagged, and the long tail is substracted from the timeline.  This procedure is not perfect, and there are residuals from the potentially biased errors in the fit, and the undetected glitches under the threshold of $3.2\,\sigma$ of the TOI noise rms. They could in principle produce a non-Gaussian signal in the final map.
In addition, these residuals could interact with the mapmaking procedure at the destriping level, since the error on the offset determination could be non-Gaussian due to undetected glitches or a possible bias in the errors of the removal of tails. This is important, because in more than $95\,\%$ of the TOI data, tails have been subtracted.

To estimate the effect of these residuals on the determination of NG, we created two sets of simulations (one including glitches and the other not) for every bolometer of the 143~GHz channel.  We generated Gaussian CMB maps, and applied the full TOI processing with a realistic instrumental noise \citep{planck2014-a08}. 
In the simulations \textit{with glitches}, we added glitches at the TOI level, following the properties measured in the data, and cleaned them with the procedure applied to the data. For the simulations \textit{without glitches}, we have the same CMB and noise realization, but no glitches added at the TOI level. 

We estimated the bias caused by glitches on the measurement of $f_{\rm NL}$ using the binned bispectrum estimator. The bias on $f_{\rm NL}$ induced by the glitch residuals $g$ on a map $T$, including noise and CMB is given by $\langle \hat{f}_{\rm NL}(T+g) -\hat{f}_{\rm NL}(T)\rangle$, where the noise in the weighting of the estimator is determined from the simulations \textit{with glitches} (as it would be for the data). Results are shown in Table~\ref{Tab_glitches}.  For most shapes, we detect no significant bias. The higher signal and high dispersion for the local shape might be due to a mis-calibration of the linear correction.  
In any case, for all shapes the bias due to glitches is a negligible correction to the value of $\fnl$, given its error bars, and we will not take it into
account in the remainder of the paper.

\begin{table*}[tb]                 
\begingroup
\newdimen\tblskip \tblskip=5pt
\caption{Results on the impact of cosmic ray residuals on the estimation of $f_\mathrm{NL}$ at 143\;GHz, determined using the binned bispectrum estimator. We produced 10 simulations. We report the mean of the bias defined in the text, and the error on this mean. We also show the Fisher error bars on $\fnl$ for these simulations.}
\label{Tab_glitches}
\nointerlineskip
\vskip -3mm
\footnotesize
\setbox\tablebox=\vbox{
   \newdimen\digitwidth 
   \setbox0=\hbox{\rm 0} 
   \digitwidth=\wd0 
   \catcode`*=\active 
   \def*{\kern\digitwidth}
   \newdimen\signwidth 
   \setbox0=\hbox{+} 
   \signwidth=\wd0 
   \catcode`!=\active 
   \def!{\kern\signwidth}
   %
%
\halign{\hbox to 1.1in{#\leaderfil}\tabskip 5em&
\hfil#\hfil\tabskip 3em&
\hfil#\hfil&
\hfil#\hfil&
\hfil#\hfil&
\hfil#\hfil\tabskip 0pt\cr
\noalign{\doubleline\vskip 2pt}
\hfill & Local & Equilateral & Orthogonal & Diffuse PS ($\times 10^{29}$) & Lens-ISW \cr
\noalign{\vskip 4pt\hrule\vskip 6pt}
\itT-only \hfill \cr
  bias mean &    1.1 $\pm$ 0.6&  !0.8 $\pm$ 1.6&  $-$1.0 $\pm$ *0.7& 0.5* $\pm$ 0.2*& !0.01 $\pm$ 0.01\cr 
 $\sigma_{f_{\rm NL}}$ & 5.2& 64& 34 & 2&  0.2\cr
\noalign{\vskip 4pt\hrule\vskip 6pt}
\itE-only \hfill \cr
  bias mean &  2.4 $\pm$ 5.8&  $-$9.6 $\pm$ 8.7&  $-$7.1 $\pm$ 14.8& 0.0* $\pm$ 0.1*& $-$3.0* $\pm$ 1.4*\cr 
 $\sigma_{f_{\rm NL}}$ & 38   & 157  & 90  & 0.6 & 7.8 \cr 
\noalign{\vskip 4pt\hrule\vskip 6pt}
\itT+\itE \hfill \cr
  bias mean &   1.8 $\pm$ 1.0&  $-$5.1 $\pm$ 2.2&  !0.1 $\pm$ *1.5& 0.01 $\pm$ 0.04& !0.01 $\pm$ 0.01 \cr 
 $\sigma_{f_{\rm NL}}$ & 4.4  & 43 &  22 & 0.3 &  0.2 \cr 
\noalign{\vskip 3pt\hrule\vskip 4pt}}}
\endPlancktablewide                 
\endgroup
\end{table*}                        

%% file: A19_Section5.tex
During the work for the 2013 release, culminating
in the NG results of \cite{planck2013-p09a}, the advantage of having multiple
independent bispectrum estimator implementations was amply
demonstrated. This allows for very useful cross-checking of results,
both during development and for the final analysis, thus greatly improving
the robustness of and confidence in the final results. For this new
release we followed the same procedure, with the same three principal
bispectrum estimators: KSW; binned; and modal, all of which had
their pipelines updated to handle polarization data in addition to
temperature.

Beyond the usefulness of cross-checking, the three estimators complement
each other and have different strengths. The KSW estimator can treat 
separable bispectrum templates without approximation, but it is more work
to add new templates and non-separable templates cannot be handled at all. 
The binned and modal estimators can reconstruct
the full bispectrum (smoothed in different domains), while the skew-$C_\ell$
extension of the KSW estimator can be used to investigate the
bispectrum beyond $f_\mathrm{NL}$. The binned bispectrum estimator
is the fastest on a single map or a set of unrelated maps, but becomes slower 
than the other two on a large set of realizations based on the same settings, 
because the linear correction term cannot be precomputed. The modal estimator 
can investigate a wide selection of oscillating or otherwise rapidly changing 
bispectrum templates that would be difficult to bin, while the binned 
bispectrum estimator can quickly implement and determine the $f_\mathrm{NL}$ 
of an additional template or the effect of a different cosmology if the binned
bispectrum of the maps has already been computed. The binned estimator gets 
the dependence of $f_\mathrm{NL}$ on $\ell$ for free with its results, while
the modal estimator allows for a statistical investigation of the mode
coefficients.

In this Section we show some of the validation tests, in particular for polarization.
In Sect.~\ref{Sec_valid_est_1} we investigate the agreement between estimators, map-by-map, on sets of successively more realistic
maps. In Sect.~\ref{Sec_valid_est_2} we show that the estimators are unbiased
in the presence of a non-zero $f_\mathrm{NL}$. Finally, in 
Sect.~\ref{Sec_valid_est_3} we show that the estimators are essentially optimal on a
set of the most realistic \Planck\ simulations available, which are those
used to compute the error bars on our final results.

\subsection{Agreement between estimators on a map-by-map basis}
\label{Sec_valid_est_1}

The maps used in this subsection are realistic simulations of the CMB
(at resolution $N_\mathrm{side}=2048$) but without any
foregrounds. They do not contain any primordial NG, but do include ISW-lensing. Since the final FFP8 simulations were not yet available, the main
goal was to make sure that the estimators
agreed with each other, not only on average, but also on a map-by-map
basis. For this purpose it was enough to look at only 49
maps. Establishing optimality of the estimators requires a larger
number of maps, and is shown on the FFP8 simulations in
Sect.~\ref{Sec_valid_est_3}.

In our first test we include the effect of the 143~GHz beam, but in other respects the simulations are ideal (no noise, and no mask).
The analysis used $\ell_\mathrm{max}=2000$ for both \itT\ and \itE.\ The results
for the average over the maps for the KSW, binned, and both modal estimators, as well as for 
the difference between each estimator and KSW, are shown in 
Table~\ref{Tab_valid_ideal}. 
The shapes are assumed to be
independent in this analysis, which means that the bias on the local
shape due to the ISW-lensing effect is clearly visible.
Results are shown for \itT-only, \itE-only, and the full combined 
\itT+\itE\ analysis. Note that the second modal
implementation cannot compute results for \itE\ alone.
One clearly sees that the results agree very well. It is also interesting to 
note that in this ideal noiseless case, one can actually determine 
$f_\mathrm{NL}$ more accurately from polarization alone than from temperature 
alone. This is due to the narrower transfer function in polarization, 
so that the primordial bispectrum is less smoothed in its projection
to two-dimensional harmonic space.

\begin{table*}[tb]                 
\begingroup
\newdimen\tblskip \tblskip=5pt
\caption{Results from the different $\fnl$ estimators for the set 
of CMB simulations described in Sect.~\ref{Sec_valid_est_1} in
the ideal case without noise or mask.
Both the results for the estimators individually and for the differences with
KSW are given, for \itT-only, \itE-only, and the full combined \itT+\itE\  
analysis. 
The shapes are assumed independent.}
\label{Tab_valid_ideal}
\nointerlineskip
\vskip -3mm
\footnotesize
\setbox\tablebox=\vbox{
   \newdimen\digitwidth 
   \setbox0=\hbox{\rm 0} 
   \digitwidth=\wd0 
   \catcode`*=\active 
   \def*{\kern\digitwidth}
   \newdimen\signwidth 
   \setbox0=\hbox{+} 
   \signwidth=\wd0 
   \catcode`!=\active 
   \def!{\kern\signwidth}
   \newdimen\dotwidth 
   \setbox0=\hbox{.} 
   \dotwidth=\wd0 
   \catcode`^=\active 
   \def^{\kern\dotwidth}
\halign{\hbox to 1.5in{#\leaderfil}\tabskip 1em&
\hfil#\hfil\tabskip 1em&
\hfil#\hfil&
\hfil#\hfil&
\hfil#\hfil&
\hfil#\hfil&
\hfil#\hfil&
\hfil#\hfil\tabskip 0pt\cr
\noalign{\doubleline\vskip 2pt}
\omit&\multispan7\hfil $f_{\rm NL}$\hfil\cr
\omit&\multispan7\hrulefill\cr
Shape\hfill & KSW & Binned & Modal\ 1 & Modal\ 2 & ** B $-$ KSW & M1 $-$ KSW & M2 $-$ KSW \cr
\noalign{\vskip 4pt\hrule\vskip 6pt}
\itT\ Local & *!7.6 $\pm$ *5.4& *!7.4 $\pm$ *5.6& *!7.4 $\pm$ *5.1& !*7.2 $\pm$ *5.7& ** $-$0.3 $\pm$ *0.6& $-$0.2 $\pm$ 0.4& $-$0.5 $\pm$ *2.2\cr
\itT\ Equilateral & *!7^* $\pm$ 53^*& !*5^* $\pm$ 58^*& !*6^* $\pm$ 53^*& !*8^* $\pm$ 56^*& ** $-$2^* $\pm$ 12^*& $-$1.0 $\pm$ 8.4& !0^* $\pm$ 17^*\cr
\itT\ Orthogonal & $-$22^* $\pm$ 27^*& $-$22^* $\pm$ 28^*& $-$22^* $\pm$ 27^*& $-$17^* $\pm$ 30^*& ** !0.5 $\pm$ *9.4& $-$0.2 $\pm$ 4.2& !5^* $\pm$ 11^*\cr
\noalign{\vskip 5pt}
\itE\ Local & *$-$0.9 $\pm$ *4.1& *$-$1.3 $\pm$ *3.4& *$-$0.9 $\pm$ *3.7&\dots& ** $-$0.3 $\pm$ *2.9& !0.1 $\pm$ 0.5& \dots\cr
\itE\ Equilateral & *$-$9^* $\pm$ 42^*& $-$10^* $\pm$ 42^*& $-$10^* $\pm$ 40^*& \dots& ** $-$1^* $\pm$ 11^*& $-$0.7 $\pm$ 9.2& \dots\cr
\itE\ Orthogonal & !*4^* $\pm$ 13^*& !*5^* $\pm$ 13^*& !*4^* $\pm$ 12^*& \dots& ** !0.1 $\pm$ *3.8& $-$0.3 $\pm$ 2.7& \dots\cr
\noalign{\vskip 5pt}
\itT+\itE\ Local & !*2.2 $\pm$ *3.1& !*1.5 $\pm$ *2.5& !*2.1 $\pm$ *2.8& !*2.0 $\pm$ *3.3& ** $-$0.6 $\pm$ *1.0& !0.0 $\pm$ 0.8& $-$0.2 $\pm$ *1.9\cr
\itT+\itE\ Equilateral & !*0^* $\pm$ 20^*& !*2^* $\pm$ 22^*& !*3^* $\pm$ 21^*& *!0^* $\pm$ 23^*& ** !1.4 $\pm$ *5.8& !2.3 $\pm$ 7.3& !0^* $\pm$ 12^*\cr
\itT+\itE\ Orthogonal & *$-$4^* $\pm$ 10^*& *$-$4^* $\pm$ *9^*& *$-$6^* $\pm$ *9^*& *$-$5^* $\pm$ 12^*& ** !0.3 $\pm$ *2.2& $-$1.1 $\pm$ 3.1& $-$1.0 $\pm$ *7.1\cr
\noalign{\vskip 3pt\hrule\vskip 4pt}}}
\endPlancktablewide                 
\endgroup
\end{table*}                        

The second test is identical to the first, except that we add realistic
anisotropic noise realizations to the full-sky maps, based on the
143~GHz channel.  The estimators now require the use of the linear
correction term, and results are shown in Table~\ref{Tab_valid_noise}.
The agreement is still very good, although slightly worse than in the ideal case, as expected.
The fact that the error bars for the \itT-only local case here are actually a bit smaller
than in the ideal case is an artefact of the small number of maps; i.e.,
the error bars have not completely converged yet. On the other hand, the fact
that the error bars for \itE-only are much larger than in the ideal case is
a real effect; the \Planck\ single-frequency polarization maps are 
noise-dominated.

\begin{table*}[tb]                 
\begingroup
\newdimen\tblskip \tblskip=5pt
\caption{
As Table~\ref{Tab_valid_ideal}, but with noise and no mask.}
\label{Tab_valid_noise}
\nointerlineskip
\vskip -3mm
\footnotesize
\setbox\tablebox=\vbox{
   \newdimen\digitwidth 
   \setbox0=\hbox{\rm 0} 
   \digitwidth=\wd0 
   \catcode`*=\active 
   \def*{\kern\digitwidth}
   \newdimen\signwidth 
   \setbox0=\hbox{+} 
   \signwidth=\wd0 
   \catcode`!=\active 
   \def!{\kern\signwidth}
   \newdimen\dotwidth 
   \setbox0=\hbox{.} 
   \dotwidth=\wd0 
   \catcode`^=\active 
   \def^{\kern\dotwidth}
\halign{\hbox to 1.5in{#\leaderfil}\tabskip 1em&
\hfil#\hfil\tabskip 1em&
\hfil#\hfil&
\hfil#\hfil&
\hfil#\hfil&
\hfil#\hfil&
\hfil#\hfil&
\hfil#\hfil\tabskip 0pt\cr
\noalign{\doubleline\vskip 2pt}
\omit&\multispan7\hfil $f_{\rm NL}$\hfil\cr
\omit&\multispan7\hrulefill\cr
Shape\hfill & KSW & Binned & Modal\ 1 & Modal\ 2 & ** B $-$ KSW & M1 $-$ KSW & M2 $-$ KSW \cr
\noalign{\vskip 4pt\hrule\vskip 6pt}
\itT\ Local & !*6.7 $\pm$ **4.8& !*6.4 $\pm$ **5.2& !*6.7 $\pm$ **4.7& !*7.0 $\pm$ *5.3& ** *$-$0.3 $\pm$ *1.0& !0.1 $\pm$ *0.4& !0.3 $\pm$ *1.2\cr
\itT\ Equilateral & !11^* $\pm$ *61^*& !12^* $\pm$ *65^*& !*9^* $\pm$ *63^*& !12^* $\pm$ 62^*& ** !*1^* $\pm$ 15^*& $-$1.9 $\pm$ *9.6& !1^* $\pm$ 12^*\cr
\itT\ Orthogonal & $-$19^* $\pm$ *31^*& $-$18^* $\pm$ *34^*& $-$20^* $\pm$ *32^*& $-$18^* $\pm$ 35^*& ** !*1^* $\pm$ 12^*& $-$1.3 $\pm$ *5.1& !0.8 $\pm$ *8.8\cr
\noalign{\vskip 5pt}
\itE\ Local & *$-$2^* $\pm$ *29^*& *$-$4^* $\pm$ *28^*& *$-$1^* $\pm$ *29^*& \dots& ** *$-$2^* $\pm$ 12^*& !0.4 $\pm$ *5.5& \dots\cr
\itE\ Equilateral & !*1^* $\pm$ 191^*& $-$18^* $\pm$ 195^*& *$-$6^* $\pm$ 200^*& \dots& ** $-$19^* $\pm$ 47^*& $-$7^* $\pm$ 23^*& \dots\cr
\itE\ Orthogonal & *$-$6^* $\pm$ 101^*& *!0^* $\pm$ 107^*& *$-$6^* $\pm$ 102^*& \dots& ** !*6^* $\pm$ 25^*& $-$0.3 $\pm$ 10^*& \dots\cr
\noalign{\vskip 5pt}
\itT+\itE\ Local & !*4.9 $\pm$ **4.2& !*4.5 $\pm$ **4.4& !*5.0 $\pm$ **4.2& *!4.9 $\pm$ *4.9& ** *$-$0.4 $\pm$ *1.2& !0.1 $\pm$ *1.5& $-$0.0 $\pm$ 1.2\cr
\itT+\itE\ Equilateral & !13^* $\pm$ *46^*& !11^* $\pm$ *49^*& !*9^* $\pm$ *48^*& !13^* $\pm$ 47^*& ** *$-$2^* $\pm$ 10^*& $-$4^* $\pm$ 13^*& $-$0.3 $\pm$ 7.0\cr
\itT+\itE\ Orthogonal & $-$11^* $\pm$ *22^*& $-$11^* $\pm$ *24^*& $-$13^* $\pm$ *22^*& $-$11^* $\pm$ 24^*& ** *!0.0 $\pm$ *7.3& $-$1.3 $\pm$ *7.1& !0.7 $\pm$ 4.5\cr
\noalign{\vskip 3pt\hrule\vskip 4pt}}}
\endPlancktablewide                 
\endgroup
\end{table*}                        

\begin{figure*}[!t]
\centering
\includegraphics{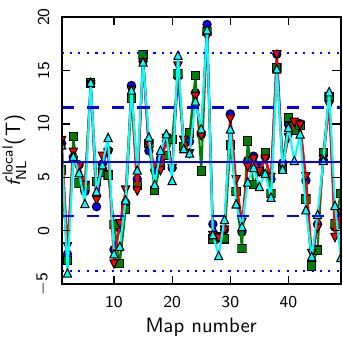}
\includegraphics{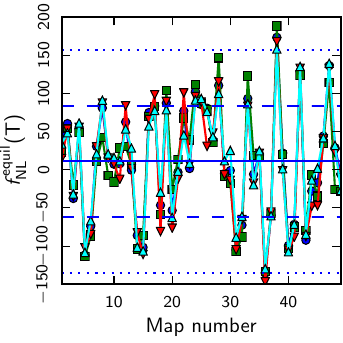}
\includegraphics{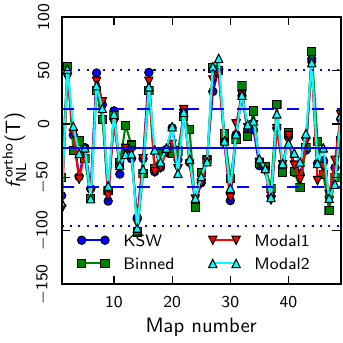}

\includegraphics{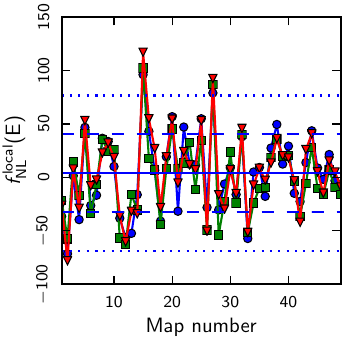}
\includegraphics{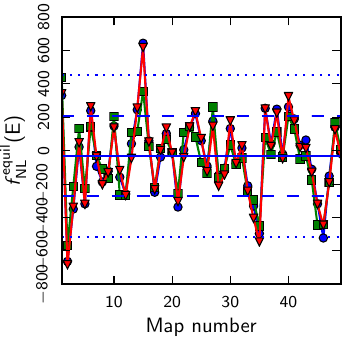}
\includegraphics{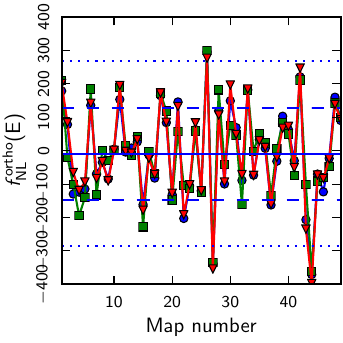}

\includegraphics{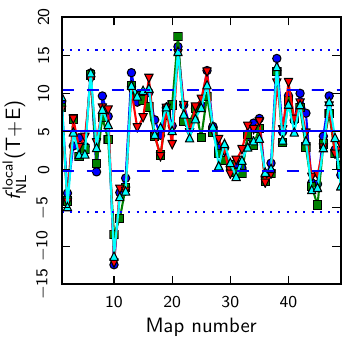}
\includegraphics{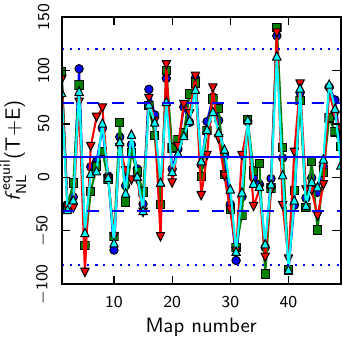}
\includegraphics{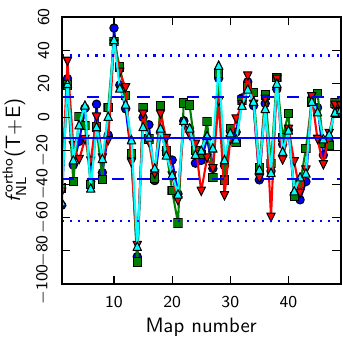}
\caption{Map-by-map comparison of the results from the different estimators 
for local (left), equilateral (centre), and orthogonal (right) $f_\mathrm{NL}$ 
(taking the shapes to be independent), for the third set of simulations 
described in Sect.~\ref{Sec_valid_est_1}, 
including both noise and a mask. Results are shown for \itT-only (top),
\itE-only (centre), and the full combined \itT+\itE\ case (bottom).
The legend for the estimators can be found in the top right figure. 
The horizontal solid line is the average value of all maps for KSW, and the
dashed and dotted horizontal lines correspond to $\pm 1\sigma$ and $\pm 2\sigma$
deviations, respectively.}
\label{Fig_valid_mask}
\end{figure*}

Finally, the third test is identical to the second, except that we now also
add a mask. The mask chosen is realistic, based on the union of the
confidence masks provided by the \SMICA, \NILC, \SEVEM, and \Commander\ methods
for this particular set of simulations.
It contains both a Galactic and a point source part. The temperature mask 
leaves 79\,\% of the sky unmasked, while the polarization mask leaves 76\,\%.
The results are shown in Table~\ref{Tab_valid_mask}, while the map-by-map
comparison is given in Fig.~\ref{Fig_valid_mask}.
From the table we see that the agreement between the different bispectrum 
estimators is still very good and only slightly degraded
when compared to the previous case.  The typical discrepancy between the bispectrum estimators, even 
in this most realistic case, is less than about a third of the uncertainty on 
$f_\mathrm{NL}$. This is apparent in the map-by-map comparison of Fig.~\ref{Fig_valid_mask}.

\begin{table*}[tb]                 
\begingroup
\newdimen\tblskip \tblskip=5pt
\caption{
As Table~\ref{Tab_valid_ideal}, but with noise and a mask.
}
\label{Tab_valid_mask}
\nointerlineskip
\vskip -3mm
\footnotesize
\setbox\tablebox=\vbox{
   \newdimen\digitwidth 
   \setbox0=\hbox{\rm 0} 
   \digitwidth=\wd0 
   \catcode`*=\active 
   \def*{\kern\digitwidth}
   \newdimen\signwidth 
   \setbox0=\hbox{+} 
   \signwidth=\wd0 
   \catcode`!=\active 
   \def!{\kern\signwidth}
   \newdimen\dotwidth 
   \setbox0=\hbox{.} 
   \dotwidth=\wd0 
   \catcode`^=\active 
   \def^{\kern\dotwidth}
\halign{\hbox to 1.5in{#\leaderfil}\tabskip 1em&
\hfil#\hfil\tabskip 1em&
\hfil#\hfil&
\hfil#\hfil&
\hfil#\hfil&
\hfil#\hfil&
\hfil#\hfil&
\hfil#\hfil\tabskip 0pt\cr
\noalign{\doubleline\vskip 2pt}
\omit&\multispan7\hfil $f_{\rm NL}$\hfil\cr
\omit&\multispan7\hrulefill\cr
Shape\hfill & KSW & Binned & Modal\ 1 & Modal\ 2 & ** B $-$ KSW & M1 $-$ KSW & M2 $-$ KSW \cr
\noalign{\vskip 4pt\hrule\vskip 6pt}
\itT\ Local & !*6.5 $\pm$ **5.1& !*6.1 $\pm$ **5.3& !*6.4 $\pm$ **5.0& !*6.0 $\pm$ *5.3& ** *$-$0.4 $\pm$ *1.5& $-$0.1 $\pm$ *0.7& $-$0.5 $\pm$ *1.3\cr
\itT\ Equilateral & !11^* $\pm$ *73^*& !*9^* $\pm$ *75^*& !*6^* $\pm$ *76^*& !11^* $\pm$ 70^*& ** *$-$2^* $\pm$ 19^*& $-$5^* $\pm$ 14^*& !0^* $\pm$ 12^*\cr
\itT\ Orthogonal & $-$22^* $\pm$ *37^*& $-$21^* $\pm$ *37^*& $-$23^* $\pm$ *36^*& $-$20^* $\pm$ 37^*& ** *!2^* $\pm$ 14^*& $-$0.9 $\pm$ *6.1& !2.6 $\pm$ *9.2 \cr
\noalign{\vskip 5pt}
\itE\ Local & !*4^* $\pm$ *36^*& **0^* $\pm$ *35^*& !*5^* $\pm$ *37^*& \dots& ** *$-$4^* $\pm$ 16^*& !1^* $\pm$ 13^*& \dots\cr
\itE\ Equilateral & $-$32^* $\pm$ 242^*& $-$49^* $\pm$ 209^*& $-$38^* $\pm$ 246^*& \dots& ** $-$17^* $\pm$ 88^*& $-$6^* $\pm$ 34^*& \dots\cr
\itE\ Orthogonal & *$-$9^* $\pm$ 138^*& *$-$7^* $\pm$ 139^*& *$-$7^* $\pm$ 142^*& \dots& ** !*2^* $\pm$ 45^*& !2^* $\pm$ 19^*& \dots\cr
\noalign{\vskip 5pt}
\itT+\itE\ Local & !*5.1 $\pm$ **5.3& !*4.2 $\pm$ **5.1& !*4.8 $\pm$ **5.0& !*4.5 $\pm$ *5.2& ** *$-$1.0 $\pm$ *1.7& $-$0.3 $\pm$ *1.7& $-$0.6 $\pm$ *1.3\cr
\itT+\itE\ Equilateral & !19^* $\pm$ *50^*& !16^* $\pm$ *50^*& !15^* $\pm$ *53^*& !16^* $\pm$ 45^*& ** *$-$3^* $\pm$ 14^*& $-$4^* $\pm$ 19^*& $-$3.2 $\pm$ *9.8\cr
\itT+\itE\ Orthogonal & $-$12^* $\pm$ *25^*& $-$11^* $\pm$ *26^*& $-$13^* $\pm$ *25^*& $-$11^* $\pm$ 23^*& ** *!1.9 $\pm$ *8.7& $-$1.0 $\pm$ *9.9& !1.4 $\pm$ *5.9\cr
\noalign{\vskip 3pt\hrule\vskip 4pt}}}
\endPlancktablewide                 
\endgroup
\end{table*}                        

\subsection{Validation of estimators in the presence of primordial 
non-Gaussianity}
\label{Sec_valid_est_2}

After the map-by-map comparison of the previous section, we next want
to make sure that the estimators are unbiased. For this purpose we
prepared a different set of 100 \itT\ and
\itE\ CMB simulations, still with cosmological parameters as determined by
\Planck. This time ISW-lensing is not present, but there is a nonzero local $f_\mathrm{NL}=12$. To these maps we add the same
beam, anisotropic noise, and mask as before. 
We again take $\ell_\mathrm{max}=2000$, and the results are given in Table~\ref{Tab_valid_fnl12}.

We see that all the estimators correctly recover the input value, both in 
temperature and in polarization. The results for the equilateral and orthogonal
shapes are consistent with the fact that those templates have a non-zero 
correlation with the local shape (the Table gives the results
for an analysis where all shapes are assumed independent). For example, a 
{\em joint} analysis of the \itT+\itE\  binned estimator gives
$f_\mathrm{NL}^\mathrm{local}=11.5 \pm 6.4$,
$f_\mathrm{NL}^\mathrm{equil}=-7.5 \pm 51$, and
$f_\mathrm{NL}^\mathrm{ortho}=-0.4 \pm 29$.
Except for the first modal estimator in \itE-only (due to
an insufficient number of maps in the linear correction term),
we also find that the error bars for the bispectrum-based estimators are very 
close to the Fisher errors.
Note that a slight increase in the error bars compared to Fisher estimates is expected for the local shape in \itT-only and in \itT+\itE, due to the signal
being significantly different from zero there (the Fisher error bars for
the local case are 5.8 for \itT-only, 26 for \itE-only, and 5.0 for \itT+\itE).
Hence the estimators are effectively optimal, as will be illustrated 
in more detail in the next section.

\begin{table}[tb]                 
\begingroup
\newdimen\tblskip \tblskip=5pt
\caption{Results from the different estimators for $f_\mathrm{NL}$ for the set 
of CMB simulations with $f_\mathrm{NL}^\mathrm{local}=12$ described in 
Sect.~\ref{Sec_valid_est_2}.
Results are given for \itT-only, \itE-only, and the full combined \itT+\itE\ analysis. 
The shapes are assumed independent (see the main text for a discussion of
this point).}
\label{Tab_valid_fnl12}
\nointerlineskip
\vskip -3mm
\footnotesize
\setbox\tablebox=\vbox{
   \newdimen\digitwidth 
   \setbox0=\hbox{\rm 0} 
   \digitwidth=\wd0 
   \catcode`*=\active 
   \def*{\kern\digitwidth}
   \newdimen\signwidth 
   \setbox0=\hbox{+} 
   \signwidth=\wd0 
   \catcode`!=\active 
   \def!{\kern\signwidth}
   \newdimen\dotwidth 
   \setbox0=\hbox{.} 
   \dotwidth=\wd0 
   \catcode`^=\active 
   \def^{\kern\dotwidth}
\halign{\hbox to 0.6in{#\leaderfil}\tabskip 0.4em&
\hfil#\hfil\tabskip 0.2em&
\hfil#\hfil\tabskip 0.2em&
\hfil#\hfil\tabskip 0.2em&
\hfil#\hfil\tabskip 0pt\cr
\noalign{\doubleline\vskip 2pt}
\omit&\multispan4\hfil $f_{\rm NL}$\hfil\cr
\omit&\multispan4\hrulefill\cr
Shape\hfill & KSW & Binned & Modal\ 1 & Modal\ 2 \cr
\noalign{\vskip 4pt\hrule\vskip 6pt}
\omit\hfil \itT\ \hfil &&\cr
Local & !11.2 $\pm$ **6.7& !10.9 $\pm$ **6.3& !11.9 $\pm$ **6.6& !11.6 $\pm$ *6.6\cr
Equilateral & !26^* $\pm$ *78^*& !24^* $\pm$ *77^*& !31^* $\pm$ *82^*& !27^* $\pm$ 76^*\cr
Orthogonal & $-$33^* $\pm$ *34^*& $-$33^* $\pm$ *35^*& $-$34^* $\pm$ *36^*& $-$33^* $\pm$ 36^*\cr
\noalign{\vskip 5pt}
\omit\hfil \itE\ \hfil &&\cr
Local & !11^* $\pm$ *29*^& !12^* $\pm$ *26^*& *!9^* $\pm$ *36^*& \dots\cr
Equilateral & !34*^ $\pm$ 182*^& !32^* $\pm$ 153*^& !10^* $\pm$ 241^*& \dots\cr
Orthogonal & $-$37^* $\pm$ 110^*& $-$28^* $\pm$ 115^*& $-$31^* $\pm$ 143^*& \dots\cr
\noalign{\vskip 5pt}
\omit\hfil \itTpE\ \hfil &&\cr
Local & !11.3 $\pm$ **5.5& !11.2 $\pm$ **5.0& !11.1 $\pm$ **5.8& !11.0 $\pm$ *5.4\cr
Equilateral & !29^* $\pm$ *52^*& !24^* $\pm$ *50^*& !28^* $\pm$ *54^*& !24^* $\pm$ 50^*\cr
Orthogonal & $-$29^* $\pm$ *26^*& $-$28^* $\pm$ !23^*& $-$30^* $\pm$ *28^*& $-$26^* $\pm$ 23^*\cr
\noalign{\vskip 3pt\hrule\vskip 4pt}}}
\endPlancktablewide                 
\endgroup
\end{table}                        

\subsection{Validation of estimators on realistic {\it Planck} simulations}
\label{Sec_valid_est_3}

As a final validation test, we ran our estimators on a large set of the
most realistic simulations available. These are the FFP8 simulations
\citep{planck2014-a14} using \SMICA for foreground separation. They are
the same simulations we use to determine the error bars on our final \SMICA\ 
results in Sect.~\ref{sec:Results}. They contain the Collaboration's best 
estimates of the CMB sky and of \Planck's noise and beam effects, and 
have been cleaned by \SMICA\ in the same way as the real sky map. 
The mask used is the same common mask defined for the real data analysis.
For this test the estimators were all processed with the same settings used for 
the final data analysis.

Here we take 159 of these maps, and process these using all the estimators. By contrast, for the final results in Sect.~\ref{sec:Results}, the convergence of the error bars of each estimator was carefully checked, using more maps if required. This explains why there are some small differences between the error bars in Sect.~\ref{sec:Results} and the ones presented here.

\begin{table*}[tb]                 
\begingroup
\newdimen\tblskip \tblskip=5pt
\caption{Results from the different estimators for $f_\mathrm{NL}$ for the set 
of \SMICA\ simulations based on FFP8 described in 
Sect.~\ref{Sec_valid_est_3}.
Both the results for the estimators individually and for the differences with
KSW are given, for \itT-only, \itE-only, and the full combined \itT+\itE\ analysis. 
The shapes are assumed independent and the lensing-ISW bias has not been
subtracted.}
\label{Tab_valid_SmicaDX11sims}
\nointerlineskip
\vskip -3mm
\footnotesize
\setbox\tablebox=\vbox{
   \newdimen\digitwidth 
   \setbox0=\hbox{\rm 0} 
   \digitwidth=\wd0 
   \catcode`*=\active 
   \def*{\kern\digitwidth}
   \newdimen\signwidth 
   \setbox0=\hbox{+} 
   \signwidth=\wd0 
   \catcode`!=\active 
   \def!{\kern\signwidth}
   \newdimen\dotwidth 
   \setbox0=\hbox{.} 
   \dotwidth=\wd0 
   \catcode`^=\active 
   \def^{\kern\dotwidth}
\halign{\hbox to 0.7in{#\leaderfil}\tabskip 0.6em&
\hfil#\hfil\tabskip 0.5em&
\hfil#\hfil\tabskip 0.5em&
\hfil#\hfil\tabskip 0.5em&
\hfil#\hfil\tabskip 0.5em&
\hfil#\hfil\tabskip 0.4em&
\hfil#\hfil\tabskip 0.7em&
\hfil#\hfil\tabskip 0.7em&
\hfil#\hfil\tabskip 0.7em&
\hfil#\hfil\tabskip 0pt\cr
\noalign{\doubleline\vskip 2pt}
\omit&\multispan9\hfil $f_{\rm NL}$\hfil\cr
\omit&\multispan9\hrulefill\cr
Shape\hfill & KSW & Binned & Modal\ 1 & Modal\ 2 & Mink.F. & ** B $-$ KSW & M1 $-$ KSW & M2 $-$ KSW & MF $-$ KSW \cr
\noalign{\vskip 4pt\hrule\vskip 6pt}
\omit\hfil \itT\ \hfil &&\cr
Local & !*7.1 $\pm$ **5.5& !*7.0 $\pm$ **5.4& !*6.2 $\pm$ **5.5& !*6.3 $\pm$ *6.2& 7 $\pm$ 12& ** $-$0.1 $\pm$ *1.1& $-$0.9 $\pm$ *1.9& $-$0.8 $\pm$ *1.9& !0 $\pm$ 11^*\cr
Equilateral & !*2^* $\pm$ *67^*& !*4^* $\pm$ *67^*& *$-$4^* $\pm$ *73^*& !*5^* $\pm$ 66^*& \dots& ** !2^* $\pm$ 19^*& $-$6^* $\pm$ 32^*& !3^* $\pm$ 18^*& \dots\cr
Orthogonal & $-$23*^ $\pm$ *32*^& $-$24^* $\pm$ *33^*& $-$24^* $\pm$ *33*^& $-$20^* $\pm$ 36^*& \dots& ** $-$1^* $\pm$ 11^*& $-$0.9 $\pm$ *9.1& !3^* $\pm$ 14^*& \dots\cr
\noalign{\vskip 5pt}
\omit\hfil \itE\ \hfil &&\cr
Local & !*0.5 $\pm$ *32^*& *!0^* $\pm$ *35^*& !*1^* $\pm$ *30^*& \dots& 0 $\pm$ 49& ** $-$0.8 $\pm$ *8.3& !0.7 $\pm$ *8.3& \dots& !0 $\pm$ 37\cr
Equilateral & !*7^* $\pm$ 144*^& *!7^* $\pm$ 143^*& !*9^* $\pm$ 152^*& \dots& \dots& ** !0^* $\pm$ 37^*& !2^* $\pm$ 35^*& \dots&  \dots\cr
Orthogonal & !*5^* $\pm$ *72^*& !*7^* $\pm$ *75^*& !*4^* $\pm$ *73^*& \dots& \dots& ** !2^* $\pm$ 22^*& $-$1^* $\pm$ 17^*& \dots& \dots\cr
\noalign{\vskip 5pt}
\omit\hfil \itTpE\ \hfil &&\cr
Local & !*5.6 $\pm$ **5.1& !*5.0 $\pm$ **4.9& !*4.7 $\pm$ **4.8& !*4.3 $\pm$ *5.3& 5 $\pm$ 11& ** $-$0.6 $\pm$ *1.2& $-$0.9 $\pm$ *1.5& $-$1.3 $\pm$ *1.7& $-$1 $\pm$ 11\cr
Equilateral & !*3^* $\pm$ *46^*& !*5^* $\pm$ *44^*& !*3^* $\pm$ *46^*& !*4^* $\pm$ 43^*& \dots& ** !2^* $\pm$ 14^*& !0^* $\pm$ 14^*& !1.0 $\pm$ *9.7& \dots\cr
Orthogonal & $-$10^* $\pm$ *22^*& *$-$9^* $\pm$ *22^*& *$-$9^* $\pm$ *21^*& *$-$7^* $\pm$ 22^*& \dots& ** !0.8 $\pm$ *7.0& !0.8 $\pm$ *7.3& !2.7 $\pm$ *7.7& \dots\cr
\noalign{\vskip 3pt\hrule\vskip 4pt}}}
\endPlancktablewide                 
\endgroup
\end{table*}                        

The results are shown in Table~\ref{Tab_valid_SmicaDX11sims}. Note that these
are the results from an independent analysis, without subtracting the 
ISW-lensing bias. We also show the results from Minkowski
functionals (for the local case only).\footnote{Since the
Minkowski-functional pipeline automatically subtracts the ISW-lensing bias, 
the theoretical value for the bias as computed from the Fisher matrix has
been added to its results, to make a direct comparison possible.}
We see that there is very good agreement between the 
bispectrum estimators even on this most complex and realistic set of 
simulations. The standard deviation of the difference between bispectrum 
estimators generally stays below one third of the error bar on $f_\mathrm{NL}$, 
the only exception being the \itT-only equilateral result for the Modal 1 pipeline, 
which is still smaller than one half of the error bar.
We see that the results from Minkowski functionals are consistent,
but clearly suboptimal  for $f_\mathrm{NL}$. 
They are however a valuable, independent check.

The exact Fisher error bars for the nine shapes considered in the table are, 
in the same order as the table: 5.4, 69, 35; 31, 131, 74; 4.7, 43, 21.
Taking into account the relative error in the standard deviation of
$1/\sqrt{2(n-1)}$, which is $5.6\,\%$ for 159 maps, we see that all bispectrum
estimators are effectively optimal on all shapes, except for the \itE-only 
equilateral case where they appear slightly suboptimal. The small suboptimality
of the Modal 2 pipeline for the local shape seen here disappears
once more maps are used (see the results in Sect.~\ref{sec:Results}).

In conclusion, all these validation tests show that we have very good
agreement between the results from the different bispectrum estimators, not just on average, but also on a map-by-map basis.
In addition we see that, despite the approximations made in the pipelines,
and the simple treatment of the masked part of the maps 
(diffusive inpainting method and $f_\mathrm{sky}$ factor), the bispectrum
estimators we use are all essentially optimal.

%% file: A19_Section6.tex
\subsection{Constraints on local, equilateral, and orthogonal $f_{\rm NL}$}
\label{fnl_loc_eq_ort_results}

In this section we investigate the local, equilateral, and orthogonal
primordial templates. These are now established as the standard 
shapes to study first when investigating the bispectrum (see 
Sect.~\ref{sec:models} for a theoretical
motivation and description of these shapes). However, they represent only
the tip of the bispectral iceberg, and many more shapes are investigated
in Sect.~\ref{sec:Other_shapes}, while full model-independent reconstructions 
of the bispectrum are presented in Sect.~\ref{Bisp_Rec}.

For a complete description of the \Planck\ data set and the bispectrum estimator configurations 
we have used, we refer the reader in particular to Sect.~\ref{sec:dataset} and Sect.~\ref{sec:settings}.  
To summarize the overall analysis methodology, we have employed four independent bispectrum estimators on the
full mission \Planck\ temperature and polarization maps obtained from the four 
different component separation pipelines, \SMICA, \SEVEM, \NILC\, and \Commander.   
The bispectrum estimators are the KSW estimator with its skew-$C_\ell$ extension
using exact separable templates (Sect.~\ref{sec:KSW_est}), the Binned estimator using fixed
multipole bins (Sect.~\ref{sec:binned_est}), and the Modal 1 and Modal 2 estimators, 
which both use separable eigenmode expansions (Sect.~\ref{sec:modal_est}).  
Temperature is analysed over the multipole range $\ell_{\rm min} =2$ to $\lmax=2000$ or above 
and polarization is analysed from  $\ell_{\rm min}=40$ to $\lmax=1500$ or above (Sect.~\ref{sec:settings}). By employing 
inpainting and a linear term, all these estimators essentially achieve optimality (as shown 
by comparison with Fisher matrix forecasts).  The linear term in Eq.~\eqref{eq:diagcovestimator}
and the uncertainties are determined using the FFP8 simulations (Sect.~\ref{sec:settings}), also 
processed through the  different foreground-separation pipelines.   
Our thorough validation campaign for these estimators is presented in Sect.~\ref{sec:Validation}.

The results of the analysis of the four cleaned maps with the four estimators,
for \itT-only, \itE-only, and full \itTpE, are shown in Table~\ref{tab:fNLsmicah},
which is one of the main results of this paper. Results are determined
while assuming all shapes to be independent, and are shown both with
and without subtraction of the ISW-lensing bias (see 
Sect.~\ref{subsec:lensingISW} for more details about ISW-lensing). This bias
is most important (relative to the size of the error bars) for the local shape,
but also non-negligible for the orthogonal shape.
Results here have not been marginalized over the point source contributions.
While Sect.~\ref{sec:point_sources} shows that there is still a significant
contamination by unclustered point sources in the cleaned maps, the correlation
with the primordial templates is so small that this has no impact on the
results reported here (as checked explicitly).

While Table~\ref{tab:fNLsmicah} is the main result of this Section, in order
to simplify the use of the \Planck\ results by the wider scientific community, we also 
present in Table~\ref{Tab_KSW+SMICA} the results that can be considered 
the final \Planck\ 2015 results for the local, equilateral, and 
orthogonal shapes.
As in 2013, we select the combination of the KSW estimator and the
\SMICA\ map for this. The \SMICA\ map consistently performs well in 
all data validation tests, which are discussed in detail in Sect.~\ref{sec:Sec_valid_data}. 
The KSW estimator, while unable to deal with 
non-separable templates, treats
separable templates exactly, and the local, equilateral, and orthogonal 
template are all separable. On the other hand, the binned and modal estimators
can deal with non-separable shapes and have other advantages as well
(like full bispectrum reconstruction), but at the price of using approximations
for the templates. However, they have all been optimized in such a way that
the correlation with the exact templates for the three primordial shapes is 
close to perfect, so that in the end the results by the different estimators 
are statistically equivalent.
Compared to the corresponding values in Table~\ref{tab:fNLsmicah}, the 
difference in the numbers in the last column of Table~\ref{Tab_KSW+SMICA}
is due to the fact that in the latter the equilateral and orthogonal $\fnl$ have 
been determined jointly.

Focusing on the results for temperature-only and the full 
temperature plus polarization (\textit{T+E}) results, we see that there is no evidence for
a non-zero bispectrum with any of these three primordial shapes (local, equilateral, and orthogonal).
After ISW-lensing subtraction, all $\fnl$ values are consistent with $0$ at $68\,\%$ 
CL. The temperature results are all very similar to the ones from
the nominal mission data published in 2013 \citep{planck2013-p09a}, with
very minor improvements in the error bars due to the additional temperature
data. We also see that results are quite consistent when including polarization,
with error bars shrinking by about 15\,\% for local, 35\,\% for equilateral,
and 40\,\% for orthogonal. 

Table~\ref{tab:fNLsmicah} displays very good agreement between the
results from the different estimators, at the level expected from
the validation tests in Sect.~\ref{sec:Validation}. We also note 
how the error bars, which are determined using the FFP8 
simulations, are statistically indistinguishable from the optimal Fisher expectation.

Different component separation methods also show a good level of agreement 
when looking at temperature-only and combined temperature plus polarization results.
The accuracy of this statement will be shown and quantified in detail in Sect.~\ref{sec:Sec_valid_data}.
However, in the same section, we will also show how the agreement between $\fnl$ extracted 
from 
different cleaned maps becomes significantly degraded when considering {\em polarization-only} 
results\footnote{The \itE-only $\fnl$ agreement is still at a reasonable $1\,\sigma_{\fnl}$ level in most cases. However 
this is larger than expectations from simulations, as described in Sect.~\ref{sec:Sec_valid_data}.}. 
The reasons behind this loss of internal consistency are not fully understood at present. 
Polarization data are, however, much noisier than temperature data, implying that the \textit{EEE} bispectra 
have a close to negligible weight in the final combined measurement, which is dominated 
by the \textit{TTT} and \textit{TTE} configurations. In fact, as just mentioned above, the combined measurement 
looks perfectly self-consistent: local, equilateral and orthogonal $\fnl$ measurements in the \textit{T+E}
column of Table~\ref{tab:fNLsmicah} pass {\em all} our tests of robustness.   

We can thus conclude that, while highly challenging from a technical
point of view, the inclusion of polarization in our estimator pipelines has
been a success, allowing for a significant tightening of the constraints on the
three standard primordial bispectrum shapes. On the other hand, in light of the outstanding 
issues in \itE-only results, we present 
our results conservatively, and recommend the reader to consider all $\fnl$ constraints 
that make use of polarization data throughout this paper as {\em preliminary} at the current stage.
We stress again that this is a conservative choice, which is made despite the fact that {\em no test} 
to date shows any evidence of leakage of the issues in \textit{EEE} bispectra into the \textit{T+E} measurements. 
A detailed description of all the data validation tests, which lead to the robustness-related assessments 
summarized here, can be found in Sect.~\ref{sec:Sec_valid_data} (for readers less interested in the technical details, 
the main results and conclusions of all these tests are summarized in Sect.~\ref{sec:data_valid_summary}).

\begin{table*}[tb]                 
\begingroup
\newdimen\tblskip \tblskip=5pt
\caption{Results for the $f_{\rm NL}$ parameters of the primordial local, 
equilateral, and orthogonal shapes, determined by the KSW, binned and modal 
estimators from the \SMICA, \NILC, \SEVEM, and \Commander\ foreground-cleaned 
maps. Results have been determined using an independent single-shape analysis
and are reported both without and with subtraction of the ISW-lensing bias; 
error bars are $68\,\%$ CL.}
\label{tab:fNLsmicah}
\nointerlineskip
\vskip -3mm
\footnotesize
\setbox\tablebox=\vbox{
   \newdimen\digitwidth
   \setbox0=\hbox{\rm 0}
   \digitwidth=\wd0
   \catcode`*=\active
   \def*{\kern\digitwidth}
   \newdimen\signwidth
   \setbox0=\hbox{+}
   \signwidth=\wd0
   \catcode`!=\active
   \def!{\kern\signwidth}
\newdimen\dotwidth
\setbox0=\hbox{.}
\dotwidth=\wd0
\catcode`^=\active
\def^{\kern\dotwidth}
\halign{\hbox to 0.7 in{#\leaderfil}\tabskip 0.1 em&
\hfil#\hfil\tabskip 0.01em&
\hfil#\hfil&
\hfil#\hfil&
\hfil#\hfil&
\hfil#\hfil&
\hfil#\hfil&
\hfil#\hfil&
\hfil#\hfil\tabskip 0pt\cr
\noalign{\doubleline\vskip 2pt}
\omit&\multispan8\hfil $f_{\rm NL}$\hfil\cr
\omit&\multispan8\hrulefill\cr
\omit&\multicolumn{4}{c}{\hfil Independent\hfil}&
\multicolumn{4}{c}{\hfil **ISW-lensing subtracted\hfil}\cr
\noalign{\vskip 2pt}
Shape\hfill&KSW&Binned&Modal 1&Modal 2&**KSW&Binned&Modal 1&Modal 2\cr
\noalign{\vskip 4pt\hrule\vskip 6pt}
\omit**{\bf\SMICA\ } \itT\hfil&&\cr
Local&  *!10.2 $\pm$ **5.7&  !**8.7 $\pm$ **5.4&  !**6.8 $\pm$  **5.5&  !**7.8 $\pm$ **6.0&
** !**2.5 $\pm$ **5.7& !**1.3 $\pm$ **5.4& !**0.5 $\pm$ **5.5& !**1.7 $\pm$ **6.0\cr
Equilateral&  *$-$13^* $\pm$ *70^*& *$-$26^* $\pm$ *66^*& *$-$16^* $\pm$ *67^*& *$-$12^* $\pm$ *68^*&
** *$-$11^* $\pm$ *70^*& *$-$27^* $\pm$ *66^*& *$-$12^* $\pm$ *67^*& *$-$13^* $\pm$ *68^*\cr
Orthogonal & *$-$56^* $\pm$ *33^*& *$-$41^* $\pm$ *33^*& *$-$47^* $\pm$ *33^*& *$-$63^* $\pm$ *36^*&
** *$-$34^* $\pm$ *33^*& *$-$14^* $\pm$ *33^*& *$-$20^* $\pm$ *33*^& *$-$44^* $\pm$ *36^*\cr
\omit**{\bf\SMICA\ } \itE\hfil&&\cr
Local & *!26*^ $\pm$ *32^*& *!35*^ $\pm$ *34^*& *!20^* $\pm$ *30^*& \dots&
** *!26*^ $\pm$ *32^*& *!34^* $\pm$ *34^*& *!20^* $\pm$ *30^*& \dots\cr
Equilateral & !144^* $\pm$ 141^*& !156^* $\pm$ 143^*& !$147$^* $\pm$ 159^*& \dots&
** !144^* $\pm$ 141^*& !155^* $\pm$ 143^*& !147^* $\pm$ 159^*& \dots\cr
Orthogonal & $-$128^* $\pm$ *72^*& $-$128^* $\pm$ *75^*& $-$137^* $\pm$ *73^*& \dots&
** $-$128^* $\pm$ *72^*& $-$126^* $\pm$ *75^*& $-$137^* $\pm$ *73^*& \dots\cr
\omit**{\bf\SMICA\ }\itTpE\hfil&&\cr
Local & !**6.5 $\pm$ **5.0& !**5.8 $\pm$ **4.9& !**4.0 $\pm$ **4.8& !**4.8 $\pm$ **4.9&
** !**0.8 $\pm$ **5.0& !**0.7 $\pm$ **4.9& **$-$0.6 $\pm$ **4.8& !**0.7 $\pm$ **4.9\cr
Equilateral & !**3^* $\pm$ *43^*& !*12^* $\pm$ *44^*& !**5^* $\pm$ *48^*& !**6^* $\pm$ *42^*&
** !**3^* $\pm$ *43^*& !**9^* $\pm$ *44^*& !**3^* $\pm$ *48^*& !**5^* $\pm$ *42^*\cr
Orthogonal & *$-$36^* $\pm$ *21^*& *$-$34^* $\pm$ *22^*& *$-$30^* $\pm$ *21^*& *$-$37^* $\pm$ *21^*&
** *$-$25^* $\pm$ *21^*& *$-$24^* $\pm$ *22^*& *$-$21^* $\pm$ *21^*& *$-$30^* $\pm$ *21^*\cr
\noalign{\vskip 6pt}
\omit**{\bf\SEVEM\ } \itT\hfil&&\cr
Local &  !*11.3 $\pm$ **5.7& !**9.7 $\pm$ **5.4& !**8.1 $\pm$ **5.8& !**9.3 $\pm$ **6.0&
** !**3.6 $\pm$ **5.7& !**2.3 $\pm$ **5.4& !**1.4 $\pm$ **5.8& !**3.1 $\pm$ **6.0\cr
Equilateral & **$-$3^* $\pm$ *69^*& *$-$16^* $\pm$ *66^*&  *$-$11^* $\pm$ *75^*& **$-$6^* $\pm$ *68*^&
** **$-$2^* $\pm$ *69^*& *$-$18^* $\pm$ *66^*& *$-$12^* $\pm$ *75^*& **$-$7^* $\pm$ *68^*\cr
Orthogonal & *$-$59^* $\pm$ *33^*& *$-$47^* $\pm$ *33^*& *$-$49^* $\pm$ *34^*& *$-$66^* $\pm$ *36^*&
** *$-$36^* $\pm$ *33^*& *$-$20^* $\pm$ *33^*& *$-$23^* $\pm$ *34^*& *$-$48^* $\pm$ *36^*\cr
\omit**{\bf\SEVEM\ } \itE\hfil&&\cr
Local & !*60^* $\pm$ *42^*& !*62^* $\pm$ *42^*& !*44^* $\pm$ *38^*&  \dots&
** *!60^* $\pm$ *42^*& !*61^* $\pm$ *42^*& *!44^* $\pm$ *38^*& \dots\cr
Equilateral & !292^* $\pm$ 167^*& !320^* $\pm$ 154^*& !302^* $\pm$ 183^*&  \dots&
** !292^* $\pm$ 167^*& !318^* $\pm$ 154^*& !302^* $\pm$ 183^*& \dots\cr
Orthogonal & $-$184^* $\pm$ *91^*& $-$156^* $\pm$ *93^*& $-$172^* $\pm$ *91^*& \dots&
** $-$183^* $\pm$ *91^*& $-$154^* $\pm$ *93^*& $-$172^* $\pm$ *91^*& \dots\cr
\omit**{\bf\SEVEM\ } \itTpE\hfil&&\cr
Local &  !**9.3 $\pm$ **5.2& !**8.3 $\pm$ **4.9& !**6.4 $\pm$ **5.0& !**7.9 $\pm$ **5.0&
** !**3.3 $\pm$ **5.2& !**2.8 $\pm$ **4.9& !**2.1 $\pm$ **5.0& !**3.5 $\pm$ **5.0\cr
Equilateral & !**9^* $\pm$ *47^*& !*21^* $\pm$ *48^*& !*15^* $\pm$ *52^*& !**5^* $\pm$ *45^*&
** !**8^* $\pm$ *47^*& !*17^* $\pm$ *48^*& !*14^* $\pm$ *52^*&  !**4^* $\pm$ *45^*\cr
Orthogonal & *$-$50^* $\pm$ *23^*& *$-$46^* $\pm$ *23^*& *$-$44^* $\pm$ *23^*& *$-$55^* $\pm$ *22^*&
** *$-$39^* $\pm$ *23^*& *$-$35^* $\pm$ *23^*& *$-$33^* $\pm$ *23^*&  *$-$47^* $\pm$ *22^*\cr
\noalign{\vskip 6pt}
\omit**{\bf \NILC\ } \itT\hfil&&\cr
Local & !*10.5 $\pm$ **5.6& !**8.7 $\pm$ **5.4& !**6.4 $\pm$ **5.6& !**8.0 $\pm$ **6.2&
** !**3.0 $\pm$ **5.6& !**1.4 $\pm$ **5.4& !**0.3 $\pm$ **5.6& !**2.2 $\pm$ **6.2\cr
Equilateral & *$-$28^* $\pm$ *69^*& *$-$45^* $\pm$ *66^*& *$-$31^* $\pm$ *75^*& *$-$15^* $\pm$ *66^*&
** *$-$28^* $\pm$ *69^*& *$-$47^* $\pm$ *66^*& *$-$30^* $\pm$ *75^*& *$-$17^* $\pm$ *67^*\cr
Orthogonal & *$-$67^* $\pm$ *33^*& *$-$48^* $\pm$ *33^*& *$-$50^* $\pm$ *33^*& *$-$63^* $\pm$ *35^*&
** *$-$45^* $\pm$ *33^*& *$-$22^* $\pm$ *33^*& *$-$28^* $\pm$ *33^*& *$-$44^* $\pm$ *35^*\cr
\omit**{\bf \NILC\ } \itE\hfil&&\cr
Local &  **!0^* $\pm$ *33*^& !*18^* $\pm$ *36^*& **$-$1^* $\pm$ *30^*& \dots&
** **$-$1^* $\pm$ *33*^& !*17^* $\pm$ *36^*& **$-$2^* $\pm$ *30*^& \dots\cr
Equilateral & !*75^* $\pm$ 140^*& *!97^* $\pm$ 141^*& !*64^* $\pm$ 162^*& \dots&
** !*75^* $\pm$ 140^*& !*96^* $\pm$ 141^*& !*64^* $\pm$ 162^*& \dots\cr
Orthogonal & *$-$79^* $\pm$ *76^*& *$-$96^* $\pm$ *81^*&  *$-$78^* $\pm$ *77^*& \dots&
** *$-$78^* $\pm$ *76^*& *$-$94^* $\pm$ *81^*& *$-$78^* $\pm$ *77^*& \dots\cr
\omit**{\bf \NILC\ } \itTpE\hfil&&\cr
Local & !**6.9 $\pm$ **5.1& !**6.1 $\pm$ **4.9&  !**3.3 $\pm$ **4.9& !**5.3 $\pm$ **5.2&
** !**1.2 $\pm$ **5.1& !**0.9 $\pm$ **4.9& **$-$2.4 $\pm$ **4.9& **!4.4 $\pm$ **5.2\cr
Equilateral & **$-$9^* $\pm$ *44^*& **$-$4^* $\pm$ *44^*& *$-$15^* $\pm$ *50^*& !**8^* $\pm$ *42^*&
** **$-$9*^ $\pm$ *44^*& **$-$7^* $\pm$ *44^*& *$-$16^* $\pm$ *50^*& !**4^* $\pm$ *42^*\cr
Orthogonal & *$-$35^* $\pm$ *21^*& *$-$31^* $\pm$ *22*^& *$-$27^* $\pm$ *23*^&  *$-$32^* $\pm$ *21^*&
** *$-$25^* $\pm$ *21^*& *$-$21*^ $\pm$ *22*^& *$-$16^* $\pm$ *23^*& *$-$26^* $\pm$ *21^*\cr
\noalign{\vskip 6pt}
\omit**{\bf\Commander\ } \itT\hfil&&\cr
Local &  **!9.6 $\pm$ **6.1&  **!9.4 $\pm$ **5.7&  **!6.4 $\pm$ **6.6& **!7.9 $\pm$ **6.3&
** !**4.0 $\pm$ **6.1& **!2.4 $\pm$ **5.7& !**1.4 $\pm$ **6.6& **!3.3 $\pm$ **6.3\cr
Equilateral & *$-$19^* $\pm$ *71^*& *$-$36^* $\pm$ *68^*&  **$-$3^* $\pm$ *77^*& *$-$14^* $\pm$ *70^*&
** *$-$20^* $\pm$ *71^*& *$-$38^* $\pm$ *68^*& **$-$4^* $\pm$ *77*^& *$-$18^* $\pm$ *70*^\cr
Orthogonal & *$-$49^* $\pm$ *35^*& *$-$38^* $\pm$ *34^*& *$-$49^* $\pm$ *36^*& *$-$45^* $\pm$ *37^*&
** *$-$29^* $\pm$ *35^*& *$-$12*^ $\pm$ *34^*& *$-$25*^ $\pm$ *38*^& *$-$28^* $\pm$ *37^*\cr
\omit**{\bf\Commander\ } \itE\hfil&&\cr
Local &  !*33^* $\pm$ *39^*& !*56*^ $\pm$ *40^*& !*28^* $\pm$ *37^*& \dots&
** !*33^* $\pm$ *39^*& *!55^* $\pm$ *40^*& *!28*^ $\pm$ *37*^& \dots\cr
Equilateral & !327*^ $\pm$ 165*^& !369*^ $\pm$ 157^*& !278^* $\pm$ 178^*& \dots&
** !327^* $\pm$ 165*^& !368^* $\pm$ 157^*& !278*^ $\pm$ 178^*& \dots\cr
Orthogonal & *$-$52^* $\pm$ *88*^& *$-$70^* $\pm$ *88*^& *$-$56*^ $\pm$ *81^*& !\dots&
** *$-$52^* $\pm$ *88^*& *$-$67*^ $\pm$ *88*^& *$-$56*^ $\pm$ *81^*& \dots\cr
\omit**{\bf\Commander\ } \itTpE\hfil&&\cr
Local & !**7.7 $\pm$ **5.2& !**7.9 $\pm$ **5.0& !**5.2 $\pm$ **5.4& !**6.8 $\pm$ **5.2&
** !**3.7 $\pm$ **5.2& **!3.0 $\pm$ **5.0& **!1.6 $\pm$ **5.4& !**3.7 $\pm$ **5.2\cr
Equilateral & !*16^* $\pm$ *46^*& !*26^* $\pm$ *45^*& !*30^* $\pm$ *50*^& !*29^* $\pm$ *46^*&
** !*14*^ $\pm$ *46*^& !*23*^ $\pm$ *45^*& !*28^* $\pm$ *50^*& *!26^* $\pm$ *46^*\cr
Orthogonal & *$-$37^* $\pm$ *22^*& *$-$37^* $\pm$ *23^*& *$-$39^* $\pm$ *23*^& *$-$35*^ $\pm$ *22^*&
** *$-$29^* $\pm$ *22*^& *$-$27*^ $\pm$ *23*^& *$-$30^* $\pm$ *23*^& *$-$28^* $\pm$ *22*^\cr
\noalign{\vskip 3pt\hrule\vskip 4pt}}}
\endPlancktablewide                 
\endgroup
\end{table*}                        

\begin{table}[tb]                 
\begingroup
\newdimen\tblskip \tblskip=5pt
\caption{Results for the $f_{\rm NL}$ parameters of the primordial local, 
equilateral, and orthogonal shapes, determined by the KSW estimator from the
\SMICA\ foreground-cleaned map. Both independent
single-shape results and results with the ISW-lensing bias subtracted are 
reported; error bars are $68\,\%$ CL. The difference between the last column
in this table and the corresponding values in the previous table is that
in the second column here the equilateral and orthogonal shapes have been
analysed jointly. The final reported results of the paper are shown in bold.}
\label{Tab_KSW+SMICA}
\nointerlineskip
\vskip -6mm
\footnotesize
\setbox\tablebox=\vbox{
   \newdimen\digitwidth
   \setbox0=\hbox{\rm 0}
   \digitwidth=\wd0
   \catcode`*=\active
   \def*{\kern\digitwidth}
   \newdimen\signwidth
   \setbox0=\hbox{+}
   \signwidth=\wd0
   \catcode`!=\active
   \def!{\kern\signwidth}
\newdimen\dotwidth
\setbox0=\hbox{.}
\dotwidth=\wd0
\catcode`^=\active
\def^{\kern\dotwidth}
\halign{\hbox to 1in{#\leaderfil}\tabskip 1em&
\hfil#\hfil\tabskip 1em&
\hfil#\hfil\tabskip 0pt\cr
\noalign{\vskip 10pt\doubleline\vskip 2pt}
\omit&\multispan2\hfil $f_{\rm NL}$(KSW)\hfil\cr
\omit&\multispan2\hrulefill\cr
Shape and method\hfill&\hfil Independent\hfil&
\hfil ISW-lensing subtracted\hfil\cr
\noalign{\vskip 2pt}
\noalign{\vskip 4pt\hrule\vskip 6pt}
\omit\hfil \SMICA\,\, (\itT) \hfil&\cr
Local& !10.2!$\pm$!*5.7&
{\bf*!2.5!$\pm$!*5.7}\cr
Equilateral& $-$13*^!$\pm$!70^*&
{\bf$-$16*^!$\pm$!70^*}\cr
Orthogonal& $-$56*^!$\pm$!33^*&
{\bf$-$34^*!$\pm$!33^*}\cr
\noalign{\vskip 5pt}
\omit\hfil {\SMICA\,\,(\itTpE) }\hfil&\cr
Local& *!6.5!$\pm$!*5.0&
{\bf*!0.8!$\pm$!*5.0}\cr
Equilateral& *!3^*!$\pm$!43^*&
{\bf*$-$4^*!$\pm$!43^*}\cr
Orthogonal& $-$36^*!$\pm$!21^*&
{\bf$-$26^*!$\pm$!21^*}\cr
\noalign{\vskip 3pt\hrule\vskip 4pt}}}
\endPlancktable                    
\endgroup
\end{table}                        

\subsection{Bispectrum reconstruction}
\label{Bisp_Rec}
\subsubsection{Modal bispectrum reconstruction}

\begin{figure*}
\centering
\includegraphics[width=.49\linewidth]{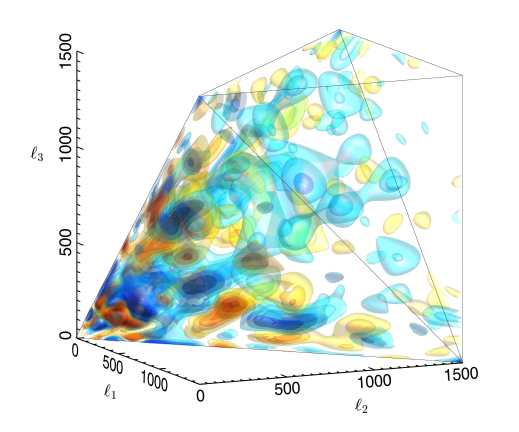}
\includegraphics[width=.49\linewidth]{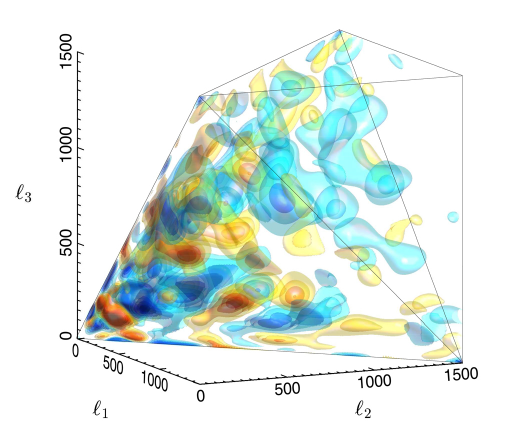}
\includegraphics[width=.75\linewidth]{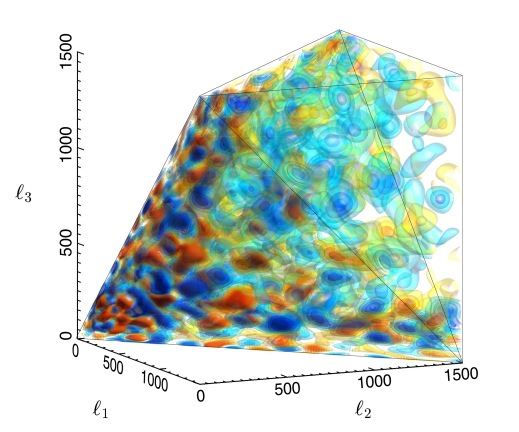}
\caption{Modal bispectrum reconstruction for \Planck\ 2013 (top left) and 2015 (top right) temperature-only data, both using the \SMICA\ maps.  Here, we restrict the 2015 resolution to the same as 2013, using  similar polynomials with $n_{\rm max} = 601$.  The two bispectra are very close to being in complete agreement in the signal-dominated regime shown up to $\ell_{\rm max}= 1500$.  In the lower panel, we show the \Planck\ 2015temperature bispectrum at high resolution using the full $n_{\rm max} = 2001$ polynomial modes.  Large-scale features in the top panels become subdivided but the main 2013 signals remain, notably a stronger measurement of the ISW-lensing signal (the regular oscillations in the squeezed limit).}
\label{fig:reconstruct}
\end{figure*}

\begin{figure*}
\centering
\includegraphics[width=.49\linewidth]{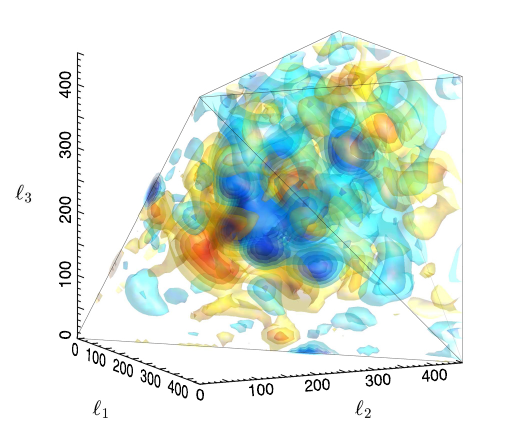}
\includegraphics[width=.49\linewidth]{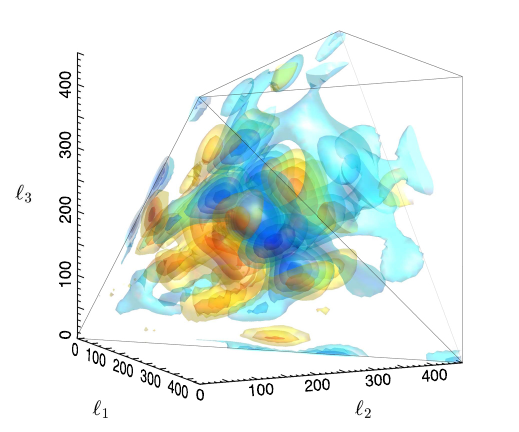}
\caption{Modal reconstruction for the WMAP-9 bispectrum (left) and the Planck \SMICA\ 2015 \itT-only bispectrum (right) plotted for the domain $\ell \le 450$, using identical isosurface levels.  Here, we employed the full 2001 eigenmodes for both the \Planck\ analysis at $\lmax =2000$ and for WMAP-9 analysis at $\lmax =600$, but for comparison purposes we have only used the first 600 eigenmodes in order to obtain a comparable resolution.   The main features in the WMAP-9 bispectrum have counterparts in the \Planck\ version, revealing an oscillatory pattern in the central region, as well as features on the tetrapyd surface.   The WMAP-9 bispectrum has a much larger noise signal beyond $\ell =350$ than the more sensitive \Planck\ experiment, leading to apparent residuals in this region.}
\label{fig:WMAPvsPlanck}
\end{figure*}

The starting point for modal bispectrum estimation is the robust extraction of the modal coefficients $\beta_n$ from each of the full mission foreground-separated maps, that is, \SMICA, \SEVEM, \NILC, and \Commander .   The $\beta_n$-coefficients are obtained for each of the temperature, polarization, and mixed bispectrum components, \textit{TTT, TTE, TEE}, and \textit{EEE}.   Their cross-correlation between cleaning methods is an important validation of their accuracy, as we shall discuss in the next section, with excellent correspondence for temperature and some differences remaining in polarization.      The modal basis number $\nmax =2001$ for the full mission analysis has been substantially increased offering a higher effective resolution when compared to the 2013 \Planck\ Data Release where $\nmax=601$ modes were used.   Several different basis functions have been used, including trigonometric functions, sinlog basis functions, and polynomials (closely related to Legendre functions), with the latter chosen because of excellent convergence in the squeezed and flattened limits. 

We can reconstruct the full three-dimensional \Planck\ bispectrum, obtained using these basis functions, to visualize its main properties and to determine robustness.   A comparison between the temperature-only bispectra from the Nominal Mission and full mission at the same $\nmax=601$ modal resolution is shown in Fig.~\ref{fig:reconstruct}.   Note the excellent agreement with all the main features replicated in the new data.   In Fig.~\ref{fig:reconstruct} in the third bispectrum, we also demonstrate the much higher bispectrum resolution achieved with the full $\nmax=2001$ modes.   The tetrapyd shape reflects the constraints on the wavenumbers $\ell_1, \ell_2$, and $\ell_3$, with the squeezed configuration appearing on the axes that lie along one $\ell_i=0$. The expected ISW-lensing bispectrum is an oscillating signal in the squeezed limit along the tetrapyd edges; it is now measured with a significance of $3.0\,\sigma$ (see Sect.~\ref{subsec:lensingISW}).   This ISW-lensing signal sets an interesting benchmark or threshold against which to compare the other strong features observed in the bispectrum and now defined with greater precision.  The original ``plus-minus" feature, with a large positive red peak around $\ell \approx 150$ followed by a larger negative peak near $\ell \approx 250$, remains though with more substructure, together with a broad  negative peak in the equilateral limit around $\ell\approx 900$, which can be associated with the third acoustic peak from the transfer functions.  Oscillatory models, which can connect these three peaks, achieve higher significance. The apparent signal observed in the flattened limit remains, with a distinct pattern of blue and red features on the surface of the tetrapyd.  

We also include a comparison with WMAP-9 in Fig.~\ref{fig:WMAPvsPlanck}, where we have restricted the reconstructions to $\ell_{\rm max} =600$ for comparison with $n_{\rm max} = 601$ modes.  These plots, using identical isosurfaces, show the same bispectrum structure including the ``plus-minus" feature clearly bisecting the main $\ell=200$ peak and the first oscillation of the ISW-lensing bispectrum visible along the lower tetrapyd edges.  The WMAP-9 reconstruction only shows significant differences from \Planck\ in the top right region, where the higher noise levels in WMAP-9 make its reconstruction less reliable.

\begin{figure*}
\centering
\includegraphics[width=.5\linewidth]{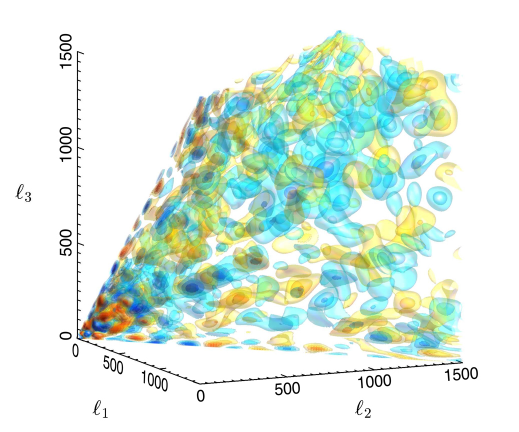}\includegraphics[width=.5\linewidth]{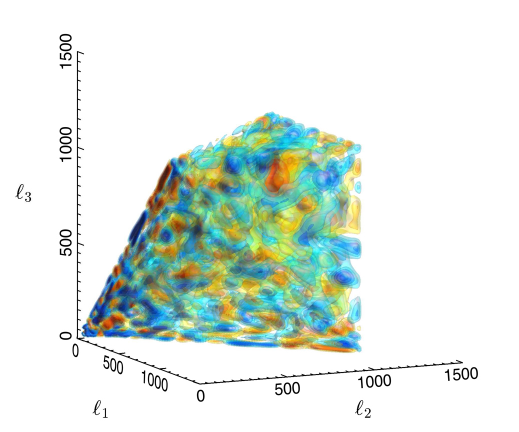}
\includegraphics[width=.5\linewidth]{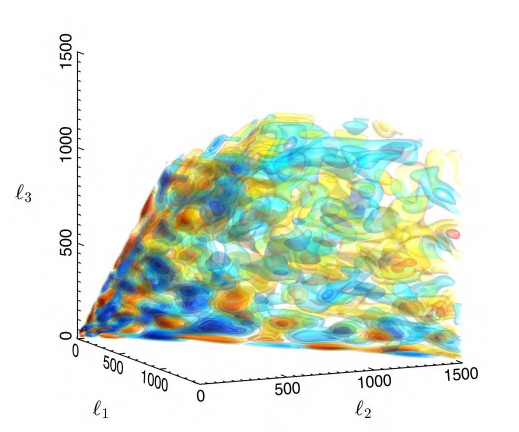}\includegraphics[width=.5\linewidth]{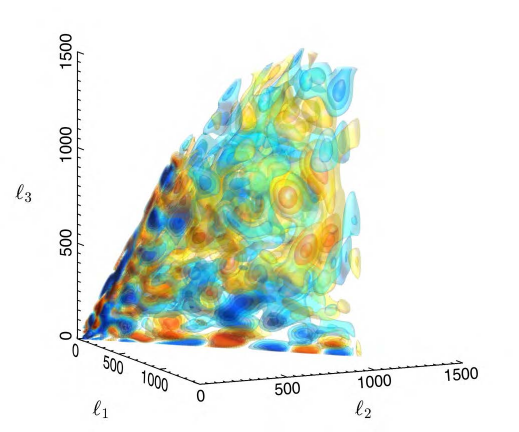}
\caption[]{\small   CMB temperature and polarization bispectrum reconstructions for \Planck\ \SMICA\ maps using the full set of  polynomial modes with $n_{\rm max} = 2001$ and with signal-to-noise weighting. The top  bispectra are the symmetric pure temperature \textit{TTT} (left) plotted with $\ell \le 1500$ and \itE-mode polarization \textit{EEE} (right) shown for $30\le \ell \le 1100$. Below are the mixed temperature/polarization  bispectra with \textit{TTE} on the left (with \itE\ multipoles in the $z$-direction) and \textit{TEE} on the right (with \itT\ multipoles in the $z$-direction).  All S/N thresholds are the same. }
\label{fig:reconstructTE}
\end{figure*}

\begin{figure*}
\centering
\includegraphics[width=.33\linewidth]{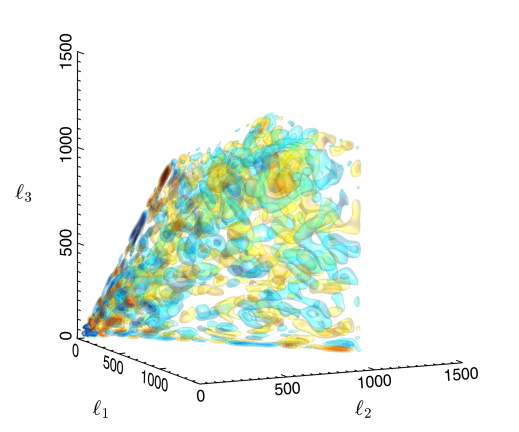}
\includegraphics[width=.33\linewidth]{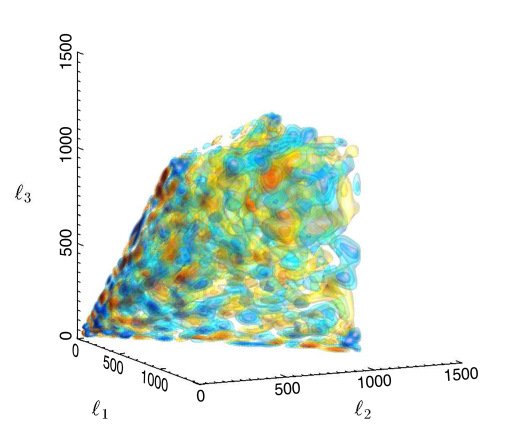}
\includegraphics[width=.33\linewidth]{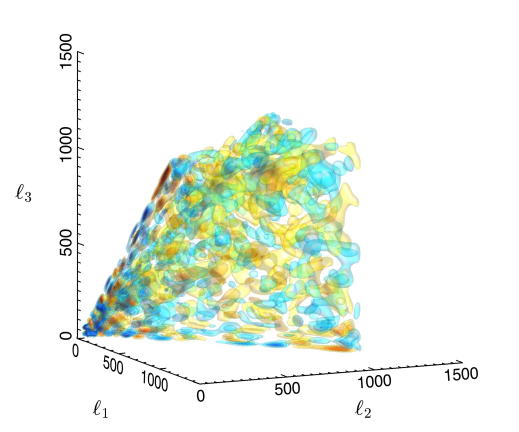}
\caption[]{\small   Comparison of CMB polarization bispectrum \textit{EEE} reconstructions for \Planck\ \NILC, \SEVEM, and \Commander\ foreground-separated maps with signal-to-noise weighting.  Note that these results are not as internally consistent between the four methods, also comparing \SMICA\ shown in Fig.~\ref{fig:reconstructTE}, which is closest to \NILC.  We will compare the underlying modal coefficients below to demonstrate these differences quantitatively.}
\label{fig:reconstructEEE}
\end{figure*}

All four components of the temperature and polarization bispectrum reconstruction obtained from  \SMICA\ are shown in Fig.~\ref{fig:reconstructTE}.  A direct comparison of the \textit{EEE} polarization bispectrum for \SEVEM, \NILC\ and \Commander, is shown in Fig.~\ref{fig:reconstructEEE}, where we note that these are orthogonalized \itE-mode contributions (see the Modal 2 discussion in Sect.~\ref{sec:SEP}).  It is interesting to observe patterns of features evident in the polarization bispectra from the different foreground-cleaned maps, which, although inherently noisier, have qualitative similarities.  At a quantitative level, however, the polarization bispectra modes from different methods are less correlated in polarization than in temperature, as we discuss in Sect.~\ref{sec:Sec_valid_data}.

\begin{figure*}
\begin{center}
\includegraphics[width=1.75in]{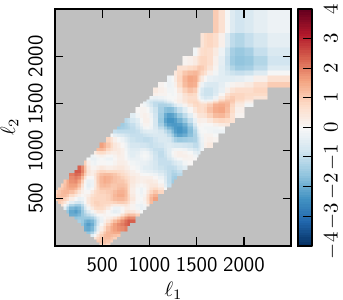}
\includegraphics[width=1.75in]{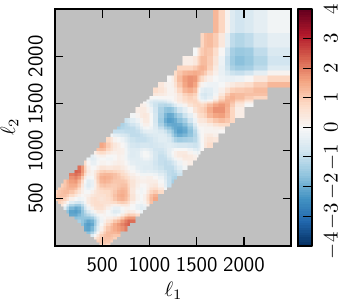}
\includegraphics[width=1.75in]{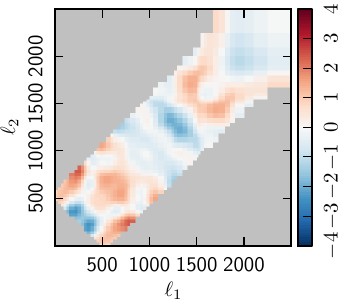}
\includegraphics[width=1.75in]{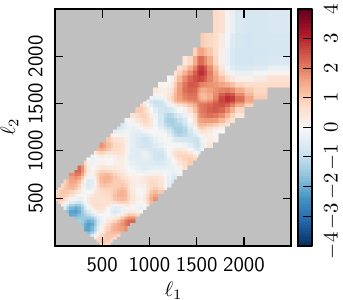}
\includegraphics[width=1.75in]{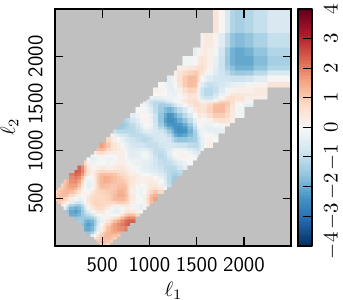}
\includegraphics[width=1.75in]{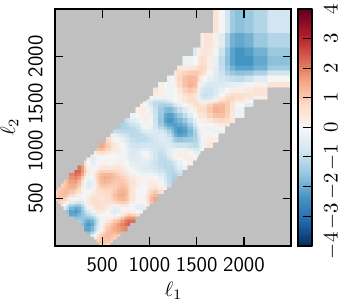}
\includegraphics[width=1.75in]{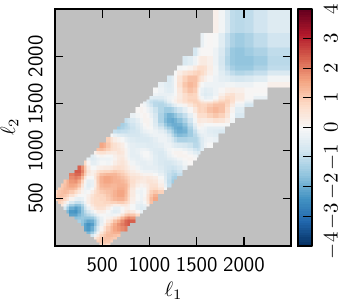}
\includegraphics[width=1.75in]{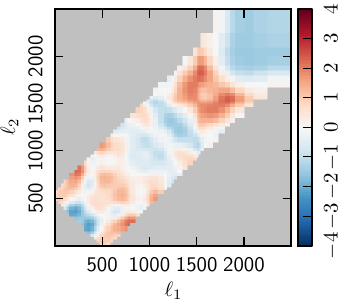}
\includegraphics[width=1.75in]{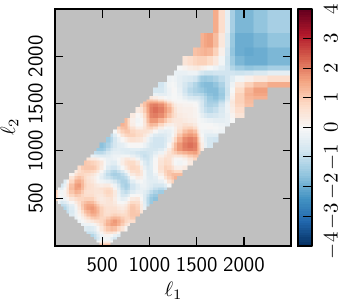}
\includegraphics[width=1.75in]{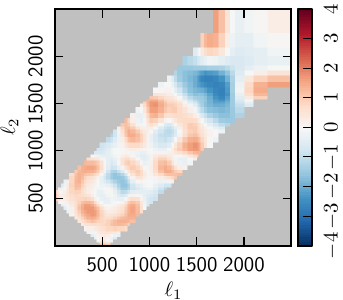}
\includegraphics[width=1.75in]{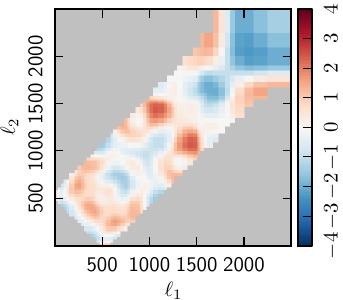}
\includegraphics[width=1.75in]{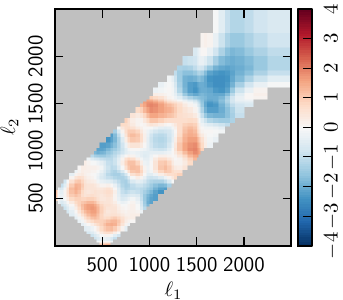}
\includegraphics[width=1.75in]{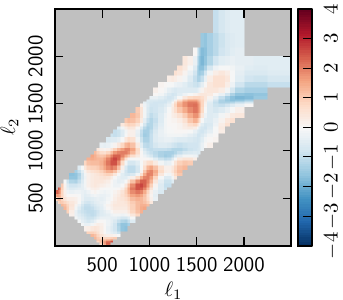}
\includegraphics[width=1.75in]{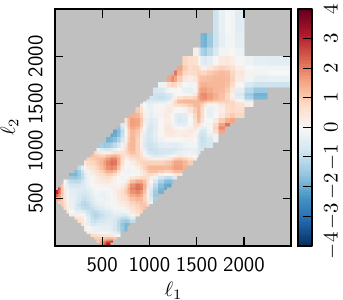}
\includegraphics[width=1.75in]{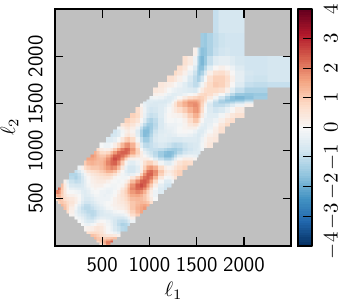}
\includegraphics[width=1.75in]{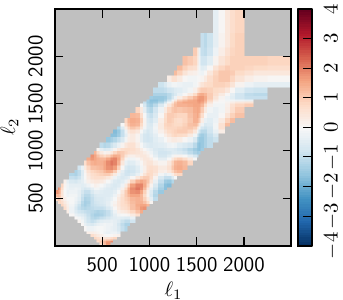}
\includegraphics[width=1.75in]{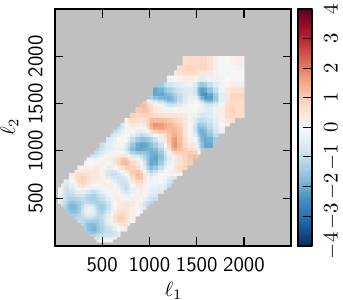}
\includegraphics[width=1.75in]{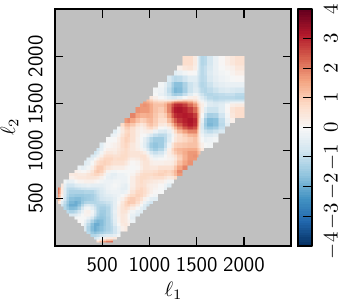}
\includegraphics[width=1.75in]{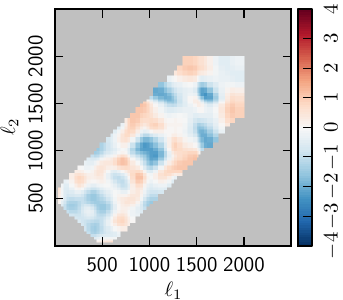}
\includegraphics[width=1.75in]{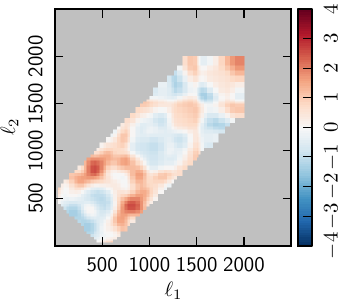}
\end{center}
\caption{Smoothed binned signal-to-noise bispectrum $\mathcal{B}$ for the
\Planck\ 2015 cleaned sky map, as determined
  with the binned estimator, as a function of $\ell_1$ and $\ell_2$ for a
fixed $\ell_3$-bin [518, 548]. 
From left to right results are shown for the
four component separation methods \SMICA, \SEVEM, \NILC, and \Commander.
  From top to bottom are shown: \textit{TTT, TTT} cleaned from radio
  and CIB point sources; \textit{T2E, TE2}; and \textit{EEE}. The colour
  range is in signal-to-noise from $-4$ to $+4$. 
  The light grey regions are where the bispectrum is not defined, either
because it is outside the triangle inequality or because of the cut
$\ell_\mathrm{max}^E = 2000$.}
\label{fig:smoothed_1}
\end{figure*}


\subsubsection{Binned bispectrum reconstruction}

The (reconstructed) binned bispectrum of a given map is a natural
product of the binned bispectrum estimator code (see Sect.~\ref{sec:binned_est}). 
To test if any bin has a significant NG signal, we study the binned 
bispectrum divided by its expected standard deviation, a quantity for which 
we will use the symbol $\mathcal{B}_{i_1 i_2 i_3}$. With the
binning used in the estimator, the pixels are dominated by noise. We
thus smooth in three dimensions with a Gaussian kernel of a certain width 
$\sigma_{\rm  bin}$. To avoid edge effects, due to the sharp boundaries of the
domain of definition of the bispectrum, we renormalize the smoothed
bispectrum, so that the pixel values would be normal-distributed for a
Gaussian map.  

\begin{figure*}
\begin{center}
\includegraphics[width=1.75in]{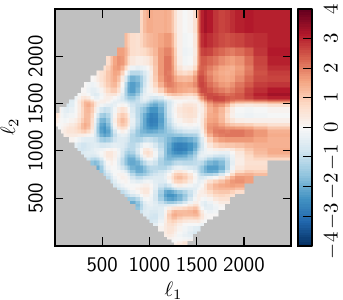}
\includegraphics[width=1.75in]{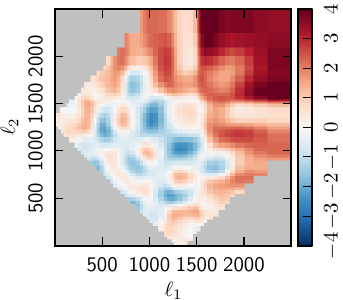}
\includegraphics[width=1.75in]{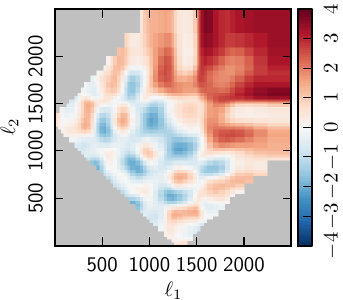}
\includegraphics[width=1.75in]{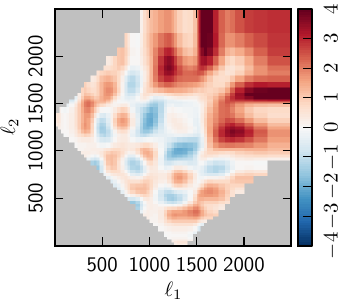}

\includegraphics[width=1.75in]{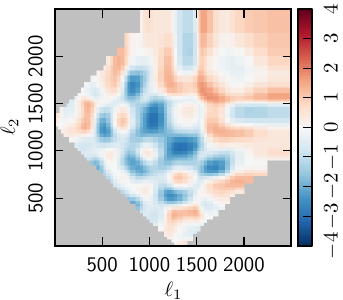}
\includegraphics[width=1.75in]{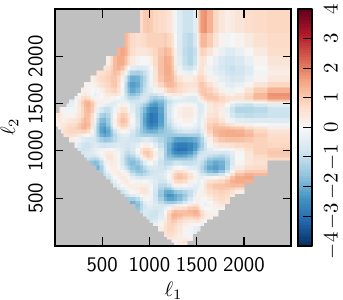}
\includegraphics[width=1.75in]{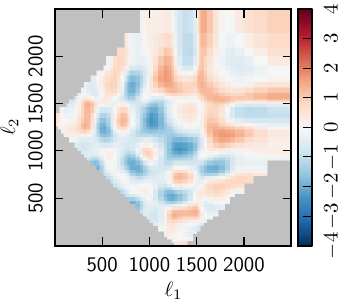}
\includegraphics[width=1.75in]{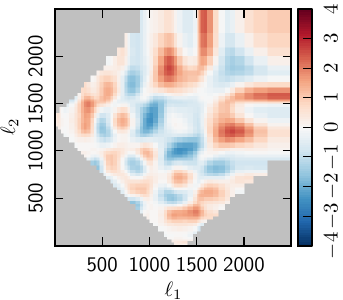}

\includegraphics[width=1.75in]{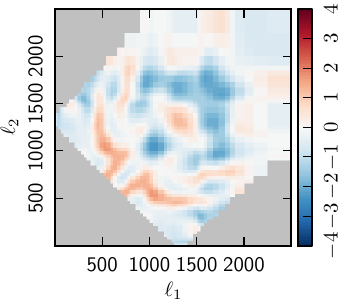}
\includegraphics[width=1.75in]{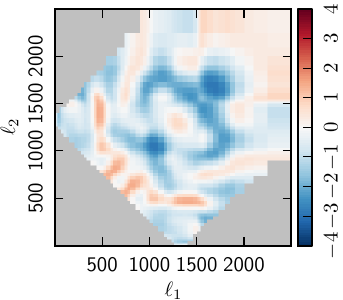}
\includegraphics[width=1.75in]{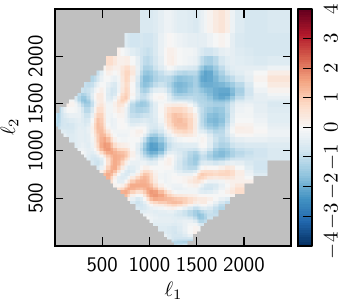}
\includegraphics[width=1.75in]{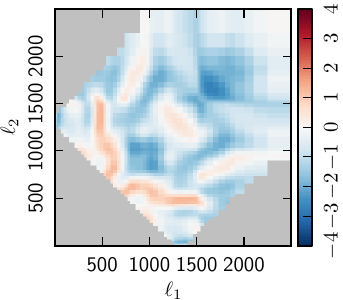}

\includegraphics[width=1.75in]{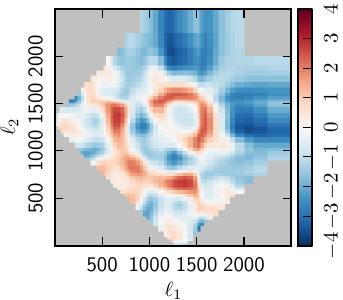}
\includegraphics[width=1.75in]{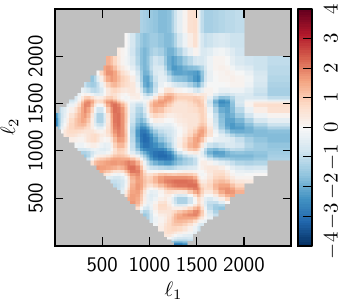}
\includegraphics[width=1.75in]{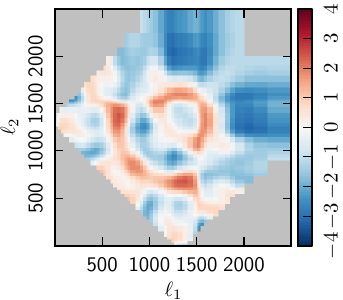}
\includegraphics[width=1.75in]{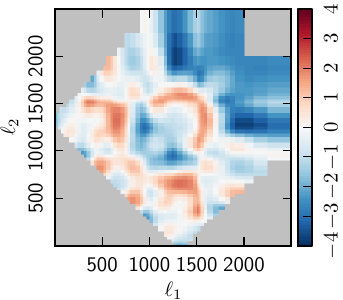}

\includegraphics[width=1.75in]{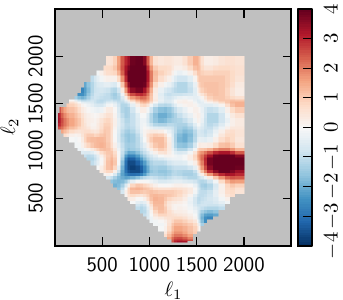}
\includegraphics[width=1.75in]{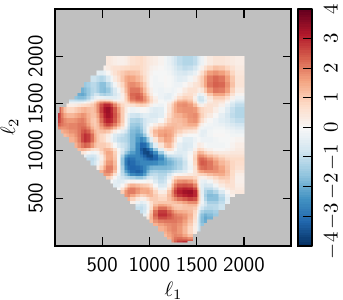}
\includegraphics[width=1.75in]{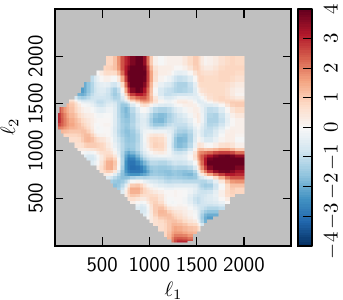}
\includegraphics[width=1.75in]{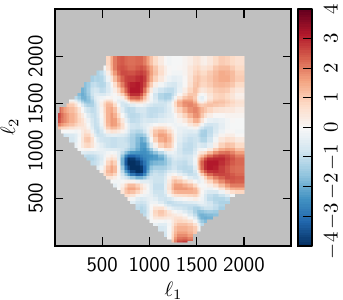}

\end{center}
\caption{Similar to Fig.~\ref{fig:smoothed_1}, but with $\ell_3 \in [1291, 1345] $. }
\label{fig:smoothed_2}
\end{figure*}

In Figs.~\ref{fig:smoothed_1} and
\ref{fig:smoothed_2}, we show slices of this smoothed binned signal-to-noise
bispectrum $\mathcal{B}_{i_1 i_2 i_3}$ with a Gaussian smoothing of 
$\sigma_{\rm bin} = 2$, as a function of $\ell_1$ and $\ell_2$. 
Very red or very blue regions correspond to a significant NG
of any type.
The two figures only differ in the value chosen for the $\ell_3$-bin, 
which is $[518, 548]$ for the first figure, and $[1291, 1345]$ for the second.
We have defined two cross-bispectra here: $B_{i_1 i_2 i_3}^{T2E} \equiv
B_{i_1 i_2 i_3}^{TTE} + B_{i_1 i_2 i_3}^{TET} + B_{i_1 i_2 i_3}^{ETT}$; and 
$B_{i_1 i_2 i_3}^{TE2} \equiv B_{i_1 i_2 i_3}^{TEE} + B_{i_1 i_2 i_3}^{ETE}
+ B_{i_1 i_2 i_3}^{EET}$. These two cross-bispectra are then divided by their 
respective
standard deviations (taking into account the covariance terms) to produce the
corresponding $\mathcal{B}_{i_1 i_2 i_3}^{T2E}$ and $\mathcal{B}_{i_1 i_2 i_3}^{TE2}$.
Those three different permutations are not equal a priori due to the condition
$i_1 \leq i_2 \leq i_3$, which is implemented in the code to reduce computations
by a factor of six. However, part of the smoothing procedure involves adding the other 
five identical copies, so that in the end the plots are symmetric under
interchange of $\ell_1$ and $\ell_2$ (and $\mathcal{B}_{i_1 i_2 i_3}$ is symmetric
under interchange of all its indices). The grey areas in the plots are regions
where the bispectrum is not defined, either because it is outside of the 
triangle inequality, or because of the limitation $\ell_\mathrm{max}^E = 2000$.
Given that in both plots $\ell_3$ is fixed at less than 2000, 
$\mathcal{B}_{i_1 i_2 i_3}^{TE2}$ is not defined if both $\ell_1$ and $\ell_2$ are 
larger than 2000, while $\mathcal{B}_{i_1 i_2 i_3}^{EEE}$ is undefined if either 
$\ell_1$ or $\ell_2$ (or both) are larger than 2000.

Results are shown for the four component separation methods \SMICA, \SEVEM,
\NILC, and \Commander, and for \textit{TTT, T2E, TE2}, and \textit{EEE}. In 
addition we show on the second line of each figure the result for \textit{TTT} 
with the radio (unclustered) and CIB (clustered) point
source bispectra subtracted according to their jointly measured amplitudes.
It is clear, in particular in the second figure, that at higher $\ell$ there
is a very significant point source contamination in the cleaned \textit{TTT} bispectra,
in agreement with the results of Table~\ref{Table:bps_and_ACIB}.
However, after removing it we do not see a clear signal of any other
residual NG. Of course this is for the moment only a 
qualitative statement; more quantitative tools for studying the amount
of NG in these smoothed bispectra are in development.

Looking at the polarized bispectra in the high-$\ell_3$ slices, in particular 
\textit{TE2} and \textit{EEE}, we do see some bluer and redder regions that
might indicate residual NG. This agrees with statements made
earlier, and discussed in greater detail in the next section, that the 
\Planck\ polarized bispectrum is for the moment not as clean and 
well-understood as the temperature one.
We also see a very good qualitative agreement between the four component 
separation methods in temperature, which worsens somewhat when mixing
in more and more polarization; in particular \SMICA\ and \NILC\ give very
similar results.

\subsection{Primordial Curvature Reconstruction}
In this section, we compress the information in $T$ and $E$ maps into maps of projected primordial curvature fluctuations, $\zeta$. To the extent that the primary CMB temperature and polarization are Gaussian fields and the early universe $\zeta$ fluctuations are presumed Gaussian, all of the information in $T,E$ is encoded in the mean field (Wiener-filtered map) plus fluctuations, characterized by a covariance matrix; it is just a different compression/re-expression of the original statistics. Thus, one could look for non-Gaussian deviations by directly evaluating the 3-point and higher statistics of this re-expression. Since fluctuations as well as the mean are given, a full description of the errors follows. That a Gaussian assumption is made does not mean that the Wiener-filtered map is in fact Gaussian, since the constraining data may drive it from a map consistent with Gaussianity. Hence one can search in the statistics of the Wiener-filtered map for evidence for non-Gaussianity. 

The weighting of the temperature and $E$-mode polarization associated with the Wiener filter is an optimal one (inverse-total-covariance weighted), and estimators for non-Gaussianity constructed using this expression are nearly optimal. However, since a (fiducial) primordial power spectrum is assumed for the Wiener filter (usually a uniform $n_s$ $\zeta$-spectrum, using the best fit Planck 2015 scalar spectral index and power amplitude $A_s$),  this approach is best suited for perturbative non-Gaussianity of the sort we treat in this paper. In practice, the estimators used here for non-Gaussianity act directly on temperature and polarization data, rather than through the intermediary of optimal $\zeta$-maps and their fluctuations. 

\begin{figure}[!t]
\centering
\includegraphics[width=0.96\linewidth]{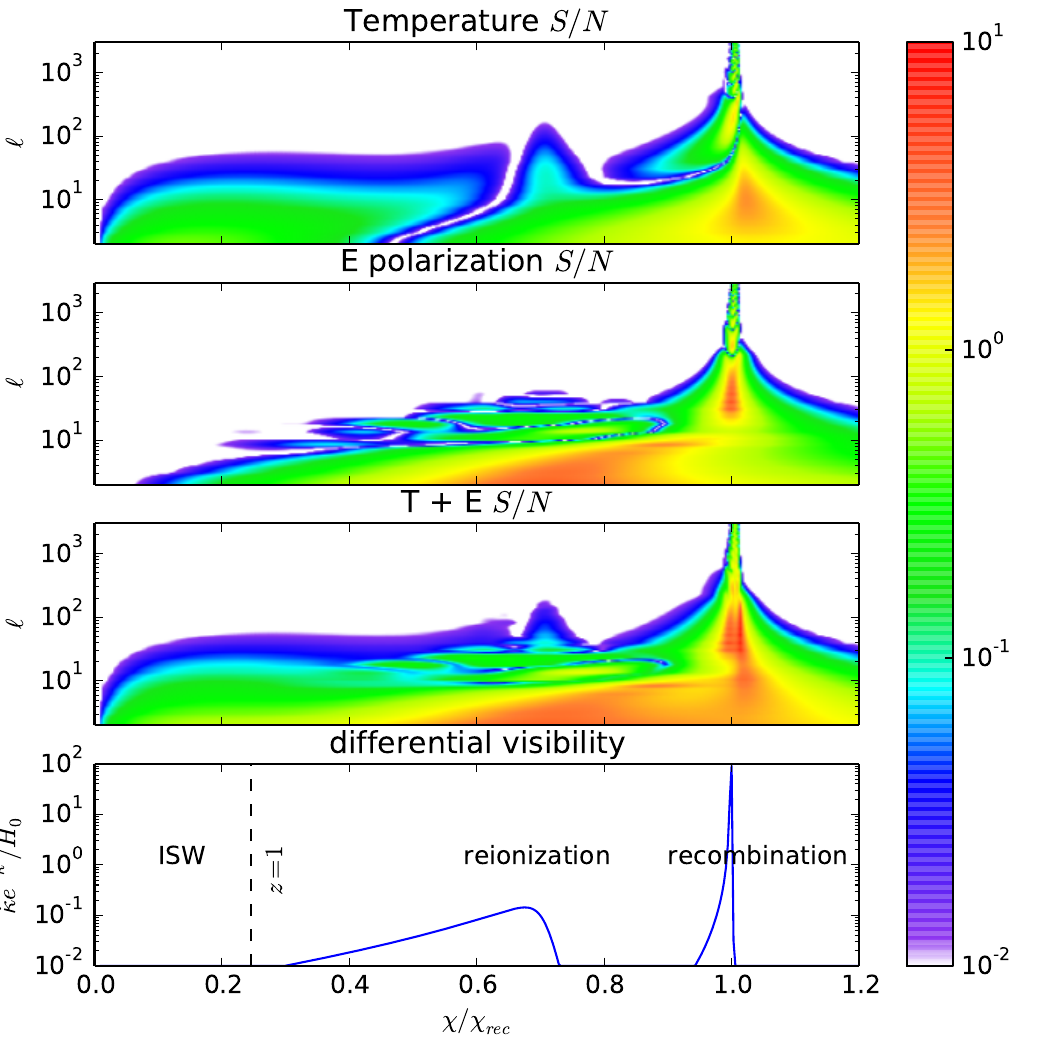}
\includegraphics[width=0.96\linewidth]{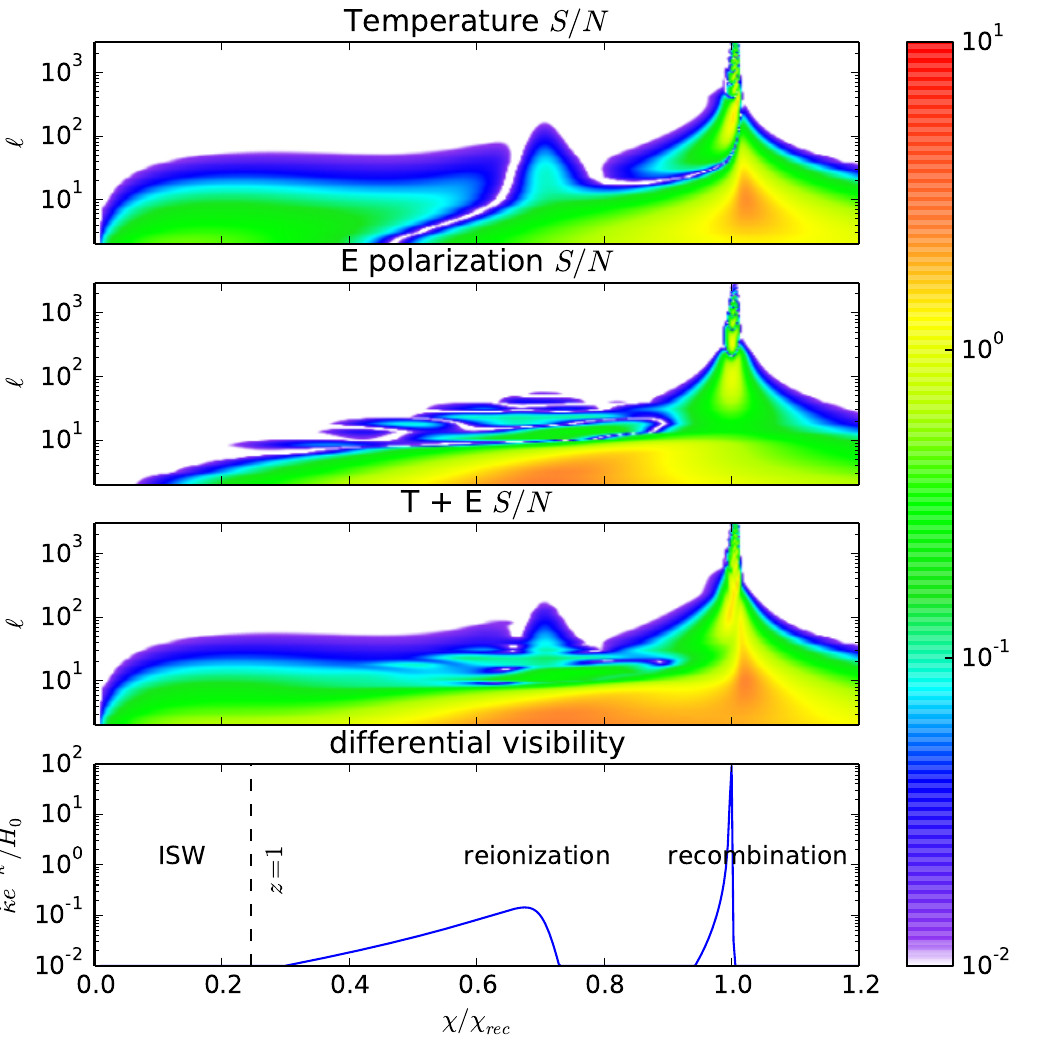}
\caption{Signal-to-noise for $T$ and $E$ as a function of multipole $L$ and comoving distance $\chi$ for an ideal cosmic-variance limited experiment (upper panels) and for the Planck experiment, with noise determined from FFP8 simulations (lower panels). The complementary nature of the information provided by $T$ and $E$ is evident. The differential visibility is shown for comparison.}\label{fig:zetaS2N} 
\end{figure}

 The scalar fluctuations can be expressed in terms of the curvature variable $\zeta({\bf x}) =\ln a ({\bf x})$ on uniform total density hypersurfaces, where $a$ is the inhomogeneous expansion factor, or, equivalently, by its wavenumber transform   $\zeta({\bf k})$. In turn, $\zeta({\bf k})$ can be  expanded in multipoles,  $\zeta_{LM\kappa} (k)$. Instead of the magnitude of the wavenumber $k$, a mixed representation gives a multipole expansion at each comoving distance from our location, $\chi$, $\zeta_{LM\kappa} (\chi )$, $M=0,. . . ,L$. Here $\kappa =c$ or $s$, with $\kappa =c$ referring to the real part of $\zeta_{LM}$ and $\kappa =s$ to the imaginary part. For $M=0$ there is no $s$ component, only $c$. 
The mean of the $\zeta$ field given the temperature field $T$ and its covariance describing allowed fluctuations about that mean are 
\begin{eqnarray}
\langle\zeta_{LM\kappa} (\chi ) \,\vert\,  a_{LM\kappa}^{T} \rangle & =& C_L^{\zeta T} (\chi) \left[ C_L^{TT} \right]^{-1}a_{LM\kappa}^T  \,  \label{eq:zetaT} \\
\langle \delta \zeta_{LM\kappa} (\chi) \delta \zeta_{L^\prime M^\prime \kappa^\prime} (\chi^\prime )\,\vert\, T \rangle & =&  \delta_{LL^\prime}  \delta_{MM^\prime} \delta_{\kappa \kappa^\prime} \times \\ 
\bigg\{ C_L^{\zeta \zeta}(\chi , \chi^\prime )  &-&    C_L^{\zeta T} (\chi  ) \left[C_L^{TT}\right]^{-1}  C_L^{T\zeta } ( \chi^\prime )\bigg\} \, . \label{eq:varzetaT} \nonumber
 \end{eqnarray}
In these expressions, $C^{TT}$ is the total covariance for the temperature, including both signal and noise variances. The specific forms assume the $C^{TT}$  matrix is diagonal in multipole space, depending only on $C_L^{TT}$, but if the noise is inhomogeneous it will have off-diagonal components and the equations become matrix equations.  In this section, we assume the noise is diagonal. Eq.~(\ref{eq:zetaT})  shows the unsurprising result that for each $LM\kappa$ only one mode is determined by $T$. Replacing $T$ by $E$ gives $\langle \zeta_{LM\kappa} (\chi ) \,|\,  a_{LM\kappa}^{E} \rangle$ and $\langle \delta \zeta \delta \zeta^\dagger \vert E \rangle$. Although $T$ and $E$ are correlated, the uncorrelated part of $E$ delivers a different mode for $\zeta$ from the one given by $T$. Thus  when $\zeta$ is constrained by both $T$ and $E$, the two modes deliver substantially more information than for $T$ and $E$ alone, and the fluctuations about the mean are thereby diminished. This is further helped by the acoustic oscillations of polarization being out of phase with those for $T$, so when $T$ is down, $E$ is not, and vice versa. The interplay of the two modes is quantified by signal-to-noise in Fig.~\ref{fig:zetaS2N}. 

When both $T$ and $E$ are included as constraints, a two by two matrix appears which includes the $ C^{TE}$ correlation in the off-diagonal as well as the $ C^{TT}$  and $ C^{EE}$  along the diagonal: 
\begin{eqnarray}
&& \langle \zeta_{LM\kappa} (\chi ) \,\vert\,  a_{LM\kappa}^{T} , a_{LM\kappa}^{E} \rangle  = \nonumber\\ 
\label{eq:zetaTE}
&& \left[\begin{array}  {cc} C_L^{\zeta T} (\chi) & C_L^{\zeta E} (\chi ) \end{array} \right]^\dagger 
\begin{pmatrix} C_L^{TT} & C_L^{TE}  \\[6pt] C_L^{ET} & C_L^{EE} \end{pmatrix}^{-1} \left[\begin{array} {c} a_{LM\kappa}^T  \\[6pt] a_{LM\kappa}^E  \end{array}\right] \, . \\ 
&& \langle \delta \zeta_{LM\kappa} (\chi) \delta \zeta_{L^\prime M^\prime \kappa^\prime} (\chi^\prime ) \,\vert\, T, E \rangle   =   \delta_{LL^\prime}  \delta_{MM^\prime}\delta_{\kappa \kappa^\prime } \Bigg\{ C_L^{\zeta \zeta}  (\chi , \chi^\prime )   \nonumber\\
&&-  \left[\begin{array} {cc} C_L^{\zeta T} (\chi) & C_L^{\zeta E} (\chi )\end{array} \right]^\dagger 
  \begin{pmatrix} C_L^{TT} & C_L^{TE} \\[6pt] C_L^{ET} & C_L^{EE} \end{pmatrix}^{-1} \left[ \begin{array} {c}  C_L^{T\zeta } (\chi^\prime ) \\[6pt] C_L^{E\zeta } (\chi^\prime )\end{array}\right] \, \Bigg\}. 
 \end{eqnarray}

\begin{figure*}[!t]
\centering
\includegraphics[width=1\linewidth]{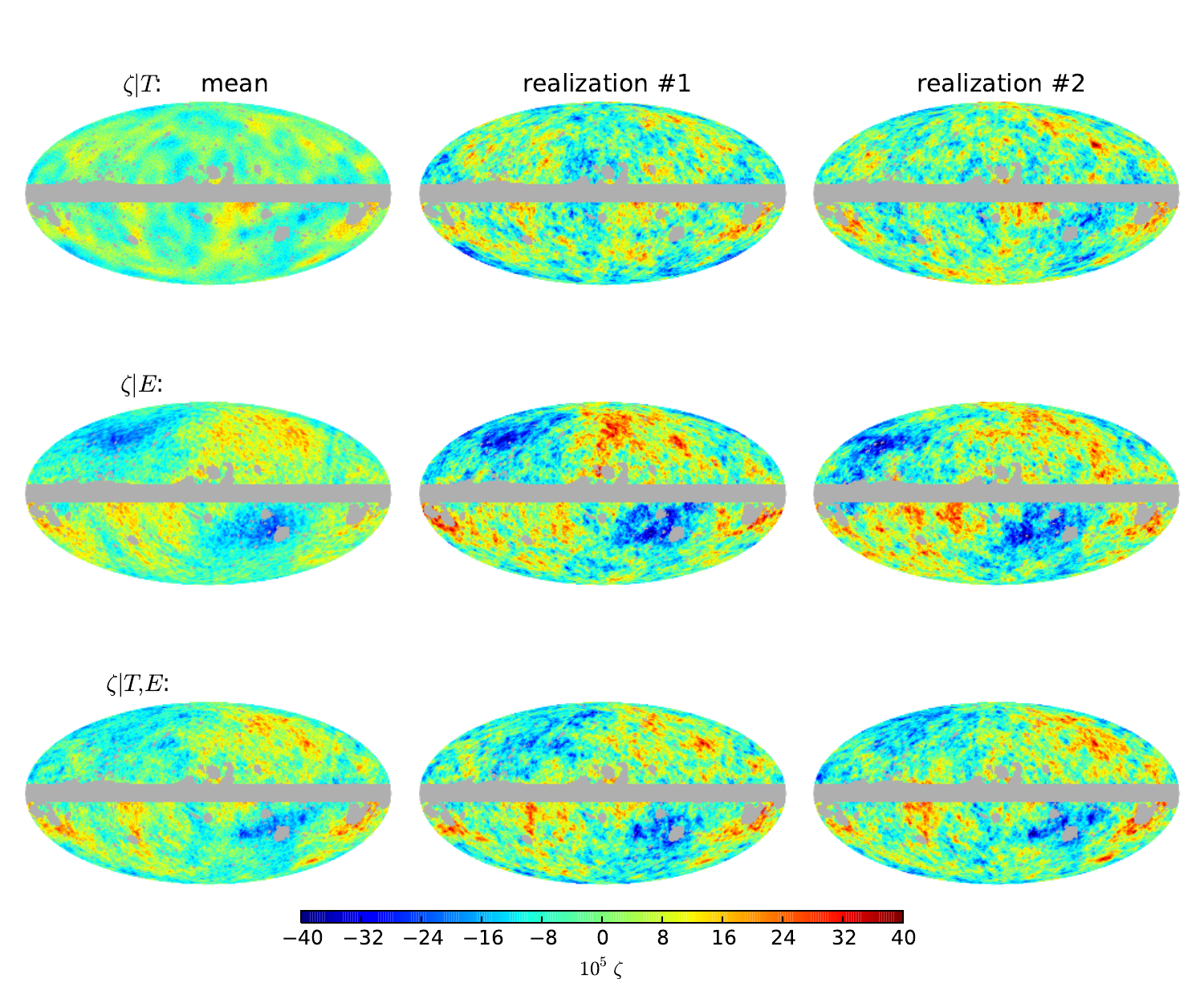}
\caption{Mean field and fluctuations for differential visibility projected $\zeta$, as described in the text. The filter used in these maps is a $40^\prime$ FWHM Gaussian. The grey regions are masked out. }\label{fig:zeta_vis} 
\end{figure*}

\begin{figure*}[!t]
\centering
\includegraphics[width=1\linewidth]{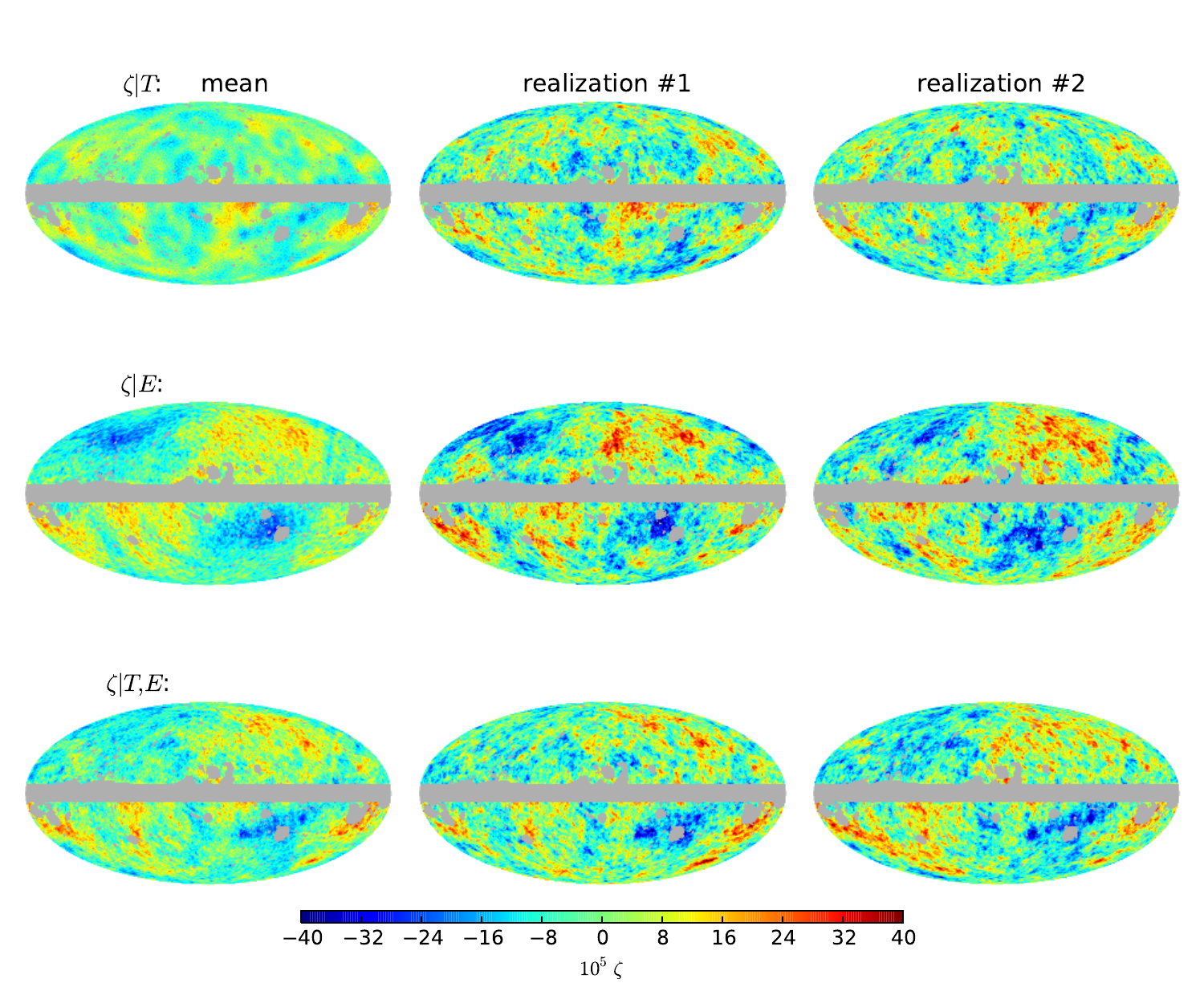}
\caption{Mean field and fluctuations for a $\chi_{\rm dec}$ slice of  $\zeta$, as described in the text. The filter used in these maps is a $40^\prime$ FWHM Gaussian. The grey regions are masked out. }\label{fig:zeta_slice} 
\end{figure*}

Fig.~\ref{fig:zeta_vis} shows an all-sky mean field (Wiener-filtered) map of the curvature variable $\zeta$ constructed in (densely-packed) shells from the multipoles $\zeta_{LM\kappa} (\chi )$. The figure actually shows $\zeta$  projected onto differential visibility, $w=de^{-\tau}/d\chi$. The map  $\zeta_w (\theta , \phi )$ is the spherical transform of $\zeta_{LM\kappa w} = \int w(\chi )d\chi  \zeta_{LM\kappa }(\chi )$. The top left map is $\langle\zeta_w \vert T \rangle$ using the all-sky maps \SMICA\ DX11. The middle left map is  $\langle \zeta_w \vert E \rangle$, based on just the polarization information. (The E-maps did not have high pass filtering imposed, so that an indication of the full results obtainable with E alone can be seen.) The bottom left panel shows  $\langle \zeta_w \vert T,E \rangle$. Visually the combined two-mode constraint has more detail than either of the one-mode constrained maps, enhancing the estimation of $\zeta_w$. In each of the three cases, two sample realizations are shown which include allowed fluctuations about the mean map. These illustrate the level of uncertainty in the patterns found in the mean maps, necessary ingredients to the full statistical description. Note that the fluctuations about the mean are larger for $T$ or $E$ alone than for the combined $T,E$. 

In all cases, the fluctuations become very large indeed if we allow for arbitrary projections, not conditioned by the visibility. Another approach similar to the differential visibility projection is to make a slice at $\chi =\chi_{\rm dec}$; i.e., making the projection a delta-function in $\chi$ at the peak where $dw/d\ln \chi =0$, which defines $\chi_{\rm dec}$.  This single slice case is shown in Fig.~\ref{fig:zeta_slice}. The figures look quite similar in structure to the projected differential visibility plots. Although $\chi \sim \chi_{\rm dec}$ is where most information is, the fluctuations of a single slice are somewhat  larger than for the differential visibility projection.  The differential visibility includes a subdominant ``reionization bump", with a small weight focused at low $L$ where sample variance is large; restricting the projection of $\zeta$ to be just over the reionization region results in a low $L$ mean map that is largely swamped by the allowed fluctuations about it. The fluctuations about the mean become much larger if we project over all $\chi$, since there are vast terrains in $\chi$ in which there is very little CMB temperature or polarization information; in particular between recombination and reionization. For that uncharted territory, a realization of the reconstructed $\zeta$-map reverts to a realization of the fiducial $\zeta$-power. 

Fig.~\ref{fig:zetaS2N} quantifies where the information resides in $L$--$\chi$ space. Here, the signal variance is the ensemble average of the square of the mean field, assuming the $\zeta$ are drawn from a Gaussian with fiducial covariance, 
\begin{equation}
\langle  \langle \zeta_{LM\kappa} (\chi)\,\vert\,  T \rangle \langle \zeta_{LM\kappa } (\chi^\prime) \,\vert\,  T \rangle \rangle = C_L^{\zeta T} (\chi  ) [C_L^{TT}]^{-1}  C_L^{T\zeta } ( \chi^\prime ), 
\end{equation} 
and the noise (i.e., fluctuation) variance is the covariance matrix, as given by Eq.~(\ref{eq:varzetaT}) for $T$ and the equivalent for $E$. Fig.~\ref{fig:zetaS2N} plots the signal-to-noise at each individual $\chi$ slice (hence ignores the components off-diagonal in $\chi$): 
\begin{equation} 
[S/N]_L^T ( \chi ) = \frac{\rho}{ \sqrt{(1-\rho^2)}}\, ; \ \rho \equiv \frac{C_L^{\zeta T} (\chi  )}{ \sqrt{ C_L^{TT}  C_L^{\zeta \zeta}(\chi , \chi) }}\,  
\end{equation} 
for $T$, with an equivalent expression for $E$.  
 The signal-to-noise structure in $\chi$ can be contrasted with the differential visibility plotted in the bottom panel:  Fig.~\ref{fig:zetaS2N} shows that the $L$--$\chi$ information has a reach beyond the differential visibility structure, especially for low multipoles, because the associated waves can straddle the last scattering surface. The ``reionization bump" in signal-to-noise seen in Fig.~\ref{fig:zetaS2N} is not that prominent. The Integrated Sachs Wolfe impact on the S/N   is evident, but is relatively low, with the consequence that we cannot draw out high significance results from the ISW effect for $\zeta$ reconstruction (or equivalently for gravitational potential reconstruction). 

The top panels in Fig.~\ref{fig:zetaS2N} are for an ideal experiment, with no noise (apart from the cosmic variance ``noise" in the Wiener map fluctuations). The bottom panels of Fig.~\ref{fig:zetaS2N} include realistic \Planck\  noise, as estimated from the FFP8 simulations: the high signal to noise in $E$ in the top panels is noticeably diminished around $L \sim 100$ over what a cosmic variance limited experiment would give. 

\begin{figure*}[!t]
\centering
\includegraphics[width=1\linewidth]{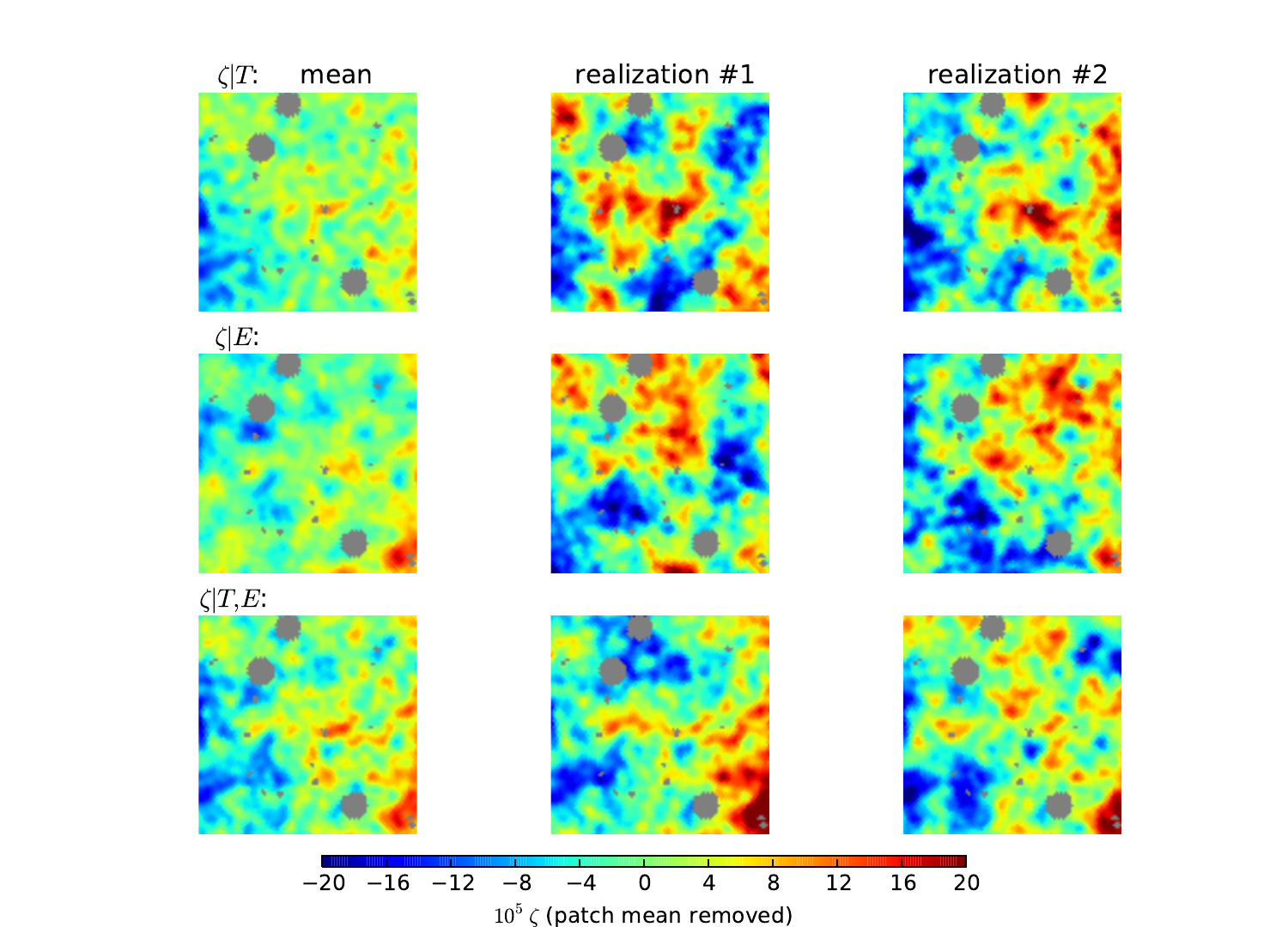}
\caption{Mean field and fluctuations for differential visibility projected $\zeta$, for a 20 degree by 20 degree patch, as described in the text. The filter used in these maps is a $20^\prime$ FWHM Gaussian. Sources in grey  are masked out. }\label{fig:zeta_vis_highres} 
\end{figure*}

\begin{figure*}[!t]
\centering
\includegraphics[width=1\linewidth]{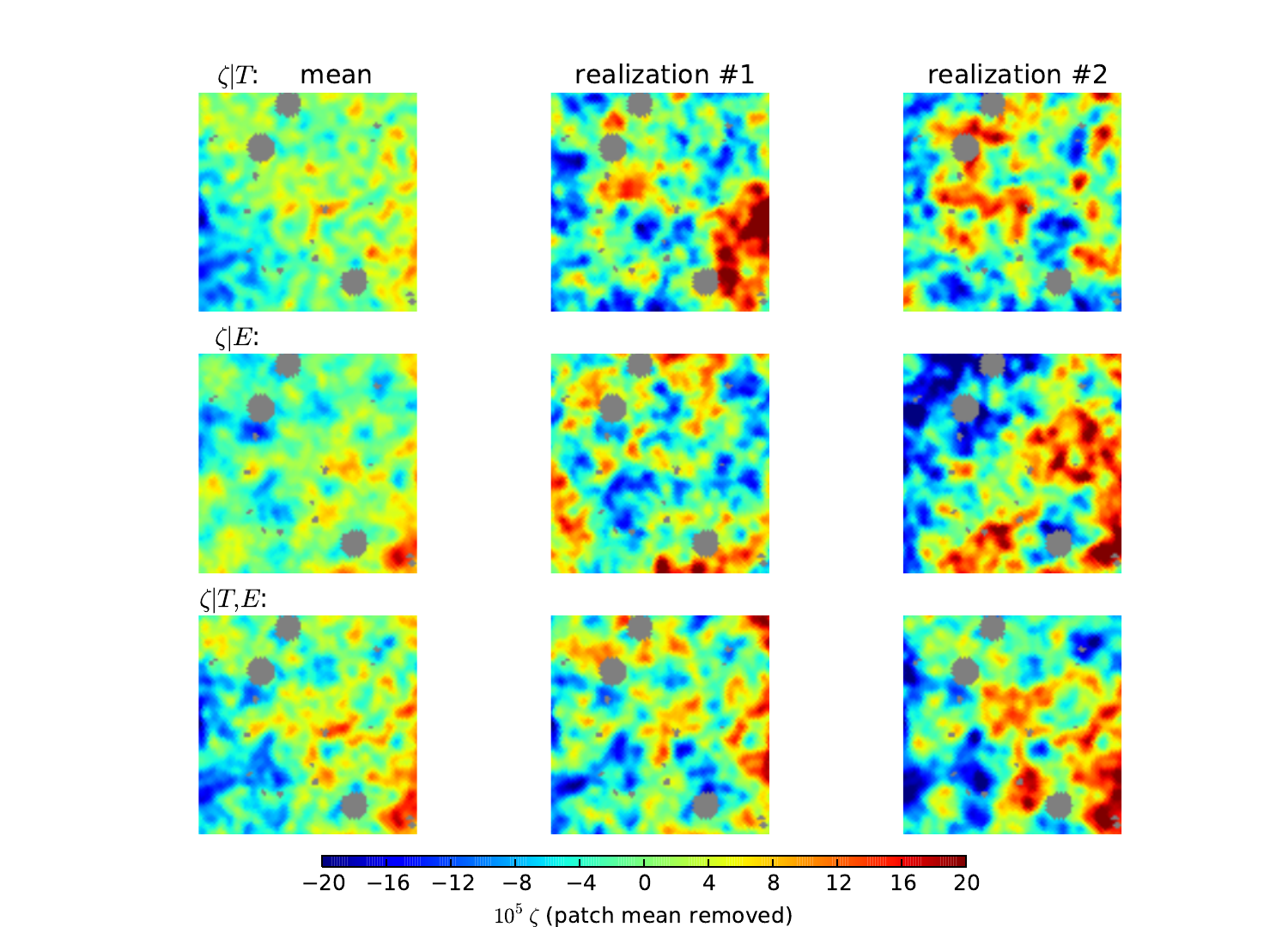}
\caption{Mean field and fluctuations for a $\chi_{\rm dec}$ slice of  $\zeta$, for a 20 degree by 20 degree patch, as described in the text. The filter used in these maps is a $20^\prime$ FWHM Gaussian. Sources in grey are masked out. }\label{fig:zeta_slice_highres} 
\end{figure*}

Figs.~\ref{fig:zeta_vis} and \ref{fig:zeta_slice} are Gaussian-filtered on a relatively large $L~\sim 200$ scale ($40^\prime$ FWHM). To see what happens at higher resolution,  Figs.~\ref{fig:zeta_vis_highres} and \ref{fig:zeta_slice_highres} zoom in on a typical $20^\circ \times 20^\circ $ patch, with long waves removed using a filter $W_L = \sin^2 (L-L_c)/\Delta L$ for $L_c < L <  L_c +  \Delta L$. The specific choices  are $L_c=20 $ and $\Delta L = 20$, the values used for the \Planck\  high pass for polarization maps. To allow direct comparison, we have just done the same high pass filtering for the $T$ map. We also removed the means of the maps. The resolution is $L~\sim 400$  ($20^\prime$ FWHM).  Figs.~\ref{fig:zeta_vis_highres} and \ref{fig:zeta_slice_highres} illustrate that  the fluctuations play a larger role in this higher resolution regime. Note that the $\chi_{\rm dec}$ slice has about the same fluctuation level as the differential visibility projection. 

 Lensing effects are not taken into account in these $\zeta$-maps. In principle one could de-lens the temperature and polarization maps before forming the Wiener filter. In practice this would be a highly noisy operation with current \Planck\  data. Hence the contaminating influence of lensing on these $\zeta$-reconstructions would be treated by comparing simulated $\zeta$-maps with and without lensing. Such corrections are expected to be a subdominant bias, as it has been modelled in the rest of the paper. 

%% file: A19_Section7.tex
In this section, we perform a battery of tests aimed at verifying the 
robustness of the results obtained in the previous section. Table 
\ref{tab:fNLsmicah} shows excellent agreement with our 
2013 analysis of nominal mission data \citep{planck2013-p09a}. The agreement 
using different component separation methods in temperature 
is also generally very good. Our focus here is thus on polarization bispectra. Redundancy is perhaps the most 
important element in our analysis, as far as robustness is concerned. We devote considerable attention to comparing the outcomes of different estimators 
and component separation pipelines, and assess their level of internal 
consistency.  We also verify the stability of our results in the harmonic 
and pixel domains, by considering different sky cuts and multipole intervals. 
Given the large computational requirements of these tests, and since results from 
different optimal estimators agree very well, as 
shown in previous sections, we will variously use the KSW, binned, or modal 
pipeline for different tests. In doing this we will also exploit the complementarity 
of different decompositions, which might make some of them more suited for 
different tests than others (for example, the binned pipeline directly works with 
 a harmonic space decomposition of the bispectrum, thus making it perfectly suited 
for tests of $\ell$-dependence, the modal pipeline can perform quick model-independent 
tests by working on a relatively small subset of bispectrum modes, and so on).

\subsection{Dependence on foreground cleaning method}
\label{sec:dep_methods}

\subsubsection{Comparison between $\fnl$ measurements}

In Table \ref{tab:fNLsmicah} we show  
$f_\mathrm{NL}$ results for the local, equilateral and orthogonal shapes, 
using four different optimal estimators, and four different foreground 
cleaning pipelines. The agreement between different estimators, on a given map, is within a 
fraction of a standard deviation, in line
with theoretical expectations and simulations, as reported in Sect.~
\ref{sec:Validation}. This level of agreement applies to all of the \textit{T+E}, (\textit{TTT}), and (\textit{EEE}) bispectra.

The overall picture becomes more complex when comparing 
outputs across different foreground cleaning methods and estimators. Whereas for \textit{TTT} and \textit{T+E} results the agreement 
also seems quite good in this case (being at the level of half a sigma or better, 
for nearly all combinations of cleaned maps and shapes), larger discrepancies 
are present in the \textit{EEE} bispectrum measurements. The most notable differences 
are found for the equilateral shape, where {\SMICA} and {\NILC} find values of  
$f_\mathrm{NL}$ consistent with $0$ within $1\, \sigma$, while {\SEVEM} and {\Commander}
 measure a roughly $ 2\, \sigma$ deviation from Gaussianity. The largest discrepancy is found 
for the pair {\Commander}--{\NILC}, using the binned pipeline (see table  
\ref{tab:fNLsmicah}). This estimator recovers 
$f_\mathrm{NL}^{\rm equil} = 369 \pm 160$ 
using the {\Commander} \itE-map and $f_\mathrm{NL}^{\rm equil} = 97 \pm 141$ using the {\NILC} \itE-map.
Other pipelines, and different choices of component separation methods,  
show slightly smaller but similar discrepancies, at a level of about $1.5 \sigma$. 
The same shape and estimator analysis of temperature maps shows good agreement: {\Commander} recovers $f_\mathrm{NL}^{\rm equil} = -36 \pm 73$,  
while {\NILC} gives $f_\mathrm{NL}^{\rm equil} = -45 \pm 71$.  The combined \textit{T+E} measurement, still for the same 
modal pipeline and equilateral shape, yields $f_\mathrm{NL}^{\rm equil} = 26 \pm 50$ for {\Commander} 
and $f_\mathrm{NL}^{\rm equil} = -4 \pm 46$ for {\NILC}, corresponding to about half a standard deviation
difference. This general trend is seen for other shapes and estimators.

\begin{table*}[t]                 
\begingroup
\newdimen\tblskip \tblskip=5pt
\caption{Comparison between local, equilateral, and orthogonal $\fnl$ results, obtained using  
the four different component separation pipelines. For each pair of cleaning methods, and for each 
NG model, we compute the difference in the measured $\fnl$. The quoted error bar is the standard deviation 
of the same difference, extracted from a set of $100$ realistic Gaussian simulations per method, not including 
foregrounds. These results have been obtained using the low resolution modal pipeline. See main text for comments 
and further details.}
\label{tab:methodsdiff}
\nointerlineskip
\vskip -3mm
\footnotesize
\setbox\tablebox=\vbox{
   \newdimen\digitwidth
   \setbox0=\hbox{\rm 0}
   \digitwidth=\wd0
   \catcode`*=\active
   \def*{\kern\digitwidth}
   \newdimen\signwidth
   \setbox0=\hbox{+}
   \signwidth=\wd0
   \catcode`!=\active
   \def!{\kern\signwidth}
\newdimen\dotwidth
\setbox0=\hbox{.}
\dotwidth=\wd0
\catcode`^=\active
\def^{\kern\dotwidth}
\halign{\hbox to 0.8 in{#\leaderfil}\tabskip 1.em&
\hfil#\hfil\tabskip 1.em&
\hfil#\hfil&
\hfil#\hfil&
\hfil#\hfil&
\hfil#\hfil&
\hfil#\hfil&
\hfil#\hfil&
\hfil#\hfil\tabskip 0pt\cr
\noalign{\doubleline\vskip 2pt}
\omit&\multispan6\hfil $f_{\rm NL}\left({\rm method}_1 \right)  - f_{\rm NL}\left({\rm method}_2 \right)$\hfil\cr
\omit&\multispan6\hrulefill\cr
\omit&\multicolumn{3}{c}{\hfil $f_{\rm sky} = 0.74$ \hfil}&
\multicolumn{3}{c}{\hfil ******\hfil $f_{\rm sky} = 0.64$ \hfil}\cr
\noalign{\vskip 2pt}
Methods\hfill&Local&Equilateral&Orthogonal&*********Local&Equilateral&Orthogonal\cr
\noalign{\vskip 4pt\hrule\vskip 6pt}
\omit\hfil \SMICA--\SEVEM\hfil&&\cr
\itT  &  *$-$1.2 $\pm$ *0.9&  **$-$6.0 $\pm$ **8.7& !**1.5 $\pm$ *4.8&
*****  \dots&  \dots&  \dots\cr
\itE  &  $-$19^* $\pm$ 21^*& $-$155^* $\pm$ *86^*&  !*34^* $\pm$ 57^*&
***** *!5^* $\pm$ 22^*& *$-$82^* $\pm$ *90^* & $-$11^* $\pm$ 66^*\cr
\itTpE  &  *$-$2.4 $\pm$ *1.6& *$-$10^* $\pm$ *18^*& !*13.5 $\pm$ *9.4&
***** *$-$1.5 $\pm$ *1.7& $-$12^* $\pm$ *18^*& !13^* $\pm$ 10^*\cr
\noalign{\vskip 6pt}
\omit\hfil \SMICA--\NILC \hfil&&\cr
\itT  &  !*0.4 $\pm$ *1.0& !*14.5 $\pm$ **8.9&  !**2.5 $\pm$ *4.7&
*****  \dots&  \dots&  \dots\cr
\itE &  !26^* $\pm$ 11^*& !*83^* $\pm$ *52^*& *$-$59^* $\pm$  27^*&
***** !26^* $\pm$ 13^*& !*32^* $\pm$ *56^*& $-$96^* $\pm$ 28^*\cr
\itTpE & *$-$0.7 $\pm$ *0.9& !*20.0 $\pm$ **8.2&  **$-$3.3 $\pm$  *3.8&
***** !*0.6 $\pm$ *0.9& !*18.4 $\pm$ **8.4& *$-$4.5 $\pm$ *4.0\cr
\noalign{\vskip 6pt}
\omit\hfil \SMICA--\Commander\hfil&&\cr
\itT &  !*0.4 $\pm$ *3.5&  *$-$14^* $\pm$ *23^*&  !**1.7 $\pm$ 14^*&
*****  \dots&  \dots&  \dots\cr
\itE & *$-$3^* $\pm$ 16^*&  $-$130^* $\pm$ *77^*& *$-$81^* $\pm$ 42^*&
***** $-$13^* $\pm$ 17^*& $-$117^* $\pm$ 100^*& $-$59^* $\pm$ 40^*\cr
\itTpE &  *$-$1.3 $\pm$ *3.2& *$-$25^* $\pm$ *18^*&  !**9^* $\pm$ 10^*&
***** *$-$1.4 $\pm$ *3.3& *$-$26^* $\pm$ *18^*& !13^* $\pm$ 10^*\cr
\noalign{\vskip 6pt}
\omit\hfil \SEVEM--\NILC \hfil&&\cr
\itT &  !*1.6 $\pm$ *1.0&  !*20^* $\pm$ *12^*&  !**1.0 $\pm$ *4.5&
*****  \dots&  \dots&  \dots\cr
\itE &  !45^* $\pm$ 26^*& !239^* $\pm$ *94^*& *$-$94^* $\pm$ 69^*&
***** !30^* $\pm$ 29^*& !114^* $\pm$ 105^*& $-$86^* $\pm$ 79^*\cr
\itTpE &  !*3.1 $\pm$ *1.8& !*30^* $\pm$ *18^*&  *$-$17^* $\pm$ 10^*&
***** !*2.2 $\pm$ *1.9& !*30^* $\pm$ *18^*& $-$18^* $\pm$ 10^*\cr
\noalign{\vskip 6pt}
\omit\hfil \SEVEM--\Commander \hfil&&\cr
\itT &  !*1.6 $\pm$ *3.4&  **$-$8^* $\pm$ *22^*&  !**0^* $\pm$ 14^*&
*****  \dots&  \dots&  \dots\cr
\itE &  !16^* $\pm$ 22^*&  !*25^* $\pm$ 112^*&  $-$116^* $\pm$ 59^*&
***** $-$18^* $\pm$ 25^*& *$-$35^* $\pm$ 121^*& $-$48^* $\pm$ 64^*\cr
\itTpE & !*1.2 $\pm$ *3.3&  *$-$14^* $\pm$ *21^*&  **$-$5^* $\pm$ 11^*&
***** !*0.2 $\pm$ *3.4& *$-$14^* $\pm$ *20^*& !*0^* $\pm$ 11^*\cr
\noalign{\vskip 6pt}
\omit\hfil \NILC--\Commander \hfil&&\cr
\itT & !*0.0 $\pm$ *3.0& *$-$28^* $\pm$ *22^*& **$-$1^* $\pm$ 12^*&
***** \dots&  \dots&  \dots\cr
\itE & $-$29^* $\pm$ 21^*& $-$213^* $\pm$ *84^*&  *$-$22^* $\pm$ 54^*&
***** $-$39^* $\pm$ 23^*& $-$149^* $\pm$ 108^*& !38^* $\pm$ 55^*\cr
\itTpE &  *$-$1.9 $\pm$ *3.1& *$-$45^* $\pm$ *18^*& !*12^* $\pm$ 11^*&
***** *$-$2.0 $\pm$ *3.2& $-$44^* $\pm$ 17^*& !18^* $\pm$ 11^*\cr
\noalign{\vskip 3pt\hrule\vskip 4pt}
}}
\endPlancktablewide                 
\endgroup
\end{table*}                        

Simulations were used to give insight into the expected level of disagreement. For each of the four component separation methods, 
we generated $100$ FFP8-based Gaussian simulations with realistic beams and noise levels. These simulations start from the same initial single frequency realizations, and are processed through 
the four different foreground cleaning pipelines. The starting maps {\em do not} include any foreground component (the same map generation procedure is used in the Monte Carlo determination of error bars). The differences in final simulations are thus generated only by the different data filtering and coadding operations performed either in pixel, harmonic, or needlet domains by the various 
foreground cleaning methods, and by additional manipulations of the maps which are required for $f_\mathrm{NL}$ estimation, such as inpainting. 
Therefore, the average scattering in $f_\mathrm{NL}$, measured from these realizations,
provides us with a baseline assessment of the expected discrepancies between different methods 
when foreground residuals and other spurious sources of NG are negligible. 
We can then compare them with differences observed on the data 
to establish whether they are consistent with expectations, or are too large.  The latter would raise 
the concern that foreground contamination, or other systematics, might be affecting the results.

Results are shown in Table 
\ref{tab:methodsdiff}, 
for \textit{EEE} and \textit{T+E} and two different sky coverages. 
The scatter between $f_\mathrm{NL}$ values from simulations is about $ 0.5\, \sigma$ for both \textit{T+E} and \textit{EEE}.  This is smaller than
the differences in the \Planck\  $\fnl$ values obtained from \textit{EEE} analysis of  different foreground-cleaned maps,  especially for the equilateral shape.  However, for the final combined \textit{T+E} measurement, observed differences are in good agreement with expectations from simulations for the majority of cases.  
Another important point is that the consistency shown in Table \ref{tab:methodsdiff} for \textit{T+E} measurements is stable to the
change of sky coverage (in polarization) from $f_{\rm sky}=0.74$ to $f_{\rm sky}=0.64$.  This will be confirmed by additional tests later in this section. For the {\SMICA}--{\SEVEM} pair we also verified stability using an even larger mask with $f_{\rm sky}=0.52$. 

Residual foregrounds may be responsible for at least some of
the observed excess of scatter in \textit{EEE}-derived $\fnl$ between different cleaning algorithms.  This is supported by the fact 
that several \textit{EEE} results for this test change significantly for different masks, and that discrepancies are alleviated by using a larger mask,
especially for equilateral shapes 
(see e.g., $\SEVEM$--$\NILC$ and $\SMICA$--$\SEVEM$ in Table \ref{tab:methodsdiff}). However, modal coefficients and their correlations are stable to a change of mask (see below), as are values of $\fnl$ for a given component separation method (see Table \ref{tab:fnlchangemask}). 

Another possible contributor is a mismatch between the noise model (used to build the estimator normalization, weights and linear term), and the actual noise in the data. 
Polarization data are very noisy, and it is a known problem that the model assumed underestimates the true noise. 
This means that the error bars for \textit{EEE} $\fnl$ results, 
quoted in Table \ref{tab:fNLsmicah}, are somewhat underestimated, which does not seem to be  a problem for the final \textit{T+E} results, since the weight of the \textit{EEE} bispectrum in the final combined measurement is very low.  
This is confirmed by the results of this test.   Indeed, we investigate \textit{EEE} in detail because it is a useful and sensitive indicator of 
various systematics in the polarized maps (which could eventually leak into 
the TTE and TEE bispectra), rather than for its statistical weight in the final measurement.
It is then fair to say that issues in the \textit{EEE} bispectra, and related $\fnl$ measurements
 are not yet fully understood and will require further investigation. Even though the \textit{T+E} are consistent, we recommend that results that include polarization data are regarded as {\em preliminary} at this stage.

\begin{table*}[t!]                 
\begingroup
\newdimen\tblskip \tblskip=5pt
\caption{For each of the four foreground cleaned maps, we compute $\fnl$ for the local, equilateral, and orthogonal modes 
using two different polarization masks, one with $f_{\rm sky} = 0.74$ and the other with $f_{\rm sky} = 0.64$, while 
for temperature we use a single mask with $f_{\rm sky} = 0.76$. 
We then calculate the difference between the two measurements and compare with expectations 
from simulations, obtained in the following way: firstly, we generate realistic Gaussian realizations 
for each component separation pipeline, not including foregrounds; then, for each simulated map and NG model, 
we measure $\fnl$ using the two masks in turn; finally, we calculate the standard deviation on $100$ 
realizations. See the main text for more details and a discussion of these results, which were obtained using the 
low resolution modal pipeline.}
\label{tab:fnlchangemask}
\nointerlineskip
\vskip -3mm
\footnotesize
\setbox\tablebox=\vbox{
   \newdimen\digitwidth 
   \setbox0=\hbox{\rm 0} 
   \digitwidth=\wd0 
   \catcode`*=\active 
   \def*{\kern\digitwidth}
   \newdimen\signwidth 
   \setbox0=\hbox{+} 
   \signwidth=\wd0 
   \catcode`!=\active 
   \def!{\kern\signwidth}
   \newdimen\dotwidth 
   \setbox0=\hbox{.} 
   \dotwidth=\wd0 
   \catcode`^=\active 
   \def^{\kern\dotwidth}
\halign{\hbox to 0.5in{#\leaderfil}\tabskip 1em&
\hfil#\hfil\tabskip 0.8em&
\hfil#\hfil&
\hfil#\hfil&
\hfil#\hfil&
\hfil#\hfil&
\hfil#\hfil&
\hfil#\hfil&
\hfil#\hfil&
\hfil#\hfil\tabskip 0pt\cr
\noalign{\doubleline\vskip 2pt}
\omit&\multicolumn{3}{c}{\hfil $f_{\rm sky} = 0.74$ \hfil}&
\multicolumn{3}{c}{$f_{\rm sky} = 0.64$ \hfil}&\multicolumn{3}{c}{\hfil\hfil\hfil Difference }\cr
\noalign{\vskip 6pt}
\omit&Local&Equilateral&Orthogonal&Local&Equilateral&Orthogonal&Local&Equilateral&Orthogonal\cr
\noalign{\vskip 4pt\hrule\vskip 6pt}
\omit\hfil \SMICA \hfil&&\cr
\itT  & !*6.8 $\pm$ *5.4& *$-$17 $\pm$ *66& *$-$48 $\pm$ 33&  \dots&  \dots&   \dots&  \dots&  \dots&   \dots\cr
\itE  & !25^* $\pm$ 30^*& !147 $\pm$ 159& $-$137 $\pm$ 73& 48^* $\pm$ 31^*& 220 $\pm$ 168&  $-$180 $\pm$ 81& $-$23^* $\pm$ 16^*& *$-$73 $\pm$ 68&  !43^* $\pm$ 34^*\cr
\itTpE & !*4.0 $\pm$ *4.8& !**5 $\pm$ *46& *$-$30 $\pm$ 21& *4.6 $\pm$ *5.2& *19 $\pm$ *55&  *$-$37 $\pm$ 22& *$-$0.7 $\pm$ *1.2& *$-$14 $\pm$ 14& !*6.7 $\pm$ *7.7\cr
\noalign{\vskip 6pt}
\omit\hfil \SEVEM \hfil&&\cr
\itT    & !*8.1 $\pm$ *5.8& *$-$11 $\pm$ *75& *$-$49 $\pm$ 34&  \dots&  \dots&   \dots&  \dots&  \dots&   \dots\cr
\itE   & !44^* $\pm$ 38^*& !302 $\pm$ 183& $-$172 $\pm$ 91& 43^* $\pm$ 39^*& 303 $\pm$ 191&  $-$170 $\pm$ 96& !*1^* $\pm$ 19^*& !**0 $\pm$ 76&  *$-$2^* $\pm$ 49^*\cr
\itTpE  & !*6.4 $\pm$ *5.0& *!15 $\pm$ *52& *$-$44 $\pm$ 23& *6.2 $\pm$ *5.3& *31 $\pm$ *54&  *$-$50 $\pm$ 25& !*0.2 $\pm$ *1.3& *$-$16 $\pm$ 15& !*6.3 $\pm$ *8.8\cr
\noalign{\vskip 6pt}
\omit\hfil \NILC \hfil&&\cr
\itT  & !*6.4 $\pm$ *5.8& *$-$31 $\pm$ *76& *$-$50 $\pm$ 33&  \dots&  \dots&   \dots&  \dots&  \dots&   \dots\cr
\itE    & *$-$1^* $\pm$ 30^*& !*64 $\pm$ 162& *$-$78 $\pm$ 77& 22^* $\pm$ 30^*& 190 $\pm$ 162&  *$-$84 $\pm$ 77& $-$23^* $\pm$ 16^*& $-$124 $\pm$ 67&  !*6^* $\pm$ 37^*\cr
\itTpE  & !*3.3 $\pm$ *4.9& *$-$15 $\pm$ *50& *$-$27 $\pm$ 23& *4.0 $\pm$ *5.3& **1 $\pm$ *56&  *$-$33 $\pm$ 23& *$-$0.7 $\pm$ *1.3& *$-$16 $\pm$ 13&  !*5.4 $\pm$ *7.5\cr
\noalign{\vskip 6pt}
\omit\hfil \Commander \hfil&&\cr
\itT  & !*6.4 $\pm$ *6.6& **$-$3 $\pm$ *77& *$-$49 $\pm$ 36&   \dots&  \dots&   \dots&  \dots&  \dots&   \dots\cr
\itE   & !28^* $\pm$ 37^*& !278 $\pm$ 178& *$-$56 $\pm$ 81& 61^* $\pm$ 38^*& 337 $\pm$ 188&  $-$122 $\pm$ 91& $-$32^* $\pm$ 20^*& *$-$60 $\pm$ 92&  !66^* $\pm$ 47^*\cr
\itTpE  & !*5.2 $\pm$ 5.4^*& *!30 $\pm$ *50& *$-$39 $\pm$ 23& *6.0 $\pm$ *5.7& *45 $\pm$ *55&  *$-$51 $\pm$ 25& *$-$0.7 $\pm$ *1.5& *$-$15 $\pm$ 14&  !11.5 $\pm$ *8.9\cr
\noalign{\vskip 3pt\hrule\vskip 4pt}
}}
\endPlancktablewide                 
\endgroup
\end{table*}                        

\begin{table*}[t]                 
\begingroup
\newdimen\tblskip \tblskip=5pt
\caption{Correlation coefficients between pairs of bispectrum modes, extracted using two different component-separated maps. For both \textit{TTT} and \textit{EEE} we compare correlations measured from data with averages over $100$ Gaussian 
realizations. The simulations are  
processed through the different component separation pipelines in the same way as the data, but they do {\em not} include 
any foregrounds. The correlation is clearly lower for \textit{EEE} bispectra than for \textit{TTT}. However, this is seen not only 
in data but also in simulations, indicating that it is not due to foreground residual contamination or other unaccounted for systematics.
The results presented in this table are obtained using the low resolution modal pipeline, with $610$ modes; results on data 
have also been cross-checked with the high-resolution modal estimator, using $2001$ modes, and they are stable.}
\label{tab:betacorr_methods}
\nointerlineskip
\vskip -3mm
\footnotesize
\setbox\tablebox=\vbox{
   \newdimen\digitwidth 
   \setbox0=\hbox{\rm 0} 
   \digitwidth=\wd0 
   \catcode`*=\active 
   \def*{\kern\digitwidth}
   \newdimen\signwidth 
   \setbox0=\hbox{+} 
   \signwidth=\wd0 
   \catcode`!=\active 
   \def!{\kern\signwidth}
   \newdimen\signwidth 
   \setbox0=\hbox{.} 
   \signwidth=\wd0 
   \catcode`^=\active 
   \def^{\kern\signwidth}
\halign{\hbox to 1.3in{#\leaderfil}\tabskip 1em&
\hfil#\hfil\tabskip 1em&
\hfil#\hfil&
\hfil#\hfil&
\hfil#\hfil&
\hfil#\hfil&
\hfil#\hfil&
\hfil#\hfil&
\hfil#\hfil&
\hfil#\hfil\tabskip 0pt\cr
\noalign{\doubleline\vskip 2pt}
\omit&\multispan4\hfil $f_{\rm sky} = 0.74$ \hfil& \multispan2\hfil $f_{\rm sky} = 0.64$ \hfil\cr
\omit&\multispan4\hrulefill&\multispan2\hrulefill\cr
\omit&\multicolumn{2}{c}{\textit {TTT}}&\multicolumn{2}{c}{\textit {EEE}}&\multicolumn{2}{c}{\textit {EEE}}\cr
\noalign{\vskip 6pt}
Methods \hfill&**Data     &**Simul.     &**Data    &**Simul.     &**Data     &**Simul.\cr
\noalign{\vskip 4pt\hrule\vskip 6pt}
\SMICA--\SEVEM&**0.97 &**0.97&**0.61 &**0.62&**0.60&**0.61\cr 
\noalign{\vskip 6pt}
\SMICA--\NILC&**0.97 &**0.97 &**0.95 &**0.95&**0.95&**0.95\cr
\noalign{\vskip 6pt}
\SMICA--\Commander&**0.78 &**0.81 &**0.70 &**0.70&**0.73&**0.73\cr
\noalign{\vskip 6pt}
\SEVEM--\NILC&**0.96 &**0.97 &**0.54 &**0.55&**0.54&**0.54\cr
\noalign{\vskip 6pt}
\SEVEM--\Commander&**0.81 &**0.83 &**0.69&**0.67&**0.70&**0.70\cr
\noalign{\vskip 6pt}
\NILC--\Commander&**0.85 &**0.86 &**0.64&**0.63&**0.66&**0.66\cr
\noalign{\vskip 3pt\hrule\vskip 4pt}
}}
\endPlancktablewide                 
\endgroup
\end{table*}                        

\subsubsection{Comparison between reconstructed bispectra}\label{sec:betacomp}

\begin{figure}[!t]
\centering
\includegraphics[width=1\linewidth]{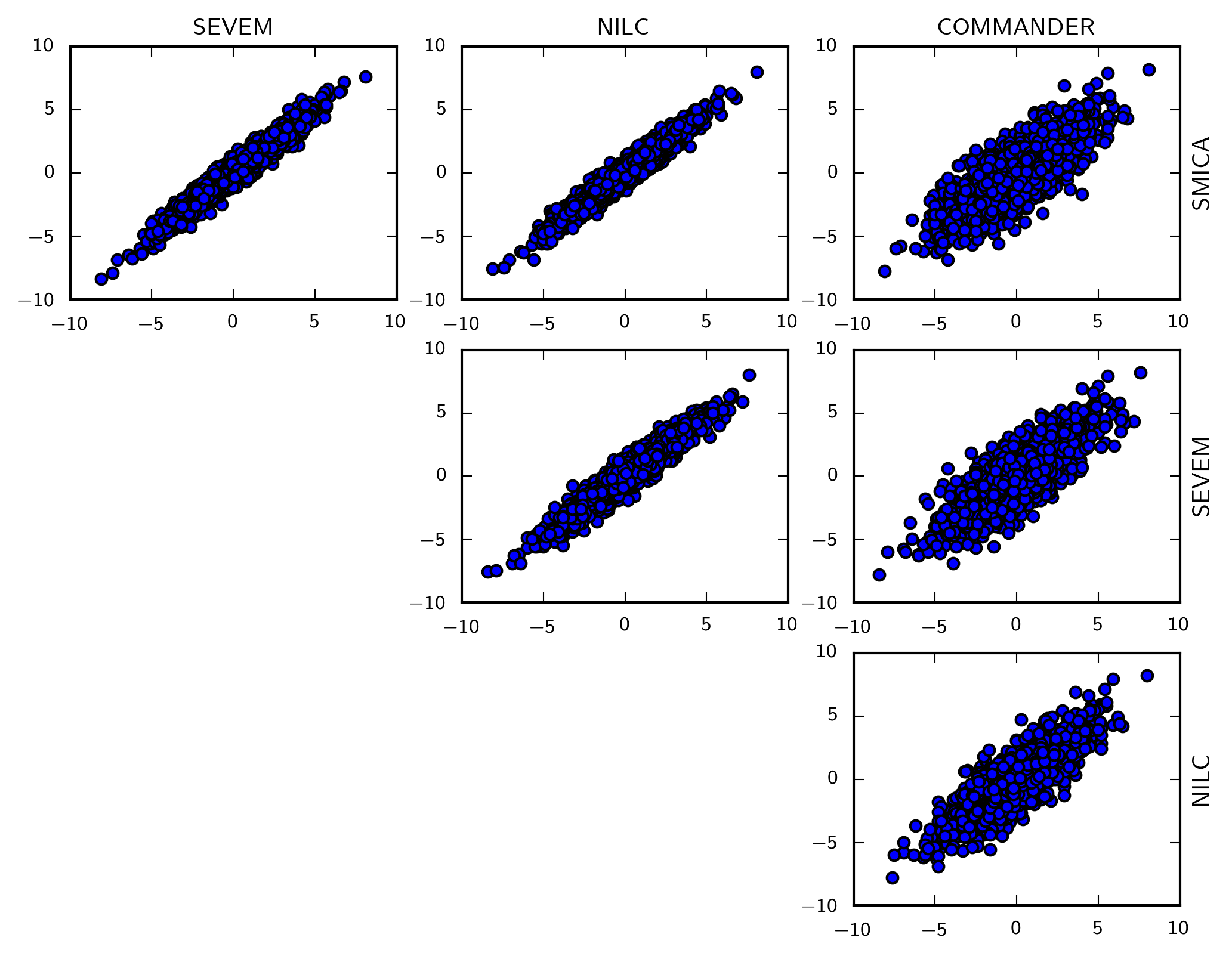}
\includegraphics[width=1\linewidth]{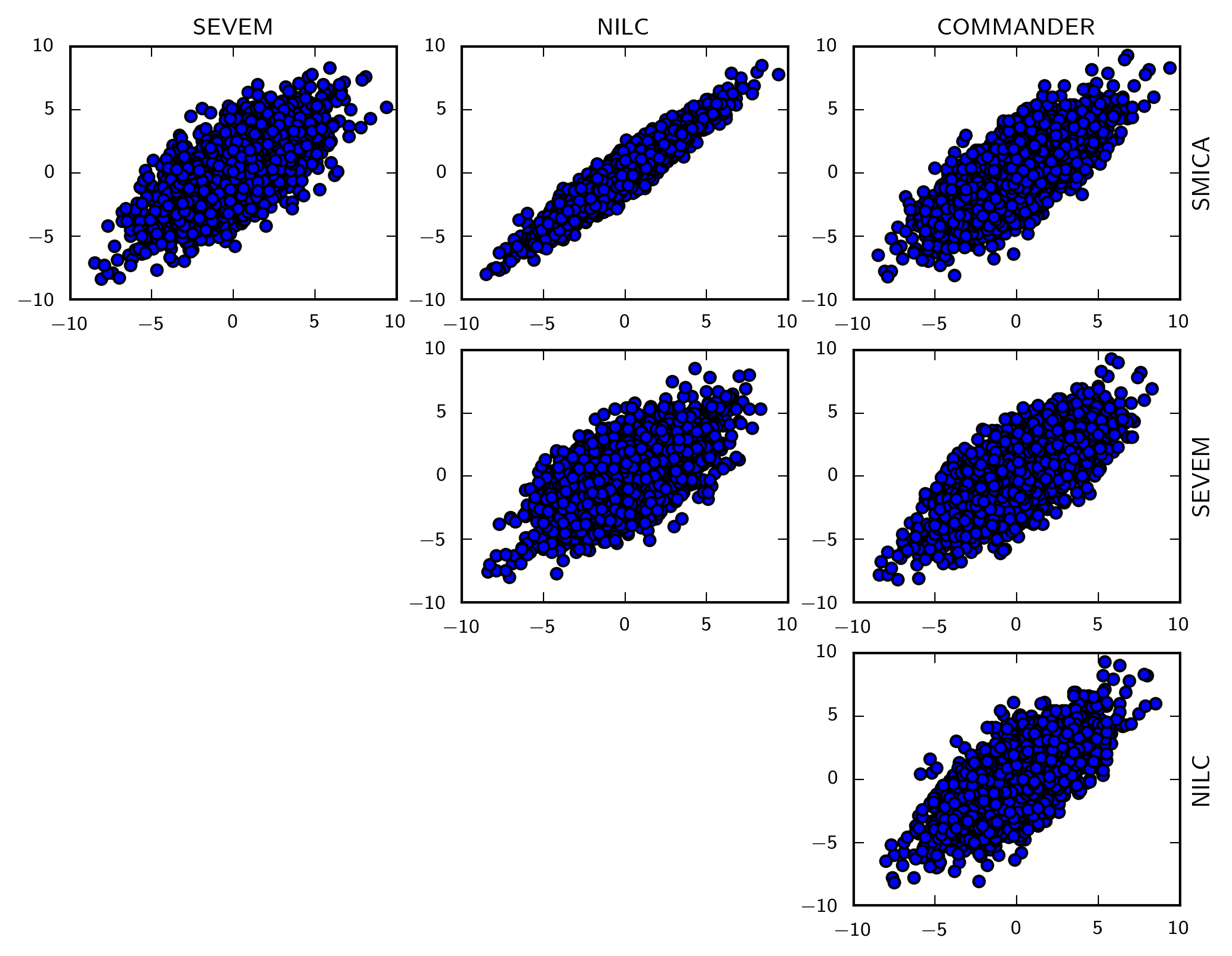}
\caption{Scatter plots showing correlations between bispectrum modes extracted from the different \Planck\
foreground-cleaned maps, for all possible pairs of component separation methods. {\em{Top:}} \textit{TTT} bispectrum 
modes. {\em{Bottom:}} \textit{EEE} bispectrum modes. While temperature shows a strong correlation, the loss of correlation in polarization 
between the different methods is evident in these plots, as discussed in the text and quantified in 
Tables \ref{tab:betacorr_methods} and \ref{tab:betacorr_table}. Results here and in Table~\ref{tab:betacorr_table} have been obtained
using the 
high resolution modal pipeline ($2001$ modes), while results in Table \ref{tab:betacorr_methods} have been 
obtained with the low resolution modal pipeline. By construction, the high resolution pipeline is measuring not the 
full \textit{EEE} bispectrum of the map, but the component of \textit{EEE} that is orthogonal to \textit{TTT}. For this reason, $r^2$ measured 
by the two pipelines for \textit{EEE} will not be identical. With this caveat in mind, the agreement between the two modal 
approaches is very good.}\label{fig:betacorr_methods} 
\end{figure}

It is important to stress that the conclusions reached at the end of the previous 
subsection refer to the three main bispectra in our 
analysis, defined by the standard scale-invariant local, equilateral, and orthogonal primordial shapes. 
These shapes select a specific subset of configurations in the overall bispectrum domain (essentially squeezed, 
equilateral, and flattened triangles). Therefore, testing 
consistency between methods  for these shapes does not guarantee that results for 
the many other NG models considered in this work (such as, for example, the oscillatory bispectra of Sect.~\ref{sec:Other_shapes}) will display 
the same level of agreement. For this reason we decided to perform a model-independent test of consistency 
between methods, based on comparisons between the $\beta_n$ eigenmodes used for bispectrum reconstruction in Sect.~\ref{Bisp_Rec}. We also reconstruct the bispectrum starting from a binned $\ell$-decomposition, and this will be used in Sect.~\ref{sec:ldep} to study stability of the results in the harmonic domain.
For the $\beta_n$ study we consider a simple test based on measuring the correlation coefficient between modes extracted 
from different foreground-cleaned maps. The correlation is defined, as usual, by
\begin{equation}\label{eq:beta_rsquared}
r_{ij}^2 = \frac{{\rm cov} \left(\beta_n^i, \beta_n^j \right)^2 }{(\sigma^2_n)_i (\sigma^2_n)_j} \; ,
\end{equation}
and we measure it for each combination of the $\SMICA$, $\SEVEM$, $\NILC$, and 
{\Commander} maps, labelled by the indices $i$,$j$.
Results are given in Table \ref{tab:betacorr_methods} and \ref{tab:betacorr_table} for the two modal pipelines and are illustrated in Fig.~\ref{fig:betacorr_methods}.   These results show 
an excellent degree of correlation between different maps in temperature (especially for 
{\SMICA}, {\SEVEM}, and {\NILC}), which is reduced when polarization is considered. In fact the correlation for polarization is not much lower than temperature for $\SMICA$ and $\NILC$, while it reduces the correlation for the pairs  $\SMICA$--$\SEVEM$, and $\NILC$--$\SEVEM$, and for {\Commander} when paired with any other method. 
This is consistent with previous findings 
of our $\fnl$-based test. To test if these results are due to foreground residuals (or other effects that are not included in the simulations),   we evaluate 
the same mode-mode correlations on the same sets of $100$ realistic, foreground-free, Gaussian simulations as previously used, and processed  through each of the 
different component separation pipelines. For this analysis we consider \textit{TTT} and \textit{EEE} bispectra, expanded 
via the low-resolution modal estimator. Our results are reported in Table \ref{tab:betacorr_methods}, in the simulation column, and they 
 clearly show that the trend in the simulations is consistent with what we see in the \Planck\ data.  In particular, \textit{EEE} results show a lower degree of correlation in
simulated maps, for the same pairs of methods. The observed loss of correlation in polarization does not seem to come from unresolved foregrounds or other unaccounted systematics, but rather something intrinsic to the foreground-removal algorithms.  They are substantially different, as $\SMICA$ and $\NILC$ both perform the cleaning in harmonic space, at the level of \itE\ and \textit{B} multipoles,  whereas $\SEVEM$ is essentially a pixel-space template fitting method, performing the subtraction on \textit{Q} and \textit{U} maps, or inpainted before $\fnl$ estimation. These issues will be studied in greater detail in future work, using Wiener-filtered, as well as inpainted maps for $\fnl$ estimation. However, we have already seen that the larger scatter between modes from different foreground cleaning methods does not have a serious impact on $\fnl$ estimation, at least for the standard local, equilateral, and orthogonal shapes.  The non-standard shapes need to be analysed separately to check robustness of NG polarization results. This is the approach we will take 
for the various non-standard NG models. 

\begin{table}[!t]
\begingroup
\newdimen\tblskip \tblskip=5pt
\caption{The $r^2$ statistic, Eq.\eqref{eq:beta_rsquared}, showing the degree of correlation between measured bispectrum $\beta$ coefficients for the component separation methods shown in Fig.~\ref{fig:betacorr_methods} (upper three rows for temperature, lower for polarization). Correlation between  \SMICA, \NILC, and \SEVEM\ is excellent in temperature; however, it declines markedly for the latter in polarization.  Note that results are from the high resolution Modal 2 pipeline using the orthogonal \textit{EEE} component only.}\label{tab:betacorr_table} 
\nointerlineskip
\vskip -3mm
\footnotesize
\setbox\tablebox=\vbox{
 \newdimen\digitwidth 
 \setbox0=\hbox{\rm 0} 
 \digitwidth=\wd0 
 \catcode`*=\active 
 \def*{\kern\digitwidth}
 \newdimen\signwidth 
 \setbox0=\hbox{+} 
 \signwidth=\wd0 
 \catcode`!=\active 
 \def!{\kern\signwidth}
\halign{#\hfil\tabskip=0.3cm& \hfil#\hfil\tabskip=0.3cm&
\hfil#\hfil\tabskip=0.3cm&
\hfil#\hfil\tabskip=0.3cm& \hfil#\hfil\tabskip=0.cm\cr 
\noalign{\doubleline}
\noalign{\vskip -2pt}
&  ~\SEVEM~ & ~~\NILC~~ & \Commander  \cr 
\noalign{\vskip 5pt\hrule\vskip 3pt} 				
\SMICA\ (\itT)	&$	0.95	$&$	0.94	$&$	0.63	$\cr
\noalign{\vskip 2pt}
\SEVEM\ (\itT)	&$		$&$	0.92	$&$	0.66	$\cr
\noalign{\vskip 2pt}
\NILC\ (\itT)	&$		$&$		$&$	0.72	$\cr
\noalign{\vskip 5pt\hrule\vskip 3pt}
\SMICA\ (\itE)	&$	0.39	$&$	0.89	$&$	0.55	$\cr
\noalign{\vskip 2pt}
\SEVEM\ (\itE) 	&$		$&$	0.30	$&$	0.50	$\cr
\noalign{\vskip 2pt}
\NILC\ (\itE)	&$		$&$		$&$	0.43	$\cr
\noalign{\vskip 5pt\hrule\vskip 3pt}}}
\endPlancktablewide                    
\endgroup
\end{table}

\begin{table}[!t]                 
\begingroup
\newdimen\tblskip \tblskip=5pt
\caption{Correlation coefficients between pairs of \textit{EEE} bispectrum modes, extracted using two different masks for each  
of the four component-separated maps. We compare correlations measured from data 
with Monte Carlo averages over $100$ Gaussian realizations. The simulations 
were processed through the different component separation pipelines in the same way as the data, but do {\em not} include 
any foreground component. According to this test, modal expansions are stable for a change of sky coverage, with 
measured correlations in full agreement with expectations from simulations.}

\label{tab:betacorr_mask}
\nointerlineskip
\vskip -3mm
\footnotesize
\setbox\tablebox=\vbox{
   \newdimen\digitwidth 
   \setbox0=\hbox{\rm 0} 
   \digitwidth=\wd0 
   \catcode`*=\active 
   \def*{\kern\digitwidth}
   \newdimen\signwidth 
   \setbox0=\hbox{+} 
   \signwidth=\wd0 
   \catcode`!=\active 
   \def!{\kern\signwidth}
   \newdimen\signwidth 
   \setbox0=\hbox{.} 
   \signwidth=\wd0 
   \catcode`^=\active 
   \def^{\kern\signwidth}
\halign{\hbox to 0.9in{#\leaderfil}\tabskip 1em&
\hfil#\hfil\tabskip 1em&
\hfil#\hfil&
\hfil#\hfil&
\hfil#\hfil&
\hfil#\hfil\tabskip 0pt\cr
\noalign{\doubleline\vskip 2pt}
\omit&\multicolumn{2}{c}{\textit{EEE}}\cr
\noalign{\vskip 6pt}
Method \hfill&**Data     &*Simul.\cr
\noalign{\vskip 4pt\hrule\vskip 6pt}
\SMICA&0.87&0.87\cr
\noalign{\vskip 6pt}
\SEVEM&0.87&0.87\cr
\noalign{\vskip 6pt}
\NILC&0.87&0.87\cr
\noalign{\vskip 6pt}
\Commander&0.88&0.87\cr
\noalign{\vskip 3pt\hrule\vskip 4pt}
}}
\endPlancktablewide                 
\endgroup
\end{table}                        

\subsection{Dependence on sky coverage}
\label{sec:dep_mask}

For each of the four component separation methods, we have used two different 
polarization masks, namely the same polarization mask as in Sect.~\ref{sec:Results}, 
with $f_{\rm sky} = 0.74$ (defined as the polarization ``common mask'' in Sect.~\ref{sec:dataset}), 
and an extended mask with $f_{\rm sky} = 0.64$. The temperature mask 
is kept unchanged in this test, and it covers a sky fraction $f_{\rm sky} = 0.76$ (the temperature ``common mask'' of Sect.~\ref{sec:dataset}).  
We report the variation in $\fnl$ for the three standard shapes in Table \ref{tab:fnlchangemask}, 
which shows insensitivity to $f_{\rm sky}$, in agreement with earlier results on \textit{T+E}. In this case, however, the \textit{EEE} results also seem quite stable, supporting the view that foreground residuals are not affecting our local, equilateral, and orthogonal 
$\fnl$ results, especially for the final, combined \textit{T+E} measurements. 
Tests on FFP8 simulations including foregrounds (see Sect.~\ref{sec:FFP8tests}) suggest that $\fnl$ measurements obtained from the $\SMICA$ and $\SEVEM$ maps are the most accurate under the current choice of mask.
As a further check of these two methods we consider a third polarization 
mask, with $f_{\rm sky}=0.53$, and repeat the combined \textit{T+E} $\fnl$ measurement, also finding stable results. For $\SMICA$ we find $\fnllocal = 5.6 \pm 5.4$, $f_{\rm NL}^{\rm equil}= 65 \pm 58$, and $f_{\rm NL}^{\rm ortho}= -30 \pm 26$, 
while for $\SEVEM$ we obtain $\fnllocal = 9.4 \pm 5.4$, $f_{\rm NL}^{\rm equil}= 75 \pm 59$, $f_{\rm NL}^{\rm ortho} = -50 \pm 30$.

We also perform model-independent checks by  looking at the correlation coefficient between different sets of bispectrum modes, in a similar way to Sect.~\ref{sec:betacomp}, but now changing the polarization mask.
Results are reported in Table \ref{tab:betacorr_mask}, 
and confirm that the data and simulations behave similarly, and that polarization modes display a lower correlation level than temperature.
%

\begin{table*}[t!]                 
\begingroup
\newdimen\tblskip \tblskip=5pt
\caption{Comparison of component separation methods, using Gaussian FFP8 simulations. We firstly consider Gaussian, foreground-free simulations, with simulated noise for each frequency band, 
process them through each 
of the four foreground cleaning pipelines, and measure $\fnl$ for the three standard shapes (columns labelled with ``Input map''). 
We then include foregrounds and repeat the measurement, {\em before} applying the cleaning, and including realistic noise levels for each method (columns labelled with ``Input map $+$ foregrounds"); 
this step is performed in order to get an idea of the level of contamination introduced by foregrounds, before cleaning. 
Finally, we apply the different component separation methods, and again estimate $\fnl$ from the final maps (columns labelled with ``Cleaned map''). The discrepancies between $\fnl$ measured on the input 
map, and $\fnl$ extracted from the cleaned map, provide a figure of merit to assess how well foregrounds are subtracted by different methods. Results below have been obtained with the KSW estimator and the ``cleaned
map'' results were also checked with the binned estimator.}
\label{tab:FFP8test}
\nointerlineskip
\vskip -3mm
\footnotesize
\setbox\tablebox=\vbox{
   \newdimen\digitwidth 
   \setbox0=\hbox{\rm 0} 
   \digitwidth=\wd0 
   \catcode`*=\active 
   \def*{\kern\digitwidth}
   \newdimen\signwidth 
   \setbox0=\hbox{+} 
   \signwidth=\wd0 
   \catcode`!=\active 
   \def!{\kern\signwidth}
   \newdimen\dotwidth 
   \setbox0=\hbox{.} 
   \dotwidth=\wd0 
   \catcode`^=\active 
   \def^{\kern\dotwidth}
\halign{\hbox to 0.5in{#\leaderfil}\tabskip 1em&
\hfil#\hfil\tabskip 0.8em&
\hfil#\hfil&
\hfil#\hfil&
\hfil#\hfil&
\hfil#\hfil&
\hfil#\hfil&
\hfil#\hfil&
\hfil#\hfil&
\hfil#\hfil\tabskip 0pt\cr
\noalign{\doubleline\vskip 2pt}
\omit&\multicolumn{3}{c}{\hfil Input map \hfil}&
\multicolumn{3}{c}{Input map $+$ foregrounds \hfil}&\multicolumn{3}{c}{\hfil\hfil\hfil Cleaned map }\cr
\noalign{\vskip 6pt}
\omit&Local&Equilateral&Orthogonal&Local&Equilateral&Orthogonal&Local&Equilateral&Orthogonal\cr
\noalign{\vskip 4pt\hrule\vskip 6pt}
\omit\hfil \SMICA \hfil&&\cr
\itT    & !*5.2 $\pm$ *5.8& !*29 $\pm$ *71& *$-$8 $\pm$ 34& $-$107.0 $\pm$ *5.8& *$-$23  $\pm$ *71& !*27 $\pm$ 34&  !*7.8 $\pm$ *5.8& !*38 $\pm$ *71 & *$-$20 $\pm$ 34\cr
\itE   &$-$39^* $\pm$ 30^*& *$-$99 $\pm$ 133& !59 $\pm$ 69& *$-$10^* $\pm$ 30^*& $-$154 $\pm$ 133&  *$-$41 $\pm$ 69& $-$56^* $\pm$ 30^*& $-$120 $\pm$ 133&  !*65 $\pm$ 34\cr
\itTpE & !*5.9 $\pm$ *5.1& !*14 $\pm$ *45& *$-$20 $\pm$ 22& $-$118.0 $\pm$ *5.1& *$-$32 $\pm$ *45&  !**8 $\pm$ 22& *!8.3 $\pm$ *5.2& !*14 $\pm$ *45& *$-$22 $\pm$ 22\cr
\noalign{\vskip 6pt}
\omit\hfil \SEVEM \hfil&&\cr
\itT   & !*5.6 $\pm$ *5.7& !*32 $\pm$ *69& *$-$8 $\pm$ 32& $-$113.2 $\pm$ *5.7& **$-$8 $\pm$ *69& !*34 $\pm$ 32& !12.7 $\pm$ *5.7& !*35 $\pm$ *69& *$-$25 $\pm$ 32\cr
\itE   & $-$17^* $\pm$ 41^*& $-$149 $\pm$ 175& !28 $\pm$ 95& *$-$14^* $\pm$ 41*^& $-$171 $\pm$ 175&  *$-$44 $\pm$ 95& $-$22^* $\pm$ 41^*& $-$120 $\pm$ 175&  !*41 $\pm$ 95\cr
\itTpE & !*7.7 $\pm$ *5.3& !*12 $\pm$ *49& $-$37 $\pm$ 24& $-$126.0 $\pm$ *5.3& *$-$29 $\pm$ *49&  *$-$57 $\pm$ 25& !13.0 $\pm$ *5.3& !*11 $\pm$ *49& *$-$41 $\pm$ 24 \cr
\noalign{\vskip 6pt}
\omit\hfil \NILC \hfil&&\cr
\itT   & !*5.1 $\pm$ *5.7& !*32 $\pm$ *69& *$-$5 $\pm$ 31& $-$102.0 $\pm$ *5.7& *$-$14 $\pm$ *69& !*32 $\pm$ 31& !17.8 $\pm$ *5.7& !*85 $\pm$ *69& *$-$16 $\pm$ 31\cr
\itE   &$-$52^* $\pm$ 33^*& $-$157 $\pm$ 156& !72 $\pm$ 73& **$-$6*^ $\pm$ 33*^& $-$155 $\pm$ 156 &  *$-$47 $\pm$ 73& $-$76^* $\pm$ 33^*& $-$179 $\pm$ 156& !113 $\pm$ 73\cr
\itTpE & !*5.7 $\pm$ *5.0& !**7 $\pm$ *46& $-$15 $\pm$ 21& $-$117.0 $\pm$ *5.9& *$-$27 $\pm$ *46&  !*12 $\pm$ 21& !15.8 $\pm$ *5.0& *$-$20 $\pm$ *46&  **$-$7 $\pm$ 21\cr
\noalign{\vskip 6pt}
\omit\hfil \Commander \hfil&&\cr
\itT   & !*0.5 $\pm$ *6.2& **$-$5 $\pm$ *73& *$-$14 $\pm$ 36& $-$127.0 $\pm$ *6.2& *$-$25 $\pm$ *73& $-$137 $\pm$ 36 & !25.6 $\pm$ *6.2& !*67 $\pm$ *73& *$-$17 $\pm$ 36\cr
\itE   & $-$51^* $\pm$ 38^*& *$-$64 $\pm$ 160& !*93 $\pm$ 86& *$-$10^* $\pm$ 38^*& $-$153 $\pm$ 160&  *$-$45 $\pm$ 86& $-$70^* $\pm$ 38^*& *$-$78 $\pm$ 159& !138 $\pm$ 86\cr
\itTpE & !*1.6 $\pm$ *5.4& **$-$2 $\pm$ *48& *$-$21 $\pm$ 23& $-$137.0 $\pm$ *5.4& *$-$29 $\pm$ *48&  !*13 $\pm$ 23& !20.4 $\pm$ *5.4& *!28 $\pm$ *48& *$-$11 $\pm$ 23\cr
\noalign{\vskip 3pt\hrule\vskip 4pt}
}}
\endPlancktablewide                 
\endgroup
\end{table*}                        

\begin{table}[t!]                 
\begingroup
\newdimen\tblskip \tblskip=5pt
\caption{Same test as in Table \ref{tab:FFP8test}, but with a NG map as input, with $\fnllocal=8.8$. For this case, we only 
report the final value after foreground cleaning for each method. Results below have been obtained 
with the KSW estimator and double-checked with the binned estimator. ISW-lensing contributions are removed.}
\label{tab:FFP8testNG}
\nointerlineskip
\vskip -3mm
\footnotesize
\setbox\tablebox=\vbox{
   \newdimen\digitwidth 
   \setbox0=\hbox{\rm 0} 
   \digitwidth=\wd0 
   \catcode`*=\active 
   \def*{\kern\digitwidth}
   \newdimen\signwidth 
   \setbox0=\hbox{+} 
   \signwidth=\wd0 
   \catcode`!=\active 
   \def!{\kern\signwidth}
   \newdimen\dotwidth 
   \setbox0=\hbox{.} 
   \dotwidth=\wd0 
   \catcode`^=\active 
   \def^{\kern\dotwidth}
\halign{\hbox to 0.5in{#\leaderfil}\tabskip 1em&
\hfil#\hfil\tabskip 0.8em&
\hfil#\hfil&
\hfil#\hfil&
\hfil#\hfil&
\hfil#\hfil&
\hfil#\hfil&
\hfil#\hfil&
\hfil#\hfil&
\hfil#\hfil\tabskip 0pt\cr
\noalign{\doubleline\vskip 2pt}
\omit&\multicolumn{3}{c}{Cleaned map. Input $\fnllocal=8.8$}\cr
\noalign{\vskip 6pt}
\omit&Local&Equilateral&Orthogonal\cr
\noalign{\vskip 4pt\hrule\vskip 6pt}
\omit\hfil \SMICA \hfil&&\cr
\itT    &!*3.1 $\pm$ *5.8& !*47 $\pm$ *71& **$-$6 $\pm$ 34\cr
\itE   & $-$53^* $\pm$ 30^*& $-$113 $\pm$ 133& !*94 $\pm$ 69\cr
\itTpE & !*5.7 $\pm$ *5.1& *!22 $\pm$ *45& *$-$19 $\pm$ 22\cr
\noalign{\vskip 6pt}
\omit\hfil \SEVEM \hfil&&\cr
\itT   & !*8.0 $\pm$ *5.7& !*43 $\pm$ *69& *$-$11 $\pm$ 32 \cr
\itE   &$-$19^* $\pm$ 41^*& $-$112 $\pm$ 175& !*35 $\pm$ 95 \cr
\itTpE & !10.2 $\pm$ *5.3& !*19 $\pm$ *49& *$-$37 $\pm$ 24\cr
\noalign{\vskip 6pt}
\omit\hfil \NILC \hfil&&\cr
\itT   & !10.2 $\pm$ *5.7& !*84 $\pm$ *69& !**7 $\pm$ 31 \cr
\itE   &$-$76^* $\pm$ 33^*& $-$179 $\pm$ 156& !113 $\pm$ 73 \cr
\itTpE & !10.1 $\pm$ *5.0& *!20 $\pm$ *46& !**4 $\pm$ 21 \cr
\noalign{\vskip 6pt}
\omit\hfil \Commander \hfil&&\cr
\itT   & !22.2 $\pm$ *6.2& !*81 $\pm$ *73& **$-$5 $\pm$ 36 \cr
\itE   &$-$68^* $\pm$ 38^*& *$-$78 $\pm$ 160& !132 $\pm$ 86 \cr
\itTpE &!18.3 $\pm$ *5.4& *!35 $\pm$ *48& **$-$9 $\pm$ 23  \cr
\noalign{\vskip 3pt\hrule\vskip 4pt}
}}
\endPlancktablewide                 
\endgroup
\end{table}                        

\subsection{Tests on simulations}\label{sec:FFP8tests}

We consider two realistic data simulations, one of which is Gaussian,
while the other includes local NG. We start with a foreground-free realization, add foregrounds according to the Planck Sky Model, and finally process through the four component separation pipelines. By estimating 
$\fnl$ in the input foreground-free simulation, for each method, and comparing to $\fnl$ 
recovered from the cleaned maps (or with the input local $\fnl$, for the NG 
case), we can assess both the impact of foregrounds 
on our measurement before subtraction and which method gives the 
highest accuracy.  The necessity to clean is very apparent in the middle set of columns in Table~\ref{tab:FFP8test}, where no cleaning has been performed.

$\SMICA$ and $\SEVEM$  give 
the best results, both in the G and NG tests. In the G test, reported in Table 
\ref{tab:FFP8test}, $\SMICA$ results show 
an agreement between the input and the cleaned map at the level 
$\sigma_{\fnl}/2$ for all shapes, and for all of \textit{TTT}, \textit{EEE}, and \textit{T+E}. $\SEVEM$ displays a similar level of accuracy, except for $\fnllocal$, where 
the difference is larger, but within one standard deviation. $\NILC$ and $\Commander$ 
clearly perform worse for the local shape, with $\NILC$ showing a 2$\,\sigma_{\fnl}$ difference, and 
$\Commander$ even larger than that. In the NG test, reported in Table \ref{tab:FFP8testNG},
$\SEVEM$ gives the most accurate results, 
recovering the input with $\sigma_{\fnl}/2$ accuracy or better. Results for $\SMICA$ are accurate at the 1$\,\sigma_{\fnl}$ level for the \textit{TTT} constraint, and worse (about 2$\,\sigma_{\fnl}$ ) in \textit{EEE}. However, the combined \textit{T+E} measurement is again very good ($\sigma_{\fnl}/2$).
 $\NILC$ is also performing very well in \textit{TTT} and \textit{T+E}, while the \textit{EEE} 
result is more than 2$\,\sigma_{\fnl}$ off.

The test described here has several limitations, 
the main and most obvious one being that it has been performed 
on just two maps (due to lack of availability of a large sample of this type of simulation at this stage). Another clear issue is that some methods, in particular $\Commander$, seem to perform much better on actual data than on these simulations. On the other hand some important trends, observed in the data 
in previous tests,
are clearly reproduced here, like the good stability of $\SMICA$ and $\SEVEM$, especially for the combined 
\textit{T+E} results and, most notably, the fact that the clear degradation in the accuracy of the \textit{EEE} measurement 
for some methods does {\em not} seem to propagate to \textit{T+E}.

\begin{figure*}[!t]
\centering
\includegraphics[width=.28\linewidth]{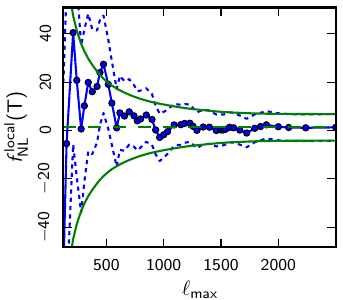}
\includegraphics[width=.28\linewidth]{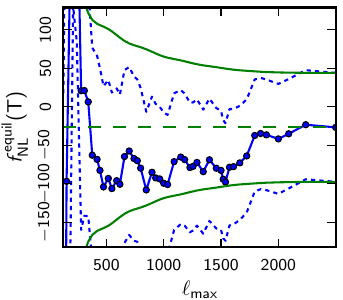}
\includegraphics[width=.28\linewidth]{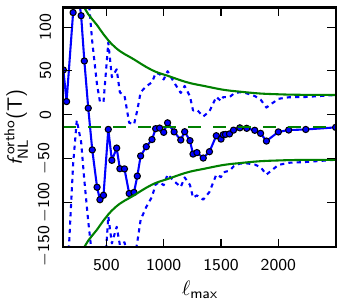}

\includegraphics[width=.28\linewidth]{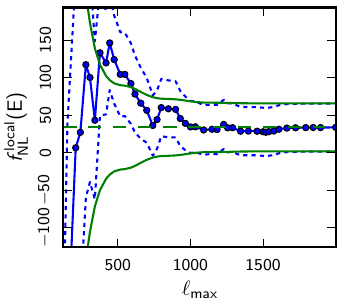}
\includegraphics[width=.28\linewidth]{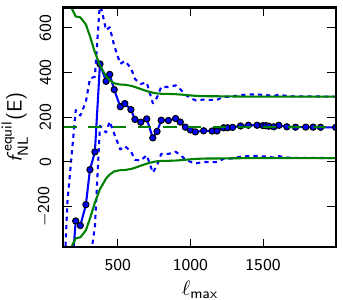}
\includegraphics[width=.28\linewidth]{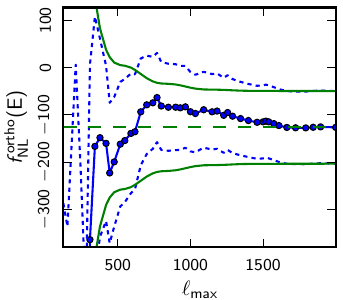}

\includegraphics[width=.28\linewidth]{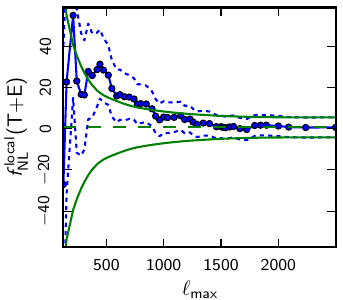}
\includegraphics[width=.28\linewidth]{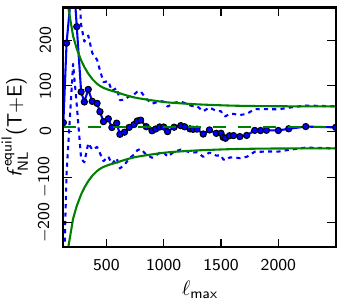}
\includegraphics[width=.28\linewidth]{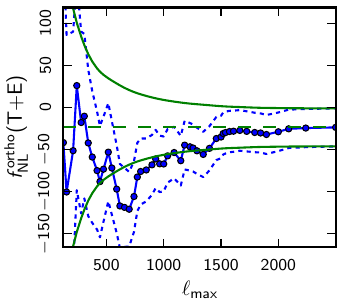}
\caption{Evolution of the $f_\mathrm{NL}$ parameters 
(solid blue line with data points) and their uncertainties (dashed lines) 
for the three primordial bispectrum templates as a function of the 
maximum multipole number $\ell_\mathrm{max}$ used in the analysis. From 
left to right the figures show, respectively local, 
equilateral, and orthogonal, while the different rows from top to bottom
show results for \itT-only, \itE-only, and full \textit{T+E}.
To indicate more clearly the evolution of the uncertainties, they are also 
plotted around the final value of $f_\mathrm{NL}$ (solid green lines without 
data points, showing the $\pm 1\,\sigma$ range around the dashed green lines).
The results are for \SMICA, and assume all shapes to be independent, and that the
ISW-lensing bias has been subtracted. They have
been determined with the binned bispectrum estimator.}
\label{Fig_lmaxdep}
\end{figure*}

\begin{figure*}[!t]
\centering
\includegraphics[width=.28\linewidth]{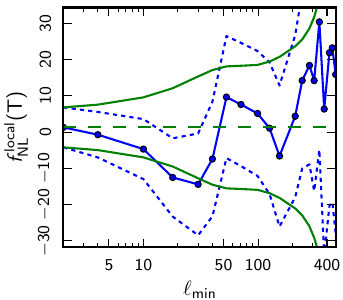}
\includegraphics[width=.28\linewidth]{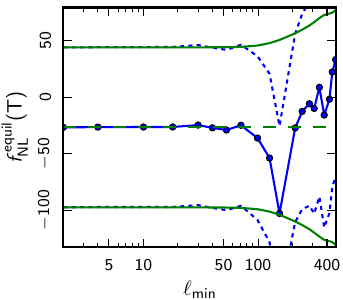}
\includegraphics[width=.28\linewidth]{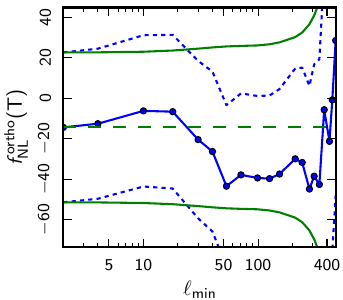}

\includegraphics[width=.28\linewidth]{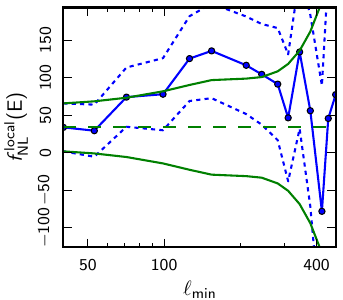}
\includegraphics[width=.28\linewidth]{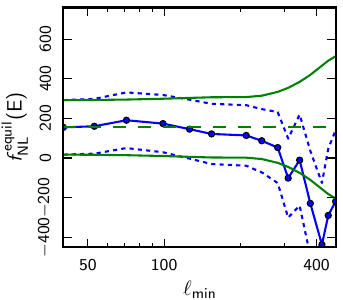}
\includegraphics[width=.28\linewidth]{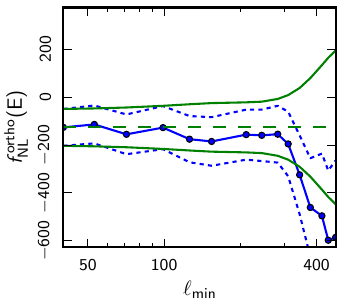}

\includegraphics[width=.28\linewidth]{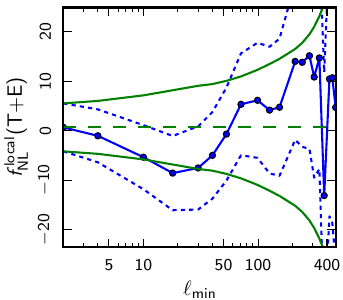}
\includegraphics[width=.28\linewidth]{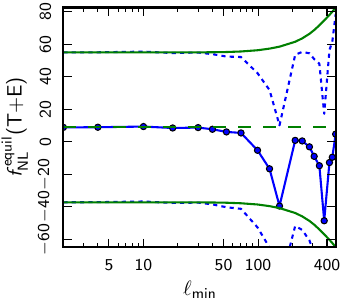}
\includegraphics[width=.28\linewidth]{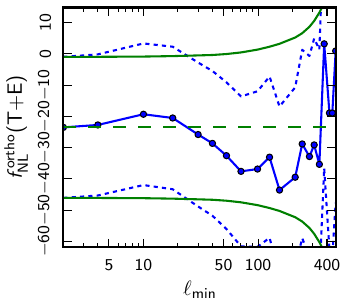}
\caption{Evolution of the $f_\mathrm{NL}$ parameters 
(solid blue line with data points) and their uncertainties (dashed lines) 
for the three primordial bispectrum templates as a function of the 
minimum multipole number $\ell_\mathrm{min}$ used in the analysis. From 
left to right the figures show respectively, local, 
equilateral, and orthogonal, while the different rows from top to bottom
show results for \itT-only, \itE-only, and full \textit{T+E}.
To indicate more clearly the evolution of the uncertainties, they are also 
plotted around the final value of $f_\mathrm{NL}$ (solid green lines without 
data points, with $\pm 1\,\sigma$ range around the dashed green line).
The results are for \SMICA, and assume all shapes to be independent, as well as that the
ISW-lensing bias has been subtracted. They have
been determined with the binned bispectrum estimator.}
\label{Fig_lmindep}
\end{figure*}

\subsection{Dependence on multipole number}\label{sec:ldep}

In this section we discuss another stability test of our results, namely
the dependence of the results for $f_\mathrm{NL}$ on the maximum and 
minimum multipole number used in the analysis. This test is most
easily performed with the binned bispectrum estimator, since it
gets the dependence of $f_\mathrm{NL}$ on $\ell$ for free with its standard 
analysis, simply by leaving out bins in the final sum when computing
$f_\mathrm{NL}$ (the binned equivalent of Eq. (\ref{eq:bispec_innerprod})).

The dependence on $\ell_\mathrm{max}$ of the results for the three standard primordial
shapes (local, equilateral, and orthogonal modes), is shown in 
Fig.~\ref{Fig_lmaxdep}, for \itT-only, \itE-only, and full \textit{T+E}. 
As mentioned in Sect.~\ref{sec:Results}, the KSW and binned estimators
use $\ell_\mathrm{max}=2500$ for temperature, while the modal estimators
use $\ell_\mathrm{max}=2000$. As can be seen in the figure, both the \itT-only
and \textit{T+E} results are basically unchanged between $\ell=2000$ and $\ell=2500$
for all three shapes, showing that this difference has no impact on the
results (as was to be expected from the excellent agreement between
estimators in Table~\ref{tab:fNLsmicah}). In fact, the values of
$f_\mathrm{NL}$ for \itT\ and \textit{T+E} are reasonably stable (given their error bars) 
down to much lower values of $\ell_\mathrm{max}$, of about 1000.

On the polarization side we can draw a similar conclusion. The binned
estimator uses $\ell_\mathrm{max}=2000$ for polarization, while the other 
estimators use $\ell_\mathrm{max}=1500$, but we see that results for \itE\ remain
basically unchanged between $\ell=1500$ and $\ell=2000$. Central values
and error bars for \itE\ for all three shapes have clearly converged by 
$\ell=1500$, and are in fact reasonably stable down to much lower values
of about 700.

As we noted in the 2013 analysis \citep{planck2013-p09a}, when going to the much
lower WMAP resolution of $\ell_\mathrm{max}\simeq 500$, we
agree with the slightly high value of $f_\mathrm{NL}^\mathrm{local}$ that the WMAP team
reported \citep{2012arXiv1212.5225B}. This value is also confirmed
when including polarization. One can clearly see the value of the
higher resolution of Planck.

The dependence on $\ell_\mathrm{min}$ is shown in Fig.~\ref{Fig_lmindep}. Here
all estimators used the same values, $\ell_\mathrm{min}^T=2$ and
$\ell_\mathrm{min}^E=40$. As explained in \cite{planck2014-a08},
not  all systematic and foreground uncertainties in the low-$\ell$ HFI polarization data have been fully characterized yet, and hence
it was decided to filter out these data.

For equilateral and orthogonal shapes the values for $f_\mathrm{NL}$ and their 
error bars are quite stable as a function of $\ell_\mathrm{min}$ up to about
$\ell = 100$ (and $\ell\simeq 300$ for \itE-only), which is not 
surprising, since these templates have little weight at low $\ell$. The local
template, on the other hand, depends very strongly on the lowest multipoles,
which is reflected in the very rapid growth of the error bars when
$\ell_\mathrm{min}$ increases. Looking at \itT-only and \textit{T+E} we see a very similar
pattern, with $f_\mathrm{NL}^\mathrm{local}$ being reasonably stable, although 
there are some jumps. The local result for \itE-only wanders a bit
more outside of the $\pm 1\,\sigma$ region, in agreement with the other tests in 
this section, which also indicate that \itE-only is not as stable as 
\itT-only and \textit{T+E}. However, that is still quite acceptable, given the small
weight of \itE-only in the full \textit{T+E} result.

One can work out quite generally that when $Y$ is a subset of a data set $X$,
and $P_X$ and $P_Y$ are the values of a parameter $P$ determined from these
two data sets, then the variance of the difference $P_Y-P_X$ is equal to
$|\mathrm{Var}(P_Y)-\mathrm{Var}(P_X)|$ 
Hence we can determine how
likely the jumps in $f_\mathrm{NL}$ as a function of $\ell_\mathrm{min}$ are.
It turns out that the jump in the \itT-only value of $f_\mathrm{NL}^\mathrm{local}$
between $\ell_\mathrm{min}=40$ and $\ell_\mathrm{min}=53$ is a $2.5\sigma$ effect
(using the values of $f_\mathrm{NL}^\mathrm{local}$ before subtraction of the 
ISW-lensing bias, which also depends on $\ell$). Similarly, the jump in the 
\itT-only value of $f_\mathrm{NL}^\mathrm{equil}$ between $\ell_\mathrm{min}=154$ and $\ell_\mathrm{min}=211$ is (by chance) also a $2.5\sigma$ effect. Given the fact
that there are 57 bins, having such a jump appears to be consistent from
a statistical point of view.

\subsection{A directional analysis with a needlet-based modal estimator}

The validation tests on simulations, described in Sect.~\ref{sec:FFP8tests}, point to 
{\SMICA} and {\SEVEM} as the best foreground cleaning methods for 
$\fnl$ estimation. Results in Table \ref{tab:fNLsmicah} show that
 {\SMICA} also has slightly smaller error bars, thus making 
it the method of choice for our final results.

As a further check of residual foreground contamination in the {\SMICA} map, 
in this section we investigate the possible directional
dependence of {\SMICA}-derived third-order statistics by means of a needlet-based
modal estimator (i.e., an estimator based on the decomposition described in Sect.~\ref{sec:modal_est}, and 
references therein, where we use cubic combinations of needlets as our basis modes). 
In other words, we analyse the behaviour of the
needlet bispectrum (see \citealt{lan.mar}, \citealt{oi.fro.mar} and \citealt{donz.mar}) on separate 
patches of the sky,
and we study the fluctuations of the corresponding residuals. 
\begin{figure}[tp]
 \includegraphics[scale=0.45]{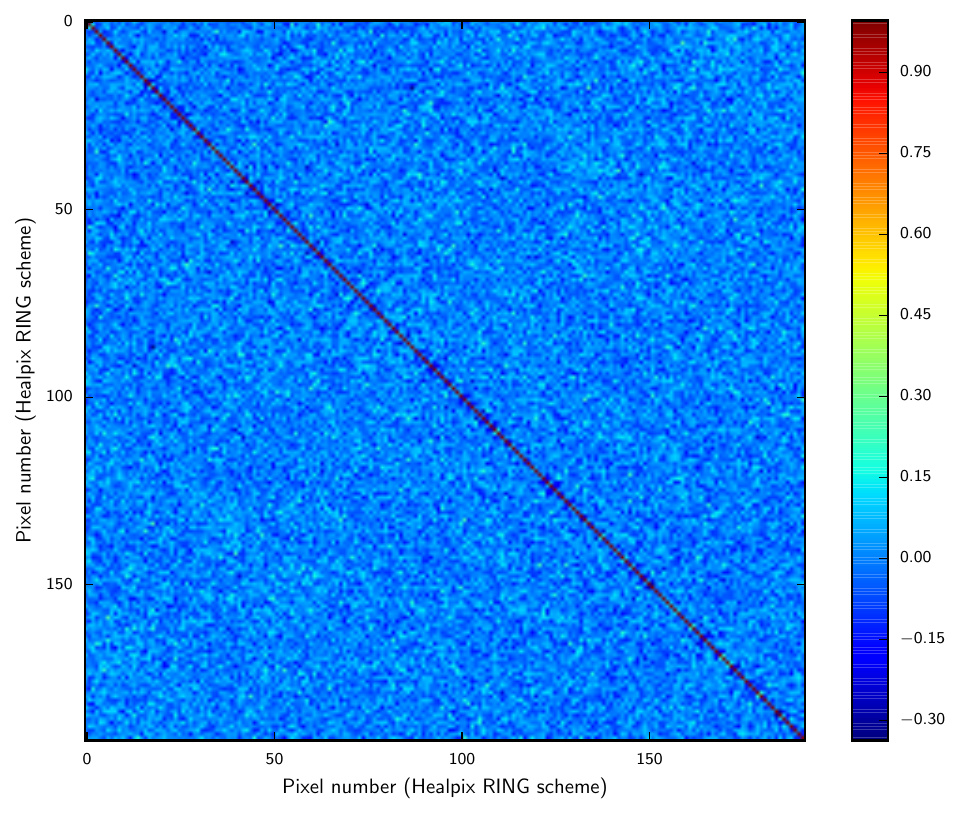}
 \caption{Temperature only, local shape, pixel correlation matrix from Monte Carlo analysis, at $N_{\rm side}=4$. The most correlated pixels are those closest to the main diagonal; 
however, these values are always lower than $34\,\%$ in the chosen case.}\label{fig:corr_4_T}
\end{figure}

Rather than assuming a specific anisotropic model, we instead calculate the 
contribution to the local $\fnl$ from different
regions of the sky and look for evidence of anisotropy in the result.\footnote{Even though we focus here on directional 
contributions to the local shape (which was typically found as one of the most sensitive to residual foreground contamination), 
this type of directional 
analysis can be done in a model-independent way, by looking separately at different needlet modes; 
we leave this for future work.}

\begin{figure*}[tp]
\includegraphics[width=0.48\textwidth]{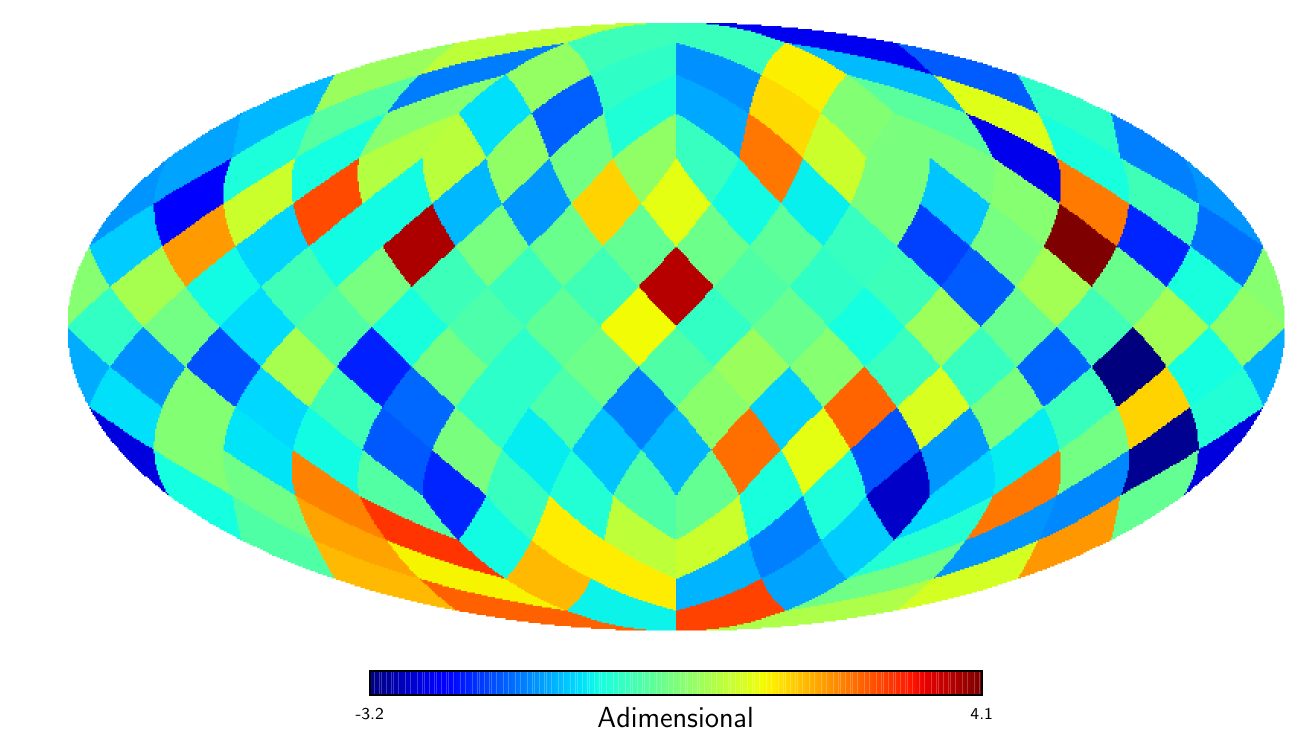}
\includegraphics[width=0.48\textwidth]{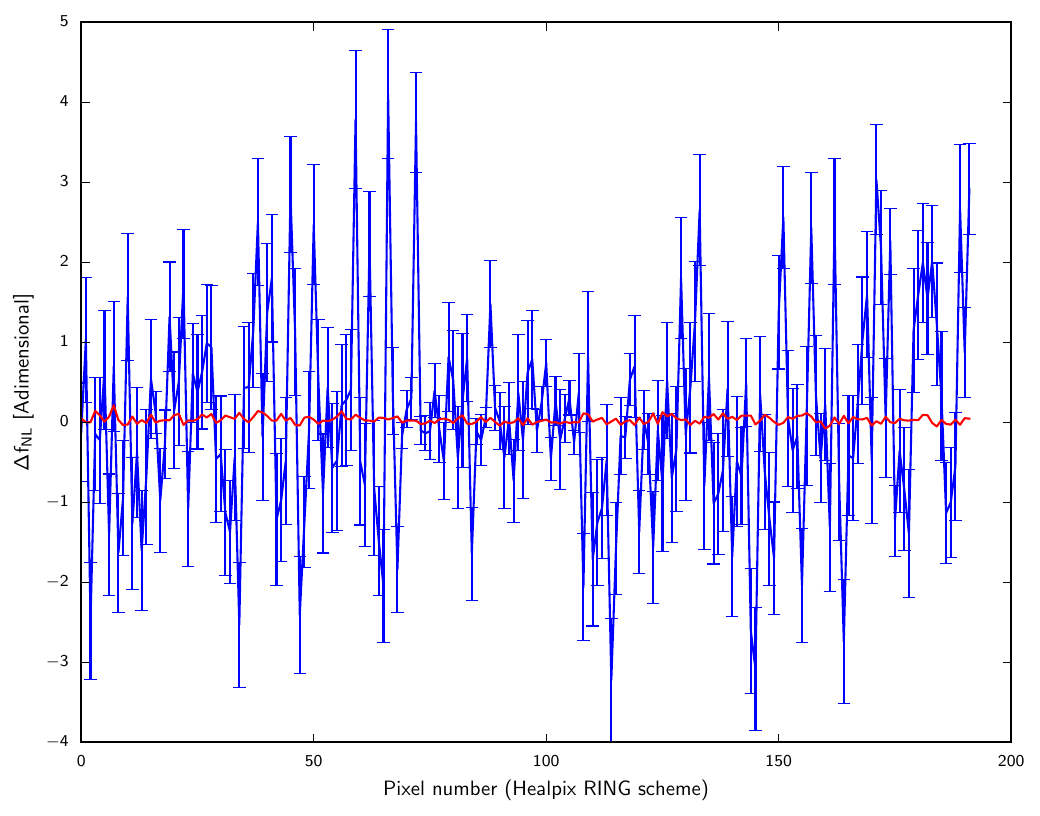}
\caption{Temperature only, local $\fnl$ directional contributions from
{\SMICA}. As explained in the text, summing over all the pixel values would
give the full sky $\fnl$ needlet estimator result. The left panel displays
the directional $\fnl$ map. On the right, the blue
points represent the $\fnl$ contibution for each direction (i.e., for each pixel in the directional map), 
with Monte Carlo error bars. The red line is the average from simulations, which is consistent
with zero. }\label{fig:smica_needlets}
\end{figure*} 

Our modal needlet estimator has been validated with 
respect to the procedures
considered in Sect.~\ref{sec:Validation}, showing excellent agreement. Since in this paper we use
the needlet estimator only in this section, and as a diagnostic 
tool, we will not explicitly report the outcome of these validation tests here, for the sake of conciseness.
The advantages of using needlet-based modal estimators have been
advocated in \cite{lan.mar}, \cite{oi.fro.mar}, \cite{2010PhRvD..82b3502F}, and \cite{2012JCAP...12..032F}; 
we refer to these papers for more
discussion and details. In short, however, they can be summarized
as follows:

\begin{enumerate}

\item it is possible to achieve a strong data compression, i.e., to
investigate cubic statistics by means of a small number of modes
(needlet frequencies) for many different bispectrum templates;

\item needlet transforms have good correlation properties in pixel
space, which allows study of the pixel contribution to the $\fnl$
signal for the templates under investigation, by treating 
different directions independently.

\end{enumerate}

In our analysis, we first divide the sky into several large ``regions'', with boundaries defined by the pixels of a {\tt{HEALPix}} grid \citep{gorski2005}. at lower resolution than the starting map (which is at $N_{\rm side}=2048$). 
For the low resolution grid we consider $N_{\rm side}$ values between $2$ and $8$. For each pixel in the coarse grid, 
we then compute the local $\fnl$ using our modal needlet estimator, and neglect contributions from external regions. 

The correlation matrices between $\fnl$ measurements in different regions were computed via Monte Carlo simulations.  $N_{\rm side}=4$ was chosen, as providing the best tradeoff between having a large number of regions for directional analysis (i.e., the total number of pixels in the low resolution grid), and a low correlation between different regions.  This is shown in Fig.~\ref{fig:corr_4_T}.
It is readily seen that, at $N_{\rm side}=4$, the correlation is largely
concentrated in one or two points near the main diagonals (where it is still low, never exceeding $34\,\%$), and falls off rapidly for all other pixels.
Note that the results here refer to temperature only. The \textit{EEE} local polarization error bars are large, even for the full sky analysis, making this directional approach uninformative for \Planck\ polarization data. We concentrate on the \textit{TTT} bispectrum here, complementing other validation tests in this section, which are mostly focused on polarization.

Having obtained our correlation matrices, and having shown that different regions are essentially uncorrelated, we can then proceed to extract $\fnl$
 for each region in the actual {\SMICA} map. As a test of directional-dependent contamination of the $\fnl$ measurement, we can also compare our results, region by region,
with the fluctuations expected by looking at the diagonal of our Monte Carlo correlation matrix.

The results of this analysis are shown in Fig.~\ref{fig:smica_needlets}. In the left panel, we represent the directional local $\fnl$ map, extracted with this method. In the right panel, we report 
the $\fnl$ values, region by region, and compare them to expectations from simulations. The red line gives the expected standard deviation, while the blue one gives estimates on the component-separated 
maps with the Monte Carlo error bars. 
Our estimator is normalized in such a way that the sum of all these contributions would yield exactly the $\fnl$ estimator for the full map, 
so these results can be viewed as a partition of the estimates along the different directions. It is readily seen that no significant fluctuation occurs, 
so that our results are consistent with the absence of directionally-dependent features (which could occur due to, for instance, residual foreground contamination).  
As an additional check, we have also investigated the possible presence of a dipole in these data, and found that our results are consistent with Gaussian isotropic simulations.

\subsection{Summary of the main validation results}\label{sec:data_valid_summary}

Throughout Sect.~\ref{sec:Sec_valid_data} we have shown a battery of tests aimed at 
evaluating the robustness of our data set, from the point of view of 
bispectrum estimators, focusing especially on the polarizaton part. 
We studied the stability 
of our results (local, equilateral, and orthogonal $\fnl$ measurements, 
model-independent bispectrum reconstruction)
under a change of sky coverage, multipole range, and choice 
 of component separation methods. We also considered simulated 
 data sets and studied the ability of different component separation 
methods to recover the input $\fnl$ after foreground subtraction. Our 
main conclusions from these tests can be summarized as follows:

\begin{itemize}

\item \textit{TTT} and \textit{T+E} results are stable both in the pixel and harmonic domains, for different 
component separation methods. For {\SMICA}, we also checked that \textit{TTT} temperature constraints on 
local $\fnl$ show no evidence of a directional variation via a needlet-based analysis. 

\item {\SMICA} and {\SEVEM} perform better than {\NILC} and {\Commander} at recovering 
 the original $\fnl$ in foreground-cleaned simulations. 
At the same time, {\SMICA} allows slightly better constraints on $\fnl$ than {\SEVEM}, due to a lower (by a small amount) 
noise level.

\item \textit{EEE} bispectra, and related $\fnl$ measurements, have some problems, and do not 
pass all the tests. Different 
component separation methods show a low level of consistency (especially 
when comparing pixel-based cleaning methods to harmonic-based cleaning methods). This disagreement 
is only partly alleviated by choosing a larger \itE-mask, so that residual foregrounds do not seem 
to fully explain all issues. 
An important caveat, already pointed out previously, is that the noise 
model in polarized FFP8 simulations is known to underestimate the actual noise level in the data, 
leading to some degree of underestimation of the error bars in our \textit{EEE} results. We stress again, however,
that this has little impact on the final \textit{T+E} constraints, due to the high noise level 
and consequent low statistical weight of \textit{EEE} bispectra. This was verified in detail both on data and 
 on simulations.

\end{itemize}

In light of the above analysis, we conclude that, as far as bispectrum estimation is concerned,
 our best cleaned map is the one produced by {\SMICA}, in line with our previous $2013$ analysis.
We also conclude that our \textit{TTT}-based $\fnl$  constraints, summarized in Tables 
 \ref{tab:fNLsmicah} and \ref{Tab_KSW+SMICA}, are robust. Joint \textit{T+E} constraints pass all our validation tests. On the other hand, in the light of the remaining issues in the \textit{EEE} bispectra and in the FFP8 polarized simulations, as we stressed at the end of Sect.~\ref{fnl_loc_eq_ort_results}, we suggest that all measurements that include polarization data in this paper should be regarded as {\em preliminary}.

%% file: A19_Section8.tex
\begin{table*}[h]                 
\begingroup
\newdimen\tblskip \tblskip=5pt
\caption{Results for local isocurvature NG, determined from the 
\SMICA\ \Planck\ 2015 map with the binned bispectrum estimator. In each case the
adiabatic mode is considered together with one isocurvature mode
(either cold dark matter, neutrino density, or neutrino velocity isocurvature).
As explained in the text this gives six different $f_\mathrm{NL}$
parameters, indicated by the different combinations of the adiabatic (a)
and isocurvature (i) modes. Results with two significant digits are shown for 
both an independent and a fully joint analysis, for \itT-only, \itE-only, and 
full \itT$+$\itE. In all cases the ISW-lensing bias has been subtracted.}
\label{Tab_isocurvNG}
\nointerlineskip
\vskip -3mm
\footnotesize
\setbox\tablebox=\vbox{
   \newdimen\digitwidth 
   \setbox0=\hbox{\rm 0} 
   \digitwidth=\wd0 
   \catcode`*=\active 
   \def*{\kern\digitwidth}
   \newdimen\signwidth 
   \setbox0=\hbox{+} 
   \signwidth=\wd0 
   \catcode`!=\active 
   \def!{\kern\signwidth}
   \newdimen\dotwidth 
   \setbox0=\hbox{.} 
   \dotwidth=\wd0 
   \catcode`^=\active 
   \def^{\kern\dotwidth}
\halign{\hbox to 0.8in{#\leaderfil}\tabskip 1em&
\hfil#\hfil\tabskip 1em&
\hfil#\hfil&
\hfil#\hfil&
\hfil#\hfil&
\hfil#\hfil&
\hfil#\hfil\tabskip 0pt\cr
\noalign{\doubleline\vskip 2pt}
\omit&\multispan6\hfil $f_{\rm NL}$\hfil\cr
\omit&\multispan6\hrulefill\cr
\omit&\multicolumn{3}{c}{\hfil Independent\hfil}&
\multicolumn{3}{c}{\hfil ******Joint\hfil}\cr
\noalign{\vskip 2pt}
Shape\hfill & Cold dark matter & Neutrino density & Neutrino velocity & ***** Cold dark matter & Neutrino density & Neutrino velocity \cr
\noalign{\vskip 4pt\hrule\vskip 6pt}
\itT\ a,aa& !***1.3 $\pm$ ***5.4& !***1.3 $\pm$ ***5.4& !**1.3 $\pm$ **5.4& ***** !***21 $\pm$ **13& **$-$27 $\pm$ **52& **$-$32 $\pm$ **48\cr
\itT\ a,ai& ***$-$2^* $\pm$ **10^*& ***$-$4^* $\pm$ **15^*& !*47^* $\pm$ *29^*& ***** ***$-$39 $\pm$ **26& !*140 $\pm$ *210& !*370 $\pm$ *350\cr
\itT\ a,ii& *!*59^* $\pm$ *910^*& *$-$130^* $\pm$ *280^*& !750^* $\pm$ 360^*& ***** !17000 $\pm$ 8200& $-$4500 $\pm$ 4500& $-$1300 $\pm$ 3800\cr
\itT\ i,aa& **!*6^* $\pm$ **50^*& !***3.0 $\pm$ ***9.0& !**1.0 $\pm$ **4.7& ***** !***96 $\pm$ *120& !**40 $\pm$ **99& **$-$27 $\pm$ **51\cr
\itT\ i,ai& ***!3^* $\pm$ **66^*& ***$-$5^* $\pm$ **22^*& !*26^* $\pm$ *21^*& ***** *$-$2100 $\pm$ 1000& !*220 $\pm$ *630& !**75 $\pm$ *170\cr
\itT\ i,ii& !**76^* $\pm$ *280^*& *$-$100^* $\pm$ *250^*& !440^* $\pm$ 230^*& ***** !*4200 $\pm$ 2000& *$-$750 $\pm$ 2400& *$-$970 $\pm$ 1400\cr
\noalign{\vskip 5pt}
\itE\ a,aa& !**34^* $\pm$ **34^*& !**34^* $\pm$ **34^*& !*34^* $\pm$ *34^*& ***** !***66 $\pm$ **50& !**51 $\pm$ *120& *$-$140 $\pm$ *150\cr
\itE\ a,ai& **$-$31^* $\pm$ *200^*& !**70^* $\pm$ *140^*& !*78^* $\pm$ *93^*& ***** **$-$380 $\pm$ *310& *$-$280 $\pm$ *640& !1100 $\pm$ *620\cr
\itE\ a,ii& $-$4200^* $\pm$ 4000^*& *$-$520^* $\pm$ 2300^*& !190^* $\pm$ 940^*& ***** *$-$8800 $\pm$ 6100& $-$6400 $\pm$ 6200& $-$9400 $\pm$ 3900\cr
\itE\ i,aa& **$-$10^* $\pm$ **87^*& !**42^* $\pm$ **42^*& !*23^* $\pm$ *27^*& ***** !***27 $\pm$ *180& !**52 $\pm$ *170& !**54 $\pm$ *120\cr
\itE\ i,ai&!**94^* $\pm$ *250^*& !**83^* $\pm$ *130^*& !*45^* $\pm$ *62^*& ***** !**910 $\pm$ *770& !*670 $\pm$ *850& *$-$190 $\pm$ *420\cr
\itE\ i,ii& !*690^* $\pm$ 2200^*& !*390^* $\pm$ 1400^*& !260^* $\pm$ 460^*& ***** *$-$6000 $\pm$ 5300& $-$4100 $\pm$ 5300& !2200 $\pm$ 1600\cr
\noalign{\vskip 5pt}
\textit{T+E} a,aa& !***0.7 $\pm$ ***4.9& !***0.7 $\pm$ ***4.9& !**0.7 $\pm$ **4.9& ***** !****5 $\pm$ **10& **$-$35 $\pm$ **27& !***2 $\pm$ **24\cr
\textit{T+E} a,ai& ***$-$2.6 $\pm$ ***9.7& ***$-$5^* $\pm$ **14^*& !*17^* $\pm$ *22^*& ***** ***$-$12 $\pm$ **20& !**74 $\pm$ **94& !*330 $\pm$ *130\cr
\textit{T+E} a,ii& !*130^* $\pm$ *450^*& *$-$130^* $\pm$ *240^*& !130^* $\pm$ 230^*& ***** *$-$1800 $\pm$ 1300& $-$3000 $\pm$ 1400& $-$3200 $\pm$ 1200\cr
\textit{T+E} i,aa& !**30^* $\pm$ **26^*& !***5.6 $\pm$ ***7.7& **$-$0.7 $\pm$ **4.1& ***** !***53 $\pm$ **47& !**51 $\pm$ **45& **$-$44 $\pm$ **24\cr
\textit{T+E} i,ai& !**26^* $\pm$ **38^*& !***2^* $\pm$ **19^*& !**6^* $\pm$ *15^*& ***** !**140 $\pm$ *170& !*170 $\pm$ *210& !**20 $\pm$ **74\cr
\textit{T+E} i,ii& !**38^* $\pm$ *170^*& **$-$26^* $\pm$ *180^*& !*85^* $\pm$ 130^*& ***** **$-$280 $\pm$ *390& *$-$390 $\pm$ *860& !*480 $\pm$ *430\cr
\noalign{\vskip 3pt\hrule\vskip 4pt}}}
\endPlancktablewide                 
\endgroup
\end{table*}                        

\begin{table*}[tb]                 
\begingroup
\newdimen\tblskip \tblskip=5pt
\caption{Similar to Table~\ref{Tab_isocurvNG}, except that we now assume that
the adiabatic and isocurvature mode are completely uncorrelated. Hence there
are only two $f_\mathrm{NL}$ parameters in this case, a purely adiabatic one
and a purely isocurvature one.}
\label{Tab_isocurvNG_uncorr}
\nointerlineskip
\vskip -3mm
\footnotesize
\setbox\tablebox=\vbox{
   \newdimen\digitwidth 
   \setbox0=\hbox{\rm 0} 
   \digitwidth=\wd0 
   \catcode`*=\active 
   \def*{\kern\digitwidth}
   \newdimen\signwidth 
   \setbox0=\hbox{+} 
   \signwidth=\wd0 
   \catcode`!=\active 
   \def!{\kern\signwidth}
   \newdimen\dotwidth 
   \setbox0=\hbox{.} 
   \dotwidth=\wd0 
   \catcode`^=\active 
   \def^{\kern\dotwidth}
\halign{\hbox to 0.8in{#\leaderfil}\tabskip 1em&
\hfil#\hfil\tabskip 1em&
\hfil#\hfil&
\hfil#\hfil&
\hfil#\hfil&
\hfil#\hfil&
\hfil#\hfil\tabskip 0pt\cr
\noalign{\doubleline\vskip 2pt}
\omit&\multispan6\hfil $f_{\rm NL}$\hfil\cr
\omit&\multispan6\hrulefill\cr
\omit&\multicolumn{3}{c}{\hfil Independent\hfil}&
\multicolumn{3}{c}{\hfil ******Joint\hfil}\cr
\noalign{\vskip 2pt}
Shape\hfill & Cold dark matter & Neutrino density & Neutrino velocity & ***** Cold dark matter & Neutrino density & Neutrino velocity\cr
\noalign{\vskip 4pt\hrule\vskip 6pt}
\itT\ a,aa& **1.3 $\pm$ ***5.4& !**1.3 $\pm$ ***5.4& **1.3 $\pm$ **5.4& ***** **1.0 $\pm$ ***5.3& !*19^* $\pm$ **12^*& **$-$0.2 $\pm$ **5.4\cr
\itT\ i,ii& *76^* $\pm$ *280^*& $-$100^* $\pm$ *250^*& 440^* $\pm$ 230^*& ***** *65^* $\pm$ *280^*& $-$840^* $\pm$ *540^*& !440^* $\pm$ 230^*\cr
\noalign{\vskip 5pt}
\itE\ a,aa& *34^* $\pm$ **34^*& !*34^* $\pm$ **34^*& *34^* $\pm$ *34^*& ***** *33^* $\pm$ **35^*& !*42^* $\pm$ **40^*& !*35^* $\pm$ *40^*\cr
\itE\ i,ii& 690^* $\pm$ 2200^*& !390^* $\pm$ 1400^*& 260^* $\pm$ 460^*& ***** 210^* $\pm$ 2200^*& $-$680^* $\pm$ 1700^*& *$-$31^* $\pm$ 540^*\cr
\noalign{\vskip 5pt}
\textit{T+E} a,aa& **0.7 $\pm$ ***4.9& !**0.7 $\pm$ ***4.9& **0.7 $\pm$ **4.9& ***** **0.5 $\pm$ ***5.0& !**3.0 $\pm$ ***7.9& **$-$0.3 $\pm$ **4.9\cr
\textit{T+E} i,ii& *38^* $\pm$ *170^*& *$-$26^* $\pm$ *180^*& *85^* $\pm$ 130^*& ***** *35^* $\pm$ *170^*& $-$120^* $\pm$ *290^*& !*87^* $\pm$ 130^*\cr
\noalign{\vskip 3pt\hrule\vskip 4pt}}}
\endPlancktablewide                 
\endgroup
\end{table*}                        

This section discusses new searches for NG beyond standard single-field inflation.  The focus here is on extensions to the analysis undertaken in \cite{planck2013-p09a} with new limits on isocurvature models, further oscillatory models over a broader frequency range, and parity-violating tensor NG.   However, we also briefly revisit all the non-standard models constrained in \cite{planck2013-p09a}, including effective field theory, non-Bunch Davies, and directionally-dependent models, in particular noting the impact of the new (preliminary) polarization data on the previous constraints.


\subsection{Isocurvature non-Gaussianity}

Here we show the results obtained for a study of the isocurvature NG in the
\Planck\ 2015 \SMICA\ map using the binned bispectrum estimator. 
As explained in Sect.~\ref{Sec:isocurv_NG_intro}, we only
investigate isocurvature NG of the local type, and in addition always consider
only one isocurvature mode (either cold dark matter, neutrino density, or
neutrino velocity isocurvature) in addition to the adiabatic mode. Note that the
baryon isocurvature mode behaves identically to the cold dark matter one, 
only rescaled by factors of $\Omega_{\rm b}/\Omega_{\rm c}$, so there is no need to
consider it separately. In that
case there are six different $f_\mathrm{NL}$ parameters: a purely adiabatic
one (a,aa), which corresponds to the result from Sect.~\ref{sec:Results}), 
a purely isocurvature one (i,ii); and four mixed ones (see 
Sect.~\ref{Sec:isocurv_NG_intro} for an explanation of the notation).

The results are given in Table~\ref{Tab_isocurvNG}\footnote{Compared to 
definitions in the literature based on $\zeta$ and $S$ (see 
e.g.,~\citealt{Langlois:2012tm}), here we adopt definitions based on 
$\Phi_\mathrm{adi}=3\zeta/5$ and $\Phi_\mathrm{iso}=S/5$, in order to make 
the link with the standard adiabatic
result more direct. Conversion factors to obtain results based on $\zeta$ 
and $S$ are 6/5, 2/5, 2/15, 18/5, 6/5, and 2/5, for the six modes, respectively.} 
and show no clear signs of any isocurvature NG. There
are a few values that deviate from zero by up to about 2.5$\,\sigma$, but such
a small deviation, in particular given the large number of results,
cannot be considered a detection. We do see that many 
constraints are tightened considerably when 
including polarization, by up to the predicted factor of about six for the 
cold dark matter a,ii, i,ai, and i,ii modes in the joint analysis.
As discussed in detail in Sect.~\ref{sec:Sec_valid_data}, results including
polarization data should be considered preliminary, and that is even more
important here, since these results depend so strongly on the additional
information from polarization.

In the results so far we allowed for a possible correlation between the 
isocurvature and adiabatic modes. However, if we assume that they are
completely uncorrelated, with a zero cross power spectrum, then there are only 
two $f_\mathrm{NL}$ parameters, the a,aa and i,ii ones. In 
Table~\ref{Tab_isocurvNG_uncorr} we give the results for this uncorrelated 
case. The independent results are the same as in the previous table,
while in the joint results one can clearly see the difference between 
the neutrino density mode (the bispectrum template of which has a large 
overlap with the adiabatic one), and the cold dark matter and neutrino 
velocity modes (with templates that are very different from the adiabatic one).
Again there is no evidence for any
isocurvature NG; the almost $2\,\sigma$ result for the neutrino
velocity isocurvature mode in the temperature-only case does not survive the addition of polarization.

\subsection{Feature models}

\begin{figure*}
\centering
~\includegraphics[width=.46\linewidth]{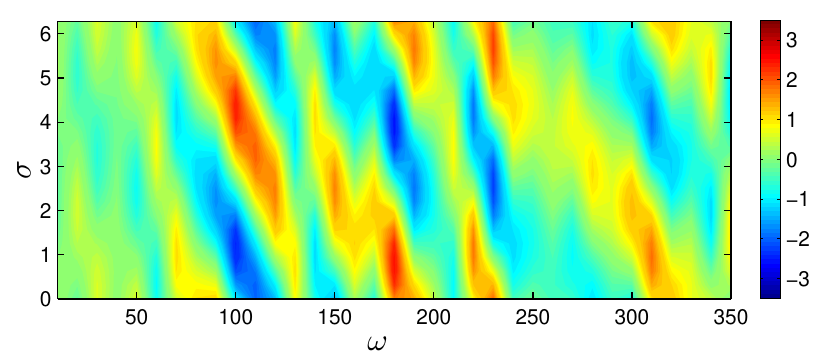}\includegraphics[width=.46\linewidth]{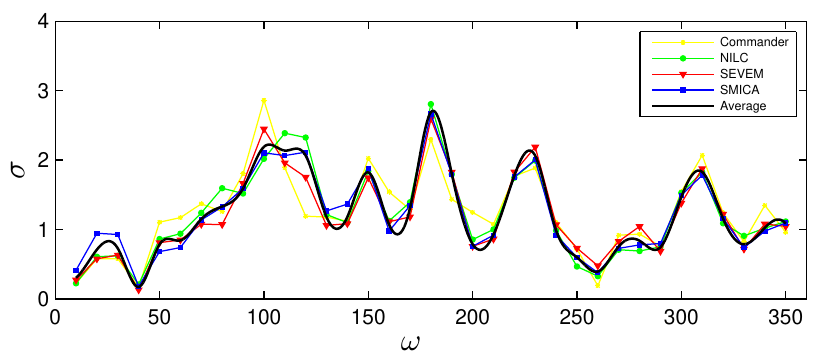}

~\includegraphics[width=.46\linewidth]{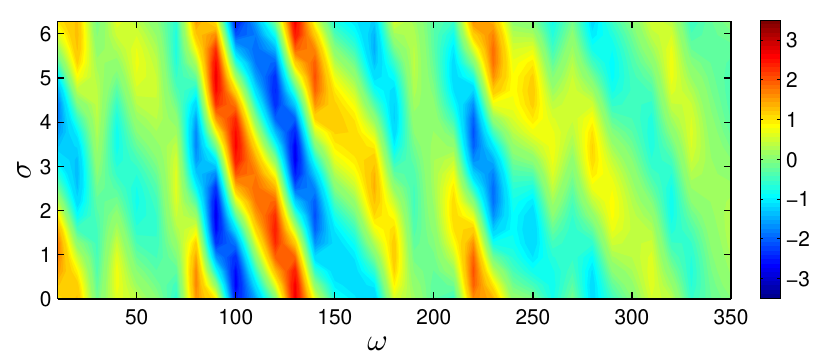} \includegraphics[width=.46\linewidth]{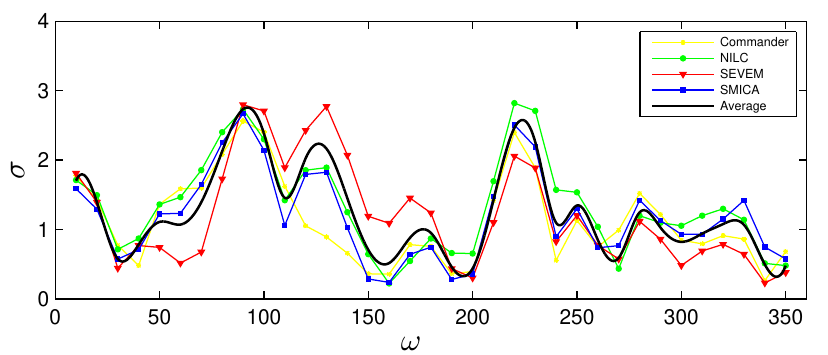}
\includegraphics[width=.95\linewidth, height=3.8cm]{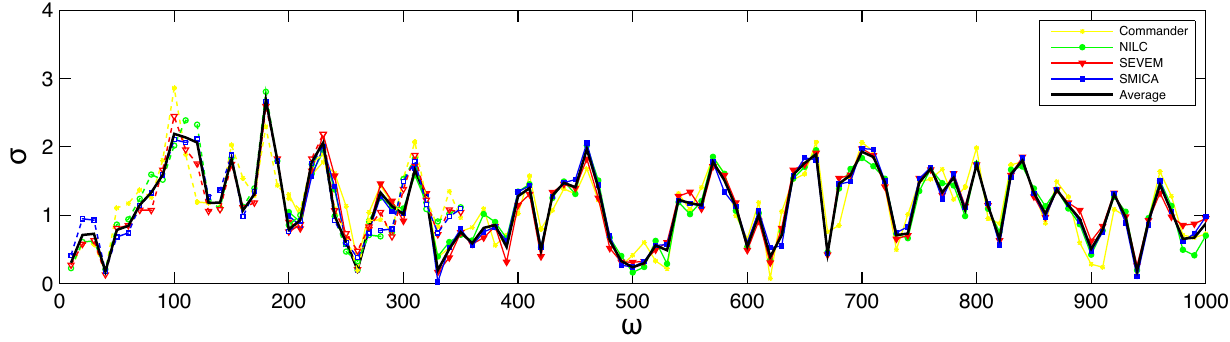} 
\includegraphics[width=.95\linewidth, height=3.8cm]{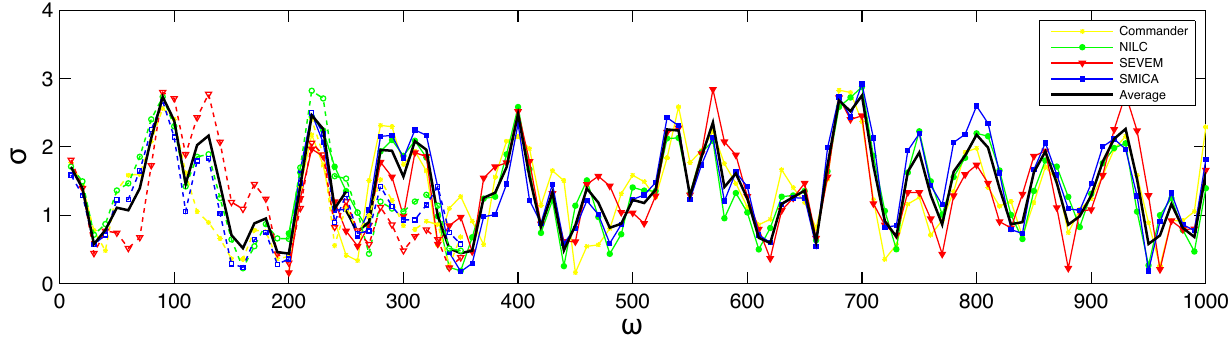} 
\includegraphics[width=.95\linewidth, height=3.8cm]{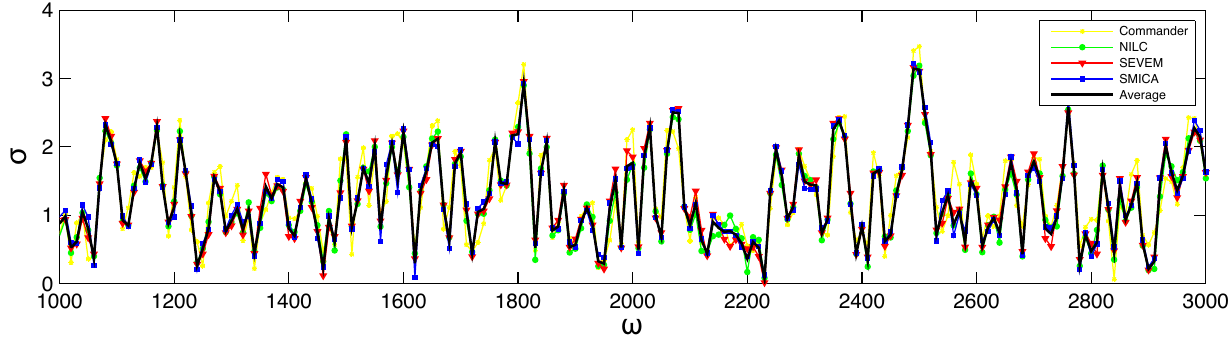} 
\includegraphics[width=.95\linewidth, height=3.8cm]{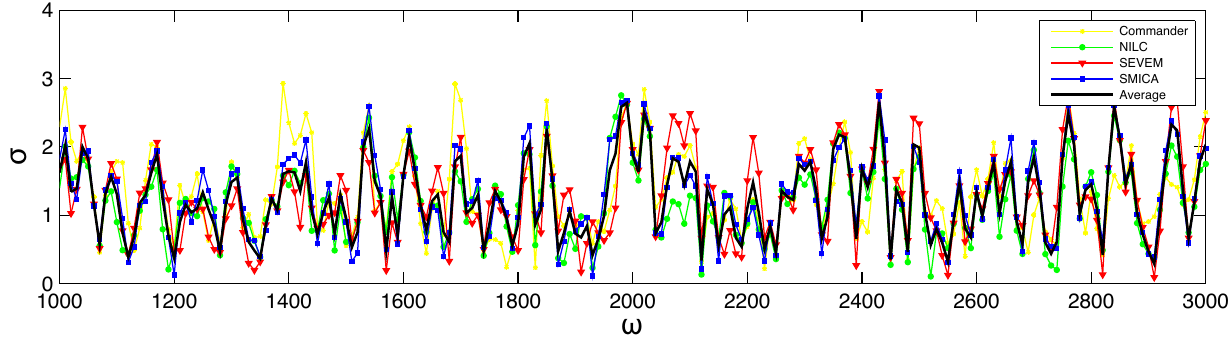} 
\caption[]{\small  Constant feature model results for both \itT-only and \textit{T+E} data across a wide frequency range.  The upper four panels show the feature signal in the Modal range  $0< \omega< 350$.  The two upper left panels show contours of the raw significance $\sigma$ obtained from the \SMICA\ map as a function of the frequency $\omega$, for \itT-only and \itT$+$\itE, respectively.   The upper right panels show the maximum signal after marginalizing over phase $\phi$ for both \itT-only and \textit{T+E}  for all foreground separation models.  The third and fourth panels show the maximum feature signal in both \itT-only and \textit{T+E}  across the frequency range $0< \omega<1000$, plotting both Modal results (dashed lines) and KSW results (solid lines for $200< \omega<1000$); these show good agreement in the overlap.  The lower two panels give the maximum KSW results for \itT-only and \textit{T+E} in the range $1000< \omega< 3000$).}
\label{fig:feat_stand}
\end{figure*}

\begin{figure*}[h]
\centering
\includegraphics[width=.49\linewidth]{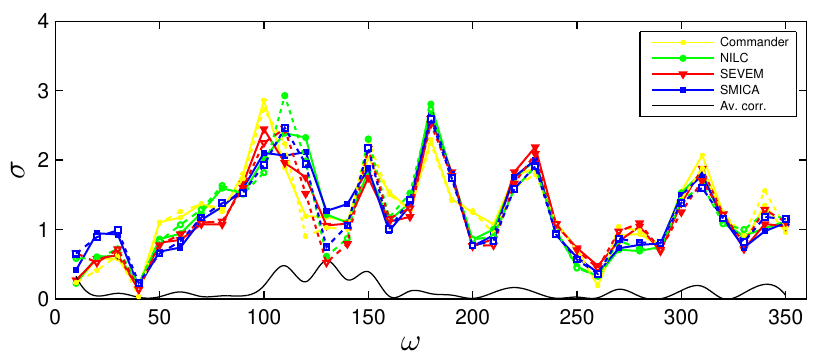} 
\includegraphics[width=.49\linewidth]{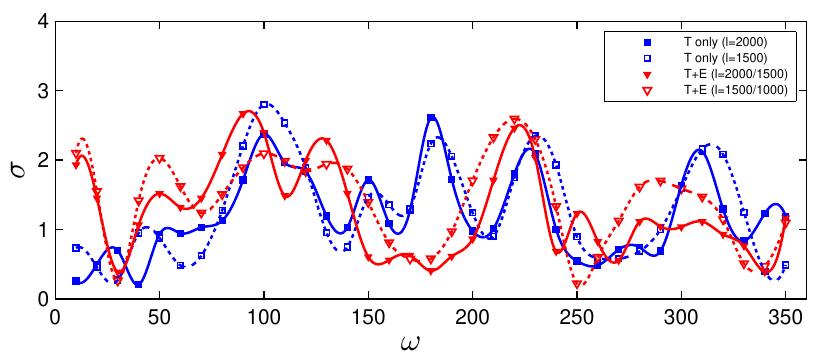}
\caption[]{\small  Constant feature ansatz validation for the Modal estimator, showing the effect of ISW-lensing and point source subtraction for $\ell < 300$ (left panel) and the impact of a lower $\ell$ maximum cutoff on the average signal (right panel), i.e., lowering $\ell_{\rm max} = 2000$ to $1500$ (\itT) and $\ell_{\rm max} = 1500$ to $1000$ (\itE). All Modal 2 results in Sect.~\ref{sec:Other_shapes} have used the extended common mask, except the validation analysis at different resolutions (right panel) which for consistency employs the common mask. }
\label{fig:feat_valid}
\end{figure*}

An important and well-motivated class of scale-dependent bispectra is the feature model, characterized by linear oscillations described by Eq.~\eqref{eq:featureBprim} and its variants in Eqs.~\eqref{eq:featequilBprim} and \eqref{eq:featflatBprim}.  In \cite{planck2013-p09a} we performed an initial search for a variety of feature models using the Modal estimator. This earlier search was limited to $\omega<200$ by the native
resolution of our implementation of the Modal estimator (using 600 modes), roughly the same range as the initial WMAP bispectrum feature model searches at lower precision (with only $50$ eigenmodes  \citep{2012JCAP...12..032F}.  Note, in the previous \Planck\ release we used wavenumber $k_{\rm c}$ in line with the theory literature, but here we switch to frequency $\omega$, in line with more recent observational power spectrum searches; the two are related by $\omega = 2\pi / 3k_{\rm c}$.
With the improved estimator resolution (now using 2001 modes) we are 
able to achieve convergence over a broader range up to $\omega=350$.   
We perform a frequency scan of 350 sampling points between $\omega=10$ and
$\omega=350$, i.e., 35 independent frequencies and 10 phases.  We also extend the number and variety of feature and resonance models that are investigated, essentially probing the resolution domain in which we have obtained a reliable Modal bispectrum reconstruction (see Fig.~\ref{fig:reconstruct}).  

\smallskip
\noindent\textit {Constant feature ansatz:}
For the constant feature shape of Eq.~\eqref{eq:featureBprim}, we can extend the frequency range much further with another approach. As the
bispectrum in Eq.~\eqref{eq:featureBprim} is separable, it allows
the construction of a KSW estimator \citep{Munchmeyer:2014nqa} for direct bispectrum estimation at any given frequency. 
The bispectrum can be written as a sum of sine and cosine components which can be estimated separately (equivalent to measuring the amplitude and phase above) and this method was used to constrain frequencies up to $\omega=3000$.  
The range where the two estimators overlap provides validation of the two methods and excellent agreement was seen (see Fig.~\ref{fig:feat_stand}).

Apart from cross-validation with two estimators, we have undertaken further tests to determine the robustness of the results to foreground and noise effects. 
In Fig.~\ref{fig:feat_valid} (left panel), we show the effect on feature model results of the subtraction of the simple point source bispectrum, as well as the ISW-lensing bispectrum.   
This study was a major motivation for adopting the more conservative ``extended" common mask, because the consistency between different component separation methods improved markedly for low frequencies, with the original common mask requiring much larger point-source subtractions (e.g., for \NILC\ subtraction the maximum raw significance has reduced from $\sigma=4.0$ to $\sigma_{\rm clean} = 2.2$ at $\omega = 110$).  After cleaning these signals, the \SMICA, \SEVEM\ and \NILC\ results are in good agreement, and also consistent with each other when polarization is included (while the \Commander\ results generally have a larger variance and so are not included in the plotted averages).  
Fortunately, the effect of subtracting ISW-lensing and point source bispectra diminishes rapidly at higher frequencies $\omega> 200$ and should be negligible; subtraction was only undertaken in the Modal region $\ell <350$.   In Fig.~\ref{fig:feat_valid} (right panel), we show the effect on the averaged significance of reducing the \Planck\ domain from the usual $\lmax=2000$ to $\lmax =1500$ ($\lmax =1500$ to $\lmax=1000$ for \itE-modes).  Despite the non-trivial change in overall signal-to-noise entailed, there is no strong evidence for an $\ell$-dependent signal, as might be expected if there was substantial NG feature contamination in the noise-dominated region.   Finally, we note that most peaks at low $\omega$ show some correlation between \itT-only and \textit{T+E}, although there are notable exceptions, such as the peak at $\omega = 180$, which is removed after inclusion of polarization (see also the phase plots in Fig.~\ref{fig:feat_stand} before marginalization).  The temperature feature peaks observed in Fig.~\ref{fig:feat_valid} at $\omega\approx 110$, $150$, and $180$ are consistent with the peaks identified previously in \cite{planck2013-p09a}.

In Fig.~\ref{fig:feat_stand}, the full set of frequency results, $0<\omega <3000$, for the constant feature model in Eq.~\eqref{eq:featureBprim} is shown for both the Modal and KSW estimators.  There is good agreement for regions where the estimators overlap and PS and ISW lensing bispectra subtraction is not necessary ($\omega>200$).   Generally, there is tighter consistency between temperature-only results, than with polarization where there is additional scatter between foreground separation methods.   Scanning across the full frequency range, there is no strong evidence for any large maximum that might indicate unequivocal evidence for a feature model signal.   The maximum peak significance obtained with either \itT-only or \textit{T+E} is consistent with expectations for a Gaussian model over this frequency range.   In particular, for the KSW estimator using the \SMICA\ data, the highest significance found in the range $200<\omega< 3000$ is 3.2$\,\sigma$ in \itT-only and 2.9$\,\sigma$ in $T$$+$$E$.   To gauge the likelihood of these results occurring randomly, realistic Gaussian \SMICA\ simulations were analysed with the KSW estimator and found to typically produce a highest peak with $3.1(\pm 0.3)\,\sigma$ over the same frequency range.   

\begin{figure*}[h]
\centering
\includegraphics[width=.45\linewidth]{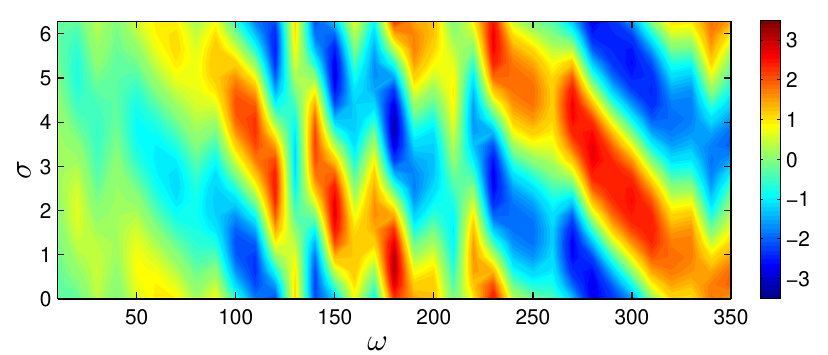} \includegraphics[width=.45\linewidth]{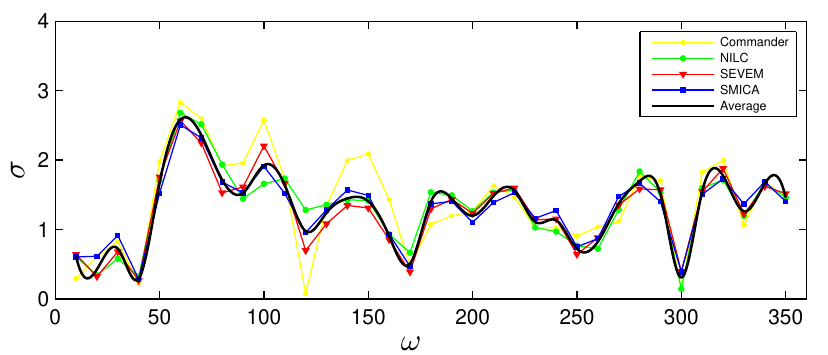}
\includegraphics[width=.45\linewidth]{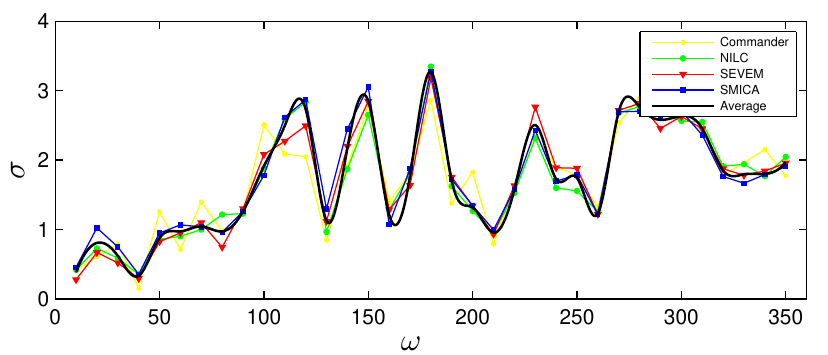} \includegraphics[width=.45\linewidth]{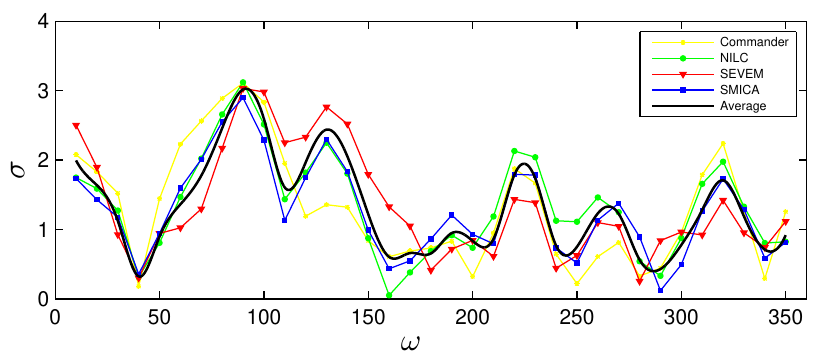}
\includegraphics[width=.45\linewidth]{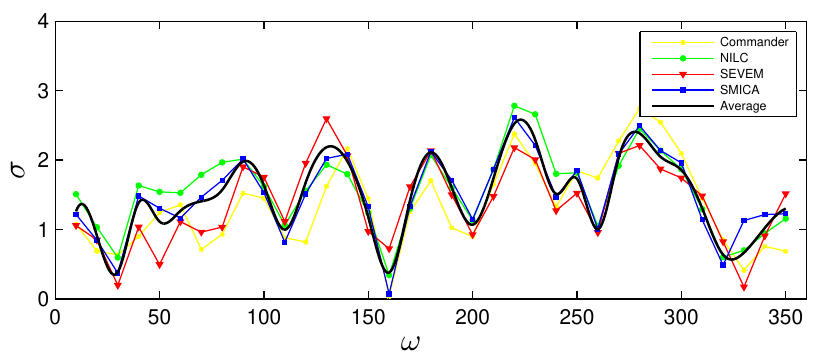} \includegraphics[width=.45\linewidth]{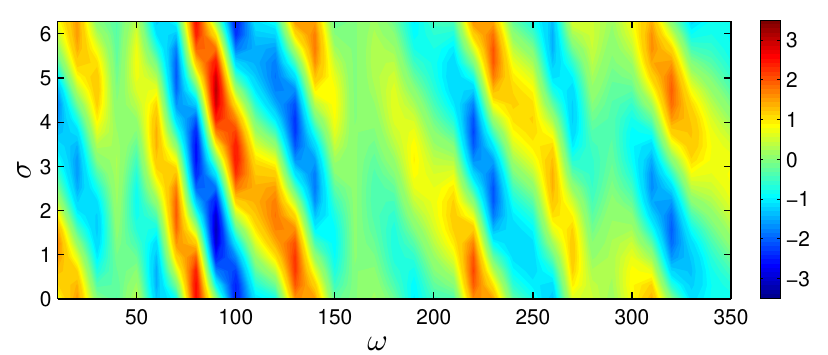}
\caption[]{\small  Generalized feature models analysed at $\ell_{\rm max} =2000$ ($E$-modes $\ell_{\rm max} =1500$) for the different \Planck\ foreground separation methods, \SMICA\ (blue), \SEVEM\ (red), \NILC\ (green), and \Commander\ (yellow), together with the SSN (\SMICA\ - \SEVEM\ - \NILC\ ) average (black).   The left three panels apply to the equilateral feature models, showing, respectively, in the top panel the full feature survey significance at each frequency and phase (temperature only), the maximum significance at each frequency for temperature only (middle), and with polarization (lower).  The right three panels apply to the flattened feature models, showing the maximum significance at each frequency for temperature only (top right) and with polarization (middle right), along with significance at each frequency and phase for temperature and polarization (right lower).  }
\label{fig:feat_equi_flat}
\end{figure*}

\begin{figure*}
\centering
\includegraphics[width=.45\linewidth]{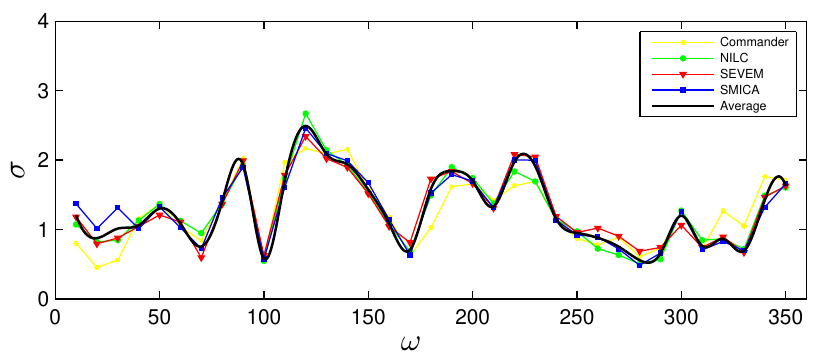}\includegraphics[width=.45\linewidth]{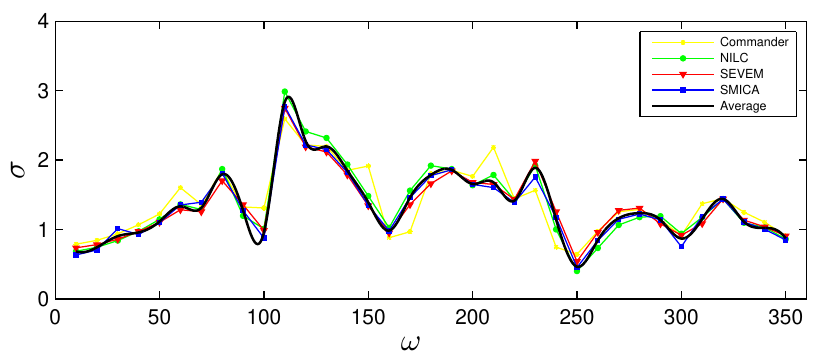}
\includegraphics[width=.45\linewidth]{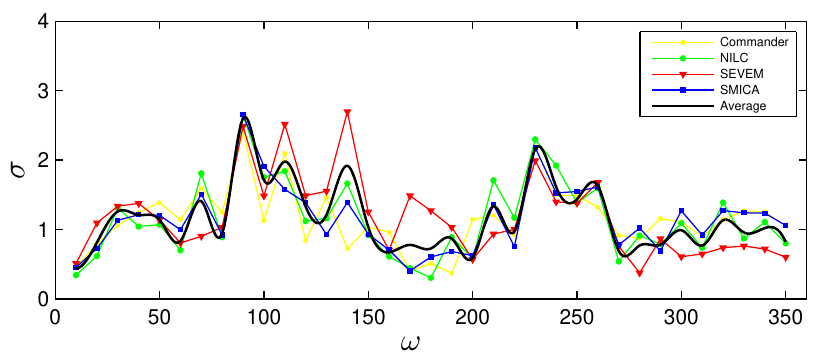} \includegraphics[width=.45\linewidth]{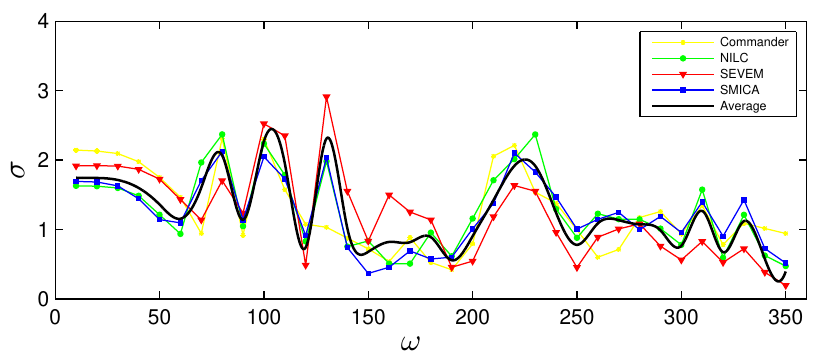}
\caption[]{\small  Single field feature model significance with a $K^2 \cos \omega K$ scaling dependence (Eq.~\ref{eq:Adsetal1}) (left panels, \itT-only upper and $T$$+$$E$ lower) or with a $K\sin \omega K$ scaling  (Eq.~\ref{eq:Adsetal2}) (right panels). To find the maximum signal, these results have been marginalized over the $\alpha$-dependent envelope function ranging from $\alpha\rightarrow0$ (no envelope) to the maximum cutoff allowed by the Modal resolution $\alpha\omega = 90$.}
\label{fig:feat_singlefields}
\end{figure*}

\begin{table*}[tb]
\begingroup
\newdimen\tblskip \tblskip=5pt
\caption{Peak statistics for the different feature models showing the \textit{Raw} peak maximum significance (for the given Modal survey domain), the corrected significance of this \textit{Single} maximum peak after accounting for the parameter survey size (the ``look-elsewhere" effect) and the \textit{Multi}-peak statistic which integrates across the adjusted significance of all peaks to determine consistency with Gaussianity.   \SMICA, \SEVEM\, and \NILC\ map analyses exhibit satisfactory bispectrum agreement for all the different models, whereas the \Commander\ results produce some anomalously large results, especially for polarization.   The significant signal for the equilateral features model  in the \itT-only multi-peak statistic is reduced when polarization is added.  The flattened feature model produces interesting results, which are reinforced with polarization to the $3\,\sigma$-level, with a high multi-peak significance. }
\label{peak_stats}
\nointerlineskip
\vskip -3mm
\footnotesize
\setbox\tablebox=\vbox{
  \newdimen\digitwidth 
  \setbox0=\hbox{\rm 0} 
  \digitwidth=\wd0 
  \catcode`*=\active 
  \def*{\kern\digitwidth}
  \newdimen\signwidth 
  \setbox0=\hbox{+} 
  \signwidth=\wd0 
  \catcode`!=\active 
  \def!{\kern\signwidth}
\halign{#\hfil\tabskip=0.3cm& \hfil#\hfil\tabskip=0.3cm&
\hfil#\hfil\tabskip=0.3cm& \hfil#\hfil\tabskip=0.3cm&
\hfil#\hfil\tabskip=0.3cm& \hfil#\hfil\tabskip=0.3cm&
\hfil#\hfil\tabskip=0.3cm& \hfil#\hfil\tabskip=0.3cm&
\hfil#\hfil\tabskip=0.3cm& \hfil#\hfil\tabskip=0.3cm&
\hfil#\hfil\tabskip=0.3cm&
\hfil#\hfil\tabskip=0.3cm& \hfil#\hfil\tabskip=0.cm\cr 
\noalign{\doubleline}
\noalign{\vskip -2pt}
&&  \SMICA\ &&& \SEVEM\hfil&&&\NILC &&& \Commander&\cr 
\noalign{\vskip 2pt\hrule\vskip 3pt} 
	&	Raw 	&	Single &	Multi	&Raw 	&	Single &	Multi	&Raw 	&	Single &	Multi	&Raw 	&	Single &	Multi	\cr							
	\noalign{\vskip 3pt\hrule\vskip 5pt}																								
Features constant \itT-only	&$	2.7	$&$	0.5	$&$	1.7	$&$	2.6	$&$	0.4	$&$	1.5	$&$	2.8	$&$	0.7	$&$	2.2	$&$	2.9	$&$	0.8	$&$	2.7	$\cr
Features constant \textit{T+E}	&$	2.7	$&$	0.5	$&$	1.9	$&$	2.8	$&$	0.7	$&$	2.5	$&$	2.8	$&$	0.7	$&$	2.4	$&$	2.6	$&$	0.4	$&$	1.5	$\cr
Features equilateral \itT-only	&$	3.3	$&$	1.5	$&$	4.0	$&$	3.2	$&$	1.3	$&$	3.5	$&$	3.3	$&$	1.6	$&$	4.1	$&$	2.9	$&$	0.9	$&$	2.5	$\cr
Features equilateral \textit{T+E}	&$	2.6	$&$	0.4	$&$	1.3	$&$	2.6	$&$	0.4	$&$	1.6	$&$	2.8	$&$	0.7	$&$	1.9	$&$	2.7	$&$	0.6	$&$	1.5	$\cr
Features flattened \itT-only	&$	2.5	$&$	0.3	$&$	1.4	$&$	2.6	$&$	0.4	$&$	1.6	$&$	2.7	$&$	0.5	$&$	2.1	$&$	2.8	$&$	0.8	$&$	2.7	$\cr
Features flattened \textit{T+E}	&$	2.9	$&$	0.9	$&$	2.9	$&$	3.0	$&$	1.1	$&$	3.5	$&$	3.1	$&$	1.2	$&$	3.8	$&$	3.1	$&$	1.2	$&$	3.8	$\cr
$K^2 \cos$ features \itT-only	&$	2.5	$&$	0.7	$&$	1.9	$&$	2.3	$&$	0.6	$&$	1.6	$&$	2.7	$&$	1.0	$&$	2.5	$&$	2.2	$&$	0.3	$&$	1.1	$\cr
$K^2 \cos$ features \textit{T+E}	&$	2.7	$&$	1.0	$&$	2.5	$&$	2.7	$&$	1.1	$&$	2.6	$&$	2.6	$&$	1.0	$&$	2.5	$&$	2.4	$&$	0.6	$&$	1.8	$\cr
$K \sin$ feature  \itT-only	&$	2.8	$&$	1.2	$&$	2.8	$&$	2.7	$&$	1.1	$&$	2.7	$&$	3.0	$&$	1.5	$&$	3.4	$&$	2.6	$&$	0.9	$&$	2.3	$\cr
$K \sin$  features  \textit{T+E}	&$	2.1	$&$	0.3	$&$	1.0	$&$	2.9	$&$	1.4	$&$	3.1	$&$	2.4	$&$	0.6	$&$	1.7	$&$	2.3	$&$	0.5	$&$	1.6	$\cr
\noalign{\vskip 5pt\hrule\vskip 3pt}}}
\endPlancktablewide                    
\endgroup
\end{table*}

\smallskip
\noindent\textit {Generalized feature models:}  We have also deployed the Modal estimator to look at (non-separable) feature models with equilateral and flattened cross-sections, as motivated by varying sound speed scenarios and those with highly excited states, respectively.   In the left panels in Fig.~\ref{fig:feat_equi_flat} we show results from the equilateral feature model of Eq.~\eqref{eq:featequilBprim}, including the frequency/phase contours before marginalization for the SMICA \itT-only map.   Multiple peaks are apparent in the temperature signal across the Modal range up to an average maximum $3.3\,\sigma$ raw significance. However,  from the lower panel it is clear that the polarization signal is not well correlated with the temperature in the equilateral case, generally reducing peak heights with the maximum now about $2.6\,\sigma$ (while eliminating the $\omega=180$ peak altogether).  This temperature peak remains present with the $\lmax =1500$ cutoff, where the signal is slightly higher, but the polarization in this case is less well correlated (using $\lmax^E =1000$).  For Gaussian noise we would not expect polarization to reinforce a high temperature signal on average.  It may also be that the equilateral temperature signal has some residual diffuse point-source contamination.  The equilateral feature model is the most affected by the removal of point sources, so the presence of a more complex correlated PS bispectrum (not removed by the constant PS template subtraction) remains for future investigation.  

Results for the flattened feature model in Eq.~\eqref{eq:featflatBprim} are shown in Fig.~\ref{fig:feat_equi_flat} (right panels), displaying more coherence between temperature and polarization.
The temperature signal with a $2.6\,\sigma$ peak between $50<\omega<150$ is reinforced by polarization and merges to make a broad 3$\,\sigma$ peak around $\omega= 90$, together with another distinct peak at $\omega\approx140$.  Such broad frequency peaks are not expected, because neighbouring feature models should be nearly uncorrelated over a range $\Delta \omega_{\rm eft} \approx 13$ (as we discuss below).  As the phase plots in Fig.~\eqref{fig:feat_equi_flat}  indicate, this breadth in frequency $\omega$ may reflect the neighbouring feature models adjusting phase $\phi$ to match an underlying NG signal of a related, but different, nature.  We note also that the frequency region for $\omega <100$ is susceptible to some degeneracy with cosmological parameters.   We shall consider a ``look-elsewhere" statistical analysis of these results below. 

\smallskip
\noindent\textit {Single-field feature solutions:}   We have also searched for the specific analytic solutions predicted for single-field inflation models with step-like potential features, as given in Eqs.~\eqref{eq:Adsetal1} and \eqref{eq:Adsetal2}, with results shown in Fig.~\ref{fig:feat_singlefields}.  The highest peaks for the $K^2$-cosine model occur around $2.5\,\sigma$ with temperature-only, then rises to 2.7$\,\sigma$ when  polarization data are included, again with peaks at other distinct frequencies apparent. The $K$-sine model shows a similar apparent signal level, with a maximum \itT-only $2.7\,\sigma$ peak, dropping to $2.4\,\sigma$ with polarization.   One further difficulty with a positive interpretation of these bispectrum results in this low frequency range is that stronger S/N counterparts in the power spectrum are predicted for these simple models \citep{2011PhRvD..84d3519A}, whereas no significant correlated oscillation signals are apparent at the relevant peak frequencies \cite{planck2014-a24}.

\smallskip
\noindent\textit {Feature model peak statistics:}  In order to determine consistency with Gaussianity for these feature model results, we can apply a number of statistical tests developed for this specific purpose \citep{Fergusson:2014hya} and, if warranted, also apply these jointly in combination with power spectrum results, as for the WMAP polyspectra analysis \citep{Fergusson:2014tza}.   When scanning across the $(\omega, \,\phi)$ parameter-dependent feature models, we are searching through independent models for which Gaussian noise, by chance, can lead to a large apparent signal. We must correct for this multiplicity of tests when determining the actual significance of results for a given model --- this is a quantitative correction for any model with free parameters, distinct from the a posteriori choice of models to test.  The simplest approach is to determine whether the maximum peak is consistent with Gaussian expectations, which can be determined from Monte Carlo simulations.  However, in \cite{Fergusson:2014hya} it was recognized that these feature models can be accurately characterized analytically with a $\chi$-distribution 
having two degrees of freedom\footnote{For the feature model with parameters $(\omega, \, \phi)$, the adjusted significance $S$ for the raw significance $\sigma$ after accounting for the ``look-elsewhere" effect is given by \citep{Fergusson:2014hya}
\eq
S = \sqrt2 \,{\rm Erf} ^{-1} [(F_{\chi,2}(\sigma))^{N_{\rm eff}}]\,,
\label{eq:lookelse}
\qe 
where $F_{\chi,2}$ is the cumulative distribution function of the $\chi$-distribution of degree two and  $N_{\rm eff}$ is the effective number of independent feature models.   This can also be used to investigate whether feature models are contributing at several frequencies. This multi-peak statistic integrates over all peak signals using the corrected significance $S$, i.e., 
\eq
S^2_I = \frac{\Delta\omega}{\Delta\omega_{\rm eff}} \sum_\omega 2 {\rm Erf} ^{-1} [F_{\chi,2}(\sigma(\omega))^{N_{\rm eff}}]^2\,, 
\label{eq:lookelsemulti}
\qe
where $\Delta\omega$ is the sampling step-size and $\Delta \omega_{\rm eff}$ is the effective correlation scale between independent models given by $\Delta \omega_{\rm eff}=(\omega_{\rm max}-\omega_{\rm min})/(N_{\rm eff}-1)$.   Essentially this sums up all significant peaks that have a non-zero adjusted significance, after accounting for the look-elsewhere effect in Eq.~\eqref{eq:lookelse}.}.  Taking $\lmax =2000$ and using this analysis, the frequency step size between models that are uncorrelated is approximately $\Delta \omega _{\rm eff} =13$, so we have an effective number of independent feature models $N_{\rm eff} \approx 27$ for the Modal frequency range (with $N_{\rm eff} \approx 230$ across the larger constant feature survey range).   Accordingly appropriate look-elsewhere corrections have been applied to find an adjusted significance for the maximum peak signal found in all the feature model searches undertaken, which is shown in Table~\ref{peak_stats}.   Given that this feature model survey is over many independent frequency models and combinations of data, even the highest raw significances above $3\,\sigma$ (``Raw" column in Table~\ref{peak_stats}) are reduced to a corrected significance below $2\,\sigma$ (``Single" column).  Hence, there appears to be no evidence from maximum peak statistics for feature model deviations from Gaussianity. 

Nevertheless, we also examine the possibility that multiple feature models are contributing to a NG signal, given the apparent emergence of several preferred frequency peaks. This integrated multi-peak statistic can also be accurately approximated analytically (Eq.~\ref{eq:lookelsemulti}) using a $\chi$-distribution; essentially we sum over all independent frequencies using the single peak significance adjusted for the ``look-elsewhere" effect (see Eq.~\eqref{eq:lookelse}).  Most of the signal surveys exhibit an unusual number of broad overlapping peaks within the accessible frequency domain, so the multi-peak statistic does yield a much higher significance, with many models above 2$\,\sigma$ after ``look-elsewhere" correction.  Notable cases are the temperature-only signal for the equilateral feature model which yields an average significance of $3.4\,\sigma$ across the  foreground-cleaned maps with concordant bispectrum results (i.e.,\ \SMICA, \SEVEM\ and \NILC); however, this interesting multi-peak significance drops to only $1.6\,\sigma$ when the polarization data are included (assuming the reliability of $E$ results).   On the other hand, the flattened feature model has an average multi-peak significance of $1.7\,\sigma$ in temperature only, which rises to $3.4\,\sigma$ with polarization included (higher at $3.7\,\sigma$ if \Commander\ data were to be included in the average).  In this case, beyond the number of peaks, it is also their width that contributes, with the main signal around $\omega\approx 90$, much broader than $\Delta \omega \approx 13$.  Finally, after look-elsewhere effects are taken into account, the $K^2$-cosine single-field solutions yield multi-peak statistics that rise with the inclusion of polarization data from $2.0\,\sigma$ to $2.5\,\sigma$, while the $K$-sine falls from $3.0\,\sigma$ ($T$) to $1.9\,\sigma$ ($T$$+$$E$).

\begin{figure*}[h]
\centering
\includegraphics[width=.45\linewidth]{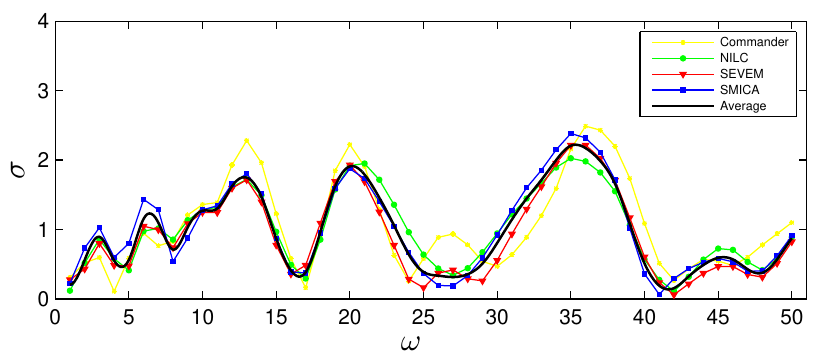} \includegraphics[width=.45\linewidth]{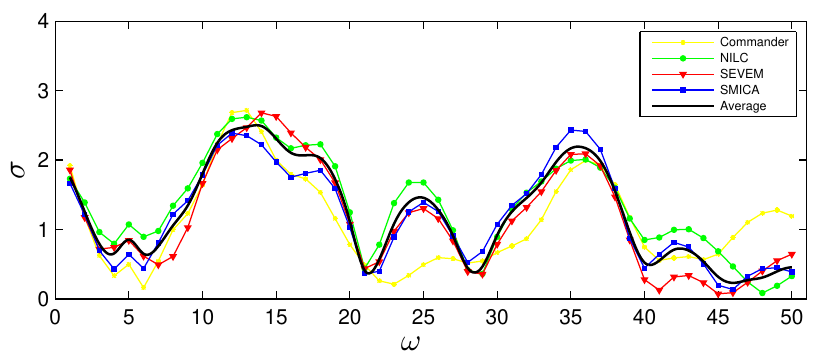}
\includegraphics[width=.45\linewidth]{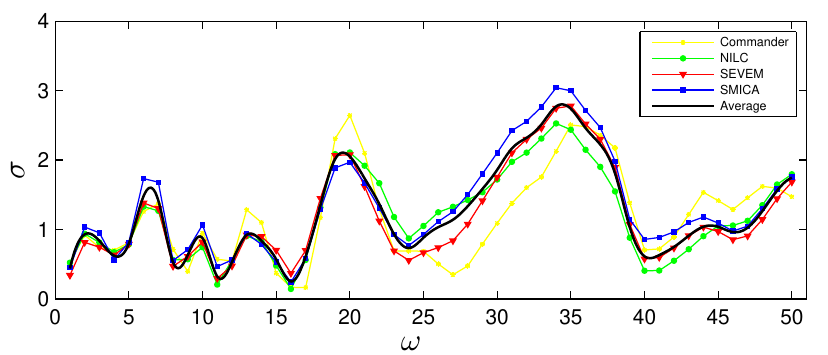} \includegraphics[width=.45\linewidth]{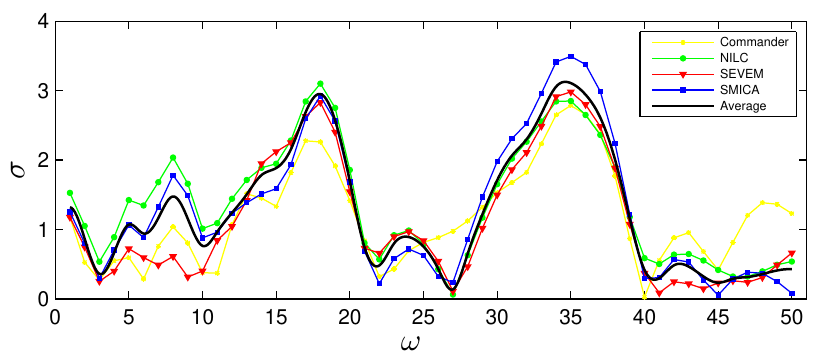}
\includegraphics[width=.45\linewidth]{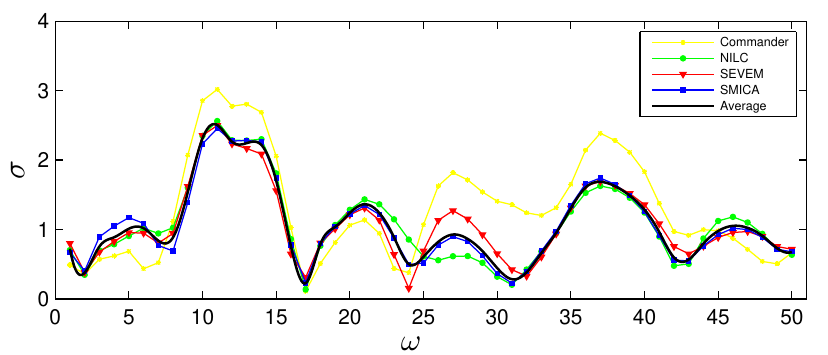} \includegraphics[width=.45\linewidth]{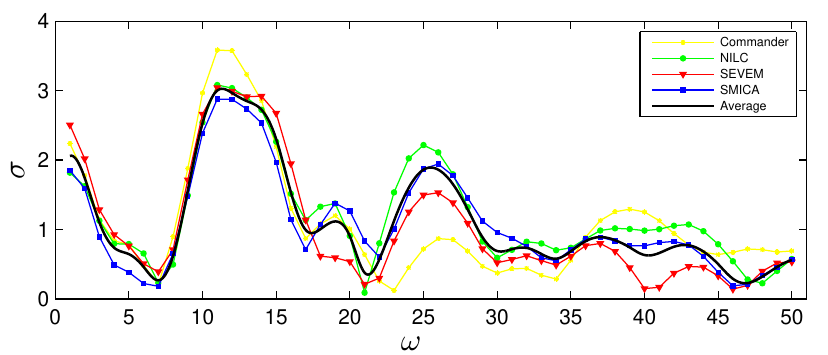}
\caption[]{\small  Generalized resonance models analysed at $\ell_{\rm max} =2000$ ($E$-modes $\ell_{\rm max} =1500$) for the different \Planck\ foreground separation methods, \SMICA~ (blue), \SEVEM~ (red), \NILC~ (green), and \Commander~(yellow), together with the SSN  (\SMICA\ - \SEVEM\ - \NILC\ ) average (black).   The upper panels apply to the constant resonance model (Eq.~\ref{eq:resBprim}), with \itT-only (left) and \textit{T+E} (right), the middle panels give results for the equilateral resonance model (Eq.~\ref{eq:resequilBprim}), and the lower panels for the flattened resonance model (Eq.~\ref{eq:resflatBprim}).   Both the equilateral and flattened resonance models produce broad peaks, which are reinforced with polarization (middle and bottom right panels). }
\label{fig:resonance_models}
\end{figure*}

An interesting, but not entirely coherent, picture emerges from these searches for non-standard models in the new \Planck\ temperature data, especially when combined with the additional (preliminary) polarization information.     In \cite{planck2013-p09a}, we noted that the feature model searches provided interesting hints of NG.   This more rigorous statistical analysis confirms this view, allowing for several alternative feature model explanations of the apparently high NG signal observed in the bispectrum reconstructions (see Fig.~\ref{fig:reconstruct}).  However, there is no strong evidence for a single large feature model at a particular frequency; rather, the high multi-peak statistics indicate signal that is spread across several broad peaks. Given the variability between different feature models and polarization component-separation methods, we note the caveat that the integrated multi-peak statistic could be sensitive to calibration issues and foreground contamination.   For this reason, we do not make strong claims for these non-standard signals at this stage, but we note that oscillatory models will continue to be investigated thoroughly over a wider frequency domain and using the more reliable polarization data available in the final \Planck\ data release.

\begin{table*}[h]
\begingroup
\newdimen\tblskip \tblskip=5pt
\caption{Peak statistics for the resonance models showing the maximum \textit{Raw} peak significance, the \textit{Single} peak significance after accounting for the parameter survey ``look-elsewhere" effect, and the \textit{Multi}-peak statistic integrating  across all peaks (also accounting for  the ``look-elsewhere" correction).   There is some evidence for  a high signal for both the equilateral and flattened resonance models, which increases when the polarization signal is added. This table does not include the results of the high frequency resonance model estimator, whose significance was assessed independently.}
\nointerlineskip
\vskip -3mm
\footnotesize
\setbox\tablebox=\vbox{
  \newdimen\digitwidth 
  \setbox0=\hbox{\rm 0} 
  \digitwidth=\wd0 
  \catcode`*=\active 
  \def*{\kern\digitwidth}
  \newdimen\signwidth 
  \setbox0=\hbox{+} 
  \signwidth=\wd0 
  \catcode`!=\active 
  \def!{\kern\signwidth}
\halign{#\hfil\tabskip=0.3cm& \hfil#\hfil\tabskip=0.3cm&
\hfil#\hfil\tabskip=0.3cm& \hfil#\hfil\tabskip=0.3cm&
\hfil#\hfil\tabskip=0.3cm& \hfil#\hfil\tabskip=0.3cm&
\hfil#\hfil\tabskip=0.3cm& \hfil#\hfil\tabskip=0.3cm&
\hfil#\hfil\tabskip=0.3cm& \hfil#\hfil\tabskip=0.3cm&
\hfil#\hfil\tabskip=0.3cm&
\hfil#\hfil\tabskip=0.3cm& \hfil#\hfil\tabskip=0.cm\cr 
\noalign{\doubleline}
\noalign{\vskip -2pt}
&&  \SMICA\ &&& \SEVEM\hfil&&&\NILC &&& \Commander&\cr 
\noalign{\vskip 2pt\hrule\vskip 3pt} 
	&	Raw 	&	Single &	Multi	&Raw 	&	Single &	Multi	&Raw 	&	Single &	Multi	&Max. 	&	Single &	Multi	\cr							
	\noalign{\vskip 3pt\hrule\vskip 5pt}	
	Sin(log) constant \itT-only	&$	2.4	$&$	0.7	$&$	1.2	$&$	2.2	$&$	0.4	$&$	0.9	$&$	2.0	$&$	0.2	$&$	0.7	$&$	2.5	$&$	0.8	$&$	1.6	$\cr
Sin(log) constant \textit{T+E}	&$	2.4	$&$	0.7	$&$	1.7	$&$	2.7	$&$	1.1	$&$	2.4	$&$	2.6	$&$	1.0	$&$	2.2	$&$	2.7	$&$	1.1	$&$	2.5	$\cr
Sin(log) equilateral \itT-only	&$	3.0	$&$	1.6	$&$	2.4	$&$	2.8	$&$	1.2	$&$	2.0	$&$	2.5	$&$	0.9	$&$	1.5	$&$	2.6	$&$	1.0	$&$	2.1	$\cr
Sin(log) equilateral \textit{T+E}	&$	3.5	$&$	2.2	$&$	3.5	$&$	3.0	$&$	1.5	$&$	2.8	$&$	3.1	$&$	1.7	$&$	3.2	$&$	2.8	$&$	1.2	$&$	2.0	$\cr
Sin(log) Êflattened \itT-only	&$	2.5	$&$	0.7	$&$	1.8	$&$	2.5	$&$	0.8	$&$	1.9	$&$	2.6	$&$	0.9	$&$	2.1	$&$	3.0	$&$	1.6	$&$	3.2	$\cr
Sin(log) Êflattened \textit{T+E}	&$	2.9	$&$	1.4	$&$	2.9	$&$	3.0	$&$	1.6	$&$	3.4	$&$	3.1	$&$	1.6	$&$	3.4	$&$	3.6	$&$	2.3	$&$	4.5	$\cr	
\noalign{\vskip 5pt\hrule\vskip 3pt}}}
\endPlancktablewide                    
\endgroup
\end{table*}

\subsection{Resonance/axion monodromy models}

\noindent\textit {Generalized resonance models:} Using the Modal expansion, we have embarked on a survey of the simplest resonance model (Eq.~\ref{eq:resBprim}), as well as the equilateral and flattened variants proposed in the literature, i.e., described by Eqs.~\eqref{eq:resequilBprim} and \eqref{eq:resflatBprim}, respectively.  The raw significance for the resonance models for both temperature-only and temperature plus polarization data are shown in Fig.~\ref{fig:resonance_models}; these are the maximal results marginalized over the phase parameter $\phi$.   Previously, the resonance model was studied in \cite{planck2013-p09a} using the Modal expansion over a narrower frequency range yielding no results above a raw significance of $1\,\sigma$.  In this extended analysis over a wider frequency range, the constant $\sin(\log)$ model  (Eq.~\ref{eq:resBprim}) produces $2.2\,\sigma$ peaks for \itT-only,  and $2.6\,\sigma$ for \itT+\itE.   The equilateral resonance model  (Eq.~\ref{eq:resequilBprim}) achieves a maximum $2.8\,\sigma$ in \itT-only at $\omega\approx35$, rising to a more impressive average $3.2\,\sigma$ for \itT+\itE. For the flattened case  (Eq.~\ref{eq:resflatBprim}) we have $2.5\,\sigma$ and $3.0\,\sigma$, respectively at $\omega\approx12$.  Qualitatively, the results shown in Fig.~\ref{fig:resonance_models}, exhibiting broad peaks, are similar to those for feature models.  

\smallskip
\noindent\textit {Resonance model peak statistics:} To determine the statistical significance of these results given the look-elsewhere effect of scanning across the parameters $(\omega,\,\phi)$, we have used the two peak statistics defined above in Eqs.~\eqref{eq:lookelse} and \eqref{eq:lookelsemulti} for feature models \citep{Fergusson:2014hya}.   In this case, the maximum peak statistic for the constant resonance model of $2.6\,\sigma$ (\textit{T+E}) is readjusted to an unremarkable ``look-elsewhere" single peak significance of $0.9\,\sigma$. Likewise the apparently significant results above $3\,\sigma$  for the equilateral and flattened models now fall below $2.0\,\sigma$ with \itTpE.  Using the single peak statistic alone, we would conclude that there is no strong evidence for any individual resonance model. Resonance models also generate oscillations in the power spectrum, and an analysis based on the 2015 temperature and polarization likelihood is presented in \citep{planck2014-a24}. A combined statistical treatment of resonance model power spectrum and bispectrum results will be reported in the future.

The multi-peak statistic in Eq.~\eqref{eq:lookelsemulti} integrates the resonance model signal across all frequencies to determine consistency with Gaussianity.  The constant resonance model has a modest multi-peak signal but, like the feature models, the equilateral and flattened resonance shapes offer stronger hints.   The multi-peak equilateral signal rose from $1.9\,\sigma$ (\itT-only) to $3.1\,\sigma$ (\itTpE) after adjusting for the ``look-elsewhere" effect, while the flattened signal went from $2.4\,\sigma$  (\itT-only) to $3.2\,\sigma$ (\itTpE).  These interesting results, reflecting those obtained for feature models, suggests the fit to any underlying NG signal might await alternative, but related, oscillatory models for a more compelling explanation. 

\begin{figure*}
\centering
\includegraphics[width=.95\linewidth]{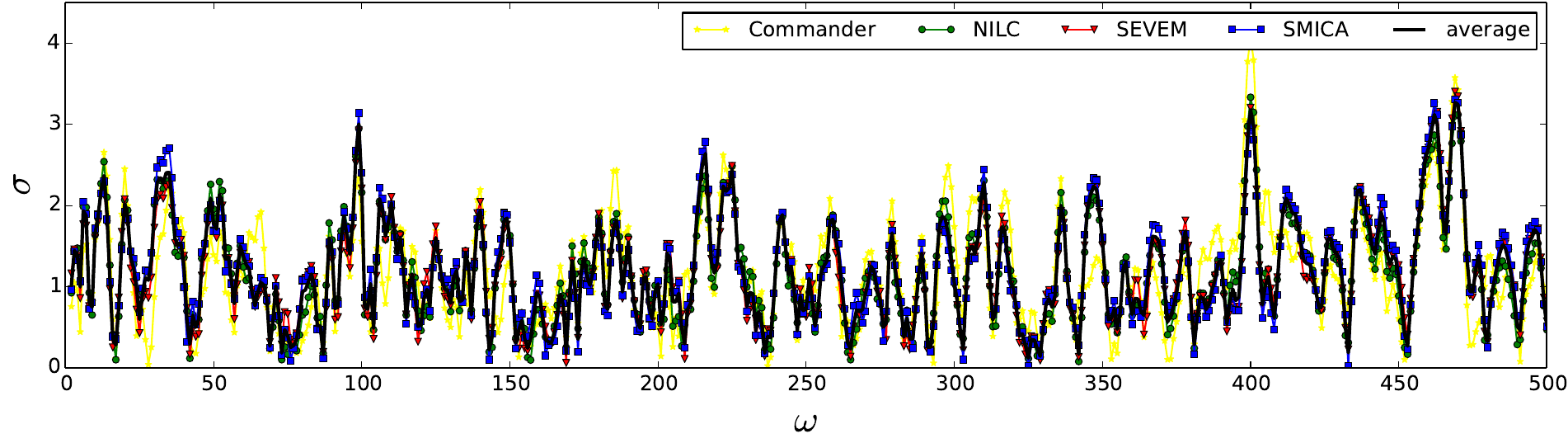} 
\includegraphics[width=.95\linewidth]{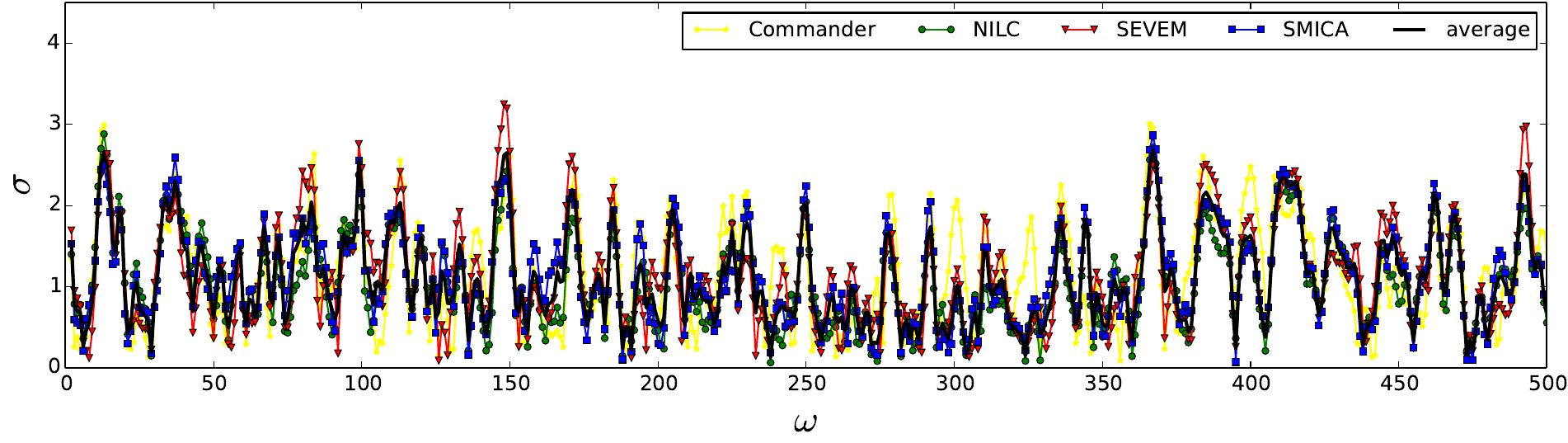} 
\includegraphics[width=.95\linewidth]{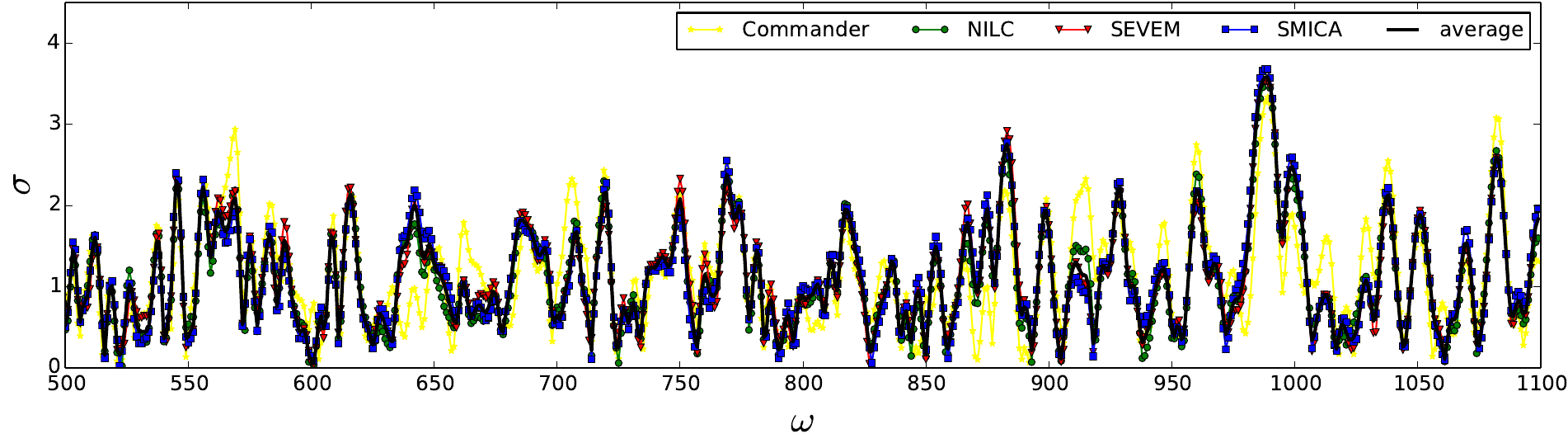} 
\includegraphics[width=.95\linewidth]{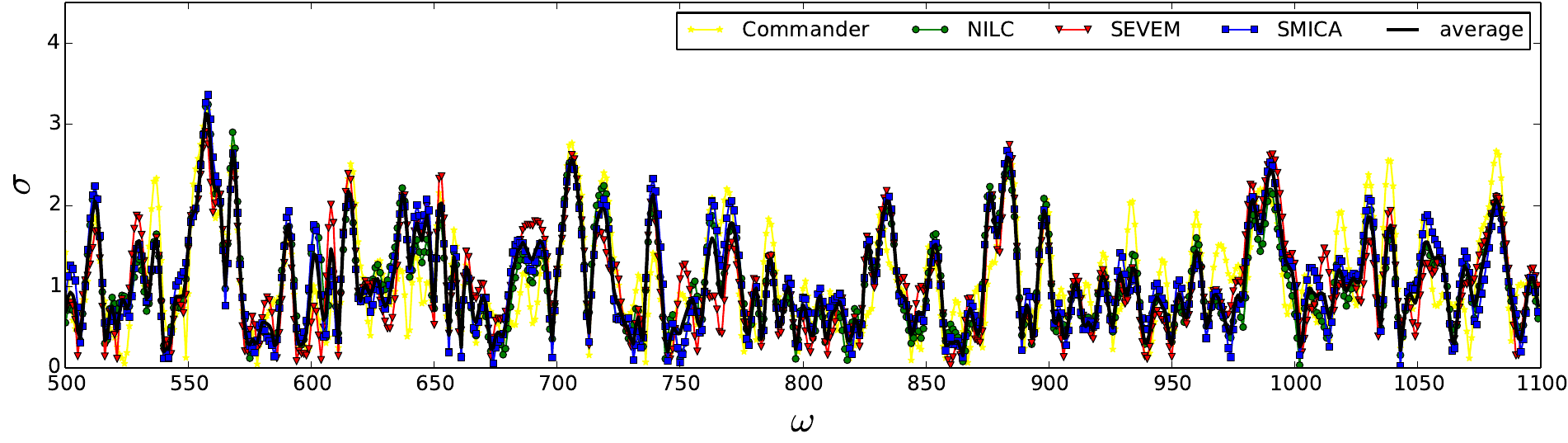} 
\caption[]{\small  Standard resonance model results for both \itT-only and \textit{T+E} across a wide frequency range using the high frequency resonance model estimator.  The first and second panels show the signal in both \itT-only (first panel) and \textit{T+E} (second panel) across the frequency range $0< \omega<500$.  The lower two panels give the results for \itT-only (third panel) and \textit{T+E} (fourth panel) in the range $500< \omega< 1100$.}
\label{fig:res_stand_Fourier}
\end{figure*}

\smallskip
\noindent\textit {High frequency resonance model estimator:}
We have further surveyed the simple resonance model (Eq.~\ref{eq:resBprim}) with a second approach. Using a model-specific expansion in terms of linear oscillation, proposed in \cite{Munchmeyer:2014cca}, it is possible to extend the frequency range of the analysis considerably. In this approach one exploits the fact that any bispectrum shape which is a function of $k_1+k_2+k_3$ can be expanded in Fourier modes of $k_1+k_2+k_3$, resulting in an effectively one-dimensional expansion, as opposed to the general Modal expansion. In the present implementation we use 800 sine and cosine modes, which cover the full frequency space of the power spectrum search, 
i.e., $0<\omega<1100$. The significances found with this method are presented in Fig.~\ref{fig:res_stand_Fourier}. As was the case for the feature model, we find that \SMICA, \SEVEM\ and \NILC\ results are in good agreement in both temperature and polarization, while \Commander\ results generally have larger variance and so are not included in the plotted averages. We find that the largest peak of the average is $3.6 \,\sigma$ in temperature and $3.1 \,\sigma$ in temperature and polarization combined. The results of the Modal expansion discussed in the previous paragraph are in good agreement with the high frequency resonance model estimator in the overlapping frequency range. 

Due to the high computational demands of this analysis, we have only exactly assessed the look-elsewhere effect for \SMICA ~\itT~data, for which we find an expected maximum peak of $3.5\,\sigma \pm 0.4\,\sigma$ in the case of Gaussian maps, to be compared to $3.7 \,\sigma$ in the \SMICA ~\itT~data, demonstrating that the results are fully consistent with Gaussianity. The expected maximum peak for Gaussian maps was calculated from the Fisher matrix with the method described in \cite{Meerburg:2015owa}. The average over component separation methods as well as the \itT+\itE~ data is even less significant. For the high frequency estimator we have assessed the significance of multiple peaks in the following way. Define the multi-peak amplitude $A_M$ as the sum of squares of the $M$ highest significances $\sigma_i$ in the frequency range, i.e., $A_M= \left( \sum_{i=1}^M \sigma_i^2 \right)^{1/2}$, where only approximately independent peaks with $\Delta \omega > 10$ are considered. One can then compare $A_M$ to its distribution in the Gaussian case and get an individual significance $\sigma_M$ for each number of peaks $M$, where we assume $M \leq 10$. The multi-peak statistic is then obtained by maximizing over $M$, leading to an additional look-elsewhere effect that we also accounted for. In this way we find that in the \SMICA ~\itT~data the raw peak maximum $3.7 \,\sigma$, is reduced to $0.5 \,\sigma$ for the single-peak statistic and to $0.6 \,\sigma$ in the multi-peak case. This large reduction in significance is due to the large number of independent frequencies as well as to the maximization over phases. One may argue that the frequency range $\omega<1100$ is too large, as EFT arguments for resonance non-Gaussianities \citep{Behbahani:2011it} limit the frequency range to $\mathcal{O}(10^2)$. As an example, we have therefore also calculated the look-elsewhere-corrected significances when we limit the analysis to $\omega<250$. In this case we find a single-peak significance of $0.6 \,\sigma$ and a multi-peak significance of $0.9 \,\sigma$. Clearly the results are fully consistent with Gaussianity.

\begin{table*}[tb]
\begingroup
\newdimen\tblskip \tblskip=5pt
\caption{Constraints on models with equilateral-type NG covering the shapes predicted by the effective field theory of inflation, together with constraints on specific non-canonical inflation models, such as DBI inflation.  See \cite{planck2013-p09a} (section 2) for further explanation of these specific models, with further implications discussed in Sect.~\ref{sec:Implications}. }
\label{tab:equilmodels}
\nointerlineskip
\vskip -3mm
\footnotesize
\setbox\tablebox=\vbox{
 \newdimen\digitwidth 
   \setbox0=\hbox{\rm 0} 
   \digitwidth=\wd0 
   \catcode`*=\active 
   \def*{\kern\digitwidth}
   \newdimen\signwidth 
   \setbox0=\hbox{+} 
   \signwidth=\wd0 
   \catcode`!=\active 
   \def!{\kern\signwidth}
   \newdimen\dotwidth 
   \setbox0=\hbox{.} 
   \dotwidth=\wd0 
   \catcode`^=\active 
   \def^{\kern\dotwidth}
\halign{#\hfil\tabskip=0.3cm& \hfil#\hfil\tabskip=0.3cm&
\hfil#\hfil\tabskip=0.3cm& \hfil#\hfil\tabskip=0.3cm&
\hfil#\hfil\tabskip=0.3cm& \hfil#\hfil\tabskip=0.3cm&
\hfil#\hfil\tabskip=0.3cm& \hfil#\hfil\tabskip=0.3cm&
\hfil#\hfil\tabskip=0.3cm& \hfil#\hfil\tabskip=0.cm\cr 
\noalign{\doubleline}
\noalign{\vskip -2pt}
&  \SMICA\ && \SEVEM\hfil&&\NILC && \Commander\cr 
\noalign{\vskip 2pt\hrule\vskip 3pt} 
Equilateral-type model & $A \pm \sigma_A$ & S/N & $A \pm \sigma_A$ & S/N & $A \pm \sigma_A$ & S/N & $A \pm \sigma_A$ & S/N\cr
\noalign{\vskip 3pt\hrule\vskip 5pt}
Constant \itT-only& !12 $\pm$ 38& !0.3& !16 $\pm$ 38& !0.4& !10 $\pm$ 37& !0.3& !*1 $\pm$ 39& !0.0\cr
Constant \textit{T+E}& !18 $\pm$ 22& !0.8& !28 $\pm$ 24& !1.2& !12 $\pm$ 23& !0.5& !26 $\pm$ 24& !1.1\cr
Equilateral \itT-only& $-$15 $\pm$ 68& $-$0.2& *$-$9 $\pm$ 68& $-$0.1& $-$19 $\pm$ 67& $-$0.3& $-$17 $\pm$ 69& $-$0.3\cr
Equilateral \textit{T+E}& !*5 $\pm$ 42& !0.1& !*4 $\pm$ 45& !0.1& *$-$2 $\pm$ 42& $-$0.1& !27 $\pm$ 45& !0.6\cr
EFT shape 1 \itT-only& *$-$3 $\pm$ 65& !0.0& !*3 $\pm$ 64& 0.0& *$-$7 $\pm$ 62& $-$0.1& *$-$9 $\pm$ 66& $-$0.1\cr
EFT shape 1 \textit{T+E}& !12 $\pm$ 39& !0.3& !15 $\pm$ 42& !0.3& !*3 $\pm$ 39& !0.1& !32 $\pm$ 42& !0.8\cr
EFT shape 2 \itT-only& !17 $\pm$ 50& !0.3& !22 $\pm$ 50& !0.5& !15 $\pm$ 47& !0.3& !*8 $\pm$ 51& !0.2\cr
EFT shape 2 \textit{T+E}& !23 $\pm$ 29& !0.8& !31 $\pm$ 31& !1.0& !15 $\pm$ 29& !0.5& !36 $\pm$ 31& !1.2\cr
DBI inflation \itT-only& !*3 $\pm$ 62& !0.0& !*9 $\pm$ 61& !0.1& *$-$1 $\pm$ 59& !0.0& *$-$4 $\pm$ 63& $-$0.1\cr
DBI inflation \textit{T+E}& !15 $\pm$ 37& !0.4& !20 $\pm$ 39& !0.5& !*7 $\pm$ 37& !0.2& !34 $\pm$ 40& !0.9\cr
Ghost inflation \itT-only& $-$50 $\pm$ 80& $-$0.6& $-$45 $\pm$ 80& $-$0.6& $-$54 $\pm$ 79& $-$0.7& $-$45 $\pm$ 82& $-$0.6\cr
Ghost inflation \textit{T+E}& $-$27 $\pm$ 50& $-$0.5& $-$37 $\pm$	54& $-$0.7& $-$31 $\pm$ 51& $-$0.6& !*1 $\pm$ 55& !0.0\cr
Inverse decay \itT-only& !17 $\pm$ 43& !0.4& !21 $\pm$ 43& !0.5& !14 $\pm$ 41& !0.3& !*4 $\pm$ 44& !0.1\cr
Inverse decay \textit{T+E}& !23 $\pm$ 25& !0.9& !32 $\pm$ 27& !1.2& !15 $\pm$ 26& !0.6& !32 $\pm$ 27& !1.2\cr 
\noalign{\vskip 5pt\hrule\vskip 3pt}}}
\endPlancktablewide                    
\endgroup
\end{table*}


\subsection{Equilateral-type models and the effective field theory of inflation}

There is considerable interest in equilateral-type models because they are physically well-motivated, through e.g., varying sound speed scenarios. There are generic predictions available from the effective field theory of inflation, notably the two specific effective field theory (EFT1 and EFT2) shapes that give rise to the equilateral and orthogonal approximations.   These models were previously constrained in \cite{planck2013-p09a} and the reader is referred to section 2 of that paper for analytic expressions for the specific shapes constrained here.   In Table~\ref{tab:equilmodels}, we list the main equilateral-type models in the literature, giving constraints for \itT-only and \textit{T+E}.  All these models correlate well with the equilateral ansatz (Eq.~\ref{eq:equilBprim}) and likewise do not show a significant signal.   However, despite this correlation, it is interesting to note the variation between models, largely due to the difference between these shapes in the flattened limit.  The implications of these results are discussed in Sect.~\ref{sec:Implications}.


\begin{table*}[tb]
\begingroup
\newdimen\tblskip \tblskip=5pt
\caption{Constraints on models with excited initial states (non-Bunch-Davies models), as well as warm inflation. See Sect. 2 for further explanation and the labelling of these classes of NBD models.  Note that the NBD, NBD1, and NBD2 models contain free parameters, so here we quote the maximum significance found over the available parameter range; the maximum for \itT\ and \itTpE\ can occur at different parameter values (on which the error bars are also dependent).}
\label{tab:fnlnonstandard}
\nointerlineskip
\vskip -3mm
\footnotesize
\setbox\tablebox=\vbox{
\setbox0=\hbox{\rm 0} 
   \digitwidth=\wd0 
   \catcode`*=\active 
   \def*{\kern\digitwidth}
   \newdimen\signwidth 
   \setbox0=\hbox{+} 
   \signwidth=\wd0 
   \catcode`!=\active 
   \def!{\kern\signwidth}
   \newdimen\dotwidth 
   \setbox0=\hbox{.} 
   \dotwidth=\wd0 
   \catcode`^=\active 
   \def^{\kern\dotwidth}
\halign{#\hfil\tabskip=0.3cm& \hfil#\hfil\tabskip=0.3cm&
\hfil#\hfil\tabskip=0.3cm& \hfil#\hfil\tabskip=0.3cm&
\hfil#\hfil\tabskip=0.3cm& \hfil#\hfil\tabskip=0.3cm&
\hfil#\hfil\tabskip=0.3cm& \hfil#\hfil\tabskip=0.3cm&
\hfil#\hfil\tabskip=0.3cm& \hfil#\hfil\tabskip=0.cm\cr 
\noalign{\doubleline}
\noalign{\vskip -2pt}
&  \SMICA\ && \SEVEM\hfil&&\NILC && \Commander\cr 
\noalign{\vskip 2pt\hrule\vskip 3pt} 
Flattened-type model & $A \pm \sigma_A$ & S/N & $A \pm \sigma_A$ & S/N & $A \pm \sigma_A$ & S/N & $A \pm \sigma_A$ & S/N\cr
\noalign{\vskip 3pt\hrule\vskip 5pt}
Flat model \itT-only& !*49^* $\pm$ *65^*& !0.8& !*57^* $\pm$ *65^*& !0.9& !*47^* $\pm$ *65^*& !0.7& !*19^* $\pm$ *65^*& !0.3\cr
Flat model \textit{T+E}& !*44^* $\pm$ *37^*& !1.2& !*70^* $\pm$ *37^*& !1.9& !*33^* $\pm$ *37^*& !0.9& !*47^* $\pm$ *37^*& !1.3\cr
Non-Bunch-Davies Ê\itT-only& !*42^* $\pm$ *82^*& !0.5& !*53^* $\pm$ *82^*& !0.6& !*26^* $\pm$ *82^*& !0.3& !*17^* $\pm$ *82^*& !0.2\cr
Non-Bunch-Davies Ê\textit{T+E}& !*61^* $\pm$ *47^*& !1.3& !*76^* $\pm$ *47^*& !1.6& !*43^* $\pm$ *47^*& !0.9& !*58^* $\pm$ *47^*& !1.2\cr
NBD sine \itT-only& !567^* $\pm$ 341^*& !1.7& !513^* $\pm$ 341^*& !1.5& !588^* $\pm$ 341^*& !1.7& !604^* $\pm$ 341^*& !1.8\cr
NBD sine \textit{T+E}& $-$387^* $\pm$ 206^*& $-$1.9& $-$485^* $\pm$ 218^*& $-$2.2& $-$425^* $\pm$ 206^*& $-$2.1& $-$417^* $\pm$ 210^*& $-$2.0\cr
NBD1 cos flattened \itT-only& *$-$10^* $\pm$ *22^*& $-$0.5& **$-$4^* $\pm$ *22^*& $-$0.2& **$-$8^* $\pm$ *22^*& $-$0.4& **$-$9^* $\pm$ *22^*& $-$0.4\cr
NBD1 cos flattened \textit{T+E}& *$-$20^* $\pm$ *19^*& $-$1.1& *$-$10^* $\pm$ *19^*& $-$0.5& *$-$19^* $\pm$ *19^*& $-$1.0& *$-$14^* $\pm$ *19^*& $-$0.8\cr
NBD2 cos squeezed \itT-only	& !*10^* $\pm$ *17^*& !0.6& !*10^* $\pm$ *17^*& !0.6& !**8^* $\pm$ *17^*& !0.5& **$-$2.5 $\pm$ *17^*& $-$0.1\cr
NBD2 cos squeezed Ê\textit{T+E}& **$-$3^* $\pm$ **5^*& $-$0.5& **$-$0.8 $\pm$ **5.5& $-$0.1& **$-$4^* $\pm$ **5^*& $-$0.8& **$-$3.8 $\pm$ **5.5& $-$0.7\cr
NBD1 sin flattened \itT-only& *$-$25^* $\pm$ *22^*& $-$1.1& *$-$27^* $\pm$ *22^*& $-$1.2& *$-$18^* $\pm$ *22^*& $-$0.8& *$-$33^* $\pm$ *23^*& $-$1.4\cr
NBD1 sin flattened \textit{T+E}& !*48^* $\pm$ *30^*& !1.6& !*49^* $\pm$ *33^*& !1.5& !*35^* $\pm$ *31^*& !1.1& !*26^* $\pm$	*34^*& !0.8\cr
NBD2 sin squeezed \itT-only& **$-$2.0 $\pm$ **1.4& $-$1.4& **$-$1.4 $\pm$ **1.4& $-$1.0& **$-$1.6 $\pm$ **1.4& $-$1.1& **$-$1.3 $\pm$ **1.4& $-$0.9\cr
NBD2 sin squeezed Ê\textit{T+E}& **$-$0.8 $\pm$ **0.4& $-$1.9& **$-$0.5 $\pm$ **0.4& $-$1.2& **$-$0.6	$\pm$ **0.4& $-$1.4& **$-$0.5 $\pm$ **0.4& $-$1.2\cr
NBD3 non-canonical Ê\itT-only ($\times1000$)& **$-$5.9 $\pm$ **6.7& $-$0.9& **$-$6.0 $\pm$ **6.8& $-$0.9& **$-$5.4 $\pm$ **6.8& $-$0.8& **$-$5.5 $\pm$ **6.7& $-$0.8\cr
NBD3 non-canonical Ê\textit{T+E} ($\times1000$)	& **$-$8.7 $\pm$ **5.0& $-$1.7& **$-$6.2 $\pm$ **5.2& $-$1.2& **$-$7.5 $\pm$ **5.2& $-$1.5& **$-$9.4 $\pm$ **5.2& $-$1.8\cr
WarmS inflation \itT-only& *$-$23^* $\pm$ *36^*& $-$0.6& *$-$26^* $\pm$ *36^*& $-$0.7& *$-$32^* $\pm$	*36^*& $-$0.9& *$-$24^* $\pm$ *36^*& $-$0.7\cr
WarmS inflation Ê\textit{T+E}& *$-$14^* $\pm$ *23^*& $-$0.6& *$-$28^* $\pm$ *23^*& $-$1.2& *$-$21^* $\pm$ *23^*& $-$0.9& *$-$17^* $\pm$ *23^*& $-$0.7\cr 
\noalign{\vskip 5pt\hrule\vskip 3pt}}}
\endPlancktablewide                    
\endgroup
\end{table*}

\subsection{Models with excited initial states (non-Bunch-Davies vacua)}

Non-Bunch-Davies (NBD) or excited initial states are models which produce flattened (or squeezed) bispectrum shapes.   The wide variety of NBD models that have been proposed are briefly classified and labelled in Sect.~\ref{sec:models}, following a more extensive overview in section 2 of \cite{planck2013-p09a} where more analytic forms and the first constraints were presented.   The latest \Planck\ constraints for these models are listed in Table~\ref{tab:fnlnonstandard}, obtained using the Modal 2 estimator with polarization,    Despite the apparent ``flattened" signal seen  in the \Planck\ bispectrum reconstructions (Fig.~\ref{fig:reconstruct}), this is generally not matched well by the specific modulation induced by the  acoustic peaks for these scale-invariant NBD models.   Tight constraints emerge for most models.   The largest signal obtained is from the NBD sinusoidal shape which gives a $1.6\,\sigma$ \itT-only raw significance, rising to $2.1\,\sigma$ for \itT+\itE; this is hardly an impressive correspondence given the number of models surveyed and the parameter freedom used in maximizing the signal.   However, an important caveat for NBD models is that the predicted shapes can be very narrow in the flattened limit, in which case solutions have been smoothed to match the current Modal resolution (though this has improved considerably since the \Planck\ 2013 NG analysis).  An improved match to the warm inflation shape means that the final constraint shown in Table~\ref{tab:fnlnonstandard} is more robust, with further implications discussed in Sect.~\ref{sec:Implications}.

\subsection{Direction-dependent primordial non-Gaussianity}
\label{cL_estimation}
We impose observational limits on direction-dependent primordial NG parametrized by Eq.~\ref{vectorBis}.
Rather than using $c_1$ and $c_2$ we instead choose to work with the non-linearity parameters $f^{L=1}_{\rm NL} = - c_1 / 4$ and $f^{L=2}_{\rm NL} = - c_2 / 16$ (chosen to match a primordial bispectrum that is equal to the equilateral shape in the equilateral limit) keeping the notation from the 2013 results.  
We estimated the $f^{L}_{\rm NL} $ values from temperature data and high-pass filtered polarization data from the four foreground-cleaned CMB maps \SMICA, \NILC, \SEVEM ,  and \Commander, where we apply
the common mask.  The details of the KSW estimator and its derivation is presented in Appendix \ref{sec:AA}.
For temperature data, we use the common mask as adopted in \cite{planck2013-p06}, which has more conservative foreground masking than the newly available mask.
We choose the more conservative foreground masking,  considering the fact that anisotropic NG is more sensitive to residual foregrounds.
We set the maximum multipole to 2000 and 1000 for temperature and polarization data, respectively.
Validating our analysis pipeline with realistic simulations, we find that the asymmetry of the \Planck\ beam, coupled with the \Planck\ scanning pattern, inflates the statistical fluctuations of the $f^{L}_{\rm NL} $ significantly. 
Noting the large angular scale of artificial anisotropy produced by the beam asymmetry, we set the minimum multipole to 101, and find that the statistical fluctuation of estimation from simulated data is close to the theoretical expectations.

These two shapes are also constrained using the Modal 2 estimator, which is not affected by the beam asymmetry and is used in the same form as elsewhere in the paper with multipoles from $2$ to $2000$ and $30$ to $1500$ being used for temperature and polarization, respectively.  The present constraints are consistent with those found for \itT-only in  \cite{planck2013-p09a}, but at higher resolution convergence has improved considerably, reflected in the lower variance. 

\begin{table*}[tb]
\begingroup
\newdimen\tblskip \tblskip=5pt
\caption{Direction-dependent NG results for both the $L=1$ and $L=2$ models.  We present results from both the KSW and Modal 2 pipelines.  The discrepancy between the central values for the $L=2$ models is due to the differing $\ell$ ranges taken for the two estimators, the key difference being the KSW $\ell_{\rm min}=101$.  As this model peaks in the squeezed configuration, a significant portion of the signal is lost, which is reflected in the increased error bars. \label{tab:cL_result}}
\nointerlineskip
\vskip -3mm
\setbox\tablebox=\vbox{
\setbox0=\hbox{\rm 0} 
   \digitwidth=\wd0 
   \catcode`*=\active 
   \def*{\kern\digitwidth}
   \newdimen\signwidth 
   \setbox0=\hbox{+} 
   \signwidth=\wd0 
   \catcode`!=\active 
   \def!{\kern\signwidth}
   \newdimen\dotwidth 
   \setbox0=\hbox{.} 
   \dotwidth=\wd0 
   \catcode`^=\active 
   \def^{\kern\dotwidth}
\halign{#\hfil\tabskip=0.3cm& \hfil#\hfil\tabskip=0.3cm&
\hfil#\hfil\tabskip=0.7cm& \hfil#\hfil\tabskip=0.7cm&
\hfil#\hfil\tabskip=0.7cm& \hfil#\hfil\tabskip=0.7cm&
\hfil#\hfil\tabskip=0.7cm& \hfil#\hfil\tabskip=0.7cm&
\hfil#\hfil\tabskip=0.7cm& \hfil#\hfil\tabskip=0.cm\cr 
\noalign{\doubleline}
\noalign{\vskip -2pt}
&  \Commander\ && \NILC\hfil&& \SEVEM&& \SMICA\cr 
\noalign{\vskip 2pt\hrule\vskip 3pt} 
& $A \pm \sigma_A$ & S/N & $A \pm \sigma_A$ & S/N & $A \pm \sigma_A$ & S/N & $A \pm \sigma_A$ & S/N\cr
\noalign{\vskip 3pt\hrule\vskip 5pt}
$L = 1$ &&&&&&&&\cr
Modal 2 \itT-only&  $-$41^* $\pm$ 43^*& $-$0.9&  $-$58^* $\pm$ 42^*& $-$1.4& $-$51^* $\pm$ 43^*& $-$1.2& $-$49^* $\pm$ 43^*& $-$1.1\cr 
KSW \itT-only&  *$-$8^* $\pm$ 46^*& $-$0.2&  $-$62^* $\pm$ 46^*& $-$1.3& $-$34^* $\pm$ 45^*& $-$0.8& $-$26^* $\pm$ 45^*& $-$0.6\cr 
Modal 2 \textit{T+E}&  $-$28^* $\pm$ 29^*& $-$1.0&  $-$30^* $\pm$ 27^*& $-$1.1&  $-$49^* $\pm$ 28^*& $-$1.7&  $-$31^* $\pm$ 26^*& $-$1.2\cr 
KSW \textit{T+E}&  $-$57^* $\pm$ 33^*& $-$1.7&  $-$62^* $\pm$ 32^*& $-$1.9&  $-$79^* $\pm$ 32^*& $-$2.5&  $-$54^* $\pm$ 32^*& $-$1.7\cr 
\noalign{\vskip 3pt\hrule\vskip 5pt}
$L = 2$&&&&&&&&\cr
Modal 2 \itT-only&  !*0.7 $\pm$ *2.8& !0.2&  !*0.8 $\pm$ *2.8& !0.4& !*1.1 $\pm$ *2.7& !0.3& !*0.5 $\pm$ *2.7& !0.2\cr 
KSW \itT$-$only&  !*1.5 $\pm$ *5.1& !0.3&  *$-$3.9 $\pm$ *5.1& $-$0.8& *$-$0.4 $\pm$ *5.1& $-$0.1& !*0.1 $\pm$ *5.0& !0.0\cr 
Modal 2 \textit{T+E}&  !*1.1 $\pm$ *2.4& !0.5&  !*0.5 $\pm$ *2.4& !0.2&  !*1.3 $\pm$ *2.4& !0.6&  *$-$0.2 $\pm$ *2.3& $-$0.1\cr 
KSW \textit{T+E}&  *$-$3.0 $\pm$ *4.1& $-$0.7&  *$-$3.6 $\pm$ *4.0& $-$0.9&  *$-$3.8 $\pm$ *4.0& $-$1.0&  *$-$1.3 $\pm$ *3.9& $-$0.3\cr 
\noalign{\vskip 5pt\hrule\vskip 3pt}}}
\endPlancktablewide                    
\endgroup
\end{table*}

We find that the ISW-lensing bispectrum and the unresolved point-sources bispectrum bias the estimation of the $f^{L}_{\rm NL}$, in particular in the analysis of temperature data.
For our final values, we subtract both these biases from our estimation.
In Table \ref{tab:cL_result}, we report the estimated value of $f^{L}_{\rm NL}$ from the foreground-cleaned CMB maps. For $L=1$ the effect of the differing $\ell$-ranges between the two estimators is not so significant and the results are quite consistent. For $L=2$, which has significant signal in the squeezed configuration, the effect of removing small scales from the KSW estimator is more pronounced, resulting in significantly enlarged error bars.  In light of this, the differences seen between the central values for the two methods is to be expected and does not indicate any inconsistencies  between the two approaches. The slight differences between the results from different foreground-cleaned temperature maps are within the likely range of statistical fluctuations, estimated from realistic simulations of CMB and noise propagated through the pipelines of foreground-cleaned map making. As seen in Table \ref{tab:cL_result}, we find that the estimated values of $f^{L}_{\rm NL}$ from \Planck\ 2015 temperature plus polarization data are consistent with zero.

 
\subsection{Parity-violating tensor non-Gaussianity}

We present observational limits on the parity-violating tensor nonlinearity parameter $f_{\rm NL}^{\rm tens}$ from the temperature and $E$-mode polarization data. Unlike the usual scalar-mode templates, the CMB bispectra sourced from the tensor NG (Eq.~\ref{eq:bis_tens_form}) are written in non-factorizable forms \citep{2013JCAP...11..051S}; hence, we use the separable Modal pipeline in our bispectrum estimations. 


The parity-violating NG under examination induces non-vanishing signals not only in parity-even configurations ($\ell_1 + \ell_2 + \ell_3 = {\rm even}$) but also in the parity-odd ones ($\ell_1 + \ell_2 + \ell_3 = {\rm odd}$) in the temperature and $E$-mode polarization bispectra \citep{2013JCAP...11..051S}. The optimal estimator, including all (even + odd) bispectrum signals, is expressed by the linear combination of the parity-even and parity-odd estimators, reading \citep{Liguori:2014}
\begin{eqnarray}
\hat{f}_{\rm NL}^{\rm all} = \frac{N^{\rm even} \hat{f}_{\rm NL}^{\rm even} + N^{\rm odd} \hat{f}_{\rm NL}^{\rm odd}}{N^{\rm even} + N^{\rm odd}} ~, 
\end{eqnarray}
where $N^{\rm even/odd}$ is the normalization factor (related to the Fisher matrix as $N^{\rm even/odd} = 6 F^{\rm even/odd}$) defined for $\ell_1 + \ell_2 + \ell_3 = {\rm even / odd}$. The parity-even estimator $\hat{f}_{\rm NL}^{\rm even}$ can be computed using the original Modal methodology \citep{2010PhRvD..82b3502F, 2012JCAP...12..032F, 2014arXiv1403.7949F, Liguori:2014}, while in computations of the parity-odd estimator $\hat{f}_{\rm NL}^{\rm odd}$, we follow the extended spin-weighted pipeline \citep{2014JCAP...05..008S, 2015JCAP...01..007S, Liguori:2014}. 
 

Our $f_{\rm NL}^{\rm tens}$ estimations (with both temperature and polarization data) are based on the resolution of $\ell_{\rm max} = 500$ and {\tt HEALPix} $N_{\rm side} = 512$, leading to feasible computational costs. This choice is not expected to change the results significantly, in comparison to the analysis at higher resolution, e.g., $\ell_{\rm max} = 2000$ and {\tt Healpix} $N_{\rm side} = 2048$, since the cosmic variance and instrumental noise are already far higher than the signals for $\ell \gtrsim 300$ \citep{2013JCAP...11..051S}. Only in the polarization data analysis is an effective $\ell_{\rm min}$ also adopted, which is motivated by the high-pass filtering process for $\ell \leq 40$ in component separation. 


Within the above $\ell$ ranges, the theoretical bispectrum templates are decomposed with the eigenbasis composed of ${\cal O}(1-10)$ polynomials and some special functions reconstructing the tensor-mode features, e.g., temperature enhancement due to the ISW effect ($\ell \lesssim 100$), and two $E$-mode peaks created by reionization ($\ell \lesssim 10$) and recombination ($\ell \simeq 100$). The resulting factorized templates are more than $95\,\%$ correlated with the original ones. The validity of our numerical computations has been confirmed through the map-by-map comparisons of $\hat{f}_{\rm NL}^{\rm even/odd}$ at very low resolution, showing the consistency between the values from the Modal methodology and those obtained by the brute-force ${\cal O}(\ell^5)$ summations like Eq.~\eqref{eq:diagcovestimator}. We have also checked that our parity-even estimator successfully leads to the constraints on $f_{\rm NL}^{\rm local}$, $f_{\rm NL}^{\rm equil}$, and $f_{\rm NL}^{\rm ortho}$ at $\ell_{\rm max} = 500$, compatible with the results from the binned estimator.


Our limits estimated from the foreground-cleaned temperature and high-pass filtered polarization data (\SMICA, \SEVEM, and \NILC) are summarized in Table~\ref{tab:fnltens}. The data and MC simulations used here, including all experimental features, i.e., beam, anisotropic noise levels and partial sky mask, have been inpainted using the identical recursive process adopted in the standard $f_{\rm NL}$ estimations (see Sect.~\ref{sec:settings}). The sky fractions of the temperature and polarization masks adopted here are, respectively, $f_{\rm sky} = 0.76$ and $f_{\rm sky} = 0.74$. Although the error bars and the linear terms have been computed using 160 MC simulations, the resulting error bars are close to the expected values, $(f_{\rm sky} F)^{-1/2}$. 

We have confirmed the stability of the \textit{T}-only constraints, and significant scatter of the \textit{E}-only constraints both in the parity-even case and in the parity-odd one, when changing $f_{\rm sky}$. Such $E$-mode instability has given insignificant effects on our \textit{T+E} constraints in the parity-even case, as they are determined almost exclusively by  \textit{TTT}, like the scalar NG analyses. In contrast, our parity-odd \textit{T+E} results vary a lot, due to the $E$-mode scatter (quantitatively speaking, only a few percent change of $f_{\rm sky}$ has shifted $f_{\rm NL}^{\rm tens}$ by more than $1\,\sigma$), because \textit{TTE} and \textit{TEE} contribute significantly to the signal-to-noise ratio in the parity-odd case \citep{2013JCAP...11..051S}. Table~\ref{tab:betacorr_fnltens} presents the correlations of the bispectra reconstructed from the component separated maps, also indicating the robustness of the \itT-only constraints and the instability of the \textit{E}-only results. We report only stable results in Table~\ref{tab:fnltens} and conclude that there is no evidence at $>2\,\sigma$ of $f_{\rm NL}^{\rm tens}$ in the parity-even, parity-odd or their whole domain. 

\begin{figure}
\begin{center}
\includegraphics[width=3.5in]{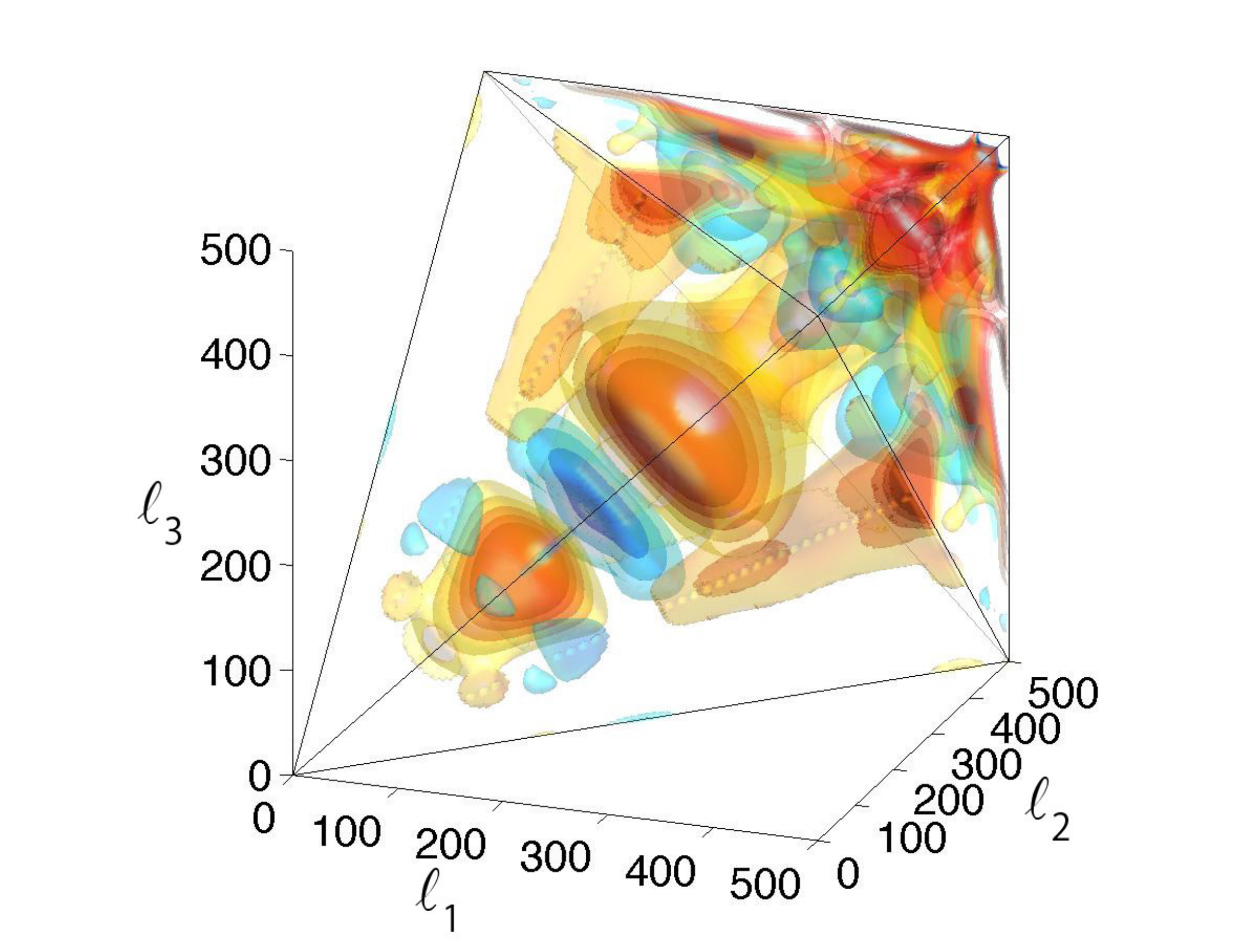}
\includegraphics[width=3.5in]{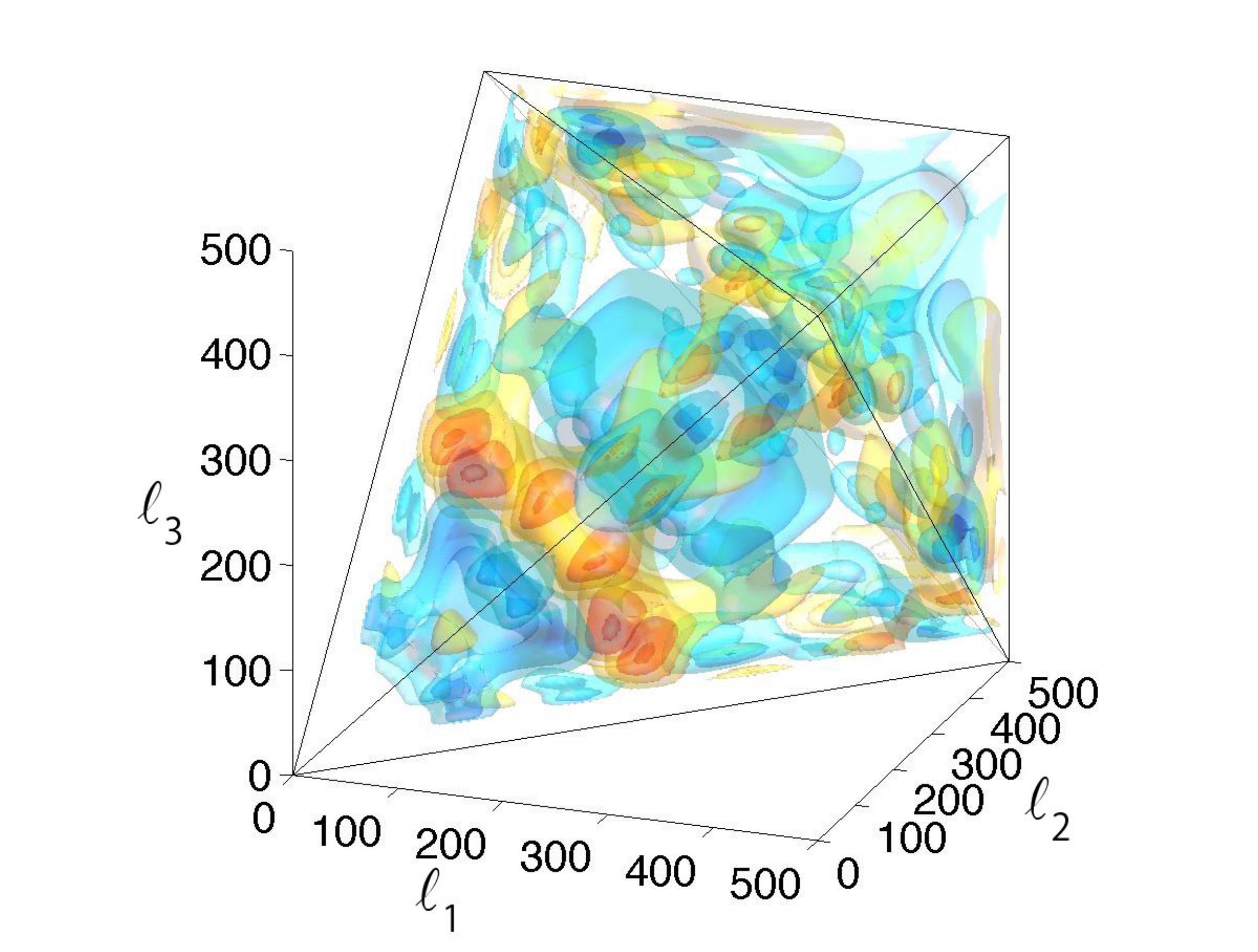}
\end{center}
\caption{Parity-odd signals ($\ell_1 + \ell_2 + \ell_3 = {\rm odd}$) of the  \textit{TTT} (top) and  \textit{EEE} (bottom) bispectra ($\ell_i \leq 500$) recovered from the \SMICA~maps by means of the Modal decomposition with 101 simple polynomial-based eigenmodes, not including any special functions fitting the CMB tensor-mode features. In the panel for  \textit{EEE}, only the signals larger than $\ell = 40$ are shown. The  \textit{TTT} and  \textit{EEE} bispectra shown here are rescaled with a constant Sachs-Wolfe weighting and signal-to-noise weighting, respectively.
}
\label{fig:reconstruct_odd}
\end{figure}

The parity-odd components of the \textit{TTT} and \textit{EEE} bispectra extracted model-independently from the \SMICA~data are visually represented in Fig.~\ref{fig:reconstruct_odd}. It is apparent from this figure that the \Planck~\textit{TTT} bispectrum has similar features to the WMAP one \citep{2015JCAP...01..007S}, e.g., distinctive signals distributed around $\ell_1 \approx \ell_2 \approx \ell_3$. As indicated by the roughly $70\,\%$ correlation between the \SMICA~and WMAP bispectra (see Table~\ref{tab:betacorr_fnltens}), the {\it Planck} \textit{T}-only limits in Table~\ref{tab:fnltens} are close to the WMAP ones (68\,\% CL): $f_{\rm NL}^{\rm tens}/10^2 = 4 \pm 16$ for parity-even; and $f_{\rm NL}^{\rm tens}/10^2 = 80 \pm 110$ for parity-odd \citep{2015JCAP...01..007S}.

\begin{table}[htb!]                 
\begingroup
\newdimen\tblskip \tblskip=5pt
\caption{
Results for the tensor nonlinearity parameter $f_{\rm NL}^{\rm tens} / 10^2$, estimated from the \SMICA, \SEVEM, and \NILC~temperature and high-pass filtered polarization maps. We separately show the central values and the errors ($68\,\%$ CL) extracted from $\ell_1 + \ell_2 + \ell_3 = {\rm even}$ (Even), $\ell_1 + \ell_2 + \ell_3 = {\rm odd}$ (Odd) and their whole domain (All). The parity-odd constraints have also been obtained from the $E$-mode data, but they are still preliminary and not currently shown. 
}
\label{tab:fnltens}
\nointerlineskip
\vskip -6mm
\footnotesize
\setbox\tablebox=\vbox{
   \newdimen\digitwidth
   \setbox0=\hbox{\rm 0}
   \digitwidth=\wd0
   \catcode`*=\active
   \def*{\kern\digitwidth}
   \newdimen\signwidth
   \setbox0=\hbox{+}
   \signwidth=\wd0
   \catcode`!=\active
   \def!{\kern\signwidth}
\newdimen\dotwidth
\setbox0=\hbox{.}
\dotwidth=\wd0
\catcode`^=\active
\def^{\kern\dotwidth}
\halign{\hbox to 1in{#\leaderfil}\tabskip 1em &
\hfil#\hfil\tabskip 1em &
\hfil#\hfil &
\hfil#\hfil\tabskip 0pt\cr
\noalign{\vskip 10pt\doubleline\vskip 2pt}\hfill & 
\hfil Even \hfil & 
\hfil Odd \hfil & 
\hfil All \hfil \cr
\noalign{\vskip 2pt}
\noalign{\vskip 4pt\hrule\vskip 6pt}
\omit\hfil \SMICA\hfil&&\cr
\itT & 2 $\pm$ 15 & 120 $\pm$ 110 & 4 $\pm$ 15 \cr
\textit{T+E} & 0 $\pm$ 13 & &  
\cr
\noalign{\vskip 5pt}
\omit\hfil \SEVEM\hfil&&\cr
\itT & 2 $\pm$ 15 & 120 $\pm$ 110 & 5 $\pm$ 15 \cr
\textit{T+E} & 4 $\pm$ 13 & &
\cr
\noalign{\vskip 5pt}
\omit\hfil \NILC\hfil&&\cr
\itT & 3 $\pm$ 15 & 110 $\pm$ 100 & 5 $\pm$ 15 \cr
\textit{T+E} & 1 $\pm$ 13 & & 
\cr
\noalign{\vskip 3pt\hrule\vskip 4pt}}}
\endPlancktable                    
\endgroup
\end{table}                        

\begin{table}[b!]                 
\begingroup
\newdimen\tblskip \tblskip=5pt
\caption{Correlation coefficients between pairs of bispectrum modes, extracted from two of the {\it Planck} component separated maps and the WMAP foreground-cleaned map at $\ell_{\rm max} = 500$ resolution. We separately present the results estimated from $\ell_1 + \ell_2 + \ell_3 = {\rm even}$ (Even) and $\ell_1 + \ell_2 + \ell_3 = {\rm odd}$ (Odd) combinations. The loss of the correlations is confirmed in the  \textit{EEE} case, like Table~\ref{tab:betacorr_methods}.}

\label{tab:betacorr_fnltens}
\nointerlineskip
\vskip -3mm
\footnotesize
\setbox\tablebox=\vbox{
   \newdimen\digitwidth 
   \setbox0=\hbox{\rm 0} 
   \digitwidth=\wd0 
   \catcode`*=\active 
   \def*{\kern\digitwidth}
   \newdimen\signwidth 
   \setbox0=\hbox{+} 
   \signwidth=\wd0 
   \catcode`!=\active 
   \def!{\kern\signwidth}
   \newdimen\signwidth 
   \setbox0=\hbox{.} 
   \signwidth=\wd0 
   \catcode`^=\active 
   \def^{\kern\signwidth}
\halign{\hbox to 0.9in{#\leaderfil}\tabskip 1em&
\hfil#\hfil\tabskip 1em&
\hfil#\hfil&
\hfil#\hfil&
\hfil#\hfil&
\hfil#\hfil\tabskip 0pt\cr
\noalign{\doubleline\vskip 2pt}
\omit & \multicolumn{2}{c}{\textit{TTT}} & \multicolumn{2}{c}{\textit{EEE}}\cr
\noalign{\vskip 6pt}
Methods \hfill  & Even & Odd & Even & Odd \cr
\noalign{\vskip 4pt\hrule\vskip 6pt}
\SMICA--\SEVEM & 1.00 & 0.99 & 0.80 & 0.80 \cr
\noalign{\vskip 6pt}
\SMICA--\NILC  & 1.00 & 1.00 & 0.90 & 0.87 \cr
\noalign{\vskip 6pt}
\SEVEM--\NILC  & 0.99 & 1.00 & 0.70 & 0.60 \cr
\noalign{\vskip 6pt}
\SMICA--WMAP   & 0.75 & 0.67 & \dots& \dots\cr
\noalign{\vskip 3pt\hrule\vskip 4pt}
}}
\endPlancktablewide                 
\endgroup
\end{table}                        

%% file: A19_Section9.tex
So far, we have considered a variety of physically motivated
possibilities for the inflationary 3-point function, or bispectrum.
A similar phenomenology exists for the 4-point function, or trispectrum.
Our constraints on the trispectrum will use CMB temperature only;
we do not use polarization in this section.
We start by briefly reviewing the inflationary physics and classifying
the signals we will search for.

First, some notation: a ``primed'' $\zeta$-trispectrum
$\langle \zeta_{\vk_1} \zeta_{\vk_2} \zeta_{\vk_3} \zeta_{\vk_4} \rangle'$
denotes the connected trispectrum without its momentum-conserving
delta function, i.e.,
\be
\langle \zeta_{\vk_1} \zeta_{\vk_2} \zeta_{\vk_3} \zeta_{\vk_4} \rangle
  = 
\langle \zeta_{\vk_1} \zeta_{\vk_2} \zeta_{\vk_3} \zeta_{\vk_4} \rangle'
  (2\pi)^3 \delta^{(3)}\Big( \sum\vk_i \Big) + \mbox{disc.},
\ee
where ``+\ disc.'' denotes the disconnected contributions to the 4-point function.

One possible signal is the ``local'' trispectrum $\gnlloc$, which arises
if the non-Gaussian adiabatic curvature $\zeta$ is of cubic-type form (see, e.g.,~\cite{2002PhRvD..66f3008O}), i.e.,
\be
\zeta(\vx) = \zeta_{\rm G}(\vx) + \frac{9}{25}\, \gnlloc \,\zeta_{\rm G}(\vx)^3
\ee
where $\gnlloc$ is a free parameter and $\zeta_{\rm G}$ is a Gaussian field.
In this model, the bispectrum is zero (since there is a $\zeta \rightarrow -\zeta$
symmetry) and the trispectrum is given by
\ba
\langle \zeta_{\vk_1} \zeta_{\vk_2} \zeta_{\vk_3} \zeta_{\vk_4} \rangle'
  = \frac{54}{25}\, \gnlloc \left[ P_\zeta(k_1) P_\zeta(k_2) P_\zeta(k_3) + \mbox{3 perm.} \right].
\label{eq:gnlloc}
\ea
Analogously to the case of the local bispectrum, $\fnlloc$, the observational
signal-to-noise for $\gnlloc$ is largest in the ``squeezed'' limit,
$k_1 \ll \min(k_2,k_3,k_4)$, and there is a consistency
relation which shows that in single-field inflation, the four-point
function is always small in the squeezed limit~(e.g.,~\citealt{Senatore:2012wy}).
Thus $\gnlloc$ can only be detectably large in multi-field models.
Conversely, there are multi-field models where $\gnlloc$ is detectable.
The main obstacle here is technical naturalness, i.e.,~ensuring that
radiative corrections do not generate an observationally larger bispectrum.
This can be the case if a large bispectrum is forbidden by a $Z_2$
symmetry, or by supersymmetry~\citep{Senatore:2010wk}.

A further category of four-point signals can be obtained by adding
quartic interactions to the inflationary action.
Following~\cite{long_trispectrum}, we will concentrate on the simplest
possibility, by considering quartic operators consistent with the symmetries
of inflation and having the lowest possible number of derivatives~\citep{2010JCAP...09..035B,2011JCAP...01..003S,Senatore:2010wk}.
There are three such operators, of the schematic form $\dot\sigma^4$,
$\dot\sigma^2 (\partial_i \sigma)^2$, and $(\partial_i \sigma)^2 (\partial_j \sigma)^2$.
By a short calculation using the in-in formalism~\citep{2003JHEP...05..013M},
the associated four-point functions are (\citealt{long_trispectrum}; see also \citealt{2006PhRvD..74l1301H}):
\ba
\langle \zeta_{\vk_1} \zeta_{\vk_2} \zeta_{\vk_3} \zeta_{\vk_4} \rangle'
   &=& \frac{9216}{25} \gnldotpi4 A_\zeta^3 \int_{-\infty}^0 d\tau_E \, \tau_E^4 
              \left( \prod_{i=1}^4 \frac{e^{k_i\tau_E}}{k_i} \right)  \nn \\
   &=& \frac{221184}{25} \gnldotpi4\, A_\zeta^3
          \frac{1}{k_1k_2k_3k_4 K^5};
\label{eq:gnldotpi4}
\ea
\ba
\langle \zeta_{\vk_1} \zeta_{\vk_2} \zeta_{\vk_3} \zeta_{\vk_4} \rangle'
  &=& -\frac{13824}{325} \gnlB A_\zeta^3 \int_{-\infty}^0 d\tau_E \, \tau_E^2 \nn \\
&& \hspace{0.5cm} \times \,
        \bigg[ \frac{(1-k_3\tau_E)(1-k_4\tau_E)}{k_1k_2k_3^3k_4^3} (\vk_3\cdot\vk_4) e^{\sum k_i\tau_E} \nn \\
&& \hspace{1.5cm} + \mbox{5 perm.} \bigg]
      \nn \\
  &=& 
    -\frac{27648}{325} 
     \gnlB A_\zeta^3 \nn \\
&& \hspace{0.5cm} \times \,
        \bigg[ \frac{K^2 + 3(k_3+k_4)K + 12k_3k_4}{k_1 k_2 k_3^3 k_4^3 K^5} (\vk_3\cdot\vk_4) \nn \\
&& \hspace{1.5cm} + \mbox{5 perm.} \bigg];
   \label{eq:gnlB}
\ea
\ba
\langle \zeta_{\vk_1} \zeta_{\vk_2} \zeta_{\vk_3} \zeta_{\vk_4} \rangle'
   &=& \frac{82944}{2575} \gnldpi4 A_\zeta^3 \int_{-\infty}^0 d\tau_E \nn \\
&& \hspace{0.5cm} \times \,
              \left[ \prod_{i=1}^4 \frac{(1-k_i\tau_E)e^{k_i\tau_E}}{k_i^3} \right] \nn \\
&& \hspace{0.5cm} \times \,
             \left[ (\vk_1\cdot\vk_2)(\vk_3\cdot\vk_4) + \mbox{2 perm.} \right]
  \nn \\
  &=& 
    \frac{165888}{2575}  \gnldpi4 A_\zeta^3 \nn \\
&& \times \,
        \left( \frac{2K^4 - 2K^2\sum k_i^2 + K \sum k_i^3 + 12 k_1 k_2 k_3 k_4}{k_1^3 k_2^3 k_3^3 k_4^3 K^5} \right) \nn \\
   && \hspace{0.5cm} \times \,
             \left[ (\vk_1\cdot\vk_2)(\vk_3\cdot\vk_4) + \mbox{2 perm.} \right].
\label{eq:gnldpi4}
\ea
Here $K = \sum k_i$ and we have introduced parameters $\gnldotpi4$, $\gnlB$, and $\gnldpi4$
to parametrize the amplitude of each trispectrum.

The normalization of the $\gnl$-parameters is chosen so that
$\langle \zeta_{\vk_1} \zeta_{\vk_2} \zeta_{\vk_3} \zeta_{\vk_4} \rangle = (216/25)\, \gnl A_\zeta^3 / k^9$
for ``tetrahedral'' configurations, with $|\vk_i| = k$ and $(\vk_i\cdot\vk_j) = -k^3/3$.
This is the analogue of the commonly-used normalization for the bispectrum,
where $\fnl$ parameters are defined so that 
$\langle \zeta_{\vk_1} \zeta_{\vk_2} \zeta_{\vk_3} \rangle = (18/5)\, \fnl A_\zeta^2 / k^6$
for equilateral configurations with $|\vk_i| = k$.

For simplicity in Eqs.~(\ref{eq:gnldotpi4})--(\ref{eq:gnldpi4})
we have assumed a scale-invariant
initial power spectrum $P_\zeta(k) = A_\zeta / k^3$.
In order to analyse {\Planck} data, we must slightly generalize this to a power-law
spectrum $P_\zeta(k) \propto k^{n_{\rm s}-4}$.
Our scheme for doing this follows Appendix~C of \cite{long_trispectrum}.

A Fisher matrix analysis shows that there is one large correlation among the
three trispectra in Eqs.~(\ref{eq:gnldotpi4})--(\ref{eq:gnldpi4}), 
so that to an excellent approximation we can treat only two of the trispectra as independent.
To quantify this, in~\cite{long_trispectrum} it is shown that the $\dot\sigma^2 (\partial\sigma)^2$
shape is $98.6\,\%$ correlated to a linear combination of the shapes $\dot\sigma^4$ and
$(\partial\sigma)^4$.
Therefore, we will only search for the parameters $\gnldotpi4$ and $\gnldpi4$.

We note that the analysis that leads to the trispectrum shapes $\gnldotpi4$
and $\gnldpi4$ is very similar to the analysis that leads to the ``standard''
bispectrum shapes $\fnlequi$ and $\fnlortho$.
However, there are some minor differences as follows.
In the bispectrum case, one considers the cubic operators $\dot\pi^3$
and $\dot\pi (\partial\pi)^2$, but it is conventional to define observables
$\fnlequi, \fnlortho$ which are related to the operator
coefficients by a linear transformation.
This is done because the two cubic operators are about $90\,\%$ correlated,
so it is convenient to orthogonalize.
In the trispectrum case the correlation is smaller (around $ 60\,\%$ for \Planck), 
and we have chosen to omit the orthogonalization step.
Another reason to omit the orthogonalization step is that the trispectrum
shape $(\partial\sigma)^4$ is a signature of multi-field inflation.
In single field inflation, the $(\partial\sigma)^4$ trispectrum is not technically 
natural; radiative corrections 
generate cubic operators of the form $\dot\pi^3$
or $\dot\pi (\partial\pi)^2$,
which generate a bispectrum with larger signal-to-noise.

There are more trispectrum shapes one might consider.
For example, classifying Galilean invariant quartic operators
leads to higher-derivative trispectra, which are not highly
correlated to the trispectra considered above~\citep{Bartolo:2013eka,Arroja:2013dya}.
We have only considered ``contact'' diagrams arising from
one power of a quartic operator, and it would be interesting to
study ``exchange'' diagrams arising from two cubic
operators and exchange of a light particle 
(e.g.,~\citealt{2009JCAP...08..008C,2009PhRvD..80d3527A,2010JCAP...04..027C,2010JCAP...08..008B,Baumann:2011nk}).
We leave these as extensions for future work.

Summarizing, we will search for the following trispectrum signals:
\be
\left\{ \gnlloc, \gnldotpi4, \gnldpi4 \right\}  \label{eq:trispectra}
\ee
defined by Eqs.~(\ref{eq:gnlloc}),~(\ref{eq:gnldotpi4}),
and~(\ref{eq:gnldpi4}) above.

\subsection{Data analysis}
\label{dataantrisp}
Turning now to data analysis, we use the machinery from~\cite{long_trispectrum}.
The first step is to represent each trispectrum as a small sum of factorizable
terms as follows.
The angular CMB trispectrum can be written
either as an integral over comoving distance $r$
(in the case of $\gnlloc$) or as a double integral over $(\tau,r)$ where
$\tau$ is conformal time during inflation (in the case of the $\dot\sigma^4$
or $(\partial\sigma)^4$ trispectra).
We approximate the integral by a finite sum, which represents the
CMB trispectrum as a sum of terms that are factorizable in a sense
defined in~\cite{long_trispectrum}.
A large number of sampling points are needed to obtain a good approximation
to the integral, leading to a large number of terms in the factorizable
representation.
However, there exists an optimization algorithm, which takes as input
a trispectrum that has been represented as a sum of many factorizable terms,
and outputs a representation with fewer terms.
The reduction can be quite dramatic, as shown in Table~\ref{tab:opt_trispectrum}.
The optimization algorithm guarantees that the output trispectrum
accurately approximates the input trispectrum, in the sense that the
two are nearly observationally indistinguishable.

\begin{table}[h!]                 
\begingroup
\newdimen\tblskip \tblskip=5pt
\caption{Number of factorizable terms $\Nin$ needed to represent each trispectrum
by direct sampling of the integral, and number of terms $\Nout$ obtained after
running the optimization algorithm from~ section VII of \cite{long_trispectrum}.}
\label{tab:opt_trispectrum}
\nointerlineskip
\vskip -6mm
\footnotesize
\setbox\tablebox=\vbox{
 \newdimen\digitwidth
   \setbox0=\hbox{\rm 0}
   \digitwidth=\wd0
   \catcode`*=\active
   \def*{\kern\digitwidth}
\halign{\hbox to 1in{#\leaderfil}\tabskip 0.5em &
\hfil#\hfil\tabskip 1em &
\hfil#\hfil &
\hfil#\hfil\tabskip 0pt\cr
\noalign{\vskip 10pt\doubleline\vskip 2pt}\hfil Trispectrum\phantom{xx}\hfil& 
\hfil $\Nin$ \hfil & 
\hfil $\Nout$ \hfil \cr
\noalign{\vskip 2pt}
\noalign{\vskip 4pt\hrule\vskip 6pt}
 $\gnlloc$ &  **436&   *17\cr
 \noalign{\vskip 2pt}
 $\dot\sigma^4$ &  *6955&  *73\cr
 \noalign{\vskip 2pt}
 $(\partial\sigma)^4$ &  20865&  192\cr
\noalign{\vskip 3pt\hrule\vskip 4pt}}}
\endPlancktable                    
\endgroup
\end{table}

Armed with ``small'' factorizable representations for each trispectrum,
the next step is to run an analysis pipeline that estimates the 
amplitude of each trispectrum from \Planck\ data.
We use the ``pure MC'' pipeline from~\cite{long_trispectrum}, which compares
the trispectrum of the data to the mean trispectrum of an
ensemble of simulations.
This pipeline requires a filtering operation $d \rightarrow \tilde a_{\ell m}$
which processes the pixel-space CMB data $d$ and generates a harmonic-space
map $\tilde a_{\ell m}$.
Our filtering operation is defined by the following steps:
\begin{enumerate}
\item Starting from the data $d$, we compute (with uniform pixel weighting)
a best-fit monopole and dipole outside the Galactic mask.  We use the temperature
``common mask'', the union of the confidence masks for the \SMICA, \SEVEM, \NILC, and \Commander\ component 
separation methods~\citep{planck2014-a11}.
\item The mask defines a few ``islands'', i.e., isolated groups of pixels that are
unmasked, but contained in a larger masked region.  We slightly enlarge the mask
so that it removes the islands.
\item We classify the components of the masked part of the sky into
``small'' masked regions with $\le 1000$ pixels 
(at {\tt{HEALPix}} resolution $N_{\rm side}=2048$),
and ``large'' regions with $> 1000$ pixels.
Small regions usually correspond to point sources, and large regions
typically correspond to areas of diffuse galactic emission.
In small regions, we inpaint the CMB
by assigning the unique map that agrees with the data on boundary pixels, 
and whose value in each interior pixel is the average of the neighboring pixels.
\item In large regions, we do not inpaint the CMB, but rather apodize the boundary
of the large region using cosine apodization with 12' radius.
\item We apply a spherical harmonic transform to the inpainted, apodized
CMB map to obtain a harmonic-space map $a_{\ell m}$ with $\ell_{\rm max}=1600$.
We then take the final filtered map $\tilde a_{\ell m}$ to be
\be
\tilde a_{\ell m} = \frac{a_{\ell m}}{b_\ell C_\ell + b_\ell^{-1} N_\ell}  \label{eq:gnl_weighting}
\ee
where $b_\ell$ is the beam,
$C_\ell$ is the fiducial CMB power spectrum,
and $N_\ell$ is the sky-averaged noise power spectrum (without beam
deconvolution).
To motivate this choice of $\ell$-weighting, we note that for an ideal
all-sky experiment with isotropic noise, we have 
$a_{\ell m} = b_\ell s_{\ell m} + n_{\ell m}$
where $s_{\ell m}, n_{\ell m}$ are signal and noise realizations.
In this case, Eq.~(\ref{eq:gnl_weighting}) weights the signal as
$s_{\ell m} / (C_\ell + b_\ell^{-2} N_\ell)$, which is optimal.
\end{enumerate}
In our pipeline, we apply this filter to the component-separated
\SMICA\ maps~\citep{planck2014-a11},
obtaining a harmonic-space map $\tilde a_{\ell m}$.  
We apply the same filter to 1000 Monte Carlo simulations
to obtain an ensemble of harmonic-space maps.
Our pipeline has the property that it always estimates the trispectrum of the
data in excess of the trispectrum in the simulations.
Since the simulations include lensing,
this means that lensing bias 
will automatically be subtracted
from our $\gnl$ estimates.

Now that the filter, data realization, and Monte Carlo simulations have
been fully specified, the details of the pipeline are described in
section IX.B of~\cite{long_trispectrum}.
For each trispectrum, the pipeline outputs an estimate of $\gnl$ and an
estimate of the statistical error.
Our basic results are:
\ba
\gnlloc &=& (-9.0 \pm 7.7) \times 10^4; \nn \\
\gnldotpi4 &=& (-0.2 \pm 1.7) \times 10^6; \label{eq:gnl_bottom_line} \\
\gnldpi4 &=& (-0.1 \pm 3.8) \times 10^5.  \nn
\ea
No deviation from Gaussian statistics is seen.
These results significantly improve the previous best constraints
on the trispectrum from 
WMAP~\citep{Vielva:2009jz,2010PhRvD..81l3007S,2010arXiv1012.6039F,hikage2012,Sekiguchi:2013hza,Regan:2013jua,long_trispectrum}
and large-scale structure~\citep{Desjacques:2009jb,Giannantonio:2013uqa,Leistedt:2014zqa}.

A constraint on $\gnlloc$ from \Planck\ 2013 data was recently reported by~\citet{Feng:2015pva},
who find $\gnlloc = (-13 \pm 18) \times 10^4$.  Our central value in Eq.~(\ref{eq:gnl_bottom_line})
agrees well with this result, but the statistical error is smaller by a factor of 2.3.
This improvement is partly due to the lower noise levels in \Planck\ 2015 data,
and partly due to the use of a better estimator.

Each line in Eq.~(\ref{eq:gnl_bottom_line}) is a ``single-$\gnl$'' constraint; i.e.,
the constraint on one $\gnl$ parameter with the other $\gnl$-parameters held fixed.  
For joint constraints, one needs to know the full covariance matrix.  
The correlation between $\gnlloc$ and the other two parameters is negligble, 
and the $\gnldotpi4$-$\gnldpi4$ correlation is:
\be
\mbox{Corr}(\gnldotpi4, \gnldpi4) = 0.61.  \label{eq:gnl_correlation}
\ee
multi-field models of inflation will generally give a linear combination
of $\dot\sigma^4$, $\dot\sigma^2 (\partial_i\sigma)^2$,
and $(\partial_i\sigma)^2 (\partial_j\sigma)^2$ trispectra.
In this case we proceed as follows.
First, if the $\dot\sigma^2 (\partial_i\sigma)^2$ coefficient is non-zero, 
we can use the near-degeneracy with a linear combination of the other two
operators to absorb it into the effective values of $\gnldotpi4$ and $\gnldpi4$.
A Fisher matrix analysis shows that the coefficients of this linear combination are
\ba
(\gnldotpi4)_{\rm eff} &=& 0.59\, \gnlB \nn \\
(\gnldpi4)_{\rm eff} &=& 0.091\, \gnlB  \label{eq:gnlB_eff}
\ea
It is convenient to define the two-component parameter vector:
\be
g_i = \left( \begin{array}{c}
  \gnldotpi4  \\
  \gnldpi4
\end{array} \right).
\ee
We also compute a two-by-two Fisher matrix $F_{ij}$, whose diagonal is given by $F_{ii} = 1/\sigma_i^2$,
where $\sigma_i$ is the single-$\gnl$ statistical error in Eq.~(\ref{eq:gnl_bottom_line}), and whose off-diagonal is
$F_{12} = r F_{11}^{1/2} F_{22}^{1/2}$, where $r$ is the correlation in Eq.~(\ref{eq:gnl_correlation}).
This procedure gives:
\be
F_{ij} = \left( \begin{array}{cc}
  3.3  &  9.2  \\
  9.2  &  68.7
\end{array} \right) \times 10^{-13}.
\ee
For a given parameter vector $g_i$, we can define a trispectrum-$\chi^2$ by
\be
\chi^2(g) = [F_{ii} \hat g_i - (Fg)_i] \, F_{ij}^{-1} \, [F_{jj} \hat g_j - (Fg)_j]\label{eq:trispectrum_chi2}
\ee
where $\hat g_i = (-0.21 \times 10^6, -0.10 \times 10^5)$ is the vector of best-fit
single-$\gnl$ values from Eq.~(\ref{eq:gnl_bottom_line}).
This definition of $\chi^2$ follows from the observation that $(F_{ii} \hat g_i)$
is an estimator with expectation value $(Fg)_i$ and covariance matrix 
$\mbox{Cov}(F_{ii}\hat g_i, F_{jj}\hat g_j) = F_{ij}$.

The inflationary implications of these trispectrum constraints are discussed
in Sect.~\ref{ssec:inflationary_trispectrum} below.

%% file: A19_Section10.tex
In this section, we present constraints on local NG at first and second order ($\fnllocal$ and $\gnlloc$) obtained with  Minkowski functionals (MFs) on temperature and polarization $E$ maps. 
MFs \citep{1994A&A...288..697M, 1997ApJ...482L...1S, 1998MNRAS.297..355S, 1998NewA....3...75W}
are a measure of fields' local morphology used to constrain their stationarity, isotropy and Gaussianity. Mostly probing general NG in a frequentist fashion in two-dimensions on CMB maps \citep{2004ApJ...612...64E, Komatsu:2003iq, 2013MNRAS.428..551M, 2010MNRAS.408.1658N,2008A&A...486..383C} or three-dimensions on LSS data \citep{2005ApJ...633...11P, 2014MNRAS.443..241W}, they have also been used to measure specific NG targets with Bayesian methods, such as $\fnllocal$ \citep{hikage2006,hikage2008,ducout2012,planck2013-p09a}, other bispectrum and trispectrum shapes \citep{hikage2012} and topological defects \citep{planck2013-p20}. New developments have been made recently, using needlets \citep{2015PhRvD..91f3501F}, neural networks \citep{2014arXiv1409.3876N} or allowing scale-dependent measurements \citep{2013MNRAS.434.2830M}.

MF-based limits are well known to be suboptimal for $\fnllocal$ and $\gnlloc$, but they provide an independent cross-check of bispectrum and trispectrum-based estimators. They are complementary to optimal estimators: they are weighted integrals of the polyspectra and are sensitive to any source of NG, including foregrounds and secondaries. While MFs are not always able to distinguish between these different sources and systematics, they allow upper bounds to be put on them. 

The most recent constraints ($1\,\sigma$) from MFs on $\fnllocal$ and $\gnlloc$ have been obtained with \WMAP\  \citep{hikage2012}
and \planck\ \citep{planck2013-p09a}:
\begin{equation}
\fnllocal = 4.2 \pm 20.5; \qquad \gnlloc = (−1.9 \pm 6.4) \times 10^5.
\end{equation}

\subsection{Method and definition of MFs}

For a smoothed two-dimensional field $\delta$ of zero mean and of variance $\sigma_0^2$, defined on the sphere, we consider an excursion set of height $\nu = \delta/\sigma_0$, i.e., the set of points where the field exceeds the threshold $\nu$. We use four functionals denoted by $V_k (\nu) (k = 0, 1, 2, 3)$. The first three correspond to MFs: $V_0$ is the fractional {\sl Area} of the regions above the threshold, $V_1$ is the {\sl Perimeter} of these regions and $V_2$ is the {\sl Genus}, defined as the total number of connected components of the excursion above the threshold minus the total number of connected components under the threshold. The fourth, $V_3$, is the {\sl Number} of clusters (also referred to as N$_{\mathrm{clusters}}$). This is the number of connected regions above the threshold for positive thresholds and below the threshold for negative thresholds. Precise definitions and formulae for the quantities $V_k$ as well as their expectation values for Gaussian fields are summarized in Appendix~\ref{sec:AB}.

We calculate the four normalized\footnote{Raw Minkowski functionals $V_k$ depend on the Gaussian part of fields through a normalization factor $A_k$, a function only of the shape of the power spectrum. We therefore normalize functionals $v_k=V_k/A_k$ to focus on NG, see Appendix~\ref{sec:AB}.} functionals $v_k(\nu)$ on $n_{\rm th}=26$ thresholds $\nu$, between $\nu_{\rm min}=-3.5$ and $\nu_{\rm max}=+3.5$.

\begin{figure}
\begin{center}
\includegraphics[width=8cm]{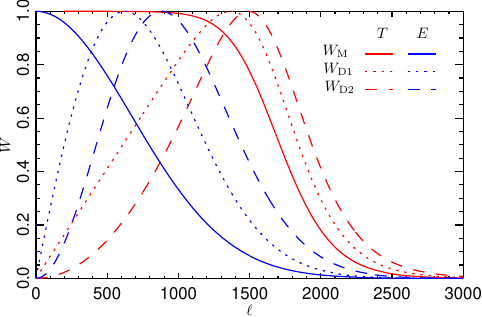}
  \caption{Filters used to optimize constraints on local NG, in harmonic space. The temperature filter $W_{\rm M}$ is a smoothed version of the true Wiener filter obtained with realistic models, while the $E$-mode $W_{\rm M}$ filter is adapted from the temperature one, with a cutoff value at $\ell\simeq 800$. The formulae for the derivative filters are given in Eq.~\eqref{eq:wiener_filters_mfs_form}.}
  \label{fig:wiener_filters_mfs}
\end{center}
\end{figure}

For this analysis, we used the same temperature and polarization $E$ data, simulations and masks described in Sect.~\ref{sec:dataset} for consistency with the bispectrum estimators. In addition, the maps are filtered to optimize constraints on local NG \citep{ducout2012}, the filters used being similar to Wiener filters for $T$ and $E$ ($W_{\rm M}$), and for the first ($W_{\rm D1}$) and second ($W_{\rm D2}$) derivatives of these fields (Fig.~\ref{fig:wiener_filters_mfs}):
\begin{eqnarray}
W_{\rm D1} & \propto & \sqrt{\ell (\ell +1) } \, W_{\rm M} \, ; \\
W_{\rm D2} & \propto & \ell (\ell +1)  \, W_{\rm M}.
\label{eq:wiener_filters_mfs_form}
\end{eqnarray}
For the temperature map, known point sources in the mask are inpainted.

We define the vector $y$ as any combination $y=\lbrace v_k^{A,W}\rbrace$ with $k=\lbrace 0,3\rbrace$, $A=\lbrace T,E\rbrace$, $W=\lbrace W_{\rm M}, W_{\rm D1}, W_{\rm D2}\rbrace$, $\hat{y}$ being the vector measured on the data.

From these measurements, we then use a Bayesian method to jointly estimate $\fnllocal$ and $\gnlloc$,
\begin{multline}
	P(\fnllocal,\, \gnlloc|\,\hat{y}) = \\
 \dfrac{P(\hat{y}\, | \fnllocal, \gnlloc) P(\fnllocal, \gnlloc)}{\int P(\hat{y}\, | \fnllocal, \gnlloc) P(\fnllocal, \gnlloc) {\rm d}\fnllocal {\rm d}\gnlloc}.  
\end{multline}
We take a uniform prior for $\fnllocal$ in the range  $-400$ to 400, and for $\gnlloc$ in the range $-4\times 10^{6}$ to  $+4\times 10^{6}$, while the evidence is just considered as a normalization. 

Assuming MFs are multi-variate Gaussian-distributed we obtain the posterior distribution for $(\fnllocal, \gnlloc)$ with a $\chi^{2}$ test 
\begin{equation}
P(\fnllocal,\, \gnlloc|\,\hat{y}) \propto \exp \left[ -\dfrac{\chi^{2}(\hat{y},\fnllocal, \gnlloc)}{2} \right] \, ,
\label{eq:postsimp}
\end{equation}
with 
\begin{multline}
\chi^{2}(\hat{y}, \fnllocal, \gnlloc)  \equiv \\
\left[ \hat{y}- \bar{y}_{\rm sim1}( \fnllocal, \gnlloc  )   \right]^{T}C_{\rm sim2}^{-1} \left[ \hat{y}- \bar{y}_{\rm sim1}( \fnllocal, \gnlloc  )  \right].
\label{eq:mychi2}
\end{multline}

For this test, we use two types of simulations to first construct a model including primordial NG $\bar{y}_{\rm sim1}( \fnllocal, \gnlloc  )$ and secondly a covariance matrix \begin{equation}C_{\rm sim2}\equiv\left\langle \left( y_{\rm sim2}- \bar{y}_{\rm sim2}\right) \left( y_{\rm sim2}- \bar{y}_{\rm sim2}\right)^{\rm T} \right\rangle,\end{equation}
with $\bar{y}_{\rm sim}\equiv\left\langle y_{\rm sim}\right\rangle_{\rm sim} $ averaged over the simulations. We now describe the details of these simulations.
\begin{itemize}
\item \textit{Simulations 1: Non-Gaussian model}\\
For the first type of simulation, we included all possible sources of NG, assuming that the total and individual levels of NG are small enough that MFs are linear with respect to those NG levels \citep{ducout2012}. The three kinds of NG we included are foreground residuals (Galactic residuals with scalable amplitude $\alpha$, as well as radio sources, CIB anisotropies, secondaries (SZ, lensing and ISW-lensing, but not SZ-lensing) and primordial NG ($\fnllocal$, $\gnlloc$):
\begin{eqnarray}
{\rm sim1}^i &=& {\rm map}^i_{\rm lensed} (\fnllocal, \gnlloc) \nonumber \\ 
& & +  \quad \rm map^{fg} (\rm radio\, sources, CIB, SZ) \nonumber \\
& &  +  \quad \alpha\times \rm map^{fg} (\rm Galactic\, residuals) \, .
\end{eqnarray}
We tried to reproduce all instrumental effects, with realistic effective beams (isotropic window functions), noise from FFP8 simulations \citep{planck2014-a14}, filtered with component separation weights. We checked the accuracy of these simulations by comparing them to FFP8 simulations, using no foreground and no primordial NG. The astrophysical models are provided by the \textit{Planck Sky Model} (PSM, \citealt{delabrouille2012}), while the primordial NG simulations are computed as in \cite{2009ApJS..184..264E}. The lensing uses {\tt LensPix}\footnote{{\tt http://cosmologist.info/lenspix}}.  The power spectrum used for these NG simulations and the lensing is the best-fit power spectrum from \Planck\ 2013+ACT/SPT+BAO \citep{planck2013-p17}. We created $n_1=200$ simulations $i$, using $n_1$ maps for the primordial NG, while we had only one astrophysical foreground simulation.

\item \textit{Simulations 2: FFP8 \citep{planck2014-a14} MC simulations} \\
Since NG is weak, the covariance matrix $C$ is computed with $n_2=10^4$ simulations, including no primordial NG and no foregrounds. These simulations reproduce realistic instrumental effects (anisotropy of beams in particular), realistic noise and component separation filtering. The only NG still present in these simulations are lensing and the ISW-lensing correlation.
\end{itemize}

\subsubsection*{Validation of the estimator}
Part of the validation for the MFs estimator is described in Sect.~\ref{Sec_valid_est_3} for $\fnllocal$, to compare the results to bispectrum estimators on realistic simulations (FFP8 MC, second item above). In addition we present in Table~\ref{tab:fnlgnl_mfs_valid} the results obtained on the same realistic simulations for $\gnlloc$, and on simulations containing primordial NG (first item above), with $\fnllocal=10$ and $\gnlloc=10^5$, with 200 simulations used in each case. These tests have been performed using the \SMICA\ method with lensing bias removed.

\begin{table}[htb!]                 
\begingroup
\newdimen\tblskip \tblskip=5pt
\caption{
Results for local NG parameters at first and second order, $\fnllocal$ and $\gnlloc$, obtained with Minkowski functionals on \SMICA\ simulations in temperature and polarization. These results are corrected for the lensing and ISW-lensing biases unless stated otherwise. Parameters are estimated jointly, and we report marginalized results, quoting 1$\,\sigma$ errors. The results are the average obtained from 200 simulations.
}
\label{tab:fnlgnl_mfs_valid}
\nointerlineskip
\vskip -6mm
\footnotesize
\setbox\tablebox=\vbox{
   \newdimen\digitwidth
   \setbox0=\hbox{\rm 0}
   \digitwidth=\wd0
   \catcode`*=\active
   \def*{\kern\digitwidth}
   \newdimen\signwidth
   \setbox0=\hbox{+}
   \signwidth=\wd0
   \catcode`!=\active
   \def!{\kern\signwidth}
\halign{\hbox to 1.5in{#\leaderfil}\tabskip 1em&
\hfil#\hfil\tabskip 1em&
\hfil#\hfil\tabskip 0pt\cr
\noalign{\vskip 10pt\doubleline\vskip 2pt}
\hfill&\hfil $\fnllocal$ \hfil& \hfil $\gnlloc$ ($\times 10^4$)  \hfil\cr
\noalign{\vskip 2pt}
\noalign{\vskip 4pt\hrule\vskip 6pt}
\omit \hfil $\fnllocal=0\;$, $\;\gnlloc=0$ \hfil&&\cr
\noalign{\vskip 5pt}
$T$**** & !* *0 $\pm$ 13 & !* $-$1 $\pm$ 19  \cr
\noalign{\vskip 5pt}
$E$**** & !* *1 $\pm$ 42 & !* !0 $\pm$ 23  \cr
\noalign{\vskip 5pt}
$T+E$** & !* *1 $\pm$ 12 & !* !0 $\pm$ 13  \cr
\noalign{\vskip 5pt}
\hline
\noalign{\vskip 5pt}
\omit\hfil $\fnllocal=10\;$, $\;\gnlloc=10^{5}$    \hfil &&\cr
\noalign{\vskip 5pt}
$T$**** & !* 10 $\pm$ 13 &  !*  !*9 $\pm$ 22 \cr
\noalign{\vskip 5pt}
$E$**** & !* 12 $\pm$ 42 &  !*  !10 $\pm$ 23 \cr
\noalign{\vskip 5pt}
$T+E$** & !* 11 $\pm$ 12 &  !* !10 $\pm$ 13  \cr
\noalign{\vskip 3pt\hrule\vskip 4pt}}}
\endPlancktable                    
\endgroup
\end{table}                        

\subsection{Results}

Results for $\fnllocal$ and $\gnlloc$ estimation with MFs on the four component separated maps in temperature and polarization are presented in Table~\ref{tab:fnlgnl_mfs}. The results for polarization $E$-only maps are not quoted, since these results were not sufficiently stable (cf. Sect.~\ref{sec:data_valid_summary}). No deviation from Gaussianity is detected. \textit{T+E} analysis generally finds higher values for $\fnllocal$, but remains consistent with Gaussianity.

\begin{figure}
\begin{center}
\includegraphics[width=9cm]{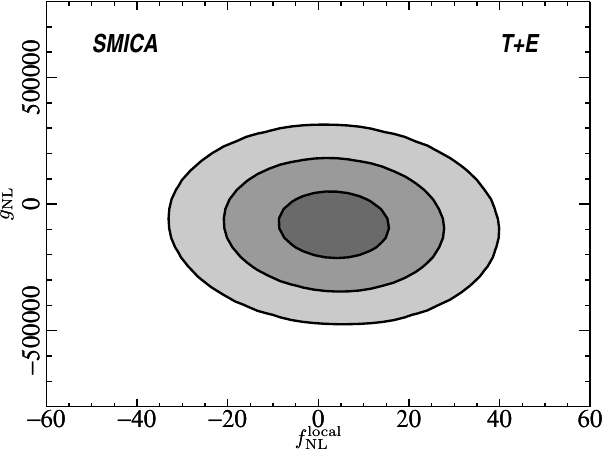}
  \caption{Joint constraint on $\fnllocal$ and $\gnlloc$ obtained with MFs. The contour lines represent 1, 2 and 3$\,\sigma$ limits of a 2D-Gaussian distribution. Constraints were obtained with \SMICA\ temperature and polarization $E$ maps.}
  \label{fig:fnlloc_vs_gnl_post2d}
\end{center}
\end{figure}

The posteriors for $\fnllocal$ and $\gnlloc$ from \SMICA\ are shown in Fig.~\ref{fig:fnlloc_vs_gnl_post2d}.
One interesting point is that the estimates of $\fnllocal$ and $\gnlloc$ are almost uncorrelated ($r<0.1$); this can be inferred when we consider the parity of MF deviations from Gaussianity, which is different for the two parameters \citep{matsubara2010}.

Foreground and secondary biases are removed from these estimates, since the NG model directly includes them. However, an estimation of their contribution in the map is reported in Table~\ref{tab:fnlgnl_biases_mfs}.

\begin{table}[htb!]                 
\begingroup
\newdimen\tblskip \tblskip=5pt
\caption{
Results for local NG parameters at first and second order,  $\fnllocal$ and $\gnlloc$, obtained with Minkowski functionals on all four component separated maps in temperature and polarization. These results are corrected for the lensing and ISW-lensing biases unless stated otherwise. Parameters are estimated jointly, and we report marginalized results, quoting 1$\,\sigma$ errors.
}
\label{tab:fnlgnl_mfs}
\nointerlineskip
\vskip -6mm
\footnotesize
\setbox\tablebox=\vbox{
   \newdimen\digitwidth
   \setbox0=\hbox{\rm 0}
   \digitwidth=\wd0
   \catcode`*=\active
   \def*{\kern\digitwidth}
   \newdimen\signwidth
   \setbox0=\hbox{+}
   \signwidth=\wd0
   \catcode`!=\active
   \def!{\kern\signwidth}
\halign{\hbox to 1in{#\leaderfil}\tabskip 1em&
\hfil#\hfil\tabskip 1em&
\hfil#\hfil\tabskip 0pt\cr
\noalign{\vskip 10pt\doubleline\vskip 2pt}
\hfill&\hfil $\fnllocal$ \hfil& \hfil $\gnlloc$ ($\times 10^4$)  \hfil\cr
\noalign{\vskip 2pt}
\noalign{\vskip 4pt\hrule\vskip 6pt}
\omit \hfil \SMICA\hfil&&\cr
$T$**** & !* *2 $\pm$ 13 & !* $-$17 $\pm$ 19  \cr
\noalign{\vskip 5pt}
$T+E$** & !* *3 $\pm$ 12 & !* *$-$8 $\pm$ 13  \cr
\noalign{\vskip 5pt}
\hline
\noalign{\vskip 5pt}
\omit\hfil \SEVEM\hfil&&\cr
$T$**** & !* *3 $\pm$ 13 &  !*  $-$23 $\pm$ 20 \cr
\noalign{\vskip 5pt}
$T+E$** & !* *7 $\pm$ 12 &  !* *$-$9 $\pm$ 13  \cr
\noalign{\vskip 5pt}
\hline
\noalign{\vskip 5pt}
\omit\hfil \NILC\hfil&&\cr
$T$**** & !* 10 $\pm$ 13 & !* $-$23 $\pm$ 20  \cr
\noalign{\vskip 5pt}
$T+E$** & !* 12 $\pm$ 12 & !* $-$15 $\pm$ 13  \cr
\noalign{\vskip 5pt}
\hline
\noalign{\vskip 5pt}
\omit\hfil \Commander\hfil&&\cr
$T$**** & !* *8 $\pm$ 13 & !* $-$30 $\pm$ 19 \cr
\noalign{\vskip 5pt}
$T+E$** & !* 10 $\pm$ 13 &  !* $-$18 $\pm$ 13 \cr
\noalign{\vskip 3pt\hrule\vskip 4pt}}}
\endPlancktable                    
\endgroup
\end{table}                        

\subsubsection*{Foreground and secondary biases}

Foreground residuals are generally negligible, in particular in the $T$ analysis. This is different from the \textit{Planck} 2013 results, where the residuals were more important; this can be explained by the beam correction applied to these previous estimates which exaggerated signals from small scales.

Lensing has a significant signature in MF estimation of $\fnllocal$, but is even stronger in $\gnlloc$ (the four-point correlation signature) and could be detected (and not treated just as a bias) with this estimator. The Wiener filters enhance the scales where lensing is dominant.

\begin{table}[htb!]                 
\begingroup
\newdimen\tblskip \tblskip=5pt
\caption{
Biases for local NG parameters at first and second order $\fnllocal$, $\gnlloc$ obtained with Minkowski Functionals on \SMICA\ in temperature and polarization. Parameters are estimated jointly, and we report marginalized results. For the corresponding error (on one map), see Table~\ref{tab:fnlgnl_mfs}.}
\label{tab:fnlgnl_biases_mfs}
\nointerlineskip
\vskip -6mm
\footnotesize
\setbox\tablebox=\vbox{
   \newdimen\digitwidth
   \setbox0=\hbox{\rm 0}
   \digitwidth=\wd0
   \catcode`*=\active
   \def*{\kern\digitwidth}
   \newdimen\signwidth
   \setbox0=\hbox{+}
   \signwidth=\wd0
   \catcode`!=\active
   \def!{\kern\signwidth}
\halign{\hbox to 1in{#\leaderfil}\tabskip 1.2em&\hfil#\hfil\tabskip 1.2em&\hfil#\hfil\tabskip 1.2em& \hfil#\hfil\tabskip 1.2em& \hfil#\hfil\tabskip 0pt\cr
\noalign{\vskip 10pt\doubleline\vskip 2pt}
\omit&\multispan2 \hfil $\Delta\fnllocal$ \hfil &\multispan2 \hfil $\Delta\gnlloc$ ($\times 10^4$) \hfil\cr
\noalign{\vskip 4pt\hrule\vskip 6pt}
\noalign{\vskip 2pt}
\hfill&\hfil $T$ & \hfil $T+E$ & \hfil $T$ & \hfil $T+E$ \cr
\noalign{\vskip 2pt}
\noalign{\vskip 4pt\hrule\vskip 6pt}
\omit \hfil \SMICA\hfil&&&&\cr
SZ &  *!0.0 &  *$-$0.3 & *!2.3 & *!1.1  \cr
\noalign{\vskip 5pt}
CIB** &  *!0.7 & *!0.5 & *$-$6.8 & *!3.4  \cr
\noalign{\vskip 5pt}
Galaxy** &  *$-$0.1 & *$-$0.2 & *$-$1.2 & *!3.1  \cr
\noalign{\vskip 5pt}
PS** & *!0.1 & *!0.2 & *!2.2 & *!1.2  \cr
\noalign{\vskip 5pt}
Lensing & !16.5 & !10.0 & !63.1 & !40.4  \cr
\noalign{\vskip 3pt\hrule\vskip 4pt}}}
\endPlancktable                    
\endgroup
\end{table}                        

%% file: A19_Section11.tex

The NG constraints obtained in this paper show consistency of \textit{Planck} data with Gaussian primordial fluctuations, thus confirming the results obtained in the 2013 release~\citep{planck2013-p09a} and improving them through the inclusion of CMB polarization data. The standard single-field slow-roll models of inflation have therefore been confirmed as a viable scenario for inflation, passing one of their most stringent tests, based on lack of measurable deviations from Gaussianity. The constraints obtained on local, equilateral, and orthogonal NG, after accounting for various contaminants, strongly limit different mechanisms proposed as alternatives to the standard inflationary models  to explain the seeds of cosmological perturbations. Measurements on deviations from Gaussianity for other primordial bispectral shapes help to shed light on more subtle effects about the detailed physics of inflation. 

As in~\cite{planck2013-p09a}, in the following we derive limits on parameters of the models from the NG constraints in the following way (unless explicitly stated otherwise): we construct a posterior based on the assumption that the sampling distribution is Gaussian (as supported by Gaussian simulations); the likelihood is approximated by the sampling distribution, but centred on the NG estimate (see \citealt{2009ApJS..184..264E}); we employ uniform or   Jeffreys' priors, over intervals of the parameters values that are physically meaningful, or as otherwise stated;  and in the cases when two or more parameters are involved, we marginalize the posterior to provide one-dimensional constraints on the parameter considered.


\subsection{General single-field models of inflation}

\noindent{\it DBI models}:  DBI models of inflation~(\citealt{2004PhRvD..70j3505S,2004PhRvD..70l3505A}), characterized by a non-standard kinetic term of the inflaton field,  predict a non-linearity parameter $f_{\rm NL}^{\rm DBI}=-(35/108) (c_\mathrm{s}^{-2}-1)$, where $c_\mathrm{s}$ is the sound speed of the inflaton perturbations~(\citealt{2004PhRvD..70j3505S,2004PhRvD..70l3505A,2007JCAP...01..002C}). The corresponding bispectrum shape is very close to the equilateral shape. Nonetheless we have constrained the exact theoretical (non-separable) shape~(see equation 7 of ~\citealt{planck2013-p09a}). The constraint we obtain $f_{\rm NL}^{\rm DBI}=2.6 \pm 61.6 $ from temperature data ($f_{\rm NL}^{\rm DBI}=15.6 \pm 37.3 $ from temperature and polarization) at $68\,\%$ CL (with ISW-lensing and point sources subtracted, see Table~\ref{tab:equilmodels}) implies 
\begin{equation}
\label{csDBIT}
c_\mathrm{s}^{\rm DBI} \geq  0.069\quad \quad  \text{95$\,\%$ CL (\itT-only)}\, ,
\end{equation}
and 
\begin{equation}
\label{csDBITE}
c_\mathrm{s}^{\rm DBI} \geq 0.087 \quad \quad  \text{95$\,\%$ CL (\textit{T+E})}\, . 
\end{equation}
In~\cite{planck2013-p09a} we constrained the so-called infrared (IR) DBI models~(\citealt{2005PhRvD..71f3506C,2005JHEP...08..045C}), which arise in string frameworks. We focused on a minimal setup, considering a regime where stringy effects are negligible and predictions for primordial perturbations are built within standard field theory. In the companion paper~\cite{planck2014-a24} we present an analysis of a more general class of IR DBI models 
which accounts for stringy signatures~(see \citealt{2008PhRvD..77b3527B}) by combining {\Planck} power spectrum and bispectrum constraints.

\medskip


\medskip 
\noindent{\it Implications for effective field theory of Inflation:}
 \label{Implications_EFT}
In this subsection we use the effective field theory approach to inflation in order to translate the contraints on $f_{\rm NL}^{\rm equil}$ and $f_{\rm NL}^{\rm ortho}$ into limits on the parameters of the Lagrangian of  general 
single-field models of inflation (of the type $P(X,\varphi)$ models). In particular we derive the most conservative bound on the sound speed of the inflaton perturbations for this class of models.

The effective field theory approach~(\citealt{2008JHEP...03..014C,2008PhRvD..77l3541W}) provides an efficient way to constrain inflationary perturbations for various classes of models that incorporate deviations from the standard single-field slow-roll scenario. In this approach the Lagrangian of the system is expanded into the (lowest dimension) operators obeying the underlying symmetries.  We consider general single-field models described by the following action
 \begin{eqnarray}
 S&=&\int d^4x \sqrt{-g} \left[ -\frac{M^2_{\rm Pl} \dot{H}}{c_\mathrm{s}^2} \left( \dot{\pi}^2-c_\mathrm{s}^2 \frac{(\partial_i \pi)^2}{a^2} \right) 
 \right. \\   
 &-& \left. M_{\rm Pl}^2 \dot{H} (1-c_\mathrm{s}^{-2}) \dot{\pi} \frac{(\partial_i \pi)^2}{a^2}  
 + \left( M_{\rm Pl}^2 \dot{H}  (1-c_\mathrm{s}^{-2})  -\frac{4}{3} M_3^4 \right) \dot{\pi}^3 \right], \nonumber 
\end{eqnarray}
where the curvature perturbation is related to the scalar field $\pi$ as $\zeta=- H \pi$. The inflaton interaction terms 
 $\dot{\pi} (\partial_i \pi)^2$ and $(\dot{\pi})^3$ generate two kind of bispectra with amplitudes, respectively, $f_{\rm NL}^{\rm EFT1}=-(85/324)(c_\mathrm{s}^{-2}-1)$ and $f_{\rm NL}^{\rm EFT2}=-(10/243)(c_\mathrm{s}^{-2}-1) \[\tilde{c}_3+(3/2) c_\mathrm{s}^2\]$, 
 where $M_3$ is the amplitude of the operator $\dot{\pi}^3$~(\citealt{2010JCAP...01..028S}, see also ~\citealt{2007JCAP...01..002C,2010AdAst2010E..72C}). These two bispectra both peak for equilateral triangles in Fourier space. Nevertheless, they are sufficiently different, and the total NG signal turns out to be a linear combination of the two, leading also to 
an orthogonal shape. We can put constraints on $c_\mathrm{s}$ and the dimensionless parameter ${\tilde c}_3 (c_\mathrm{s}^{-2}-1)=2 M_3^4 c_\mathrm{s}^2 /(\dot{H} M_{\rm Pl}^2)$~\citep{2010JCAP...01..028S}. Notice that DBI inflationary models corresponds to having $\tilde{c}_3= 3(1-c_\mathrm{s}^2)/2$, while $c_\mathrm{s}=1$ and $M_3=0$ (or ${\tilde c}_3 (c_\mathrm{s}^{-2}-1)=0)$ represent the non-interacting (vanishing NG) case. 

\begin{figure}[!t]
\includegraphics[width=\hsize]{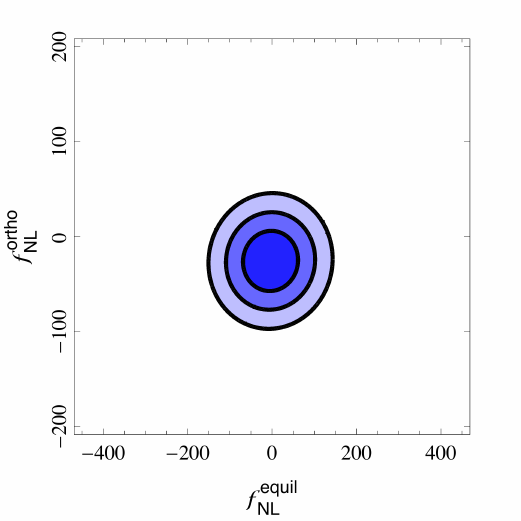}
\caption{$68\,\%$, $95\,\%$, and $99.7\,\%$ confidence regions in the parameter space $(f_\mathrm{NL}^\mathrm{equil}, f_\mathrm{NL}^\mathrm{ortho})$, defined by thresholding $\chi^2$ as described in the text.}
\label{fig:eq_ort}
\end{figure}

\begin{figure}[!t]
\includegraphics[width=\hsize]{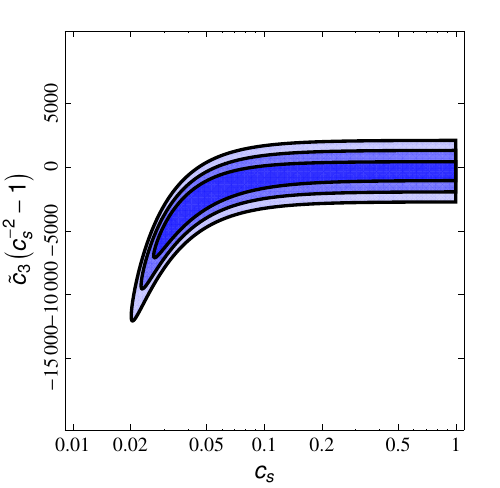}
\caption{$68\,\%$, $95\,\%$, and $99.7\,\%$ confidence regions in the single-field inflation parameter space $(c_\mathrm{s}, \tilde{c}_3)$, obtained from Fig.~\ref{fig:eq_ort} via the change of variables in Eq.~(\ref{meanfNL}).}
\label{fig:cs_c3}
\end{figure}

The mean values of the estimators for $f^{\mathrm{equil}}_{\mathrm{NL}}$ and $f^{\mathrm{ortho}}_{\mathrm{NL}} $ are expressed in terms of $c_\mathrm{s}$ and $\tilde{c}_3$ by
\begin{eqnarray}
\label{meanfNL}
f^{\mathrm{equil}}_{\mathrm{NL}} &=&\frac{1-c_\mathrm{s}^2}{c_\mathrm{s}^2} \left [-0.275 - 0.0780 c_{\rm s}^2 -  (2/3) \times 0.780  \tilde{c}_3 \right] \nonumber \\
f^{\mathrm{ortho}}_{\mathrm{NL}} &=&\frac{1-c_\mathrm{s}^2}{c_\mathrm{s}^2} \left[ 0.0159 - 0.0167 c_{\rm s}^2 - (2/3) \times 0.0167  \tilde{c}_3\right] \, .
\end{eqnarray}
Here the coefficients come from the Fisher matrix between the equilateral and orthogonal templates and the theoretical bispectra predicted by the two operators $\dot{\pi} (\nabla \pi)^2$ and $\dot{\pi}^3$. We use a 
$\chi^2$ statistic given by $\chi^2(\tilde{c}_3,c_\mathrm{s})={\vec v}^{\rm T}(\tilde{c}_3,c_\mathrm{s}) \C^{-1} {\vec v}(\tilde{c}_3,c_\mathrm{s})$,  where $v^i(\tilde{c}_3,c_\mathrm{s})=f^i(\tilde{c}_3,c_\mathrm{s})-f^i_{P}$ ($i$=\{equilateral, orthogonal\}), $f^i_{P}$ being the joint estimates of equilateral and orthogonal $f_{\rm NL}$~(see Table~\ref{Tab_KSW+SMICA}), $\C$ the covariance matrix  of the joint estimators, and  $f^i(\tilde{c}_3,c_\mathrm{s})$ is provided by Eq.~(\ref{meanfNL}).  As an example in Fig.~\ref{fig:eq_ort} we show the $68 \,\%, 95 \,\%$, and $99.7\,\%$ confidence regions for $f^{\rm equil}_{\rm NL}$ and $f^{\rm ortho}_{\rm NL}$ obtained 
from the $T+E$ constraints, requiring $\chi^2 \leq 2.28, 5.99$, and $11.62$ respectively (corresponding to a $\chi^2$ variable with two degrees of freedom). In Fig.~\ref{fig:cs_c3} we show the corresponding confidence regions in the $(\tilde{c}_3,c_\mathrm{s})$ parameter space.
Marginalizing over $\tilde{c}_3$ we find 
\begin{equation}
c_\mathrm{s} \geq 0.020 \quad \quad  \text{95$\,\%$ CL (\itT-only)}\, ,
\end{equation}
and 
\begin{equation}
c_\mathrm{s} \geq 0.024 \quad \quad  \text{95$\,\%$ CL (\textit{T+E})}\, .
\end{equation}
The constraints improve by a few percent in \itT-only and by up to $25\,\%$ by including polarization, in comparison with those of \cite{planck2013-p09a}.

\medskip

\begin{figure}
\includegraphics[width=8cm,height=8cm,keepaspectratio]{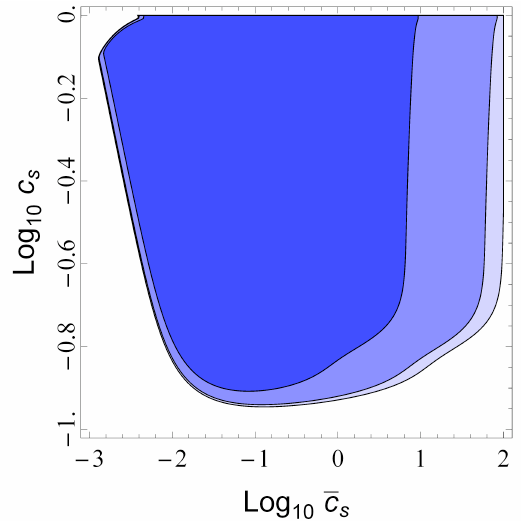}
\caption{
$68\,\%$, $95\,\%$, and $99.7\,\%$ probability contours in the Galileon models for $c_{\rm s}$ and $\bar{c}_{\rm s}$ parameters
for the $\bar{c}_{\rm s}>0$ branch (tensor spectral index $n_{\rm T}<0$).}
\label{Galileon_cs_csbar_positive}
\end{figure}

\begin{figure}
\includegraphics[width=8cm,height=8cm,keepaspectratio]{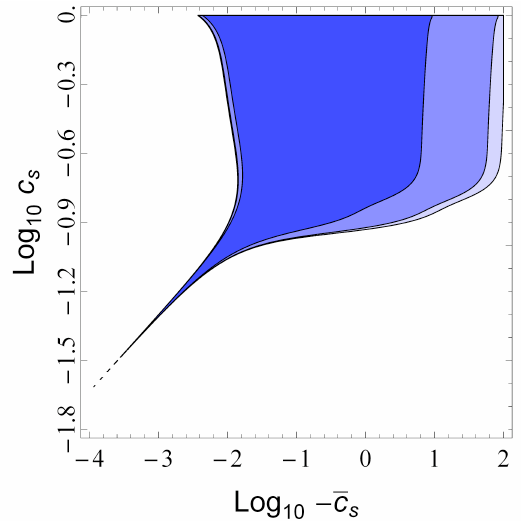}
\caption{
$68\,\%$, $95\,\%$, and $99.7\,\%$ probability contours in the Galileon models for $c_{\rm s}$ and $\bar{c}_{\rm s}$ parameters
for the $\bar{c}_{\rm s}<0$ branch (blue tensor spectral index $n_{\rm T}>0$).}
\label{Galileon_cs_csbar_negative}
\end{figure}

\noindent{\it Galileon models of inflation:}  \label{Ginflation}
Galileon models of inflation~\citep{2011JCAP...01..014B,2010PhRvL.105w1302K,2010PhRvD..82j3518M,2012JCAP...10..035O}  are well motivated models based on the so called ``Galilean symmetry''~\citep{2009PhRvD..79f4036N}. They are characterized by stability properties that are quite well understood (ghost-free, and stable against quantum corrections) and can arise naturally within fundamental physics models \citep{2010PhLB..693..334D,2010PhRvD..82d4020D}. Moreover they are an interesting example of models where gravity is modified on large scales and we focus on them also as a typical example of a more general class of modified gravity theories that are ghost-free (the so called Horndeski theories, \citealt{1974IJTP...10..363H}; see also~\cite{planck2014-a16} where these models are discussed in the context of dark energy/modifed gravity scenarios for the late time evolution of the universe). The predictions for the primordial perturbations are very rich. Bispectra can be generated with the same shapes as the ``EFT1'' and ``EFT2''  bispectra (see also discussion in~\citealt{2011JCAP...02..006C}), however the amplitude(s) scale with the fluctuation sound speed as $c_{\rm s}^{-4}$, differently from the general single-field models of inflation considered in the above subsection.     
They can be written as (at the lowest-order in slow-roll parameters) 
\begin{eqnarray}
\label{fnlG}
\fnl^{\rm EFT1}&=&\frac{17}{972}\left(-\frac{5}{c_{\mathrm{s}}^4}+\frac{30}{c_{\mathrm{s}}^2}-\frac{40}{c_{\mathrm{s}} \bar c_{\mathrm{s}}}+15 \right) 
\\
\label{fnlG2}
\fnl^{\rm EFT2}&=&\frac{1}{243} \left(\frac{5}{c_{\mathrm{s}}^4}+\frac{30/A-55}{c_{\mathrm{s}}^2}+\frac{40}{c_{\mathrm{s}} \bar c_{\mathrm{s}}}-320 \frac{c_{\mathrm{s}}}{\bar c_{\mathrm{s}}}-\frac{30}{A}+275 
\right. \nonumber\\
&-& \left.225 c_{\mathrm{s}}^2+280 \frac{c^3_{\mathrm{s}}}{\bar c_{\mathrm{s}}}\right)\, .
\end{eqnarray}
Here $A$, $\bar{c}_{\rm s}$, and $c_{\rm s}$ are dimensionless parameters of the models. In particular $c_{\rm s}$ is the sound speed of the Galileon scalar field, while $\bar{c}_{\rm s}$ is a parameter that appears to break the standard consistency relation for the tensor-to-scalar perturbation ratio ($r=16 \epsilon \bar{c}_{\rm s}=-8n_{\rm T} \bar{c}_{\rm s}$, $n_{\rm T}$ being the tensor spectral index)\footnote{For the explicit expressions of these 
parameters in terms of the coefficients of the Galileon Lagrangian see~\cite{planck2014-a24}.}.  Accordingly to Eq.~(\ref{meanfNL}) a linear combination of these two bispectra generate equilateral and orthogonal bispectra templates\footnote{
We note that we are neglecting $\mathcal{O}(\epsilon_1/c_\mathrm{s}^4)$ corrections (where $\mathcal{O}(\epsilon_1)$ means also $\mathcal{O}(\eta_{\mathrm{s}},{\mathrm{s}},...)$); see ~\citep{2011JCAP...01..014B} and \citep{2011JCAP...10..027R}. These corrections will have a different shape associated with them and they are not necessarily small when compared with some of the terms displayed, e.g., the terms $\mathcal{O}(1/c_\mathrm{s}^2)$. }. 
From the \textit{Planck} constraints on $f^{\rm equil}_{\rm NL}$ and $f^{\rm ortho}_{\rm NL}$ (see Table~\ref{Tab_KSW+SMICA}), we derive constraints on these model parameters following the procedure described at the beginning of this section. We choose log-constant priors in the ranges $10^{-4} \leq  A \leq 10^4$, and $10^{- 4} \leq  \bar c_{\mathrm{s}} \leq 10^{2}$, together with a uniform prior $10^{-4} < c_{\rm s} <1$. 
These priors have been choosen essentially on the basis of perturbative regime validity of the theory and to allow for a quite wide range of parameter values. In Fig.~\ref{Galileon_cs_csbar_positive}, as an example, probability contours are shown in the parameter space 
$(c_{\rm s}, \bar{c}_{\rm s})$ from the $T+E$ constraints, after marginalizing over the parameter $A$.  Marginalizing over both $A$ and $\bar c_{\mathrm{s}}$ we find 
\begin{equation}
c^{\rm Galileon}_\mathrm{s} \geq 0.21 \quad \quad  \text{95$\,\%$ CL (\itT-only)}\, ,
\end{equation}
and 
\begin{equation}
c^{\rm Galileon}_\mathrm{s} \geq 0.23 \quad \quad  \text{95$\,\%$ CL (\textit{T+E})}\, .
\end{equation}
Notice that interestingly enough the parameter $\bar c_{\mathrm{s}}$ can be negative in principle, corresponding to a blue spectral tilt of 
inflationary gravitational waves (without any kind of instability). We therefore also explore this branch, with a log-constant prior (for $-\bar c_{\mathrm{s}}$), $-10^{2} \leq  \bar c_{\mathrm{s}} \leq - 10^{-4}$, and the same priors for the other parameters 
as above. Fig.~\ref{Galileon_cs_csbar_negative} shows the probability contours in the $(c_{\rm s}, \bar{c}_{\rm s})$ plane, after marginalizing over the parameter $A$, for the $n_{\rm T} <0$ branch. Marginalizing over both $A$  and $\bar c_{\mathrm{s}}$ gives
\begin{equation}
c^{\rm Galileon}_\mathrm{s} \geq 0.19\quad \quad  \text{95$\,\%$ CL (\itT-only)}\, ,
\end{equation} 
and
\begin{equation}
c^{\rm Galileon}_\mathrm{s} \geq 0.21\quad \quad  \text{95$\,\%$ CL (\textit{T+E})}\, .
\end{equation}

A combined analysis of the \textit{Planck} bispectrum and power spectrum constraints on the Galileon models is presented in the companion \textit{Planck} paper on inflation~\citep{planck2014-a24}.


\subsection{Multi-field models}
\label{multifield}
\noindent{\it Curvaton models:}
The simplest adiabatic curvaton models predict local NG with an amplitude~\citep{2004PhRvD..69d3503B,2004PhRvL..93w1301B}
\begin{equation}
\label{fNLcurv}
f_\mathrm{NL}^\mathrm{local} = \frac{5}{4r_{\rm D}} - \frac{5 r_{\rm D}}{6} - \frac{5}{3}\, ,
\end{equation}
for a quadratic potential of the curvaton field~(\citealt{2002PhLB..524....5L,2003PhRvD..67b3503L,2005PhRvL..95l1302L,2006JCAP...09..008M,2006PhRvD..74j3003S}), where $r_{\rm D}=[3\rho_{\rm curvaton}/(3 \rho_{\rm curvaton}+4\rho_{\rm radiation})]_{\rm D}$ is the ``curvaton decay fraction''  at the time of the curvaton decay in the sudden decay approximation. Assuming a uniform prior, $0<r_\mathrm{D}<1$,  our constraint $f_{\rm NL}^{\rm local}=2.5 \pm 5.7$  at 68\,\% CL (see Table~\ref{Tab_KSW+SMICA}) yields 
\begin{equation}
r_{\rm D} \geq 0.16 \quad \quad  \text{95$\,\%$ CL (\itT-only)}\, , 
\end{equation}
while accounting for temperature and polarization data ($f_{\rm NL}^{\rm local}=0.8 \pm 5.0$ at 68\,\% CL) gives 
\begin{equation}
r_{\rm D} \geq 0.19 \quad \quad  \text{95$\,\%$ CL (\textit{T +E})}\, , 
\end{equation}
improving the previous  {\Planck} bound which was previously $r_{\rm D} \geq 0.15$ ($95\,\%$ CL; \citealt{planck2013-p09a}). In \cite{planck2014-a24}, assuming there is some relic isocurvature fluctuations in the curvaton field, 
a limit on $r_{\rm D}$ is derived from the bounds on isocurvature fluctuations. In this restricted case, the limit $r_{\rm D} > 0.98$  ($95 \,\%$ CL) is derived, which is consistent with the constraint given here. 

Notice that the above expression of $f_{\rm NL}^\mathrm{local}$ Eq.~\eqref{fNLcurv} is valid under the assumption that there is no significant decay of the inflaton into curvaton particles. In general one should account for such a possibility. For example, if the classical curvaton field survives and starts to dominate, then the curvaton particles produced during reheating (which have the same equation of state as the classical curvaton field) are expected to survive and dominate over other species at the epoch of their decay. The classical curvaton field and the curvaton particles decay at the same time, inevitably producing adiabatic perturbations (for a detailed discussion see \citealt{2006JCAP...04..009L}). A general formula for $f_\mathrm{NL}^\mathrm{local}$, accounting for the possibility that the inflaton field decays into curvaton particles, is provided in~\cite{2006PhRvD..74j3003S}:
\label{fNLcurvextended}
\begin{equation}
f_\mathrm{NL}^\mathrm{local} = (1+\Delta_{\rm s}^2) \frac{5}{4r_{\rm D}} - \frac{5 r_{\rm D}}{6} - \frac{5}{3}\, ,
\end{equation}
where $\Delta_{\rm s}^2=\rho_{\rm curv.\,\,particles}/\rho_{\rm curv. field}$ measures the ratio of the energy density of curvaton particles to the energy density of the classical curvaton field \citep{2006JCAP...04..009L,2006PhRvD..74j3003S} and $\rho_{\rm curvaton}$ in the expression for  $r_{\rm D}$ is given by $\rho_{\rm curvaton}= \rho_{\rm curv. particles}+\rho_{\rm curv. field}$. Using the uniform prior  $0<r_\mathrm{D}<1$ and $0 < \Delta_{\rm s}^2 <10^2$ our measurements of $f_{\rm NL}^{\rm local}$ constrain $\Delta_{\rm s}^2 \leq 8.5$ at $95\,\%$ CL (\itT) and  $\Delta_{\rm s}^2 \leq 6.9$ (\textit{T+E}).

\medskip

\medskip



\subsection{Non-standard inflation models}

\noindent{\it Directional dependence motivated by gauge fields}:
In Table ~\ref{tab:cL_result} we constrained directionally-dependent bispectra (Eq.~\ref{vectorBis}). This kind of NG 
is generated by inflationary models characterized by the presence of gauge fields.
An actual realization of this type of scenario can be 
obtained with a coupling between the inflaton and the gauge field(s), via the  
kinetic term of the field(s), i.e., $\mathcal{L}=-I^2(\phi) F^2/4$. In this formula, $F^2$ represents the strength of the gauge field, while 
$I(\phi)$ is a function of the inflaton field with an appropriate time dependence (see, e.g.,~\citealt{1992ApJ...391L...1R}). In  
this type of scenario, vector fields can be generated during inflation, and this in turn determines the excitation of 
$L=0$ and $2$ modes in the bispectrum, with $f_{\rm NL}^L=X_L (|g_*|/0.1)\, (N_{k_{3}}/60)$, 
where $X_{L=0}=(80/3)$ and $X_{L=2}=-(10/6)$ \citep{Barnaby:2012tk,Bartolo:2012sd,Shiraishi:2013vja}. 
The parameter $g_*$, appearing in the equations above, represents the amplitude of a quadrupolar 
anisotropy in the power spectrum (see, e.g.,~\citealt{2007PhRvD..75h3502A}), while $N$ defines the number of e-folds, before 
the end of inflation, when the relevant scales exit the horizon. 
It is thus clear that these models predict both a degree of statistical anisotropy in the power spectrum, and a potentially non-negligible 
bispectrum, as well as a direct relation between the two. 

Starting from our \SMICA~constraints from \itT (\textit{T+E}) in Table~\ref{tab:cL_result}, marginalizing over a uniform prior $50 \leq N \leq 70$, and 
assuming uniform priors on $-1 \leq g_* \leq 1$, we obtain the limits 
$- 0.050 < g_* < 0.050 $ ($- 0.040 < g_* < 0.040$), and $- 0.31 < g_* < 0.31$ ($- 0.29 < g_* < 0.29$), from the $L=0$, $L=2$ modes, respectively ($95 \,\%$ CL) (considering $g_*$ as scale-independent).
We note that these constraints refer to all models in which curvature perturbations are sourced by an $I^2(\phi) F^2$ term (see references in ~\citealt{Shiraishi:2013vja}). 
The constraints we obtain are consistent with the tighter (model-independent) limits obtained in~\cite{planck2014-a24}  for the case of a scale-independent $g_*$, from an analysis of quadrupolar anisotropies in the CMB power spectrum.

\medskip

\noindent{\it NG from gauge field production during axion inflation:}
We have also constrained the inverse decay NG of Eq.~(\ref{invdec}) arising  typically in models where the inflaton field is a pseudoscalar axion that couples to a gauge field. Using the modal estimator we get the following constraints (removing ISW-lensing bias):
\begin{equation}
f^{\rm inv.dec}_{\rm NL}= 17 \pm 43 \quad \quad  \text{68$\,\%$ CL}\, ,
\end{equation}
for temperature only; and 
\begin{equation}
\label{invdecconst2}
f^{\rm inv.dec}_{\rm NL}= 23 \pm 26 \quad \quad  \text{68$\,\%$ CL}\, ,
\end{equation}
from temperature+polarization (see Table~\ref{tab:equilmodels}).
The NG amplitude is given by $f_\mathrm{NL}^{\mathrm{inv.dec}} = f_3(\xi_*,1,1) \mathcal{P}_*^3 e^{6\pi\xi_*}/ \mathcal{P}_{\mathcal{\zeta}}^2(k_*)$, where $\mathcal{P}^{1/2}=H^2/(2\pi|\dot\phi|)$ 
is the power spectrum of vacuum-mode curvature perturbations (i.e., the power spectrum predicted without the coupling to gauge fields), $\mathcal{P}_{\mathcal{\zeta}}^2(k_*)$ is the dimensionless scalar power spectrum of curvature perturbations (a star denoting evaluation at the pivot scale). The NG parameter is exponentially sensitive to the strength of the coupling between the axion and the gauge field. From Eq.
(\ref{invdecconst2}) we limit the strength of the coupling to $\xi\leq 2.5 \quad (95\,\%~\mathrm{CL})$. The details together with constraints on the axion decay constant can be found in \cite{planck2014-a24}, where an overview of various observational limits on axion (monodromy) models of inflation is presented. This limit is in agreement with the one that can be derived from tensor non-Gaussianities (see below).  

\medskip

\noindent{\it Tensor NG and pseudoscalars}: 
In inflationary scenarios associated with a pseudoscalar coupling to a U(1) gauge field, the tensor bispectrum generated via the gravitational interaction with the gauge field is expressed by Eq.~\eqref{eq:bis_tens_form}, and the amplitude $f_{\rm NL}^{\rm tens}$ depends on the following: the coupling strength of the pseudoscalar field to the gauge field ($\xi$); a slow-roll parameter for the inflaton ($\epsilon$); and the power spectrum of vacuum-mode curvature perturbations (${\cal P}$).  The expression is  $f_{\rm NL}^{\rm tens} \approx 6.4 \times 10^{11} {\cal P}^3 \epsilon^3 e^{6\pi\xi} / \xi^9$ \citep{2013JCAP...11..047C, 2013JCAP...11..051S}. The constraints on $f_{\rm NL}^{\rm tens}$ presented in Table~\ref{tab:fnltens} can then be used to constrain the model parameters. Clearly there are strong degeneracies, but if we marginalize over a uniform prior $1.5 \times 10^{-9} \leq {\cal P} \leq 3 \times 10^{-9}$ and set $\epsilon = 0.01$, then assuming a uniform prior $0.1 \leq \xi \leq 7$, from the \SMICA~(\itT-only or \textit{T+E}) limit, we obtain $\xi < 3.3$ (95\,\% CL).

\medskip

\noindent{\it Warm inflation}:
We update the constraints on warm inflation models in the strongly dissipative regime, when dissipative effects are relevant. In this regime $f_\mathrm{NL}^\mathrm{warm} = -15 \ln \left(1 + r_\mathrm{d}/14 \right) - 5/2$
\citep{2007JCAP...04..007M} with a large dissipation parameter $r_\mathrm{d} = \Gamma/(3H)$ (where $\Gamma$ is a friction term for the inflaton evolution describing the energy transfer from the inflaton field to radiation).. The limit from the 2013 {\Planck} release is  $\log_{10} r_\mathrm{d} \leq 2.6$ ($95 \,\%$ CL) \citep{planck2013-p09a}. Assuming a constant prior $0 \leq \log_{10} r_\mathrm{d} \leq 4$, the new \SMICA~constraint $f_\mathrm{NL}^\mathrm{warmS} =  
-23 \pm 36$
at $68\,\%$ CL from \itT\  
($f_\mathrm{NL}^\mathrm{warmS} =  
 -14 \pm 23$
 from \textit{T+E}), see Table~\ref{tab:fnlnonstandard}, yields a limit on the dissipation parameter of $\log_{10} r_\mathrm{d} \leq 3.3$ ($\log_{10} r_\mathrm{d} \leq 2.5$) at $95 \,\%$ CL, with the \textit{T+E} constraints (in brackets), slightly improving the $2013$ {\Planck} limits. Values of $r_\mathrm{d} \gtrsim 2.5$ (strongly-dissipative regime) are still allowed; however, the {\it Planck} constraint puts the model in a regime 
where there might be an overproduction of gravitinos (see~\citealt{2008JCAP...01..027H} and references therein). Unlike the strong dissipative regime, in the intermediate and weak dissipative regimes $(r_{\rm D} \leq 1)$ the NG level strongly depends on the microscopic parameters ($T/H$ and $r_{\rm D}$), giving rise to a new additional bispectrum shape (for details see~\citealt{2014JCAP...12..008B}).

\subsection{Alternatives to inflation}
Ekpyrotic/cyclic models have been proposed as an alternative to inflation~(for a review, see~\citealt{2010AdAst2010E..67L}). Local NG is generated from the conversion of ``intrinsic'' non-Gaussian entropy perturbation modes into curvature fluctuations. Models based on a conversion taking place during the ekpyrotic phase (the so called ``ekpyrotic conversion mechanism'') are already ruled out~\citep{Koyama:2007if,planck2013-p09a}. 
Ekpyrotic models where ``kinetic conversion''  occurs after the ekpyrotic phase predict a local bispectrum with $f_{\rm NL}^\mathrm{local} = (3/2) \, \kappa_3 \sqrt{\epsilon} \pm 5$, where the sign depends on the details of the conversion process~\citep{Lehners:2007wc,2010AdAst2010E..67L,Lehners:2013cka}, where $\epsilon \simeq 50$ or greater are typical values. If we take $\epsilon \simeq 100$ and a uniform prior on $-5 < \kappa_3 <5$ the constraints on $f_{\rm NL}^\mathrm{local}$ from \itT-only (see Table~\ref{Tab_KSW+SMICA}), yield $-0.91 < \kappa_3 <  0.58$  and $-0.25 < \kappa_3 <  1.2$ at 95\,\% CL, for the plus and minus sign in $f_{\rm NL}^\mathrm{local}$ respectively. From the \textit{T+E}  $f_{\rm NL}^\mathrm{local}$ constraints (Table~\ref{Tab_KSW+SMICA}) we obtain $-0.94 < \kappa_3 <  0.38$  and $-0.27 < \kappa_3 <  1.0$ at 95\,\% CL, for the plus and minus sign in $f_{\rm NL}^\mathrm{local}$ respectively.
If we consider  $\epsilon \simeq 50$ we derive the following limits: $-1.3 < \kappa_3 <  0.81$  and $-0.35 < \kappa_3 <  1.8$ at 95\,\% CL from \itT-only; $-1.3 < \kappa_3 <  0.53$  and $-0.38 < \kappa_3 <  1.5$ at 95\,\% CL  from \textit{T+E}. 
Another variant of the ekpyrotic models has been investigated in~\citep{Qiu:2013eoa,Li:2013hga,Fertig:2013kwa}, where the intrinsic NG is zero and NG is generated only by non-linearities in the conversion mechanism, reaching a value of  $f_{\rm NL}^\mathrm{local} \simeq \pm 5$.


\subsection{Inflationary interpretation of CMB trispectrum results}
\label{ssec:inflationary_trispectrum}

We briefly interpret the trispectrum constraints in an inflationary context.
First, consider the case of single field inflation.
The action for the Goldstone boson $\pi$ is highly constrained by residual
diffeomorphism invariance (see, e.g.,~\cite{long_trispectrum}).
To lowest order in the derivative expansion, the most general action is:
\ba
S_{\pi} &=& \int d^4 x \sqrt{- g}\, \bigg\{ -M^2_{\rm Pl}\dot{H} \left(\partial_\mu \pi\right)^2 \nn \\
&& \hspace{1cm}
  + 2 M^4_2 \left[\dot\pi^2+\dot{\pi}^3-\dot\pi\frac{(\partial_i\pi)^2}{a^2}+(\partial_\mu\pi)^2(\partial_\nu\pi)^2 \nn \right] \nn \\
&& \hspace{1cm}
  - \frac{M_3^4}{3!} \left[ 8\,\dot{\pi}^3+12 \dot\pi^2(\partial_\mu\pi)^2+\cdots \right] \nn \\
&& \hspace{1cm}
  + \frac{M_4^4}{4!} \left[ 16\,\dot\pi^4+32\dot\pi^3(\partial_\mu\pi)^2 + \cdots \right]+\cdots \bigg\},
\ea
where the parameter $M_4$ is related to the trispectrum by
\be
\gnldotpi4 = \frac{25}{288} \frac{M_4^4}{H^4}\, A_\zeta \, c_{\rm s}^3 \ .
\ee
The $\gnldotpi4$ constraint in Eq.~(\ref{eq:gnl_bottom_line}) translates to the following
parameter constraint in single field inflation:
\be
- 9.70 \times 10^{14} < \frac{M_4^4}{H^4 c_{\rm s}^3} < 8.59 \times 10^{14}\hspace{1cm} \mbox{(95\,\% CL).}
\ee
This constraint is a factor 1.8 better than WMAP.

Turning now to multifield inflation, we consider an action of the form
\ba
S_\sigma &=& \int d^4x\, \sqrt{-g} \,\, \bigg[ \frac{1}{2} (\partial_\mu\sigma)^2 
  + \frac{1}{\Lambda^4_1} \dot\sigma^4 \nn \\
&& \hspace{1cm}
  + \frac{1}{\Lambda^4_2} \dot\sigma^2 (\partial_i\sigma)^2
  + \frac{1}{\Lambda^4_3} (\partial_i\sigma)^2 (\partial_j\sigma)^2
\bigg],  \label{eq:S_multi}
\ea
where $\sigma$ is a spectator field that acquires quantum fluctuations
with power spectrum $P_\sigma(k) = H^2/(2k^3)$ and converts to adiabatic curvature
via $\zeta = (2 A_\zeta)^{1/2} H^{-1} \sigma$.
The trispectrum in this model is
\ba
\gnldotpi4 A_\zeta
  &=& \frac{25}{768}\frac{H^4}{\Lambda_1^4}, \nn \\
\gnlB A_\zeta
  &=& -\frac{325}{6912}\frac{H^4}{\Lambda_2^4}, \\
\gnldpi4 A_\zeta 
  &=& \frac{2575}{20736} \frac{H^4}{\Lambda_3^4}, \nn
\ea
so we can constrain its parameters by thresholding the $\chi^2$
defined in Eq.~(\ref{eq:trispectrum_chi2}).
For example, if we consider the Lorentz invariant model
\be
S = \int d^4x\, \sqrt{-g}\,
  \left[ \frac{1}{2} (\partial_\mu \sigma)^2 + \frac{1}{\Lambda^4} (\partial_\mu \sigma)^2 (\partial_\nu \sigma)^2 \right],  \label{eq:S_lorentz}
\ee
so that the parameters $\Lambda_i$ of the more general action in Eq.~(\ref{eq:S_multi})
are given by $\Lambda_1^4 = -2 \Lambda_2^4 = \Lambda_3^4 = \Lambda^4$, then by thresholding
at $\Delta\chi^2 = 4$ (as appropriate for one degree of freedom), 
we obtain the following constraint on the parameter $\Lambda$:
\be
-0.26 < \frac{H^4}{\Lambda^4} < 0.20 \hspace{1cm} \mbox{(95\,\% CL).}
\ee
Constraints in other parameter spaces can also be obtained by
thresholding the $\chi^2$ defined in Eq.~(\ref{eq:trispectrum_chi2}).
For example we show $68\,\%$ and $95\,\%$ confidence regions in the
$(\gnldotpi4, \gnldpi4)$-plane, obtained by thresholding at $\chi^2=2.28$ and
$\chi^2=5.99$ as appropriate for a $\chi^2$ random variable with two degrees
of freedom.

\begin{figure}
\begin{center}
\includegraphics[width=9cm]{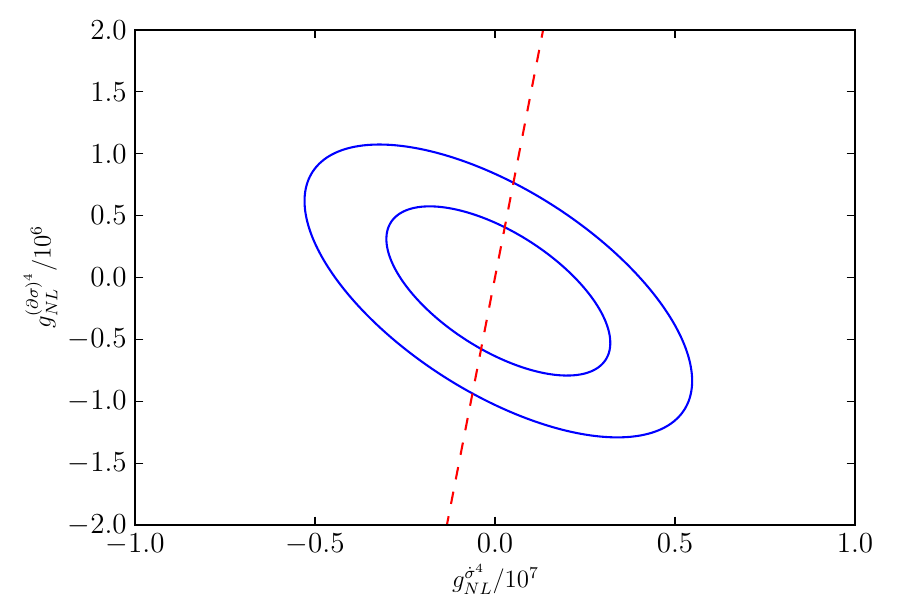}
\caption{$68\,\%$ and $95\,\%$ confidence regions in the $(\gnldotpi4, \gnldpi4)$ plane,
with the Lorentz invariant model in Eq.~(\ref{eq:S_lorentz}) shown as the dashed line.}
\end{center}
\label{fig:gnl_plane}
\end{figure}

\medskip

\noindent{\it DBI Trispectrum:} 
The trispectrum constraint on the shape $\dot\sigma^4$ in Eq.~(\ref{eq:gnl_bottom_line}) can also be used to obtain a lower bound on the DBI model 
sound speed. This is because, in the small sound speed limit \citep{2009JCAP...08..008C,2009PhRvD..80d3527A}, the dominant contribution to 
the contact interaction trispectrum \citep{2006PhRvD..74l1301H} has this 
shape. The corresponding non-linearity parameter is 
$g_\mathrm{NL}^{\dot\sigma^4}=-25/(768\, c_\mathrm{s}^4)$. 
We follow the same procedure as described at the beginning of this section~\ref{sec:Implications} and, assuming a uniform prior in the range $0\leq c_\mathrm{s}\leq1/5$, we can 
derive a constraint on $c_\mathrm{s}$ as 
\begin{equation} 
c_\mathrm{s}^{\rm DBI} \geq0.021,\quad \quad  \text{95$\,\%$ CL}\, . 
\end{equation} 
This constraint is consistent with the ones derived from the bispectrum 
measurements (see Eqs. \ref{csDBIT} and \ref{csDBITE})) and it is only a factor 
of about three worse. Notice, however, that in this case we are ignoring the scalar exchange 
contribution, which is of the same order in $c_\mathrm{s}$.

\noindent{\it Curvaton trispectrum:} 
For the simplest curvaton scenario, the trispectrum non-linearity parameter $g_\mathrm{NL}^\mathrm{local}$ 
prediction is \citep{2006PhRvD..74j3003S}
\begin{equation} 
g_\mathrm{NL}^\mathrm{local}=\frac{25}{54}\left(-\frac{9}{r_{\rm D}}+\frac{1}{2}+10r_{\rm D}+3r_{\rm D}^2\right). 
\end{equation} 
Following the procedure described at the beginning of Sect.~\ref{sec:Implications}, we use the observational constraint obtained in Sect. \ref{sec:tau_gnl} (Eq.~\ref{eq:gnl_bottom_line}), and the 
same prior ($0<r_\mathrm{D}<1$) as in~\ref{multifield}, to obtain a lower bound on the curvaton decay fraction as 
\begin{equation} 
r_{\rm D} \geq0.05\quad \quad  \text{95$\,\%$ CL}\,. 
\end{equation} 
This limit is consistent with the previous ones derived using the bispectrum 
measurements and it is a factor of about 3 to 4 worse.

%% file: A19_Section12.tex
In this paper we have presented the constraints on primordial NG using the full \Planck\ mission data. The results have improved compared to the \Planck\ 2013 release~\citep{planck2013-p09a} as a consequence  of including data from the full mission and taking advantage of \Planck's polarization capability --- the first time that maps of the CMB polarization anisotropies have been used to constrain primordial NG.

Using temperature data alone, the constraints on the local, equilateral, and orthogonal bispectrum templates are $f_{\rm NL}^{\rm local} = 2.5 \pm 5.7$, $f_{\rm NL}^{\rm equil} = - 16 \pm 70$, and $f_{\rm NL}^{\rm ortho} = - 34 \pm 33$.   Moving from the nominal  \Planck\ 2013 data to the full mission data yielded modest improvements of up to 15\,\% (for the orthogonal shape).  After the inclusion of full mission polarization data, our final constraints become  $f_{\rm NL}^{\rm local} = 0.8 \pm 5.0$, $f_{\rm NL}^{\rm equil} = 
- 4 \pm 43$, and $f_{\rm NL}^{\rm ortho} = - 26 \pm 21$, which represents a substantial step forward relative to  \Planck\ 2013, with error bars shrinking by 14\,\% (local), 43\,\% (equilateral), and 46\,\% (orthogonal).   These improved limits on the standard shapes enhance our understanding of different inflationary models that can potentially lead to subtle effects beyond the simplest models of inflation.

The reason that the polarization data  provide such complementary constraints on primordial curvature perturbations is due to the phase shift of the CMB polarization  transfer functions compared to the temperature transfer functions.  So, despite the relatively much lower signal-to-noise in the polarization maps, their inclusion leads to  appreciable improvements on limits on NG parameters.  Nevertheless, the full characterization of the noise properties in the polarized maps is still ongoing. In spite of the extensive testing and cross-checks validating the combined temperature and polarization results, we therefore conservatively recommend that \textit{all} results that include polarization information, not just the polarization-only results,  be taken as preliminary at this stage.

The complementary nature of the polarization information also represents an important cross-check on the analysis. The \Planck\ results based on polarization alone are statistically consistent with the results based on temperature alone, with a precision comparable to that achievable in an optimal analysis of the WMAP 3-year temperature maps \citep{WMAP3Cosmology,2007JCAP...03..005C,2008PhRvL.100r1301Y}.

Our analysis was subject to an extensive validation exercise. In addition to extensive simulation tests, including for the first time a detailed test of the impact of time-domain de-glitching, our results are supported by tests for robustness under change of estimator implementations  (KSW, binned bispectrum, and two modal estimators), and variations in sky coverage as well as upper and lower multipole cutoffs.  We also tested for possible directional dependence using a needlet estimator. These tests form the basis of our selection of \SMICA\ as the main foreground cleaning method for our headline results.

The \Planck\ 2015 analysis presented here provides constraints on a greatly extended range of template families. These  extensions include  a tenfold increase in the range of frequencies covered in feature models, giving rise to linearly oscillating bispectra,  generalized shapes for oscillating models including for logarithmic oscillations, tests for deviations from the Bunch-Davies vacuum, models of equilateral type in the context of the effective field theory of inflation, and direction-dependent primordial NG. Beyond purely scalar mode templates we also tested for parity-violating tensor NG.  

Using the full mission data with polarization, we have investigated the hints of NG reported in the \Planck\ 2013 analysis of oscillatory features.   While no individual feature or resonance model rises above our detection threshold of $3\,\sigma$ (after inclusion of the look-elsewhere effect), the results of integrated (multi-peak) statistical tests  indicate that continued investigation of oscillatory and non-scaling models is warranted.

In addition to searches for specific NG templates, we present model-independent reconstructions of the temperature and polarization bispectra using the  modal and binned bispectrum approaches.  These full mission reconstructions can achieve twice the resolution of the \Planck\ 2013 results, demonstrating excellent consistency in temperature, and good agreement  with the WMAP9 reconstruction in regions where this earlier data set is signal-dominated.

The inclusion of polarization information leads to significantly improved constraints on NG in primordial isocurvature  perturbations, providing complementary information to 2-point function results for models where the NG in isocurvature components is more easily detectable than its contribution to the power spectrum.

A significant addition to this year's analysis is the inclusion of detailed trispectrum results due to cubic NG.   The local trispectrum is constrained by \Planck\ temperature data to be $g_{\rm NL}^{\rm local} = (-9.0 \pm 7.7) \times 10^4$ and the other two shapes were also found to be consistent with Gaussianity.   
Both 3-point and 4-point constraints are consistent with the improved (though still suboptimal) constraints from Minkowski functionals, a very different estimation framework. This concordance adds confidence in our results.

We have discussed the implications of our results on the physics 
of the early Universe, showing that the $n$-point functions for 
$n > 2$ provide a significant window onto the primordial Universe 
beyond the power spectrum, constraining general-single field, multifield, 
and non-standard inflation models, as well as alternatives to inflation. Using bispectrum and 
trispectrum limits we updated results on the parameter space 
of  the inflationary models (and alternatives) already tested in 2013, and constrained the 
parameter space of other well-motivated inflationary models (e.g., Galileon-like models of inflation, and 
models where axion/pseudoscalar fields are present during inflation). 

The global picture that emerges is one of consistency with the premises of the $\Lambda$CDM cosmology, namely that the structure we observe today is the consequence of passive evolution of adiabatic, Gaussian, primordial seed  perturbations. Nevertheless, NG at some level is expected in all inflationary models, and hence we should strive to find means to reduce errors on $\fnl$ still further.

%% file: A19_acknowledgements.tex
\begin{acknowledgements}
The Planck Collaboration acknowledges the support of: ESA; CNES and CNRS/INSU-IN2P3-INP (France); ASI, CNR, and INAF (Italy); NASA and DoE (USA); STFC and UKSA (UK); CSIC, MINECO, JA, and RES (Spain); Tekes, AoF, and CSC (Finland); DLR and MPG (Germany); CSA (Canada); DTU Space (Denmark); SER/SSO (Switzerland); RCN (Norway); SFI (Ireland); FCT/MCTES (Portugal); ERC and PRACE (EU). A description of the Planck Collaboration and a list of its members, indicating which technical or scientific activities they have been involved in, can be found at 
\url{http://www.cosmos.esa.int/web/planck/planck-collaboration}.
Some of the results in this paper have been derived using the {\tt{HEALPix}} package.
Part of this work was undertaken on the STFC COSMOS@DiRAC HPC Facility at the University of Cambridge, funded by UK BIS NEI grants.
We gratefully acknowledge IN2P3 Computer Center (\url{http://cc.in2p3.fr}) for
providing a significant amount of the computing resources and services needed 
for the analysis with the binned bispectrum estimator. This research used resources of the National Energy Research
Scientific Computing Center, a DOE Office of Science User Facility
supported by the Office of Science of the U.S. Department of Energy
under Contract No. DE-AC02-05CH11231. We also acknowledge the IAP magique3 computer facilities. 
Some computations were performed on the GPC cluster at the SciNet HPC
Consortium.  SciNet is funded by the Canada Foundation for Innovation under
the auspices of Compute Canada, the Government of Ontario, and the
University of Toronto.

 \end{acknowledgements}

%% file: A19_AppendixA.tex
\section{Derivation of an estimator for $c_L$}
\label{sec:AA}
As parameterized by Eq.~(\ref{vectorBis}), we express a primordial bispectrum of direction-dependence:
\begin{eqnarray}
\lefteqn{B_\Phi(k_1,k_2,k_3)=\nonumber}\\
&&\sum_{L\ge 1} c_L \left[P_L(\vkhat_1\cdot \vkhat_2) P_\Phi(k_1)P_\Phi(k_2)+2\,\mathrm{perm.}\right]
\end{eqnarray}
where $P_L(\vkhat_1\cdot \vkhat_2)$ is a Legendre Polynomial of  order $L$.
It can be shown that such a primordial bispectrum leads to a CMB bispectrum \begin{eqnarray}
\lefteqn{\langle a^{p_1}_{\ell_1m_1} a^{p_2}_{\ell_2 m_2} a^{p_3}_{\ell_3 m_3} \rangle=}\label{alm_3point_cL}\\
&&(-\imath)^{\ell_1+\ell_2+\ell_3}(4\pi)^3 \int \frac{d^3\vec k_1}{(2\pi)^3} \,g^{p_1}_{l_1}(k_1)\,Y^*_{l_1m_1}(\hat {\vec k}_1)\times\nonumber\\
&&\int \frac{d^3\vec k_2}{(2\pi)^3} \,g^{p_2}_{l_2}(k_2)\,Y^*_{l_2m_3}(\hat {\vec k}_2)\int \frac{d^3\vec k_3}{(2\pi)^3} \,g^{p_3}_{l_3}(k_3)\,Y^*_{l_3m_3}(\hat {\vec k}_3)\times\nonumber\\
&& (2\pi)^3 \delta^{(3)}(\vec k_1+\vec k_2+\vec k_3)\, c_L \left[P_L(\hat {\mathbf k}_1\cdot \hat {\mathbf k}_2) P_\Phi(k_1)P_\Phi(k_2)+2\,\mathrm{perm.}\right]\nonumber\\
&=&\left(\begin{array}{ccc}\ell_1&\ell_2&\ell_3\\m_1&m_2&m_3\end{array}\right) \left(\sum_L\,c_L\,b^{(c_L),p_1 p_2 p_3}_{\ell_1 \ell_2 \ell_3}\right),\nonumber
\end{eqnarray}
where $p$ denotes either temperature or \itE-mode polarization, $b^{(c_L),p_1 p_2 p_3}_{\ell_1 \ell_2 \ell_3}$ is the reduced CMB bispectrum, and
the term with large parentheses denotes the Wigner-3j symbol.
The reduced CMB bispectrum $b^{(c_L),p_1 p_2 p_3}_{\ell_1 \ell_2 \ell_3}$ is given by~\citep{Shiraishi:2013vja}
\begin{eqnarray}
\lefteqn{b^{(c_L),p_1 p_2 p_3}_{\ell_1 \ell_2 \ell_3}=\frac{4\pi}{2L+1}\frac{w_{\ell_1}w_{\ell_2}w_{\ell_3}\,\imath^{\ell_1+\ell_2+\ell_3}}{h_{\ell_1 \ell_2 \ell_3}}\int^{\infty}_0 r^2 dr}\nonumber\\
&\times&\sum_{L_1 L_2 L_3}\imath^{L_1+L_2+L_3}h_{L_1 L_2 L_3}\left[h_{\ell_1 L_1 L}\,h_{\ell_2 L_2 L}\,(-1)^{L_1+L_2+L}\,\delta_{L_3,\ell_3}\right.\nonumber\\
&\times&\left.\beta^{p_1}_{\ell_1 L_1}(r)\,\beta^{p_2}_{\ell_2 L_2}(r)\,\alpha^{p_3}_{\ell_3}(r)\left\{\begin{array}{ccc}\ell_1&\ell_3&\ell_3\\L_2&L_1&L\end{array}\right\}
+\mathrm{2\:perm.}\right],\label{B_L}
\end{eqnarray}
where $\delta_{L,l}$ denotes the Kronecker delta function, $\{\ldots\}$ the Wigner 6j symbol, ``perm.'' means permutations, 
$w_\ell$ is a beam window function, and
\begin{eqnarray}
\alpha^{p}_\ell(r)&=&\frac{2}{\pi}\int k^3\,d \ln k\:\,g^p_{\ell}(k)\,j_\ell(kr),\label{alpha}\\
\beta^{p}_{\ell,\ell^\prime}(r)&=&\frac{2}{\pi}\int d k^3\ln k\:P_\Phi(k)\,g^p_{\ell}(k)\,j_{\ell^\prime}(kr).\label{beta_lL}
\end{eqnarray}
Here, $g^p_{\ell}(k)$ is the radiation transfer function for temperature or \itE-mode polarization, and
$j_\ell(x)$ is a spherical Bessel function. The $h$ symbol is 
\begin{eqnarray}
h_{\ell_1 \ell_2 \ell_3}=\sqrt{\frac{(2 \ell_1+1)(2 \ell_2+1)(2 \ell_3+1)}{4\pi}}\left(\begin{array}{ccc}\ell_1&\ell_2&\ell_3\\0&0&0\end{array}\right).
\end{eqnarray}
By maximizing the likelihood with respect to $c_L$, we obtain the KSW estimator for $c_L$:
\begin{eqnarray}
\hat c_L&=&\frac{1}{6\,\,\mathcal N_{c_L}} \sum_{p_i q_i}\sum_{\ell_im_i} \mathcal G^{m_1 m_2 m_3}_{\ell_1 \ell_2 \ell_3} b^{(c_L),p_1 p_2 p_3}_{\ell_1 \ell_2 \ell_3}\nonumber\\
&\times&\left[(C^{-1} a)^{p_1}_{\ell_1 m_1}(C^{-1} a)^{p_2}_{\ell_2 m_2} (C^{-1} a)^{p_3}_{\ell_3 m_3}\right.\nonumber\\
&&\left.-3 \langle (C^{-1} a)^{p_1}_{\ell_1 m_1} (C^{-1} a)^{p_2}_{\ell_2\,m_2} \rangle_{\mathrm{MC}}\,(C^{-1}a)^{p_3}_{\ell_3 m_3}\right],\label{cL_estimator}
\end{eqnarray}
where MC denotes that the average is over Monte Carlo simulations and $\mathcal N_{c_L}$ is a normalization constant.
The normalization constant $\mathcal N_{c_L}$ is given by
\begin{eqnarray}
{\mathcal N_{c_L}}&=&\frac{1}{6}\sum_{p_i q_i}\sum_{\ell_i} (h_{\ell_1 \ell_2 \ell_3})^2\\
&\times&b^{(c_L),p_1 p_2 p_3}_{\ell_1 \ell_2 \ell_3} (C^{-1})^{p_1 q_1}_{\ell_1}(C^{-1})^{p_2 q_2}_{\ell_2}\,(C^{-1})^{p_3 q_3}_{l_3}\,b^{(c_L),q_1 q_2 q_3}_{\ell_1 \ell_2 \ell_3}\,.\nonumber
\end{eqnarray}
Using Eq.~(\ref{alm_3point_cL}), we find that
\begin{eqnarray}
\lefteqn{\sum_{p_i \ell_i m_i} \mathcal G^{m_1 m_2 m_3}_{\ell_1 \ell_2 \ell_3} \,b^{(c_L),p_1 p_2 p_3}_{\ell_1 \ell_2 \ell_3} \,(C^{-1} a)^{p_1}_{\ell_1 m_1}(C^{-1} a)^{p_2}_{\ell_2 m_2} (C^{-1} a)^{p_3}_{\ell_3 m_3}}\nonumber\\
&=&\int r^2 dr\int d\Omega \frac{4\pi}{2 L+1}\left(\sum_{q_3\ell_3m_3} \alpha^{p_3}_{\ell_3}(r)\,(C^{-1} a)^{p_3}_{\ell_3 m_3}Y_{\ell_3m_3}(\hat {\mathbf n})\right)\times\nonumber\\
&&\sum^{L}_{M=-L}(-1)^M\left(\sum_{\ell'_1 m'_1} b^{LM}_{\ell'_1 m'_1}(r)\:Y_{\ell'_1 m'_1}(\hat {\mathbf n})\right)
\left(\sum_{\ell'_2 m'_2} b^{L\,-M}_{\ell'_2 m'_2}(r)\:Y_{\ell'_2 m'_2}(\hat {\mathbf n})\right)\nonumber\\
&&+\mathrm{2\;\;perm.},\label{cL_part}
\end{eqnarray}
where 
\begin{eqnarray}
b^{LM}_{\ell'm'}(r)=\sum_{p q\ell m}(-1)^{\ell+m}\,\imath^{\ell+\ell'}\,h_{\ell' \ell L}\,\left(\begin{array}{ccc}\ell&\ell'&L\\-m&m'&M\end{array}\right)
\beta^{p}_{\ell,\ell'}(r)\,w_{\ell}\,(C^{-1})^{pq}_l a^{q}_{\ell m},\nonumber
\end{eqnarray}
with $\beta^p_{\ell,\ell'}(r)$ being defined in Eq.~(\ref{beta_lL}).
In the derivation above, we used the identities
\begin{eqnarray}
P_L(\vkhat_1 \cdot \vkhat_2)=\frac{4\pi}{2L+1}\sum_M Y_{LM}(\vkhat_1)\,Y^*_{LM}(\vkhat_2),\nonumber
\end{eqnarray}
and
\begin{eqnarray}
\lefteqn{\delta(\vk_1 + \vk_2 + \vk_3)=8\int\,dr\,d\vec n^2\sum_{\ell_1 m_1} \imath^{\ell_1} j_{\ell_1}(k_1r) Y_{\ell_1 m_1}(\hat {\vec k}_1)\,Y^*_{\ell_1 m_1}(\hat {\vec n})}\nonumber\\
&\times&\sum_{\ell_2 m_2} \imath^{\ell_2} j_{\ell_2}(k_2r) Y_{\ell_2 m_2}(\hat {\vec k}_2)\,Y^*_{\ell_2 m_2}(\hat {\vec n})
\sum_{\ell_3 m_3} \imath^{\ell_3} j_{\ell_3}(k_3r) Y_{\ell_3 m_3}(\hat {\vec k}_3)\,Y^*_{\ell_3 m_3}(\hat {\vec n}).\nonumber
\end{eqnarray}
Applying Eq.~(\ref{cL_part}) to Eq.~(\ref{cL_estimator}), we find
\begin{eqnarray}
\lefteqn{\hat c_L=\frac{1}{\mathcal N_{c_L}}\frac{2\pi}{(2 L+1)}\sum^{L}_{M=-L} (-1)^M \int r^2 dr\int d^2\hat {\vec n}}\nonumber\\
&&[A(\hat {\vec n},r) B_{LM}(\hat {\vec n},r)B_{L,-M}(\hat {\vec n},r)-A(\hat {\vec n},r)\langle B_{LM}(\hat{\vec n},r) B_{L,-M}(\hat{\vec n},r)\rangle_{\mathrm{MC}}\nonumber\\
&&-2B_{L,-M}(\hat {\vec n},r)\langle A(\hat{\vec n},r) B_{LM}(\hat{\vec n},r)\rangle_{\mathrm{MC}}],\nonumber
\end{eqnarray}
where 
\begin{eqnarray}
A(\vnhat,r)=\sum_{\ell m} \sum_{pq} \alpha^{p}_{\ell}(r)(C^{-1})^{pq}_l\,a^{q}_{\ell m}Y_{\ell m}(\vnhat),\nonumber
\end{eqnarray}
\begin{eqnarray}
B_{LM}(\vnhat,r)=\sum_{\ell m} b^{LM}_{\ell m}(r)\:Y_{\ell m}(\vnhat).\nonumber
\end{eqnarray}
Since $b^{LM}_{\ell'm'}(r)\ne (-1)^{m'} \left[b^{LM}_{\ell',-m'}(r)\right]^*$, $B_{LM}(\vnhat,r)$ is not a real function, but a complex function. 
We estimate $B_{LM}(\vnhat,r)$ efficiently by computing the following  with {\tt{HEALPix}} \citep{gorski2005}:
 \begin{eqnarray}
\mathrm{Re}[B_{LM}(\vnhat,r)]&=&\sum_{\ell m}\mathcal R^{LM}_{\ell m}(r)\:Y_{\ell m}(\hat {\mathbf n});\nonumber\\
\mathrm{Im}[B_{LM}(\vnhat,r)]&=&\sum_{\ell m}\mathcal I^{LM}_{\ell m}(r)\:Y_{\ell m}(\hat {\mathbf n}). \nonumber
 \end{eqnarray}
 Here
 \begin{eqnarray}
 \mathcal R^{LM}_{\ell m}(r)&=&\frac{b^{LM}_{\ell m}(r)+(-1)^m [b^{LM}_{\ell\,-m}(r)]^*}{2},\nonumber\\
 \mathcal I^{LM}_{\ell m}(r)&=&\frac{b^{LM}_{\ell m}(r)-(-1)^m [b^{LM}_{\ell\,-m}(r)]^*}{2\imath}.\nonumber
 \end{eqnarray}

%% file: A19_AppendixB.tex
\section{Definition of Minkowski Functionals and theoretical expectations}
\label{sec:AB}

For a field $f(x)$ of zero average and variance $\sigma^{2}_0$ defined on the two-dimensional sphere $\mathbb{S}^{2}$, an overdense excursion set is defined as
\begin{equation} 
\Sigma \equiv \lbrace x \in \mathbb{S}^{2} \vert \: f(x) > \nu\sigma_0  \rbrace.
\end{equation}
The boundary of the excursion is
\begin{equation}
\partial  \Sigma \equiv \lbrace x \in \mathbb{S}^{2} \vert \: f(x) = \nu\sigma_0  \rbrace. 
\end{equation}
Then the three Minkowski functionals on the sphere are
\begin{equation} {\rm Area:}\ V_{0}(\nu)=\dfrac{1}{4\pi}\int_{\Sigma}{\rm d}\Omega, \end{equation}
\begin{equation} {\rm Perimeter:}\ V_{1}(\nu)=\dfrac{1}{4\pi}\dfrac{1}{4}\int_{\partial \Sigma} {\rm d}l, \end{equation}
\begin{equation} {\rm Genus:}\ V_{2}(\nu)=\dfrac{1}{4\pi}\dfrac{1}{2\pi}\int_{\partial \Sigma} \kappa \, {\rm d}l, \end{equation}
where ${\rm d}\Omega$ and ${\rm d}l$ are respectively elements of solid angles (surface) and of angle (distance), $\kappa$ is the geodesic curvature. Note that the Genus can be also expressed as the number of components\footnote{A component is a connected subset of the excursion.} in the excursion minus the number of holes in the excursion.

The fourth functional $V_3(\nu)$, is defined, for $\nu > 0$,  as the number of components in the excursion. Symmetrically, for $\nu <0$, it is the number of underdense components (or the number of components in the excursion $\lbrace x \in \mathbb{S}^{2} \vert \: f(x) < \nu\sigma_0  \rbrace$).

In the Gaussian limit, the functionals can be expressed the following way \citep[see, e.g.,][]{matsubara2010,1983rafi.book.....V}:
\begin{equation}
V_{k}(\nu)=A_{k}v_{k}(\nu),
\end{equation} 
with
\begin{eqnarray}
v_{k}(\nu) &= &\exp(-\nu^{2}/2) H_{k-1}(\nu), \quad k \leq 2 \label{eq:nuk1} \\
v_{3}(\nu) &= & \dfrac{{\rm e}^{-\nu^{2}}}{\mathrm{erfc}\left(\nu /\sqrt{2}\right)}, \label{eq:nuk2}
\end{eqnarray}
and
\begin{equation}
H_{n}(\nu)={\rm e}^{\nu^{2}/2}\left(-\dfrac{\rm d}{{\rm d}\nu}\right)^{n} {\rm e}^{-\nu^{2}/2}.
\end{equation}
The amplitude $A_k$ depends only on the shape of the power spectrum $C_{\ell}$:
\begin{eqnarray} 
A_{k} &= &\dfrac{1}{(2\pi)^{(k+1)/2}}\dfrac{\omega_{2}}{\omega_{2-k}\omega_{k}}\left( \dfrac{\sigma_{1}}{\sqrt{2}\sigma_{0}}\right)^{k},\quad k \leq 2 \\
A_{3} &= &\dfrac{2}{\pi}\left( \dfrac{\sigma_{1}}{\sqrt{2}\sigma_{0}}\right)^{2}
 \end{eqnarray}
where $\omega_{k}\equiv\pi^{k/2}/\Gamma(k/2+1)$, which gives $\omega_{0}=1$, $\omega_{1}=2$, $\omega_{2}=\pi$ and $\sigma_{0}$ and $\sigma_{1}$ are respectively the rms of the field and its first derivatives.